\documentclass[preprint]{aastex631}
\usepackage[T1]{fontenc}

\newcommand{\ldl}{$\lambda/\Delta\lambda$}
\newcommand{\teff}{$T_{\rm eff}$}
\newcommand{\logg}{$\log{g}$}
\newcommand{\kzz}{$\log{\kappa_{zz}}$}
\newcommand{\vtan}{$V_{\rm tan}$}

\accepted{5 Nov 2024}

\submitjournal{ApJ}

\shorttitle{T Subdwarfs}
\shortauthors{Burgasser et al.}
\graphicspath{{./}{figures/}}

\begin{document}

\title{New Cold Subdwarf Discoveries from Backyard Worlds and a Metallicity Classification System for T Subdwarfs}

\correspondingauthor{Adam Burgasser}
\email{aburgasser@ucsd.edu}

\author[0000-0002-6523-9536]{Adam J.\ Burgasser}
\affiliation{Department of Astronomy \& Astrophysics, UC San Diego, La Jolla, CA, USA}
%; \email{aburgasser@ucsd.edu}}

\author[0000-0002-6294-5937]{Adam C.\ Schneider}
\affiliation{US Naval Observatory, Flagstaff Station, Flagstaff, AZ, USA}

\author[0000-0002-1125-7384]{Aaron M.\ Meisner}
\affiliation{NSF's National Optical-Infrared Astronomy Research Laboratory, 950 North Cherry Avenue, Tucson, AZ 85719, USA}

\author[0000-0001-7896-5791]{Dan Caselden}
\affiliation{Department of Astrophysics, American Museum of Natural History, Central Park West at 79th Street, NY 10024, USA}

\author[0000-0002-5370-7494]{Chih-Chun Hsu}
\affil{Center for Interdisciplinary Exploration and Research in Astrophysics (CIERA), Northwestern University,
1800 Sherman, Evanston, IL 60201, USA}

\author[0000-0003-0398-639X]{Roman Gerasimov}
\affiliation{Department of Physics \& Astronomy, University of Notre Dame, Notre Dame, IN 46556, USA}

\author[0000-0003-2094-9128]{Christian Aganze}
\affiliation{Center for Astrophysics and Space Sciences, Department of Physics, UC San Diego, La Jolla, CA, USA}
\affiliation{Department of Physics \& Astronomy, Stanford University, Stanford, CA 94305, USA}
\affiliation{Stanford Science Fellow and Rubin Fellow}

\author[0000-0002-1420-1837]{Emma Softich}
\affiliation{Department of Astronomy \& Astrophysics, UC San Diego, La Jolla, CA, USA}

\author[0000-0002-1480-9041]{Preethi Karpoor}
\affiliation{Department of Astronomy \& Astrophysics, UC San Diego, La Jolla, CA, USA}

\author[0000-0002-9807-5435]{Christopher A.\ Theissen}
\affiliation{Department of Astronomy \& Astrophysics, UC San Diego, La Jolla, CA, USA}

\author[0000-0002-5253-0383]{Hunter Brooks}
\affiliation{Department of Astronomy and Planetary Science, Northern Arizona University, Flagstaff, AZ 86011, USA}

\author[0000-0003-2235-761X]{Thomas P. Bickle}
\affiliation{School of Physical Sciences, The Open University, Milton Keynes, MK7 6AA, UK}

\author[0000-0002-2592-9612]{{Jonathan Gagn\'{e}}}
\affiliation{Plan\'{e}tarium Rio Tinto Alcan, Espace pour la Vie, 4801 av.\ Pierre-de Coubertin, Montr\'{e}al, Qu\'{e}bec, Canada}

\author[0000-0003-3506-5667]{{\'{E}tienne Artigau}}
\affiliation{Institut Trottier de recherche sur les exoplan\`{e}tes, D\'{e}partement de Physique, Universit\'{e} de Montr\'{e}al, Montr\'{e}al, Qu\'{e}bec, Canada}

\author[0000-0001-8617-2425]{{Micha\"{e}l Marsset}}
\affiliation{European Southern Observatory, Alonso de Cordova 3107, 1900 Casilla Vitacura, Santiago, Chile}

\author[0000-0003-4083-9962]{Austin Rothermich}
\affiliation{Department of Astrophysics, American Museum of Natural History, Central Park West at 79th Street, NY 10024, USA}
\affiliation{Department of Physics, Graduate Center, City University of New York, 365 5th Ave., New York, NY 10016, USA}
\affiliation{Department of Physics and Astronomy, Hunter College, City University of New York, 695 Park Avenue, New York, NY, 10065, USA}

\author[0000-0001-6251-0573]{Jacqueline K.\ Faherty}
\affiliation{Department of Astrophysics, American Museum of Natural History, New York, NY, USA}

\author[0000-0003-4269-260X]{J.\ Davy Kirkpatrick}
\affiliation{IPAC, Mail Code 100-22, Caltech, 1200 East California Boulevard, Pasadena, CA 91125, USA}

\author[0000-0002-2387-5489]{Marc J.\ Kuchner}
\affiliation{Exoplanets and Stellar Astrophysics Laboratory, NASA Goddard Space Flight Center, 8800 Greenbelt Road, Greenbelt, MD 20771, USA}

\author[0000-0003-4714-3829]{Nikolaj Stevnbak Andersen}
\affiliation{Backyard Worlds: Planet 9}

\author{Paul Beaulieu}
\affiliation{Backyard Worlds: Planet 9}

\author[0000-0002-7630-1243]{Guillaume Colin}
\affiliation{Backyard Worlds: Planet 9}

\author[0000-0002-1044-1112]{Jean Marc Gantier}
\affiliation{Backyard Worlds: Planet 9}

%\author{Sam Goodman}
%\affiliation{Backyard Worlds: Planet 9}

\author[0000-0002-8960-4964]{Leopold Gramaize}
\affiliation{Backyard Worlds: Planet 9}

\author[0000-0002-7389-2092]{Les Hamlet}
\affiliation{Backyard Worlds: Planet 9}

\author[0000-0002-4733-4927]{Ken Hinckley}
\affiliation{Backyard Worlds: Planet 9}

\author[0000-0003-4905-1370]{Martin Kabatnik}
\affiliation{Backyard Worlds: Planet 9}

\author[0000-0001-8662-1622]{Frank Kiwy}
\affiliation{Backyard Worlds: Planet 9}

\author{David W.~Martin}
\affiliation{Backyard Worlds: Planet 9}

\author{Diego H.~Massat}
\affiliation{Backyard Worlds: Planet 9}

\author{William Pendrill}
\affiliation{Backyard Worlds: Planet 9}

\author[0000-0003-4864-5484]{Arttu Sainio}
\affiliation{Backyard Worlds: Planet 9}

\author[0000-0002-7587-7195]{J{\"o}rg Sch{\"u}mann}
\affiliation{Backyard Worlds: Planet 9}

\author[0000-0001-5284-9231]{Melina Th{\'e}venot}
\affiliation{Backyard Worlds: Planet 9}

\author{Jim Walla}
\affiliation{Backyard Worlds: Planet 9}

\author{Zbigniew W\k{e}dracki}
\affiliation{Backyard Worlds: Planet 9}

\collaboration{100}{The Backyard Worlds: Planet 9 Collaboration}

\begin{abstract}
We report the results of a spectroscopic survey of candidate T subdwarfs identified by the Backyard Worlds:~Planet 9 program. 
Near-infrared spectra of {31} sources with red $J-W2$ colors 
%(indicating low temperature) 
and large $J$-band reduced proper motions 
%(indicating low luminosities and high velocities) 
show varying signatures of subsolar metallicity, including strong 
collision-induced H$_2$ absorption, obscured methane and water features, and weak K~I absorption.
These metallicity signatures are supported by spectral model fits and 3D velocities, indicating thick disk and halo population membership for several sources.
We identify {three} new metal-poor T subdwarfs ([M/H] $\lesssim$ {$-$0.5}),
CWISE~J062316.19+071505.6, {WISEA~J152443.14$-$262001.8}, and CWISE~J211250.11-052925.2; and {19} new ``mild'' subdwarfs with modest metal deficiency ([M/H] $\lesssim$ {$-$0.25}).
We also identify three metal-rich brown dwarfs with thick disk kinematics.
%and a candidate L dwarf/T dwarf binary.
We provide kinematic evidence that the extreme L subdwarf 2MASS~J053253.46+824646.5 and the mild T subdwarf CWISE J113010.07+313944.7 may be part of the Thamnos population, while the T subdwarf CWISE J155349.96+693355.2 may be part of the Helmi stream.
%Combining our sample with spectra of previously-identified metal-poor brown dwarfs, 
We define a metallicity classification system for T dwarfs that adds mild subdwarfs (d/sdT), subdwarfs (sdT), and extreme subdwarfs (esdT) to the existing dwarf sequence. We also define a metallicity spectral index that  correlates with 
%metallicity classification, 
metallicities inferred from spectral model fits and iron abundances from stellar primaries of benchmark T dwarf companions.
This expansion of the T dwarf classification system supports investigations of ancient, metal-poor brown dwarfs now being uncovered in deep imaging and spectroscopic surveys.
\end{abstract}

\keywords{
Brown dwarfs (185) --- 
L subdwarfs (896) --- 
Stellar classification (1589) ---
T dwarfs (1679) --- 
T subdwarfs (1680)
}

\section{Introduction\label{sec:intro}}

Over the past 25 years, wide-field {red optical and near-infrared} surveys have dramatically expanded our census of the lowest-mass stars, brown dwarfs, and free-floating {exoplanets}, uncovering thousands of low temperature ultracool dwarfs (UCDs) that span the
%Brown dwarfs are distinguished from stars by their lack of hydrogen fusion, resulting in long-term cooling and dimming. As such, the end of the currently defined spectral sequence---
late-M, L, T, and Y dwarf spectral classes \citep{1999ApJ...519..802K,2006ApJ...637.1067B,2011ApJ...743...50C}.
%s---is a cooling sequence for a typical brown dwarf (M $\lesssim$ 0.078 M$_{\odot}$).  
This sequence is well-sampled among solar-metallicity objects down to effective temperatures {\teff} $\approx$ 300~K \citep{2014ApJ...786L..18L},
%thanks to the discovery of L and T dwarfs largely by the 2MASS, DENIS, and SDSS surveys (REFS), and the subsequent discovery of Y dwarfs in the main survey of WISE and its subsequent missions CatWISE and NEOWISE (REFS). Several thousand brown dwarfs are now known in the Solar Neighborhood, including XXX Y dwarfs, 
providing a rich sample for studying the formation, evolution, atmospheres, structures, and spatial distribution of objects that bridge the gap between the lowest mass stars and gas giant planets.

Metal-poor ultracool subdwarfs, including members of the thick disk and halo populations of the Milky Way, remain a very small fraction ($\lesssim$0.1\%) of the currently known population. The rarity of these sources can be directly attributed to the sparsity of metal-poor halo stars in the vicinity of the Sun ($\approx$0.5\%; \citealt{2008ApJ...673..864J}), the long-term cooling of brown dwarfs that makes old metal-poor objects cold and faint, and the relatively small distances ($\lesssim$30~pc) to which low-temperature brown dwarfs can be detected.  
Nevertheless, metallicity variety among the known brown dwarf population is crucial for understanding how composition influences atmospheric chemistry, thermal evolution, interior structure, and the hydrogen burning mass threshold \citep{1994ApJ...424..333S,1997AandA...327.1054B,2011ApJ...736...47B,2020RNAAS...4..214G}. In addition, since UCDs have either long hydrogen fusion lifetimes ($\gtrsim$10$^{12}$~yr; \citealt{1997ApJ...482..420L}) or never fuse, these objects record the conditions {and chemical abundances} of the earliest generations of star formation in the Milky Way \citep{2009IAUS..258..317B,2015AandA...579A..58L,2017arXiv170200091C}.

While metal-poor M-type subdwarfs have been known for over 75 years \citep{1947ApJ...105...96J,1978ApJ...220..935M,1997AJ....113..806G,2007ApJ...669.1235L},
significantly metal-poor brown dwarfs are a recent discovery.
The first serendipitous discoveries of metal-poor L-type subdwarfs 
\citep{2003ApJ...592.1186B,2004A&A...428L..25S,2009ApJ...694L.140S,2010ApJ...708L.107L} 
were made with data from
the Two Micron All-Sky Survey (hereafter 2MASS, \citealt{2006AJ....131.1163S}), 
the Sloan Digital Sky Survey (hereafter SDSS, \citealt{2000AJ....120.1579Y}), and
the UKIRT Infrared Deep Sky Survey (hereafter UKIDSS, \citealt{2007MNRAS.379.1599L}).
These sources have distinct near-infrared colors and spectra compared to solar-metallicity L dwarfs,
shaped by enhanced collision-induced H$_2$ absorption \citep{1969ApJ...156..989L,2003ApJ...592.1186B} and metallicity-induced variations in optical molecular absorption and condensate formation \citep{2006AJ....132.2372G,2007ApJ...657..494B,2021ApJ...923...19G}.
However, most L subdwarfs are low-mass stars, since brown dwarfs at the age of the Galactic thick disk and halo 
($\gtrsim$8~Gyr; \citealt{2017ApJ...837..162K}) have temperatures $\lesssim$1500~K, within the late-L and T dwarf regime
\citep{2001RvMP...73..719B,2003AandA...402..701B}.
While several modestly metal-poor T dwarfs ([M/H] $\sim$ $-$0.5) {were found early on} by wide-field surveys \citep{2002ApJ...564..421B,2014MNRAS.440..359B,2014MNRAS.437.1009P,2018AJ....155...87K} and as companions to metal-poor stars \citep{2011MNRAS.414..575M,2012MNRAS.422.1922P,2010AandA...515A..92S,2013MNRAS.433..457B,2013ApJ...777...36M},
%but these were distinguished by their own color and spectral peculiarities (Burgasser, Tinney, others). Deeper infrared surveys UKIDSS and VISTA, and WISE, expanded the census of metal-poor L and T dwarfs (Pinfield, Burningham, Kirkpatrick, Lodieu, Zhang), enabling the first spectral sequences of these objects (Burgasser, Gizis, Kirkpatrick, Zhang) including initial delineations of metallicity classes (sd/esd/usd) following the system defined for M subdwarfs by Gizis and Lepine. 
{significantly} metal-poor substellar T subdwarfs with [M/H] $\lesssim$ $-$1 were initially elusive.

The first examples of true T subdwarfs were uncovered through the deep infrared photometry, multi-epoch astrometry, and {search efforts of} citizen scientists deployed by the Backyward Worlds: Planet 9 program (BYW; \citealt{2017ApJ...841L..19K}). This project aims to build the census of low-temperature brown dwarfs in the Solar Neighborhood by providing  multi-epoch unWISE images \citep{2014AJ....147..108L,2018AJ....156...69M,2019ApJS..240...30S} to citizen scientists who visually identify faint moving sources. This approach is particularly well-suited for finding low-temperature, nearby thick disk and halo brown dwarfs, whose large velocities translate into large proper motions. 
\citet{2020ApJ...898...77S} reported the discovery of
%the two faint, high proper-motion sources,
WISEA J041451.67$-$585456.7 and WISEA~J181006.18$-$101000.5 (hereafter J0414$-$5854 and J1810$-$1010), whose near-infrared spectra are indicative of both low temperatures ({\teff} $\lesssim$ 1300~K) and {significantly} subsolar metallicities ([M/H] $\lesssim$ $-$1).
%, as confirmed through comparison to metal-poor atmosphere models. 
\citet{2020ApJ...899..123M,2021ApJ...915..120M} subsequently reported three additional T subdwarf candidates in BYW data, one (WISE~J155349.96+693355.2; hereafter J1553+6933) that exhibits spectroscopic evidence of modest subsolar metallicity ([M/H] $\approx$ $-$0.5).
\citet{2022AJ....163...47B} reported the BYW T subdwarf candidate, CWISE~J052306.42$-$015355.4 (hereafter J0523$-$0153) whose distinct $J-W2$ and $W1-W2$ colors and large proper motion suggest it is a metal-poor member of the thick disk or halo.
%($-$1.5 $\lesssim$ [Fe/H] $\lesssim$ $-$0.5).
{Finally}, WISEA~J153429.75$-$104303.3 (hereafter WISE~J1534$-$1043; aka, ``The Accident'') may be the first known metal-poor Y subdwarf \citep{2021ApJ...915L...6K}.
Candidate thick disk and halo T subdwarfs have also been identified in deep multi-band imaging and spectroscopic surveys conducted with JWST
%, with hints of modified molecular chemistry in their metal-poor atmospheres 
\citep{2023ApJ...942L..29N,2023ApJ...947L..25G,2023MNRAS.523.4534W,2024ApJ...962..177B,2024ApJ...964...66H,2024MNRAS.529.1067H}.

Low-temperature subdwarf candidates are exceptionally faint, and most lack the spectroscopic validation or astrometric follow-up necessary
% , crucial for understanding how metallicity modifies low-temperature atmospheric chemistry
% %, establishing spectral indicates of metallicity and a classification system, 
% and comparing absolute fluxes to models 
to assess physical properties.
Indeed, \citet{lodieu2022} used improved spectroscopy, photometry, and a parallax measurement for J1810$-$1010
%including a parallax distance measurement; 
%of 8.9$^{+0.7}_{-0.6}$~pc. This study 
to conclude that it is both colder ({\teff} = 800$\pm$100~K, equivalent to a T7 dwarf) and more metal-poor ([Fe/H] = $-$1.5$\pm$0.5~dex) than originally estimated.
%by \citet{2020ApJ...898...77S}.
This study also found evidence that J1810$-$1010 lacks methane absorption even in the strong 3.3~$\micron$ band, a defining trait for T dwarfs, suggesting significantly modified atmospheric chemistry.
Separately, \citet{2024ApJ...962..177B} have claimed evidence for modified PH$_3$/CO$_2$ chemistry in the distant T subdwarf UNCOVER-BD-3
identified with JWST, drawing an analogy to variations in CaH/TiO chemistry among metal-poor M subdwarfs, {although this claim has yet to be verified \citep{2024ApJ...973...60B}}.
%These temperatures align with those of early T dwarfs, yet the spectra of these targets exhibited little evidence of methane absorption, obscured by strong H$_2$ absorption.
%The unusual spectral properties of these low-temperature subdwarfs, and their potential importance in understanding the star formation conditions of the early Milky Way, motivates the search for additional examples. 

In this article, we report spectroscopic follow-up of a sample of late-L and T subdwarf candidates uncovered by Backyward Worlds: Planet 9.
% WISEA~J062316.19+071505.6 (hereafter WISE~J0623+0715), 
% WISEA~J152443.14-262001.8 (hereafter WISE~J1524$-$2620), and 
% WISEA~J211250.11-052925.2 (hereafter WISE~J2112$-$0529). 
% All three show evidence of low temperatures and subsolar metallicities, as confirmed from spectroscopic observations and atmosphere model comparison.
In Section~\ref{sec:selection} we describe the selection of candidate T subdwarfs 
%from sources identified in the BYW program by citizen scientists, 
and summarize their associated photometric and astrometric properties.
We also estimate distances for metal-poor candidates using new empirical color-magnitude relations based on late-L and T subdwarfs with parallax distance measurements.
In Section~\ref{sec:spectra} we describe new near-infrared spectroscopic observations of {24} candidates and three metal-poor L and T dwarf benchmarks. {We also} define comparison samples of solar metallicity and metal-poor brown dwarf spectra from the literature, {including sources that match our selection criteria}.
In Section~\ref{sec:analysis} we analyze these data, reporting classifications from dwarf standards and spectral indices, K~I equivalent widths, spectral model fits, and radial velocities (RVs) from forward modeling. We also use estimated distances, proper motions, and RVs to compute 3D velocities and assess Galactic population membership.
In Section~\ref{sec:individual} we assess the status of our candidates, identifying {three} new T subdwarfs, {19 new slightly metal-poor T dwarfs}, two brown dwarfs with exceptional Galactic orbits, and three apparently metal-rich brown dwarfs with thick disk kinematics.
In Section~\ref{sec:classify} we use our full spectral sample to define anchors for a two-dimensional classification scheme for T dwarfs, extending the existing dwarf sequence to include mild subdwarfs (d/sdT), subdwarfs (sdT), and extreme subdwarfs (esdT). We also construct a calibrated metallicity index that correlates with model-fit metallicities and the iron abundances of the stellar primaries of benchmark T dwarf companions.
In Section~\ref{sec:discussion} we examine metallicity dependence in the temperature scale of T dwarfs, continued challenges in modeling T subdwarf spectra, and prospects for expanding the known cold subdwarf population.
%discuss additional issues in the identification and characterization of known and future T and Y subdwarf discoveries.
%compare our spectroscopic data to subsolar-metallicity atmosphere models to infer estimates of temperature and metallicity for our candidates and benchmarks.
%use our spectral sample to define a set of metallicity indices for T dwarfs, and assess the validity of these indices using benchmarks and models.
%We also define a preliminary spectral sequence spanning mild subdwarfs, subdwarfs, and extreme subdwarfs in the T dwarf regime.
In Section~\ref{sec:summary} we summarize our results.

\section{BYW T Subdwarf Candidates} \label{sec:selection}

\subsection{Color and Reduced Proper Motion Selection}

%All of the T subdwarf candidates discussed here were identified as part of the 
The Backyard Worlds: Planet 9 program
%, which is described in further detail in \citet{2017ApJ...841L..19K}. In brief, this program 
uses multi-epoch imaging data from the {\em WISE} mission to find faint, infrared moving sources, which are identified by a community of {over 80,000} citizen scientists. 
%This program has successfully uncovered over 100 late-T and Y dwarfs in the vicinity of the Sun \citep{2020ApJ...899..123M,2021ApJS..253....7K}, including several T subdwarfs as noted above.
Objects identified as moving point sources are matched to optical and near-infrared photometry and astrometry from
2MASS, 
UKIDSS,
CatWISE2020 \citep{2021ApJS..253....8M},
Pan-STARRS DR1 \citep{2016arXiv161205560C}, 
Gaia DR3 \citep{2021AandA...650C...3G}, 
%the UKIRT Infrared Deep Sky Survey (UKIDSS; \citealt{2007MNRAS.379.1599L}), 
the UKIRT Hemisphere Survey (UHS; \citealt{2018MNRAS.473.5113D}), and
the Vista Hemisphere Survey (VHS/VIKING; \citealt{2013Msngr.154...35M}).
There are currently over 4,000 motion-confirmed cool brown dwarf candidates
in the BYW sample,
most without existing spectroscopic data. 
Of these, roughly 3,400 sources have well-measured $J$-band (2MASS, UHS/UKIRT, VHS/VIKING) and $W2$-band (WISE) photometry, as well as proper motions from CatWISE2020 or Gaia.

\begin{figure}[t]
\centering
\includegraphics[height=2.7in]{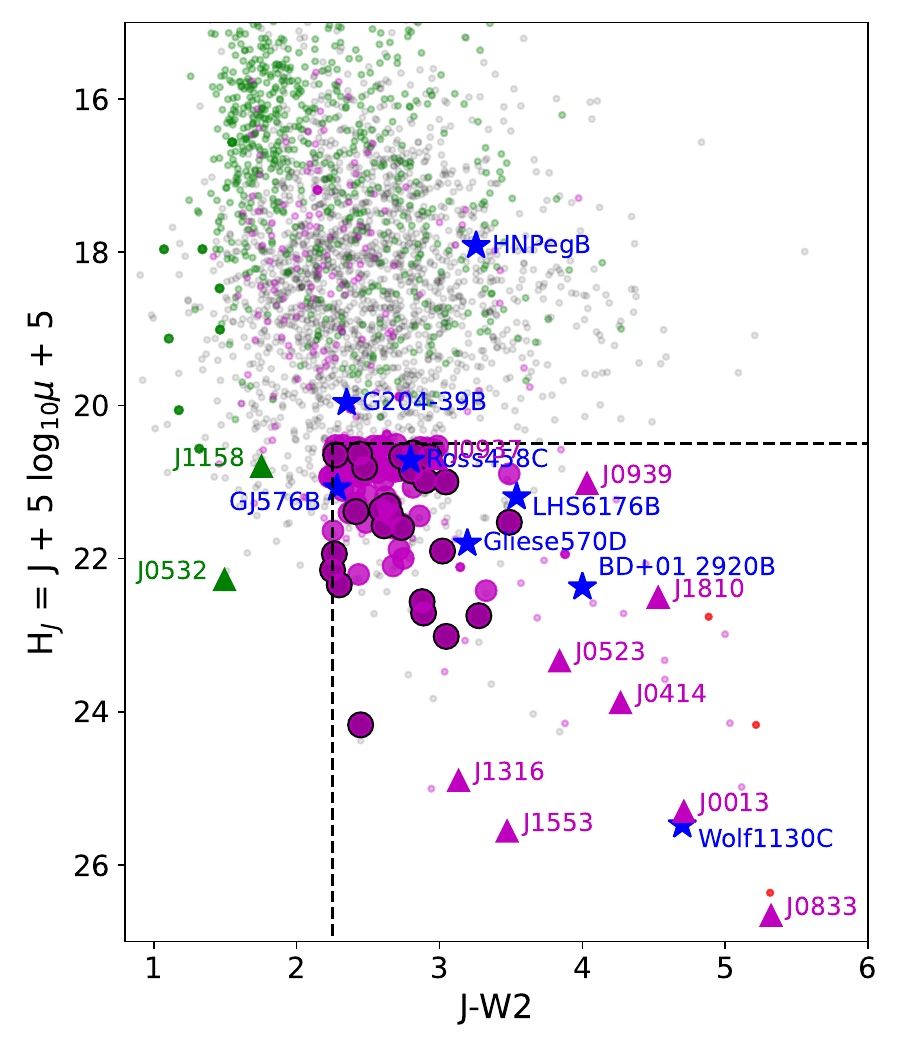}
\includegraphics[height=2.7in]{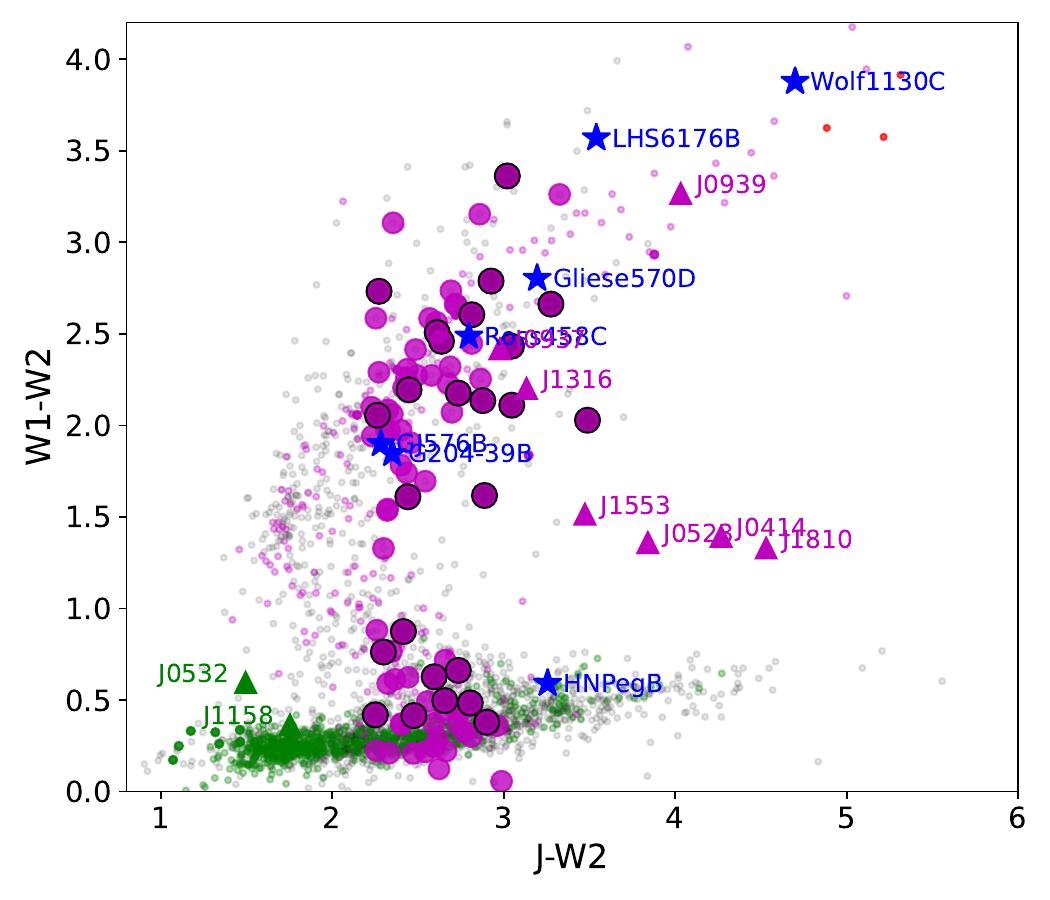}
%\plottwo{figures/rpmJ.pdf}{figures/color.pdf}
\caption{(Left) Selection of candidate T subdwarfs from BYW targets (small grey dots).
Semi-transparent circles indicate disk late-M and L dwarfs (green), T dwarfs (magenta), and Y dwarfs (red)  from the compilation of \citet{2018ApJS..234....1B}; 
while opaque circles indicate metal-poor objects from that compilation and from
\citet{2018MNRAS.479.1383Z}. 
We label previously identified late-L (green triangles) and T (magenta triangles) subdwarf comparison sources
and {some of the} benchmark companions (blue stars) from our spectroscopic sample (Tables~\ref{tab:observations} and~\ref{tab:additional}).
% and several spectroscopically-confirmed or candidate metal-poor T subdwarfs
% from the literature (large magenta triangles; \citealt{2018MNRAS.479.1383Z,2022AJ....163...47B}), 
We delineate our selection criteria of $J-W2$ $\geq$ 2.25~mag and H$_J$ $\geq$ 20.5~mag with dashed lines.
All candidates
%listed in Appendix~\ref{app:candidates} 
are indicated by large magenta circles,
while those with spectroscopic observations reported in this study are further highlighted by black outlines.
(Right): $W1-W2$ versus $J-W2$ color-color diagram of the same sample. 
%BYW candidates and previously identified sources, as labeled in the left panel. 
Note that sources without W1 detections are not shown in this plot. 
\label{fig:rpm}}
\end{figure}

Guided by the photometric and astrometric properties of previously identified L and T subdwarfs compiled in \citet{2018MNRAS.479.1383Z}, \citet{2021ApJ...915..120M}, and \citet{2022AJ....163...47B}, and ``normal'' field ultracool dwarfs compiled in \citet{2018ApJS..234....1B}, we identified a series of selection criteria to isolate high-probability T subdwarf candidates. We imposed a color limit of $J-W2 \geq 2.25$ mag to select sufficiently low temperature objects, and applied a reduced proper motion (RPM) constraint
\begin{equation}
    H_J \equiv J + 5\log_{10}\frac{\mu}{\rm arcsec/yr} + 5 = M_J + 5\log_{10}\frac{V_{tan}}{\rm 4.74~km/s} \geq 20.5~{\rm mag}
\end{equation}
to select sources with large tangential velocities, indicative of thick disk and halo objects (Figure~\ref{fig:rpm}). 
Note that unlike \citet{2021ApJ...915..120M} and \citet{2022AJ....163...47B} we do not impose a $W1-W2$ color constraint, using $H_J$ to segregate
potential T subdwarfs from contaminant field late-L and T dwarfs.
We further required $\mu/\sigma_{\mu} > 5$, $J \leq 18.5$~mag and declination $\delta \geq -35\degr$ to facilitate follow-up from Keck Observatory. 
We identified 95 candidate T subdwarfs, including 19 ``high-priority'' sources with measured $J-K < -0.5$~mag and/or estimated {\vtan} $>$ 150~km/s, the latter based on color/absolute magnitude relations from \citet{2021ApJ...915L...6K} or measured parallaxes. 
These candidates
%The full candidate sample is provided in Appendix~\ref{app:candidates}, and 
include the previously identified T subdwarfs 
J1810$-$1010 and the sdT5 CWISE~J113019.19-115811.3 (hereafter J1130-1158; \citealt{2021ApJ...915L...6K}). Here, we investigate a subset of {31} candidates with infrared spectroscopy.

\subsection{Photometric Distance Estimation}

\begin{figure}[tbp]
\centering
\includegraphics[height=2.5in]{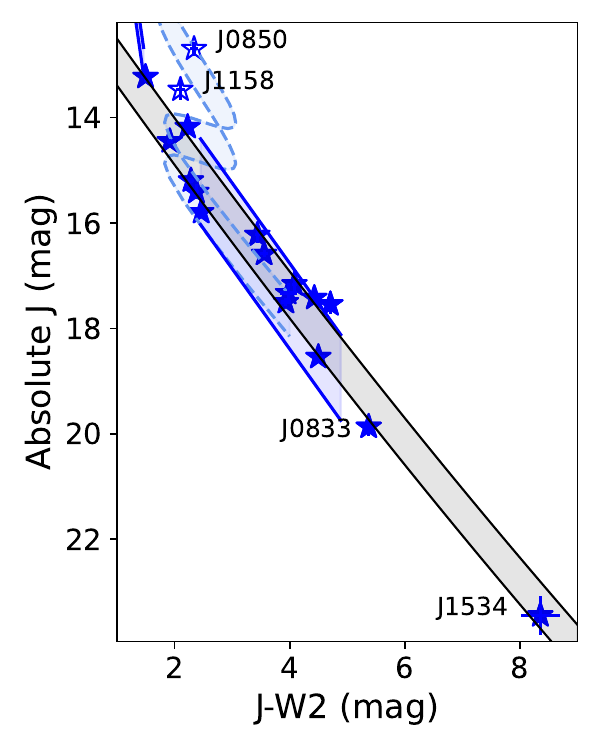}
\includegraphics[height=2.5in]{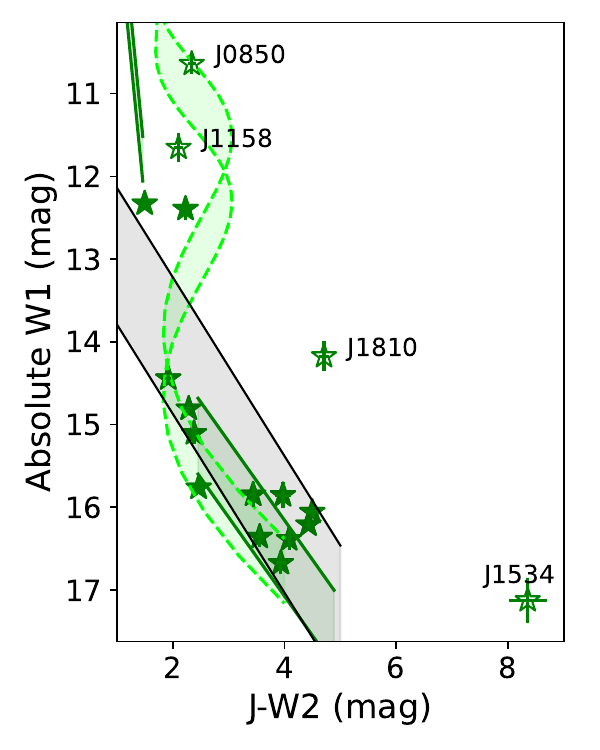}
\includegraphics[height=2.5in]{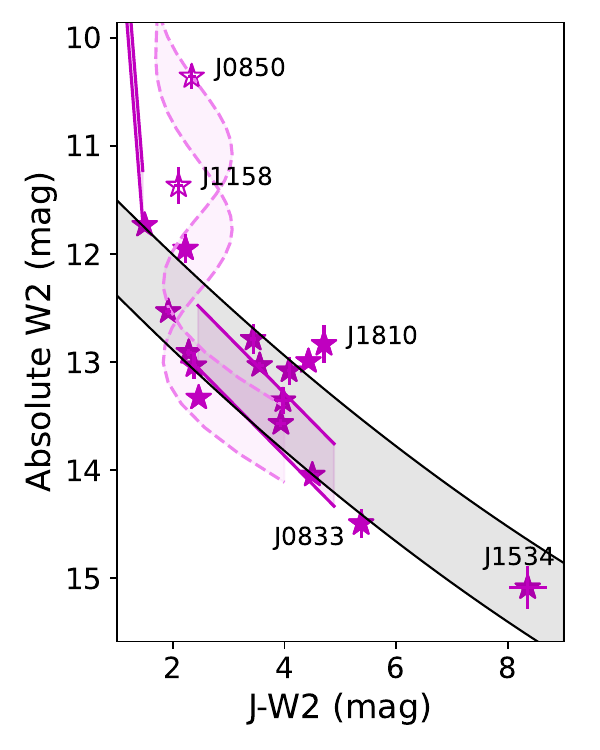} \\
\includegraphics[height=2.5in]{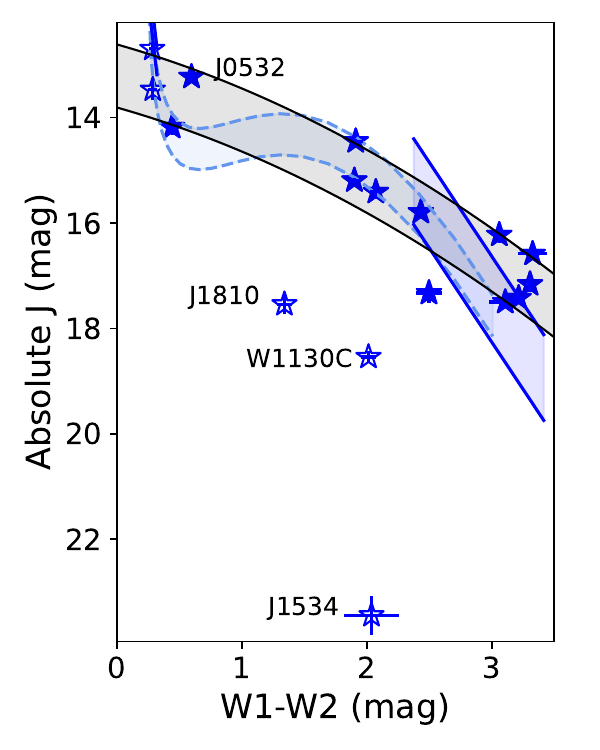}
\includegraphics[height=2.5in]{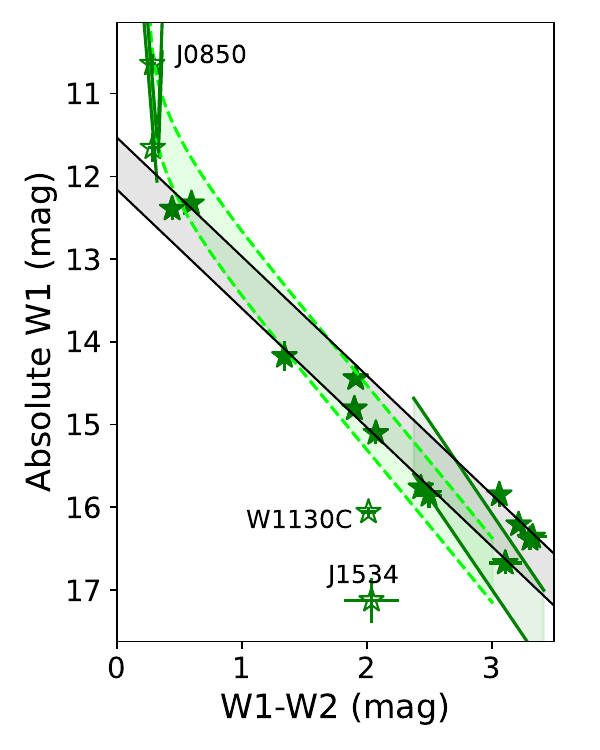}
\includegraphics[height=2.5in]{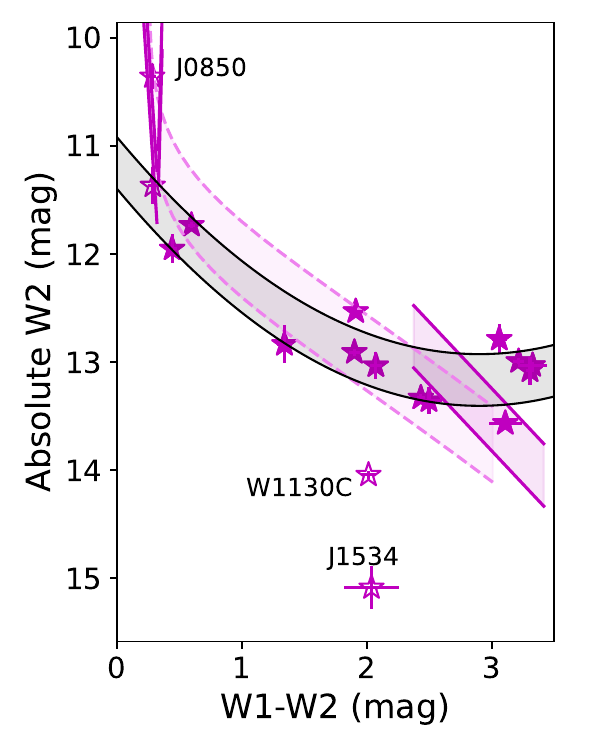}
\caption{{Absolute magnitudes in $J$ (left, in blue), $W1$ (middle, in green) and $W2$ (right, in magenta) bands
as a function of $J-W2$ (top) and $W1-W2$ (bottom) colors.
Symbols display measurements for 
metal-poor late L and T dwarfs with parallax measurements (Table~\ref{tab:absmagsample}) compared to empirical relations 
for dwarfs (light colored shaded regions with dashed lines: \citealt{2012ApJS..201...19D})
and subdwarfs (dark colored shaded regions with solid lines: \citealt{2018ApJ...864..100G,2019MNRAS.486.1260Z}).
Some individual sources are labeled.
%(Left) Absolute magnitudes versus spectral type. 
%For J0532+8246 and J1810-1010, we indicate the shift that would occur if spectral types were adjusted to esdT0 (based on CH$_4$ at 1.1~$\micron$) and esdT7 (based on effective temperature estimate).
Grey shaded regions delineate our empirical relationships computed over the ranges 1~mag $\leq$ $J-W2$ $\leq$ 8.5~mag and 0~mag $\leq$ $W1-W2$ $\leq$ 3.5~mag.
%for $M_J$ and $M_{W2}$ and 1~mag $\leq$ $J-W2$ $\leq$ 5~mag for $M_{W1}$ (excluding the proposed Y subdwarf WISEA~J1534$-$1043).
Sources not included in each color-magnitude fit are indicated by open symbols.}
\label{fig:absmag}}
\end{figure}

The majority of our sample lack trigonometric parallaxes, and assessment of physical properties (temperature, radius) and population membership (kinematics) requires a distance estimate. 
\citet{2018ApJ...864..100G} derived linear relationships between spectral type and absolute 2MASS $JHK_s$ and WISE $W1W2$ magnitudes for late-M and L subdwarfs, but these do not extend into the T dwarf regime.
\citet{2019MNRAS.486.1260Z} derived separate linear relationships between spectral type and absolute MKO $YJHK$ and WISE $W1W2$ magnitudes for L and for mid/late T subdwarfs (the latter encompassing relatively modest subsolar metallicities), but were unable to connect these due to the lack of L/T transition subdwarfs known at the time of that study.
With several more cool subdwarfs now known,  
we revisited these analyses with a sample of {18} metal-poor late-L and T dwarfs with measured parallactic distances, either of the source or its primary companion\footnote{We exclude three sources from this sample: the {esdL6} J0616-6407 and the sdT1 J0301$-$2319, whose measured parallaxes from \citet{2012ApJ...752...56F} and \citet{2020AJ....159..257B} were deemed too low quality for this analysis; and the subdwarf candidate J0055+5947, a co-moving companion to the white dwarf LSPM~J0055+5948 \citep{2020ApJ...899..123M}, which does not appear to be significantly metal-poor. We also excluded previously reported subdwarfs J0004$-$2604 and J1055+5443 as our analysis and that of \citet{2023ApJ...958...94R} indicate they are not significantly metal-poor.} (Table~\ref{tab:absmagsample}).
Figure~\ref{fig:absmag} compares absolute $J$, {$W1$}, and $W2$ magnitudes to $J-W2$ and $W1-W2$ colors for this sample, as well as empirical relationships for subdwarfs from \citet{2018ApJ...864..100G} and \citet{2019MNRAS.486.1260Z}, and for dwarfs from \citet{2012ApJS..201...19D}.
%and \citet{2019ApJS..240...19K,2021ApJ...915L...6K}. 
{There are reasonably linear trends between color and absolute magnitude for these combinations, and good overlap with prior empirical relations for most of the metal-poor T dwarfs. However, the metal-poor late L dwarfs show considerable scatter, particularly in absolute $W1$ and $W2$ magnitudes where empirical dwarf relations become nearly vertical (a large range of absolute magnitudes over nearly constant color). In addition, both the esdT J1810$-$1010 and the putative Y subdwarf J1534$-$1043 are significantly offset from the bulk of the sample in $M_{W1}$ versus $J-W2$, while
J1534$-$1043 and the T subdwarf companion Wolf~1130C are offset in all of the absolute-magnitude versus $W1-W2$ color samples.}
We therefore chose to estimate distances from $M_J$ and $M_{W2}$ versus $J-W2$ color-magnitude trends, fitting {16 sources over 1~mag $\leq J-W2 \leq$ 8.5~mag} to second order polynomials
by linear least squares regression:
\begin{equation}
    M_J = 11.4103901 + 1.55433844[J-W2] - 0.01643861[J-W2]^2 
    \label{eqn:mj}
\end{equation}
\begin{equation}
    M_{W2} = 11.41790678 + 0.53871427[J-W2] - 0.01195053[J-W2]^2. 
\label{eqn:mw2}
\end{equation}
{Both relations have a scatter of 0.44~mag, or 22\% error in distance.}
%We computed distances with both relations, finding the $M_{W2}$ consistently led to larger distances that were more often in significant disagreement ($>$3$\sigma$) with parallax distances. We therefore 
For our analysis, we used the average of both relations to estimate distances, and include in our uncertainty estimates a 0.3~mag scatter
in addition to the photometric uncertainties.
%for our subsequent analysis. 
We emphasize that these relations are based on a limited sample and require validation with a larger parallax sample of verified low-temperature subdwarfs.  

\begin{deluxetable}{llcccccccl}
\tablecaption{Late-L and T Subdwarf Absolute Magnitude Standards \label{tab:absmagsample}} 
\tabletypesize{\tiny} 
\tablehead{ 
%  & & & \multicolumn{5}{c}{Best Fit Model} \\
% \cline{4-10}
\colhead{Name} & 
\colhead{SpT\tablenotemark{a}} & 
\colhead{$J$} & 
\colhead{$J-W2$} & 
\colhead{$W1-W2$} & 
\colhead{$\pi$} & 
\colhead{$M_J$} & 
\colhead{$M_{W1}$} & 
\colhead{$M_{W2}$} & 
\colhead{Ref.} \\ 
 & 
\colhead{} & 
\colhead{(mag)} & 
\colhead{(mag)} & 
\colhead{(mag)} & 
\colhead{(mas)} & 
\colhead{(mag)} & 
\colhead{(mag)} & 
\colhead{(mag)} & \\
} 
\startdata
%{2MASS~J06164006$-$6407194} & esdL6 & 16.58$\pm$0.01 & 1.43$\pm$-0.00 & 0.44$\pm$0.02 & 19.90$\pm$6.50 & 13.08$\pm$0.71 & 12.09$\pm$0.71 & 11.65$\pm$0.71 & [1];[2] \\
2MASS~J085039.11-022154.3 & d/sdL6.5 & 15.44$\pm$0.04 & 2.34$\pm$0.05 & 0.282$\pm$0.016 & 28.2$\pm$1.3 & 12.70$\pm$0.11 & 10.64$\pm$0.10 & 10.36$\pm$0.10 & [1];[2] \\
2MASS~J11582077+0435014  & d/sdL8 & 15.61$\pm$0.06 & 2.10$\pm$0.06 & 0.286$\pm$0.023 & 37$\pm$3 & 13.47$\pm$0.16 & 11.65$\pm$0.16 & 11.37$\pm$0.16 & [1];[2] \\
2MASS~J053253.46+824646.5  & esdL8: & 15.18$\pm$0.06 & 1.50$\pm$0.06 & 0.596$\pm$0.016 & 40.71$\pm$0.15 & 13.23$\pm$0.06 & 12.328$\pm$0.013 & 11.732$\pm$0.014 & [1];[2] \\
2MASS~J06453153$-$6646120 & d/sdT0 & 15.525$\pm$0.009 & 2.227$\pm$0.017 & 0.441$\pm$0.023 & 54$\pm$3 & 14.18$\pm$0.12 & 12.39$\pm$0.12 & 11.95$\pm$0.12 & [3];[4] \\
%{2MASS~J20304235+0749358} & d/sdT1 & 14.23$\pm$0.03 & 2.02$\pm$0.03 & 0.81$\pm$0.19 & 102.8$\pm$0.8 & 14.29$\pm$0.03 & 13.08$\pm$0.19 & 12.26$\pm$0.02 & [1];[2] \\
WISEA~181006.18$-$101000.5\tablenotemark{b}  & esdT3: & 17.29$\pm$0.04 & 4.71$\pm$0.06 & 1.34$\pm$0.05 & 112$\pm$8 & 17.54$\pm$0.16 & 14.17$\pm$0.16 & 12.83$\pm$0.16 & [5];[5] \\
{HD~65486B} & d/sdT4.5 & 15.78$\pm$0.02 & 1.92$\pm$0.03 & 1.91$\pm$0.04 & 54.157$\pm$0.018\tablenotemark{c} & 14.45$\pm$0.02 & 14.44$\pm$0.04 & 12.531$\pm$0.017 & [6];[2] \\
{WISE~J044853.28$-$193548.6} & sdT5\tablenotemark{d} & 16.608$\pm$0.011 & 2.378$\pm$0.019 & 2.07$\pm$0.03 & 58$\pm$3 & 15.41$\pm$0.11 & 15.10$\pm$0.12 & 13.03$\pm$0.11 & [3];[4] \\
GJ~576B & d/sdT5.5 & 16.59$\pm$0.02 & 2.284$\pm$0.023 & 1.90$\pm$0.03 & 52.51$\pm$0.02\tablenotemark{c} & 15.192$\pm$0.016 & 14.81$\pm$0.03 & 12.908$\pm$0.017 & [7];[2] \\
2MASS~09373487+2931409  & sdT6 & 14.65$\pm$0.04 & 2.46$\pm$0.04 & 2.432$\pm$0.017 & 169.4$\pm$1.8 & 15.79$\pm$0.04 & 15.76$\pm$0.03 & 13.33$\pm$0.02 & [1];[8] \\
WISE~J030919.67$-$501614.3 & d/sdT7.5 & 17.10$\pm$0.02 & 3.437$\pm$0.021 & 3.06$\pm$0.05 & 67$\pm$4 & 16.22$\pm$0.13 & 15.85$\pm$0.13 & 12.79$\pm$0.13 & [3];[9] \\
ULAS~J141623.94+134836.3  & sdT7 & 17.26$\pm$0.02 & 4.428$\pm$0.020 & 3.21$\pm$0.03 & 107.7$\pm$0.2\tablenotemark{c} & 17.421$\pm$0.018 & 16.21$\pm$0.03 & 12.993$\pm$0.012 & [7];[2] \\
LHS~6176B & d/sdT7.5 & 18.05$\pm$0.04 & 3.55$\pm$0.04 & 3.33$\pm$0.10 & 51.00$\pm$0.03\tablenotemark{c} & 16.58$\pm$0.04 & 16.36$\pm$0.10 & 13.03$\pm$0.02 & [7];[2]\\
{HD~126053B} & sdT7.5 & 18.71$\pm$0.05 & 3.93$\pm$0.06 & 3.11$\pm$0.12 & 57.27$\pm$0.04\tablenotemark{c} & 17.50$\pm$0.05 & 16.67$\pm$0.12 & 13.57$\pm$0.03 & [10];[2] \\
{WISE~J083641.10$-$185947.0} & sdT8\tablenotemark{d} & 19.10$\pm$0.12 & 3.97$\pm$0.09 & 2.50$\pm$0.09 & 44.2$\pm$2.2 & 17.33$\pm$0.16 & 15.85$\pm$0.14 & 13.36$\pm$0.11 & [3];[4] \\
2MASS~J09393548$-$2448279  & d/sdT8 & 15.700$\pm$0.007 & 4.082$\pm$0.011 & 3.31$\pm$0.02 & 196$\pm$10 & 17.16$\pm$0.12 & 16.39$\pm$0.12 & 13.08$\pm$0.12 & [1],[7];[8] \\
Wolf~1130C & sdT6: & 19.64$\pm$0.09 & 4.50$\pm$0.09 & 2.01$\pm$0.04 & 60.30$\pm$0.03\tablenotemark{c} & 18.54$\pm$0.09 & 16.06$\pm$0.04 & 14.04$\pm$0.02 & [11];[2] \\
WISE~J083337.82+005214.1 & d/sdT8 & 20.28$\pm$0.10 & 5.37$\pm$0.10 & \nodata & 83$\pm$5 & 19.86$\pm$0.15 & \nodata & 14.49$\pm$0.12 & [12];[13] \\
WISEA~J153429.75$-$104303.3  & (e)sdY:\tablenotemark{e} & 24.5$\pm$0.3 & 8.4$\pm$0.3 & 2.04$\pm$0.21 & 61$\pm$5 & 23.4$\pm$0.3 & 17.1$\pm$0.3 & 15.09$\pm$0.19 & [14],[4] \\
\enddata
\tablecomments{$W1W2$ photometry is from CATWISE2020 \citep{2021ApJS..253....8M} {as provided by the Vizier catalog service \citep{vizier2000}}, with exceptions noted.}
\tablenotetext{a}{{Spectral type from this paper unless otherwise noted; see Table~\ref{tab:assessment}.}}
\tablenotetext{b}{Using WISE photometry as reported in \citet{lodieu2022}.}
\tablenotetext{c}{Parallax from primary.}
\tablenotetext{d}{{Based on \citet{2019MNRAS.486.1260Z}.}}
\tablenotetext{e}{{Based on \citet{2021ApJ...915L...6K}.}}
\tablerefs{The first reference is for $J$-band photometry, the second reference is for parallax;
[1] 2MASS \citep{2003yCat.2246....0C};
[2] Gaia~EDR3 \citep{2021AandA...650C...3G};
[3] VISTA VHS DR5 \citep{2013Msngr.154...35M};
[4] \citet{2021ApJS..253....7K};
[5] \citet{lodieu2022};
[6] \citet{2021AJ....161...42B};
[7] UKIDSS DR9 \citep{2013yCat.2319....0L};
[8] \citet{2012ApJ...752...56F};
[9] \citet{2019ApJS..240...19K};
[10] \citet{2012MNRAS.422.1922P};
[11] \citet{2013ApJ...777...36M};
[12] \citet{2014MNRAS.437.1009P};
[13] \citet{2019ApJS..240...19K};
[14] \citet{2023AJ....166...57M}
}
\end{deluxetable}

\section{Spectroscopic Observations} \label{sec:spectra}

\subsection{Keck/NIRES}

We obtained moderate-resolution near-infrared spectroscopy of {20} of our T subdwarf candidates with the Near-Infrared Echellette Spectrometer (NIRES; \citealt{2004SPIE.5492.1295W}) on the Keck II 10m telescope.
NIRES is a cross-dispersed spectrograph providing {\ldl} $\approx$ 2700 spectra spanning 0.9--2.45 $\micron$ over four spectral orders, with a fixed 0$\farcs$55$\times$18$\arcsec$ slit.
Data were obtained between 2019 October and {2024 January} in various conditions (Table~\ref{tab:observations}). 
All sources were observed with individual exposure times of 300~s in multiple ABBA nodding sets separated by 10$\arcsec$ along the slit for background subtraction.
One candidate, J0623+0715, resides in a crowded field, so we obtained six exposures at one slit position, and offset to measure the sky background at the start, middle, and end of the sequence. For all sources, we observed a nearby A0~V calibrator star at equivalent airmass to measure telluric and instrumental throughput. We also obtained calibration dome flat lamp exposures at the start of both nights. We observed three additional benchmarks for comparative analysis: the late L subdwarf 2MASS~J05325346+8246465 (hereafter J0532+8246; \citealt{2003ApJ...592.1186B}) and the metal-poor T dwarf benchmark companions 
GJ~576B (T5p, [Fe/H] = $-$0.37$\pm$0.09; \citealt{2010AandA...515A..92S,2019AandA...627A.138A}) and
LHS~6176B (T8p, [Fe/H] = $-$0.30$\pm$0.10; \citealt{2013MNRAS.433..457B}).

\begin{longrotatetable}
\begin{deluxetable}{llclcccclc}
\tablecaption{Spectroscopic Observations \label{tab:observations}} 
\tabletypesize{\scriptsize} 
\tablehead{ 
\colhead{Name} & 
\colhead{Designation} & 
\colhead{$J$}  & 
\colhead{Instrument}  & 
\colhead{UT Date}  & 
\colhead{Exposure}  & 
\colhead{Airmass}  & 
\colhead{Seeing}  & 
\colhead{Calibrator}  & 
\colhead{$J$ SNR} \\ 
\colhead{} & 
\colhead{(J2000)} & 
\colhead{(mag)}  & 
\colhead{}  & 
\colhead{}  & 
\colhead{(s)}  & 
\colhead{}  & 
\colhead{($\arcsec$)}  & 
\colhead{}  & 
\colhead{}  \\ 
} 
\startdata 
\hline
\multicolumn{9}{c}{Metallicity Benchmarks and Comparison Sources} \\
\hline
J0532+8246 & J053310.53+824620.5 & 15.18$\pm$0.06 & Keck/NIRES & 2022 Jan 19 & 2400 & 2.67 & 0$\farcs$7 &  HD 48049 & 277  \\
GJ~576B &  J150457.52+053759.8 & 16.59$\pm$0.02 & Keck/NIRES  & 2022 Jun 11 & 2400 & 1.03 &  0$\farcs$6 & HD 123233 & 224  \\
LHS~6176B &  J095047.39+011731.6 & 18.05$\pm$0.04 & Keck/NIRES  & 2022 Jan 20 & 3600 & 1.06 & 0$\farcs$4\tablenotemark{b} & HD 74721 & 91  \\
\hline
\multicolumn{9}{c}{Subdwarf Candidates} \\
\hline
J0045+7958 & J004556.21+795847.8 & 16.97$\pm$0.21 & APO/TSpec & 2018 Jan 21 & 3600 & 1.53 & 1$\farcs$5 &  HD 9612 & 5 \\
J0140+0150 & J014016.91+015054.7\tablenotemark{a} & 17.20$\pm$0.02 & IRTF/SpeX & 2020 Dec 24  & 1800 & 1.25 & 0$\farcs$5 & HD 1154  & 23 \\
\nodata & \nodata  & \nodata  & Keck/NIRES & 2021 Jan 02 & 2400 & 1.09 &  0$\farcs$8 & HD 18571  & 64 \\
J0411+4714 & J041102.42+471422.6 & 17.67$\pm$0.02 & Keck/NIRES & 2023 Dec 27 & 1200 & 1.34 &  0$\farcs$5 & BD+45 981  & 90 \\
J0429+3201 & J042941.12+320125.4 & 18.28$\pm$0.08 & Keck/NIRES & 2023 Nov 03 & 2400 & 1.02 &  0$\farcs$4\tablenotemark{b} & BD+43 1021  & 37 \\
J0623+0715 & J062316.20+071506.3 & 17.75$\pm$0.03 & Keck/NIRES & 2022 Jan 19 & 1800 & 1.46 & 0$\farcs$5 & HD 37887  & 46 \\
J0659+1615	& J065919.43+161552.0 & 17.97$\pm$0.06 & Keck/NIRES & 2024 Jan 30 & 2400 & 1.24 & 0$\farcs$5 & HD~47540 & 69 \\
%{J0805+5153} & J080556.14+515330.4 & 17.08$\pm$0.02 & SpeX & 2022 Mar 25 & 1800 & 1.38 & XXXX  & HD~71906 & 104 \\
J0843+2904 & J084325.37+290435.8 & 17.21$\pm$0.02 & IRTF/SpeX & 2022 Feb 22 & 1800 & 1.18 & 0$\farcs$8 & HD 71906  & 15 \\
J1110$-$1747 & J111055.13$-$174738.8 & 17.68$\pm$0.02 & Keck/NIRES & 2022 Jan 19 & 2400 & 1.27 & 0$\farcs$5 & HD 98884  & 76 \\
J1130+3139 & J113010.21+313947.0 & 17.79$\pm$0.05 & Keck/NIRES & 2022 Jan 20 & 2400 & 1.05 & 0$\farcs$5 & HD 98989 & 90 \\
{J1138+7212} & J113833.47+721207.8 & 17.08$\pm$0.23 & Gemini/GNIRS & 2013 Dec 03 & 2160 & 1.71 & 0$\farcs$7 & HIP~57348 & 13 \\
J1204$-$2359 & J120444.33$-$235927.4 & 17.91$\pm$0.03 & Keck/NIRES & 2022 Jan 19 & 2400 & 1.40 & 0$\farcs$5 & HD 110649  & 44 \\
J1304+2819 & J130440.45+281927.7 & 18.10$\pm$0.04 & Keck/NIRES & 2022 Jul 11 & 2400 & 1.30 &  0$\farcs$6 & HD 109055 & 40  \\
J1308$-$0321 & J130841.33$-$032157.8 & 18.07$\pm$0.02 & Keck/NIRES & 2022 Jan 20 & 1800 & 1.09 & 0$\farcs$5 & HD 114381 & 72  \\
J1401+4325 & J140118.31+432553.6 & 18.40$\pm$0.08 & Keck/NIRES & 2022 Jan 23 & 1800 & 1.09 & 0$\farcs$6 & HD 128039 & 36  \\
J1458+1734 & J145837.85+173450.2 & 18.39$\pm$0.07 & Keck/NIRES & 2022 Jul 11 & 3600 & 1.15 &  0$\farcs$5 & HD 124773 & 58  \\
{J1515$-$2157} & J151521.28$-$215737.3 & 17.48$\pm$0.03 & Magellan/FIRE & 2024 May 29 & 400 & 1.03 & 0$\farcs$6 & HD 139273 & 33 \\
J1524$-$2620 & J152443.15$-$262001.8 & 18.42$\pm$0.10 & Keck/NIRES & 2022 Jun 11 & 2400 & 1.45 &  0$\farcs$5 & HD 147384 & 25  \\
J1710+4537 & J171003.01+453756.4 & 18.23$\pm$0.06 & Keck/NIRES & 2022 Jun 11 & 3600 & 1.17 &  0$\farcs$7 & HD 123299 & 60  \\
J1801+4717 & J180122.59+471741.4 & 18.49$\pm$0.09 & Keck/NIRES & 2021 Aug 23 & 2400 & 1.15 & 0$\farcs$5 & HD 164899 & 33  \\
J2013$-$0326 & J201342.30$-$032643.4 & 18.24$\pm$0.12 & Keck/NIRES & 2021 Aug 23 & 2400 & 1.10 & 0$\farcs$5 & HD 192230  & 49  \\
J2021+1524 & J202130.11+152418.3 & 17.18$\pm$0.02 & Keck/NIRES & 2023 Jul 03 & 2400 & 1.01 & 0$\farcs$9 & HD 333145  & 70  \\
J2112$-$0529 & J211250.14$-$052924.7 & 17.50$\pm$0.03 & Keck/NIRES & 2022 Jun 11 & 2400 & 1.11 &  0$\farcs$5 & HD 203769 & 62  \\
J2112+3030 & J211255.59+303037.6 & 18.18$\pm$0.06 & Keck/NIRES & 2023 Jul 03 & 3600 & 1.02 & 0$\farcs$9 & HD 333145  & 42  \\
J2251$-$0740 & J225109.56$-$074037.4 & 18.47$\pm$0.08 & Keck/NIRES & 2023 Jul 10 & 2400 & 1.16 & 0$\farcs$5 & HD 203769 & 41  \\
\enddata 
\tablecomments{{All designations are from the CatWISE2020 catalog \citep{2021ApJS..253....8M} and are proper motion corrected to epoch 2015 May 28 except for noted sources.}}
\tablenotetext{a}{{Astrometry from UKIDSS DR9 \citep{2007MNRAS.379.1599L} epoch 2010.716.}}
\tablenotetext{b}{Seeing significantly less than the NIRES 0$\farcs$55 slit width, hence systematic radial velocity offsets are possible (see Section~\ref{sec:rv}).}
\end{deluxetable} 
\end{longrotatetable}

All NIRES data were reduced using a modified version of the Spextool package
\citep{2004PASP..116..362C} with standard settings. In brief, we used the dome flat lamps to rectify the spectral orders and correct for pixel-to-pixel response variations, computed the wavelength calibration from OH emission lines in deep exposures,
%(adjusted for barycentric motion), 
optimally extracted spectra from the rectified 2D traces, averaged multiple spectra for each source order-by-order with a sigma-clipped weighted mean (individual cosmic ray hits were removed by hand), and corrected the spectra for telluric absorption and instrumental response using the A0~V star spectra following the methodology of \citet{2003PASP..115..389V}. We chose to stitch the individual spectral orders into a single spectrum manually to correct for slight inter-order flux scaling variations not accounted for in the Spextool package. The final data have median signal-to-noise ratios between 23 and 275 at the 1.27~$\micron$ peak at native resolution.

\subsection{IRTF/SpeX}

Two candidates, J0140+0150 and J0843+2904, were observed with the SpeX spectrograph on the 3m NASA Infrared Telescope Facility \citep{2003PASP..115..362R}.
J0140+0150 was observed on 2020 Dec 24 (UT) in clear conditions at an average airmass of 1.25;
%{J0805+5153 was observed on 2022 Mar 25 (UT) in XXX conditions at an average airmass of 1.38}, and
J0843+2904 was observed on 2022 Feb 22 (UT) in partly cloudy conditions at an average airmass of 1.18.
Both sources were observed in prism mode with the 0$\farcs$8$\times$15$\arcsec$ slit, providing a resolution of {\ldl} $\approx$ 75 spanning 0.75--2.5~$\mu$m in a single order, with 
ten exposures of 180~s each obtained
in an ABBA nodding pattern along the slit for background subtraction. A0~V stars were observed immediately after each source, as well as internal flat field and arc lamp exposures. Data were reduced using Spextool version 4.1 with standard settings, following the procedures described above.

% {Three} candidates, J0140+0150, {J0805+5153,} and J0843+2904, were observed with the SpeX spectrograph on the 3m NASA Infrared Telescope Facility \citep{2003PASP..115..362R}.
% J0140+0150 was observed on 2020 Dec 24 (UT) in clear conditions at an average airmass of 1.25,
% {J0805+5153 was observed on 2022 Mar 25 (UT) in XXX conditions at an average airmass of 1.38}, and
% J0843+2904 was observed on 2022 Feb 22 (UT) in partly cloudy conditions at an average airmass of 1.18.
% {All three} sources were observed in prism mode with the 0$\farcs$8$\times$15$\arcsec$ slit, providing a resolution of {\ldl} $\approx$ 75 spanning 0.75--2.5~$\mu$m in a single order, with 
% ten exposures of 180~s each obtained
% in an ABBA nodding pattern along the slit for background subtraction. A0~V stars were observed immediately after each source, as well as internal flat field and arc lamp exposures. Data were reduced using Spextool version 4.1 with standard settings, following the procedures described above.

\subsection{APO/TripleSpec}

One candidate, J0045+7958, was observed on {2018 January 21 (UT)} with the TripleSpec spectrograph (TSpec; \citealt{2004SPIE.5492.1295W}) on the 3.5m Astrophysical Research Consortium Telescope at Apache Point Observatory (APO). Like NIRES, TSpec is a cross-dispersed spectrograph spanning 0.95--2.46~$\mu$m at a resolution of {\ldl} $\approx$ 3500 for a 1$\farcs$1-wide slit. Conditions were clear with 1$\farcs$5 seeing. We acquired twelve 300~s exposures in an ABBA nod pattern at an average airmass of 1.53, followed by observations of the B8~IV standard HD~9612 ($V$ = 6.57) at an airmass of 1.42. We also obtained quartz lamp flat field and NeAr arc lamp exposures illuminated on the dome at the start of the night. Data were reduced using a modified version of Spextool that incorporates instrument parameters for APO/TSpec (cf. \citealt{2012ApJ...745...14S}) following the procedures described above.

\subsection{Magellan/FIRE}

{One candidate, J1515$-$2157, was observed on 2024 May 29 (UT) with the Folded-port InfraRed Echellette (FIRE; \citealt{2013PASP..125..270S}) on the 6.5m Magellan Baade Telescope at Las Campanas Observatory. FIRE was operated in its prism-dispersed mode, providing 0.82--2.51~$\mu$m spectra at an average resolution of {\ldl} $\approx$ 300 for the deployed 0$\farcs$8-wide slit. 
%Conditions were XXXX. 
We acquired eight 50~s exposures in an ABBA nod pattern at 
an average airmass of 1.03, followed by observations of the 
A0~V standard HD~139273 ($V$ = 10.13) at the same airmass. 
We obtained quartz lamp flat field exposures reflected off the Baade flat field screen
at two settings to obtain uniform illumination for pixel calibation at the start of the night, 
and NeAr arc lamp exposures after the science observation for wavelength calibration. 
Data were reduced using a custom version of the FIREHOSE package \citep{2015zndo.....18775G}, which is based on the MASE pipeline \citep{2009PASP..121.1409B} and Spextool,
following procedures described in \citet{2024AJ....168...66B}.}

\subsection{Gemini-North/GNIRS}

{One candidate, J1138+7212, was observed on 2013 December 3 (UT) with the Gemini Near-Infrared Spectrograph (GNIRS; \citealt{2006SPIE.6269E..14E}) on the 8m Gemini-North telescope. Data were obtained as part of program GN-2013B-Q-91 in queue service mode, and were obtained through light clouds and seeing of 0$\farcs$7.
The cross-dispersed mode of GNIRS was used with the 0$\farcs$675 slit, short 0$\farcs$15~pix$^{-1}$ camera, and 31.7 lines/mm grating, yielding a 0.9-–2.4~$\mu$m spectrum at an average resolution of {\ldl} $\approx$ 800. A series of 12 exposures of 180~s each were obtained in an ABBA nod pattern at an average airmass of 1.71,
preceded by observations the A5~V star HIP~57348 (V = 10.67) at similar airmass. 
Data were reduced using the pipeline described in \citet{2008A&A...482..961D} and \citet{2011AJ....141..203A}, following similar procedures as noted above, with wavelength calibration achieved using a combination of Ar arc lamps and bright OH telluric emission lines.}

\subsection{Additional Spectra\label{sec:additional}}

In addition to these newly-observed candidates, we included in our analysis spectra of late-L and T dwarfs and subdwarfs previously published in the literature. A sample of 199 L7--T8 spectra were obtained from the SpeX Prism Library Analysis Toolkit (SPLAT; \citealt{2017ASInC..14....7B}), which excluded known resolved and candidate binaries and young brown dwarfs (Table~\ref{tab:indices_splat} in Appendix~\ref{app:splat}). 
We also included previously published spectra of 
T dwarf companions to primary stars with independent iron abundance measurements ranging over $-$0.64 $\leq$ [Fe/H] $\leq$ {$+$0.25},
and previously identified metal-poor late-L and T dwarfs (Table~\ref{tab:additional}).
For the late L subdwarf J0532+8246, we stitched our Keck/NIRES spectrum with the Keck/LRIS \citep{1995PASP..107..375O} optical spectrum reported in \citet{2003ApJ...592.1186B} by matching in the overlapping region of 0.98--1.0~$\mu$m.
For the T subdwarf J1810$-$1010, we stitched together the Palomar/TripleSpec \citep{2004SPIE.5492.1295W} near-infrared spectrum reported by \citet{2020ApJ...898...77S} with the GTC/OSIRIS \citep{2000SPIE.4008..623C} optical and GTC/EMIR \citep{2007RMxAC..29...12G} $YJ$ spectra reported by \citet{lodieu2022}
by matching in the overlapping region of 0.98--1.32~$\mu$m.

For those sources observed with cross-dispersed spectrographs (Keck/NIRES, APO/TSpec, Palomar/TripleSpec, SOAR/ARCoIRIS) and with well-measured $J$, $H$, and $K$ magnitudes, we re-evaluated the relative scaling of the spectral orders to ensure they align with photometry.
We computed synthetic spectrophotometric $J-H$ and $J-K$ colors by integrating the spectra over the relevant filter profiles from 2MASS $JHK_s$, UKIRT/WFCam $JHK$, or VISTA $JHK$\footnote{Filter profiles were obtained from the Spanish Virtual Observatory (SVO) Filter Profile Service \citep{2012ivoa.rept.1015R,2020sea..confE.182R} at \url{http://svo2.cab.inta-csic.es/theory/fps/}.} using SPLAT tools, with inter-order gaps filled in using a {BT-Settl atmosphere model \citep{2012RSPTA.370.2765A}} at an appropriate temperature, surface gravity, and metallicity (see Section~\ref{sec:modelfit}). For the majority of our sources, the inter-order flux calibration is generally consistent with the photometric uncertainties. However, for J1810$-$1010 we found a 50\% reduction in $K$-band flux is required to align the spectrum with its UKIDSS photometry \citep{2020ApJ...898...77S}. We used the appropriately scaled spectrum of this source for subsequent analysis.

% the late-L and T subdwarfs and subdwarf candidates
% J0414-5854, J1130-1158, J1553+6933, and J1810-0040;
% 2MASS~J11582077+0435014 (J1158+0435, sdL7; \citealt{2010ApJS..190..100K}),
% ULAS~J002136.00+155227.3 (J0021+1552, T4p; \citealt{2019MNRAS.486.1260Z}),
% 2MASS~J09373487+2931409 (J0937+2931, T6p; \citealt{2002ApJ...564..421B}),
% ULAS~J131610.28+075553.0 (J1316+0755, T6.5p; \citealt{2014MNRAS.440..359B}),
% ULAS~J141623.94+134836.3 (J1416+1348B, T7p; \citealt{2010MNRAS.404.1952B,2010AandA...510L...8S,2010AJ....139.2448B}), and
% 2MASS~J09393548-2448279 (J0939-2448, T8p; \citealt{2006ApJ...637.1067B});
% and the benchmark companions
% HN~Peg~B (young T2.5, [Fe/H] = $-$0.08; 
% \citealt{2007ApJ...654..570L,2018AandA...614A..55A}),
% GJ~576B (T6p, [Fe/H] = $-$0.37; \citealt{2010AandA...515A..92S,2011MNRAS.414..575M,2014MNRAS.440..359B,2019AandA...627A.138A,2019MNRAS.486.1260Z}),
% G~204-39B (T6.5, [Fe/H] = $-$0.04; \citealt{2010AJ....139..176F,2006ApJ...637.1067B,2019AandA...625A..68S}),
% Gliese~570D (T7.5, [Fe/H] = $+$0.05; \citealt{2000ApJ...531L..57B,2004AJ....127.2856B,2018AandA...614A..55A}),
% HD~3651B (T8, [Fe/H] = $+$0.143; \citealt{2006MNRAS.373L..31M,2007ApJ...654..570L,2007ApJ...660.1507L,2007ApJ...658..617B,2020ApJ...898..119R}), and
% Wolf~1130B (sdT8, [Fe/H] = $-$0.64; \citealt{2013ApJ...777...36M,2012ApJ...748...93R}).

%\adamb{add in J071121.36-573634.2 from kellogg 2018?}

\startlongtable
\begin{deluxetable}{llcllcll}
\tablecaption{Additional Spectral Data \label{tab:additional}} 
\tabletypesize{\scriptsize} 
\tablehead{ 
\colhead{Name} & 
\colhead{Designation} & 
\colhead{$J$}  & 
\colhead{Lit.}  & 
\colhead{Instrument}  & 
\colhead{$J$ SNR}  & 
\colhead{[Fe/H]}  & 
\colhead{Ref}  \\ 
\colhead{} & 
\colhead{(J2000)} & 
\colhead{(mag)}  & 
\colhead{SpT}  & 
\colhead{}  & 
\colhead{}  & 
\colhead{(dex)}  & 
\colhead{}  \\ 
} 
\startdata 
\hline
\multicolumn{8}{c}{Metallicity Benchmarks} \\
\hline
HD~3651B & J003918.91+211516.8\tablenotemark{a} & 16.31$\pm$0.03 & T8 & IRTF/SpeX & 73 & +0.14 & [1]; [2] \\ % & [Fe/H] = 0.14 \\
{HD~65486B} & J07580132$-$2538587 & 16.12$\pm$0.08 & T4.5 & IRTF/SpeX & 34 & $-$0.28 & [30]; [31],[34]\tablenotemark{d} \\ % 
{Ross~458C} & J130041.40+122114.3 & 16.69$\pm$0.02 & T8.5p & Magellan/FIRE & 67 & +0.25 & [28]; [29] \\ % & [Fe/H] = 0.14 \\
{HD~126053B} & J142320.90+011635.3 & 18.71$\pm$0.05 & T8p & Gemini/GNIRS & 24 & $-$0.38 & [25] \\ % & [Fe/H] = 0.14 \\
Gliese~570D & J145716.21$-$212216.5 & 15.32$\pm$0.05 & T7.5 & IRTF/SpeX & 37 & +0.05 & [3]; [4] \\ % & [Fe/H] = 0.05 \\
GJ~576B &  J150457.34+053757.0 & 16.59$\pm$0.02 & sdT5.5 & Gemini/GNIRS  & 56 & $-$0.37 & [5]; [6]  \\ % &  [Fe/H] = $-$0.37 \\
\nodata &  \nodata  & \nodata  & \nodata  & VLT/X-Shooter  & 21 & \nodata & [7]  \\ % &  \nodata  \\
G~204-38B & J175805.45+463319.8 & 16.15$\pm$0.09 & T6.5 & IRTF/SpeX & 75 & $-$0.04 & [8]; [9] \\ % & [Fe/H] = -0.04 \\
Wolf~1130C & J200519.70+542428.5 & 19.64$\pm$0.09 & sdT8 & Keck/NIRSPEC & 78 & $-$0.70 & [10]; [11]\\ %  & [Fe/H] = -0.64 \\
\nodata & \nodata & \nodata & \nodata & Keck/MOSFIRE & \nodata & \nodata & [10] \\ % & \nodata \\
HN~Peg~B & J214428.67+144606.0\tablenotemark{b} & 15.86$\pm$0.03 & T2.5 & IRTF/SpeX & 114 & $-$0.08 & [12]; [12] \\ % & [Fe/H] = -0.08 \\
{Wolf~940B} & J214638.70$-$001042.3 & 18.02$\pm$0.06 & T8 & Magellan/FIRE & 69 & +0.08 & [26]; [6] \\ 
\hline
\multicolumn{8}{c}{Metal-poor Comparison Sources} \\
\hline
J0004$-$1336 & J000458.59$-$133655.2 & 16.84$\pm$0.17 & T2p & IRTF/SpeX &  18 & \nodata & [19] \\
J0004$-$2604 & J000430.67$-$260403.6 & 16.49$\pm$0.13 & sdT2 & IRTF/SpeX &  40 & \nodata & [19] \\
{J0013+0634} & J001354.85+063445.4 & 19.75$\pm$0.05 & T8p & Gemini/GNIRS & 22 & \nodata & [18]  \\
J0021+1552 & J002135.97+155226.8 & 17.88$\pm$0.04 & T4p & VLT/X-Shooter & 9 & \nodata & [7]  \\
J0055+5947\tablenotemark{c} & J005559.94+594744.4 & 17.90$\pm$0.04 & T8 & Keck/NIRES & 13 & \nodata & [22] \\
J0057+2013 & J005757.95+201302.8 & 16.32$\pm$0.10 & sdL7 & IRTF/SpeX & 40 & \nodata & [24] \\
J0301$-$2319 & J030119.47$-$231921.8 & 16.64$\pm$0.14 & sdT1 & IRTF/SpeX &  25 & \nodata & [19] \\
J0309$-$5016 & J030919.92$-$501614.0 & 17.17$\pm$0.03 & T7p & Magellan/FIRE &  37 & \nodata & [19] \\
J0348$-$5620 & J034858.85$-$562016.9 & 16.52$\pm$0.15 & T3p & Magellan/FIRE & 22 & \nodata & [19] \\
J0414$-$5854 & J041451.86$-$585455.3 & 19.63$\pm$0.11 & esdT & Magellan/FIRE & 6 & \nodata & [13]  \\
J0433+1009\tablenotemark{c} & J043309.36+100902.3 & 18.00$\pm$0.03 & T8 & Keck/NIRES & 32 & \nodata & [15]  \\
J0532+8246 & J053310.53+824620.5 & 15.18$\pm$0.06 & esdL7 & Keck/LRIS & \nodata & {$-$1.6} & [14],[16]  \\
J0616$-$6407 & J061643.51$-$640719.8 & 16.35$\pm$0.01 & esdL6 & VLT/X-Shooter & 26 &{$-$1.6} & [16]  \\
J0645$-$6646 & J064529.11$-$664550.7 & 15.52$\pm$0.01 & sdL8 & VLT/X-Shooter & 47 & \nodata & [17]  \\
%J0711$-$5736 & J071121.36$-$573634.2 & \nodata & sdT0 & Magellan/FIRE & \nodata & \nodata & [24] \\
{J0833+0052} &  J083338.11+005206.2 & 20.28$\pm$0.10 & T9p & Gemini/GNIRS & 15 & \nodata & [18]  \\
J0845$-$3305\tablenotemark{c} & J084506.49$-$330532.1 & 16.85$\pm$0.01 & T7 & Magellan/FIRE & 82 & \nodata & [15]  \\
J0850$-$0221 & J085038.99$-$022155.1 & 15.44$\pm$0.04 & sdL7 & IRTF/SpeX &  83 & \nodata & [19] \\
J0911+2146\tablenotemark{c} & J091105.03+214645.4  & 17.75$\pm$0.03 & T9 & Magellan/FIRE & 29 & \nodata & [15]  \\
J0937+2931 & J093735.65+293127.2 & 14.65$\pm$0.04 & T6p & IRTF/SpeX & 128 & \nodata & [8]  \\
J0939$-$2448 & J093936.13$-$244844.0 & 15.67$\pm$0.01 & T8p & IRTF/SpeX & 17 & {$-$0.24} & [8],[35]  \\
J0953$-$0943\tablenotemark{c} & J095316.33$-$094318.8  & 17.01$\pm$0.03 & T5.5 & IRTF/SpeX & 24 & \nodata & [15],[34]\tablenotemark{d}  \\
J1019$-$3911 & J101944.47$-$391150.9 & 16.03$\pm$0.10 & sdT3 & SOAR/ARCoIRIS & 20 & \nodata & [19] \\
J1035$-$0711 & J103534.52$-$071148.5 & 16.39$\pm$0.09 & sdL7 & IRTF/SpeX  & 33 & \nodata & [19] \\
J1055+5443 & J105512.03+544328.3 & 18.87$\pm$0.07 & sdT8/Y0 & Keck/NIRES  & 29 & \nodata & [15],[20] \\
J1130$-$1158 & J113019.20$-$115811.5 & 17.22$\pm$0.03 & sdT5 & SOAR/ARCoIRIS & 11 & \nodata & [15]   \\
{J1158+0435} & J115821.40+043446.3 & 15.61$\pm$0.06 & sdL7 & IRTF/SpeX & 103 & \nodata & [32]   \\
J1307+1511 & J130710.04+151102.4 & 18.14$\pm$0.04 & sdL8 & VLT/X-Shooter & 10 & \nodata & [17]   \\
J1316+0755 & J131609.68+075553.9 & 19.29$\pm$0.12 & sdT6.5 & Gemini/GNIRS & 13 & \nodata & [5]  \\
J1338$-$0229 & J133836.94$-$022912.2 & 17.37$\pm$0.03 & sdL7 & Magellan/FIRE & 7 & \nodata & [16] \\
J1416+1348B & J141623.96+134837.3 & 17.26$\pm$0.02 & T7.5p & IRTF/SpeX & 48 & $-$0.35 & [21],[35] \\
J1553+6933 & J155347.98+693400.5 & 19.09$\pm$0.07 & sdT4 & Keck/NIRES & 19 & \nodata & [22],[33]  \\
J1810$-$1010\tablenotemark{c} & J181005.99$-$101001.8 & 17.27$\pm$0.02 & esdT &  Palomar/TripleSpec & 33 & {$-$1.5} & [13],[23]  \\
\nodata &\nodata &\nodata &\nodata &  GTC/OSIRIS+EMIR & \nodata & \nodata & [23]  \\
J2105$-$6235 & J210529.52$-$623604.6 & 16.85$\pm$0.14 & sdT1.5 & Gemini/Flamingos & 69 & \nodata & [24] \\
J2218+1146\tablenotemark{c} & J221859.41+114642.9 & 17.96$\pm$0.03 & T7p & Keck/NIRES & 51 & \nodata & [22] \\
\enddata 
\tablecomments{{All designations are from the CatWISE2020 catalog \citep{2021ApJS..253....8M} and are proper motion corrected to epoch 2015 May 28 except for noted sources.}}
\tablenotetext{a}{{Coordinates from \citet{2009AJ....137....1F}.}}
\tablenotetext{b}{{Coordinates from Pan-STARRS DR1 \citep{2016arXiv161205560C} epoch 2011 Nov 25.}}
\tablenotetext{c}{Also identified as a T subdwarf candidate by our selection criteria.}
\tablenotetext{d}{{Data re-reduced from the IRSA Legacy Archive.}}
\tablerefs{{The first reference corresponds to the source of the spectral data; the second reference corresponds to the source of the metallicity (for benchmarks):}
[1] \citet{2007ApJ...658..617B};
[2] \citet{2020ApJ...898..119R};
[3] \citet{2004AJ....127.2856B}; 
[4] \citet{2018AandA...614A..55A}; 
[5] \citet{2014MNRAS.440..359B}; 
[6] \citet{2019AandA...627A.138A}; 
[7] \citet{2019MNRAS.486.1260Z}; 
[8] \citet{2006ApJ...639.1095B}; 
[9] \citet{2019AandA...625A..68S}; 
[10] \citet{2013ApJ...777...36M}; 
{[11] \citet{2018ApJ...854..145M}}; 
[12] \citet{2007ApJ...654..570L}; 
[13] \citet{2020ApJ...898...77S};
[14] \citet{2003ApJ...592.1186B};
[15] \citet{2021ApJS..253....7K};
[16] \citet{2017MNRAS.464.3040Z};
[17] \citet{2018MNRAS.480.5447Z};
[18] \citet{2014MNRAS.437.1009P};
[19] \citet{2019AJ....158..182G}; 
[20] \citet{2023ApJ...958...94R};
[21] \citet{2010AJ....139.2448B};
[22] \citet{2020ApJ...899..123M};
[23] \citet{lodieu2022};
[24] \citet{2014ApJ...787..126L};
[25] \citet{2012MNRAS.422.1922P};
[26] \citet{2011ApJ...735..116B};
[27] \citet{2021AandA...656A.162M};
[28] \cite{2010ApJ...725.1405B};
[29] \citet{2021ApJS..255....8R};
[30] \citet{2012ApJ...755...94D};
[31] \citet{2022AandA...663A...4S};
{[32] \citet{2010ApJS..190..100K}};
{[33] \citet{2021ApJ...915..120M}};
[34] This paper;
{[35] \citet{2017ApJ...848...83L}}.
}
\end{deluxetable}

\section{Spectral Analysis\label{sec:analysis}}

\subsection{Comparison to Spectral Templates}

\begin{figure}[tbp]
\centering
\includegraphics[height=6.8in]{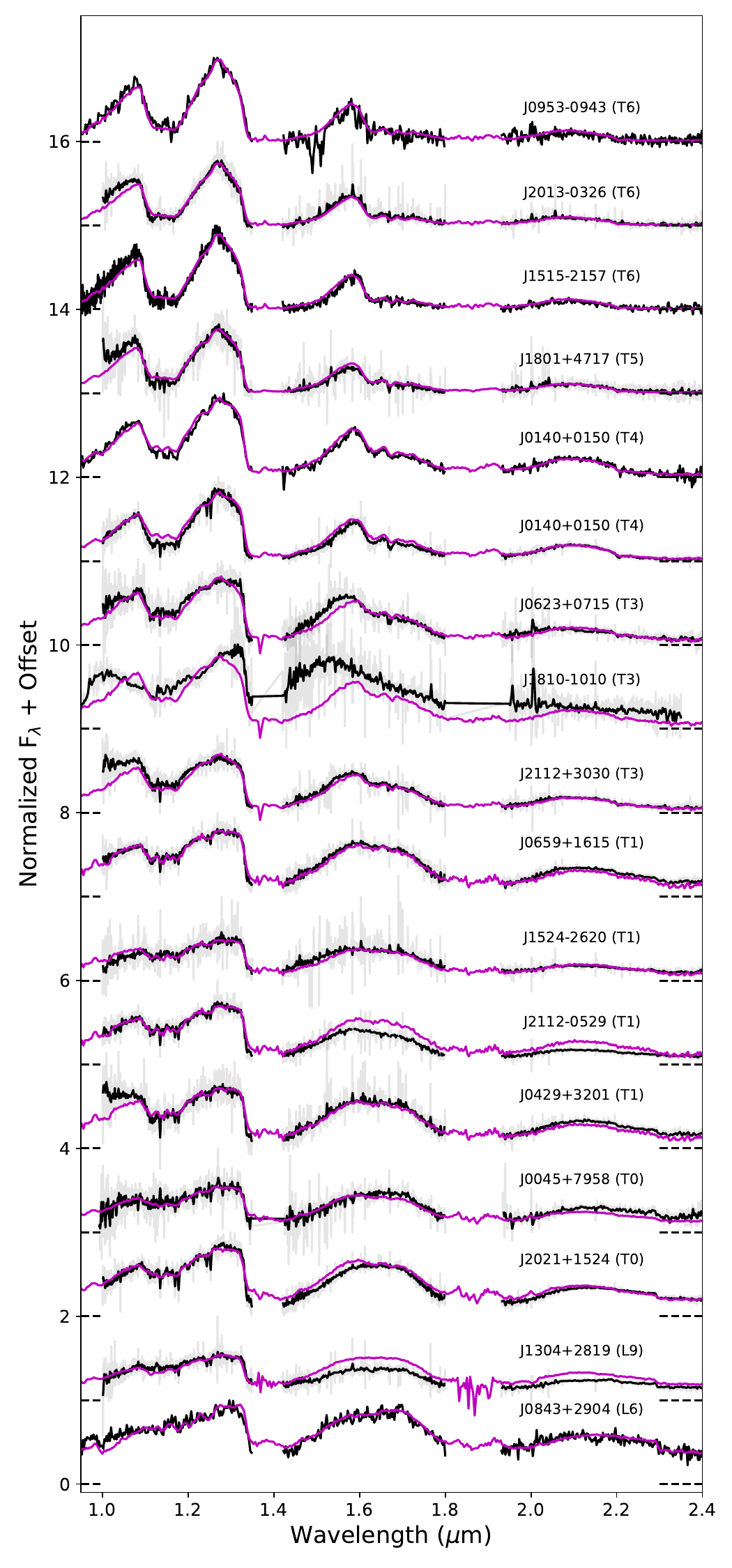}
\includegraphics[height=6.8in]{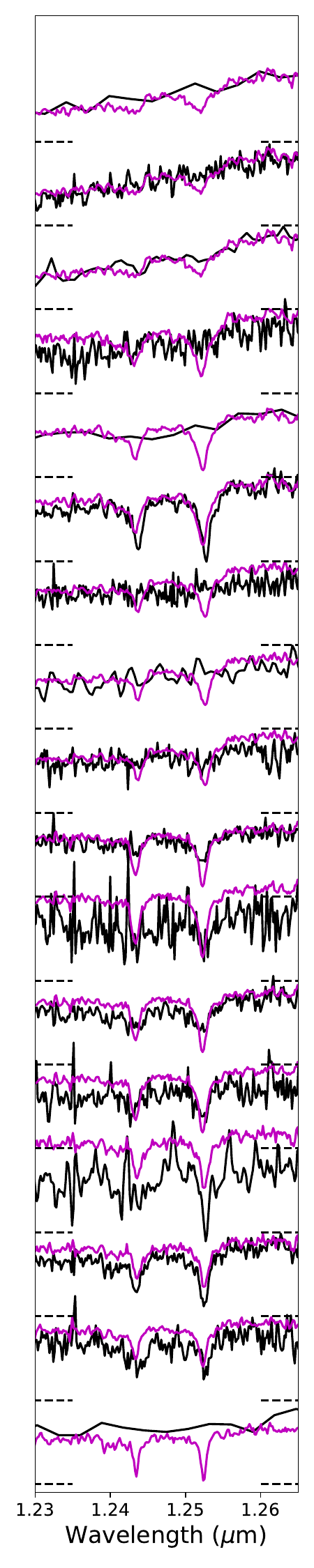}
\includegraphics[height=6.8in]{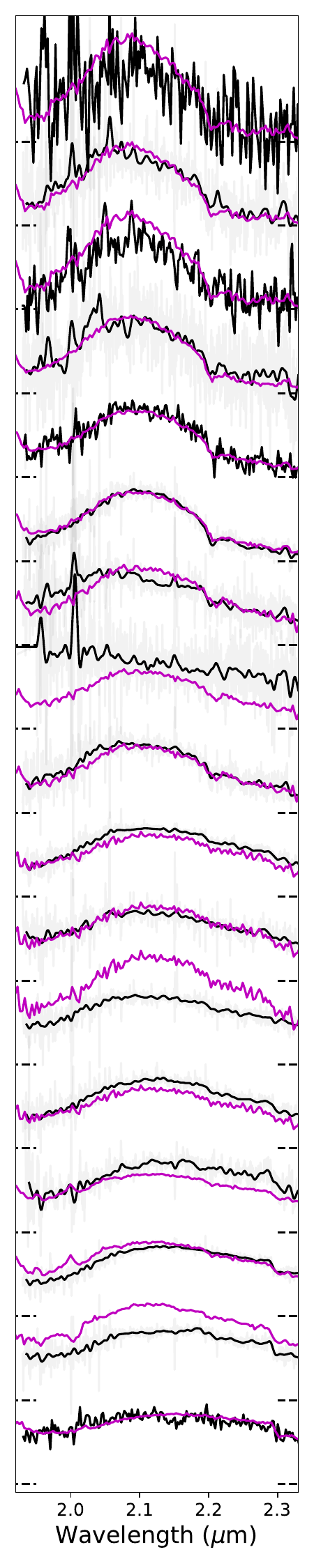}
\caption{\footnotesize 
Observed spectra of candidate late-L and T subdwarfs 
from Tables~\ref{tab:observations} and~\ref{tab:additional}. 
%compared to dwarf spectral standards (magenta lines) with corresponding types indicated next to each source label. 
Spectra are normalized at the $J$-band (1.25--1.28~$\mu$m) flux peak and offset by a constant (dashed lines on left and right edges).
Source are ordered from bottom to top by increasing $H$-band CH$_4$ band strength.
The left panel shows the full 1--2.5~$\mu$m spectral data, while
the middle and right panels {focus on} the 1.23--1.265~$\micron$ $J$-band and 2.0--2.3~$\micron$ $K$-band regions. 
In the left and right panels, the full resolution data are shown as grey lines and smoothed spectra ({\ldl} $\approx$ 125) as black lines, and are compared to
low-resolution dwarf spectral standards (magenta lines) drawn from
\citet{2010ApJS..190..100K} and \citet{2006ApJ...637.1067B},
{with corresponding types indicated next to each source label}. 
The middle panel shows only the full resolution data compared to medium-resolution spectral standards from \citet{2022RNAAS...6..151T} in magenta.
Note that IRTF/SpeX {and Magellan/FIRE} data
%for J0140+0150, J0843+2904, J0845$-$3305
J0911+2146, and J0953$-$0943 {(the latter three are in Figure~\ref{fig:spectra-all2})} 
are not smoothed as their native resolution is {{\ldl} $\approx$ 100--300}. 
The $K$-band spectra are re-scaled to highlight structure in this region.
%at center and right. 
%In all panels, data are normalized at the local flux peak and .
\label{fig:spectra-all}}
\end{figure}

\begin{figure}[tbp]
\centering
\includegraphics[height=6.8in]{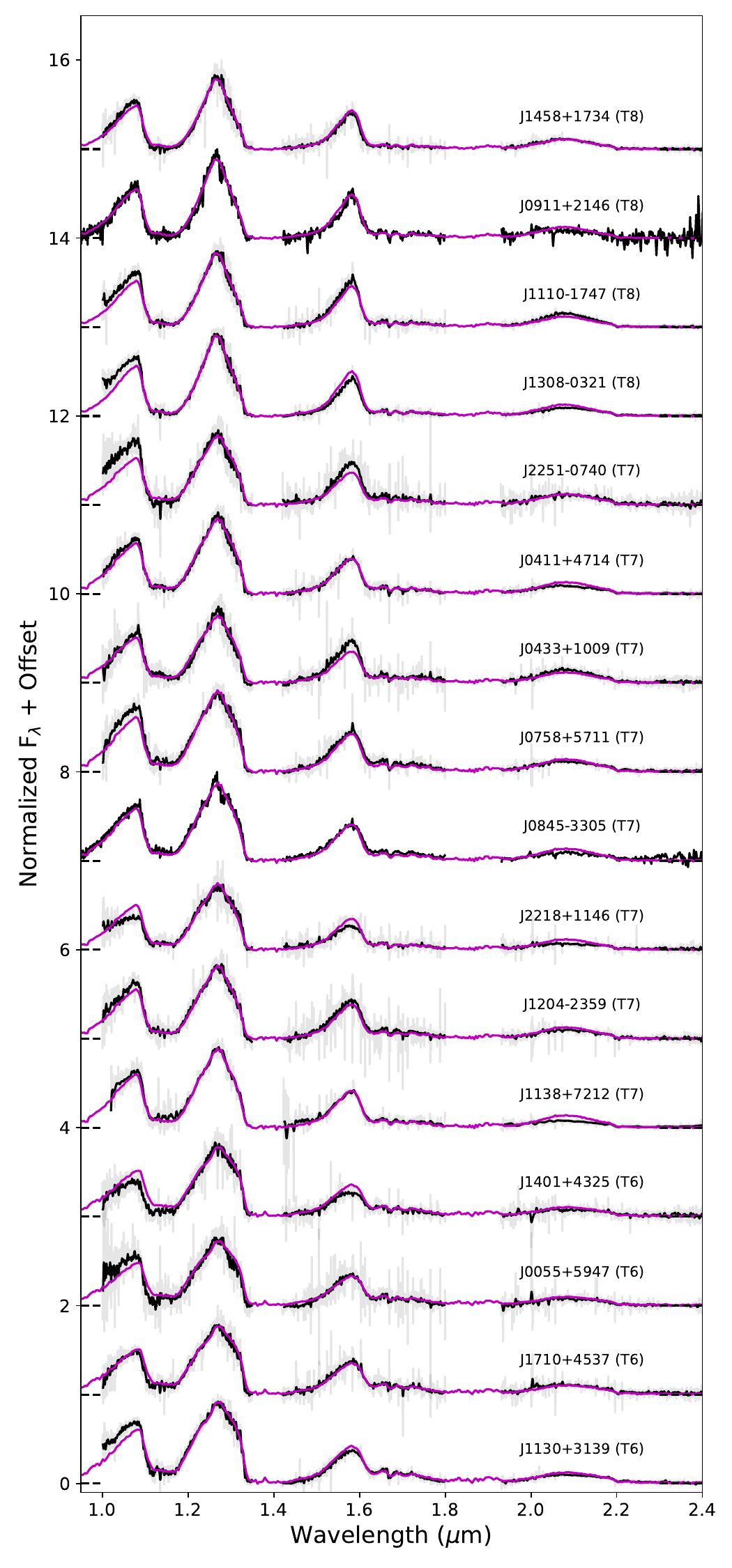}
\includegraphics[height=6.8in]{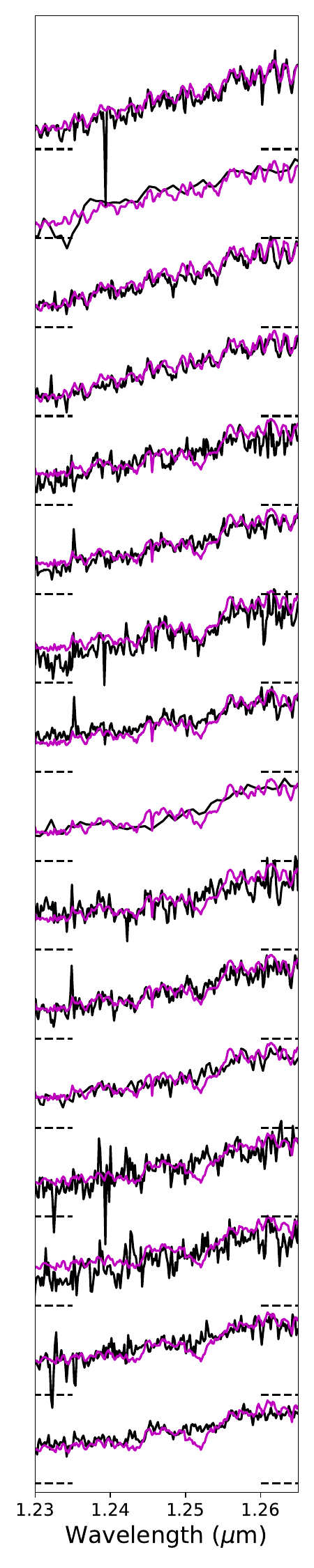}
\includegraphics[height=6.8in]{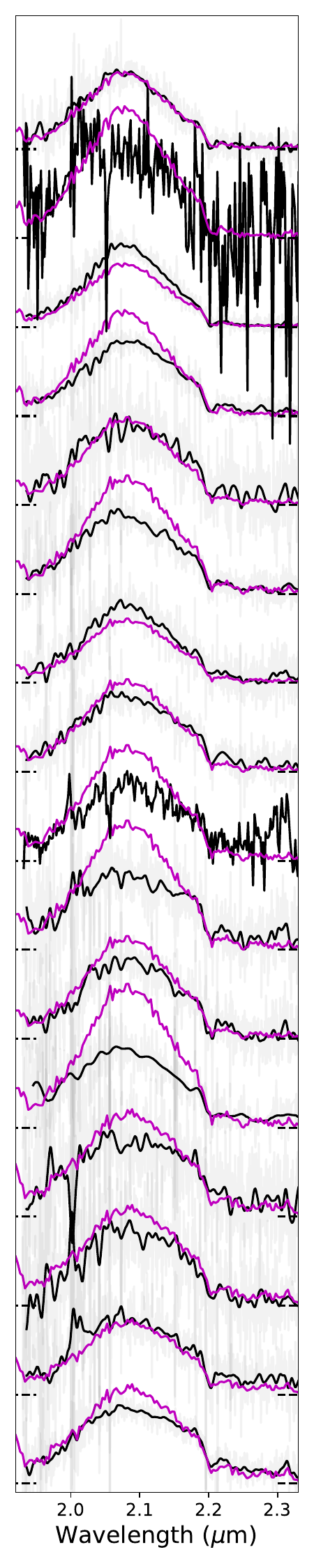}
\caption{{Continuation of Figure~\ref{fig:spectra-all} for the latest-type sources in our spectral sample.}
\label{fig:spectra-all2}}
\end{figure}

%Full resolution spectra of all our sources are provided in Appendix~\ref{appendix:spectra}.
{Figures~\ref{fig:spectra-all} and \ref{fig:spectra-all2} display} the reduced spectral data for our subdwarf candidates in both native resolution and smoothed to {\ldl} $\approx$ 125. {The smoothed spectra are compared to low resolution spectra of} L and T dwarf near-infrared spectral standards from SPLAT\footnote{Spectral data for late-L and T dwarf standards are from  
\citet{2004AJ....127.2856B,2006ApJ...637.1067B,2007ApJ...658..557B,2007AJ....134.1162L}; and
\citet{2014ApJ...794..143B}.} as defined by \citet{2010ApJS..190..100K} and \citet{2006ApJ...637.1067B}.
A best-fit standard for each source was selected based on comparison over the 0.9-1.4~$\micron$ range using
a $\chi^2$ statistic (see Eqn.~\ref{eqn:chi1}); these standard types are listed in
Table~\ref{tab:indices}.
%lists the best-match template for each source based on comparison over the 0.9-1.4~$\micron$ range (cf.\ \citealt{2010ApJS..190..100K}).
For most of the sources, including the companions GJ~576B and
LHS~6176B, we see close agreement with T6--T9 standards, with minor deviations in the $Y$-band (1~$\micron$) and $K$-band (2.1~$\micron$) peaks consistent with slightly subsolar metallicities \citep{2006ApJ...639.1095B}. 
%We argue below that these sources are ``mild'' T subdwarfs, with metallicities -0.5 $\lesssim$ [Fe/H] $\lesssim$ 0. 
A handful of sources appear to be late-type L or early-type T dwarfs based on weak or absent 1.6~$\micron$ CH$_4$ absorption, and several exhibit even more suppressed $K$-band peaks relative to their matched standards. 
For our higher resolution Keck/NIRES data, we also see {variety} in the strengths of the 1.2436~$\mu$m  and 1.2526~$\mu$m K~I doublet, in many cases weaker than the corresponding NIRES dwarf standard \citep{2022RNAAS...6..151T}. We quantify these spectral deviations below.
%We argue below that these are candidate ``early-type'' T subdwarfs comparable to WISE~J1810-1010.
%One source, WISE J1304+2819, has a distinct spectrum that we discuss separately below.

\subsection{Spectral Indices}\label{sec:indices}

For all of our spectra, we evaluated ten near-infrared spectral indices that sample molecular features (H$_2$O and CH$_4$ bands) and overall spectral shape, defined in \citet{2006ApJ...639.1095B,2006ApJ...637.1067B} and \citet{2019ApJ...883..205B}. Tables~\ref{tab:indices} and~\ref{tab:indices_splat} summarize the index values for our candidates and comparison sources, with uncertainties determined through Monte Carlo sampling of spectral flux errors.
We also list index-based spectral classifications based on the index versus spectral type relationships of \citet{2007ApJ...659..655B}
which are defined for spectral types L0--T9. Note that these relationships are defined for solar-metallicity dwarfs and are inaccurate for exceptionally metal-poor sources. 
Indeed, classifications from matched standards (Figure~\ref{fig:spectra-all}) and from indices can differ by more than 3 subtypes {(e.g., J0532+8246).} 
%\adamb{this may change} 
%We adopted final near-infrared classifications as a combination of the standard- and index-based values, indicating deviations from the standard classification with d/sd (mild), sd (significant), or esd (very significant) prefixes, as qualified below. Note that one source, J0140+0150 is given an ``r'' prefix to denote its unusual K-band brightness, also detailed below.

%\adamb{Actually these are still left in, need to replace}
%These measurements are provided in Appendix~\ref{app:spl}.

%\include{table_indices}

\begin{longrotatetable}
\begin{deluxetable}{lccccccccccll}
\tablecaption{Index Measurements for New and Literature Spectra \label{tab:indices}} 
\tabletypesize{\scriptsize} 
\tablehead{ 
\multicolumn{11}{c}{} & 
\multicolumn{2}{c}{SpT\tablenotemark{a}} \\
\cline{12-13}
\colhead{Name} & 
\colhead{[CH4$-$J]} & 
\colhead{[CH4$-$H]} & 
\colhead{[CH4$-$K]} & 
\colhead{[H2O$-$J]} & 
\colhead{[H2O$-$H]} & 
\colhead{[H2O$-$K]} & 
\colhead{[Y/J]} & 
\colhead{[K/H]} & 
\colhead{[K/J]} & 
\colhead{[H$-$dip]} & 
\colhead{Std} &
\colhead{Ind}  \\ 
\colhead{(numerator)} & 
\colhead{1.315--1.335} &
\colhead{1.635--1.675} &
\colhead{2.215--2.255} &
\colhead{1.14--1.165} &
\colhead{1.48--1.52} &
\colhead{1.975--1.995} &
\colhead{1.005--1.045} &
\colhead{2.06--2.10} &
\colhead{2.06--2.10} &
\colhead{1.61--1.64} \\
\colhead{(denominator)} & 
\colhead{1.26--1.285} &
\colhead{1.56--1.60} &
\colhead{2.08--2.12} &
\colhead{1.26--1.285} &
\colhead{1.56--1.60} &
\colhead{2.08--2.12} &
\colhead{1.25--1.29} &
\colhead{1.25--1.29} &
\colhead{1.56--1.60} &
\colhead{1.56--1.59,1.66--1.69} \\
} 
\startdata 
\hline
\multicolumn{12}{c}{Benchmark Companions} \\
\hline
HD~3651B & 0.25$\pm$0.01 & 0.14$\pm$0.01 & 0.04$\pm$0.02 & 0.04$\pm$0.01 & 0.17$\pm$0.01 & 0.32$\pm$0.02 & 0.42$\pm$0.01 & 0.29$\pm$0.01 & 0.14$\pm$0.01 & 0.25$\pm$0.02 & T8 & T7.7 \\
{HD~65486B} & 0.64$\pm$0.02 & 0.50$\pm$0.02 & 0.26$\pm$0.07 & 0.32$\pm$0.01 & 0.37$\pm$0.02 & 0.56$\pm$0.09 & 0.64$\pm$0.02 & 0.30$\pm$0.02 & 0.14$\pm$0.01 & 0.59$\pm$0.03 & T4.5 \\
LHS~6176B & 0.26$\pm$0.01 & 0.16$\pm$0.01 & 0.12$\pm$0.01 & 0.06$\pm$0.01 & 0.16$\pm$0.01 & 0.39$\pm$0.02 & 0.52$\pm$0.01 & 0.22$\pm$0.01 & 0.11$\pm$0.01 & 0.31$\pm$0.01 & T8 & T7.5 \\
{Ross~458C} & 0.19$\pm$0.01 & 0.12$\pm$0.01 & 0.07$\pm$0.01 & 0.05$\pm$0.01 & 0.19$\pm$0.01 & 0.36$\pm$0.01 & 0.35$\pm$0.01 & 0.45$\pm$0.01 & 0.26$\pm$0.01 & 0.23$\pm$0.01 & T8 & T8.1 \\
{HD~126053B} & 0.27$\pm$0.02 & 0.14$\pm$0.01 & 0.10$\pm$0.05 & 0.05$\pm$0.01 & 0.22$\pm$0.02 & 0.37$\pm$0.05 & 0.54$\pm$0.03 & 0.11$\pm$0.01 & 0.05$\pm$0.01 & 0.40$\pm$0.03 & T7 & T7.5 \\
GJ~570D & 0.28$\pm$0.01 & 0.13$\pm$0.01 & 0.11$\pm$0.02 & 0.06$\pm$0.01 & 0.20$\pm$0.01 & 0.34$\pm$0.02 & 0.41$\pm$0.01 & 0.24$\pm$0.01 & 0.11$\pm$0.01 & 0.34$\pm$0.02 & T8 & T7.5 \\
GJ~576B (GNIRS) & 0.43$\pm$0.01 & 0.37$\pm$0.01 & 0.21$\pm$0.05 & 0.16$\pm$0.01 & 0.30$\pm$0.02 & 0.47$\pm$0.04 & 0.52$\pm$0.01 & 0.22$\pm$0.01 & 0.10$\pm$0.01 & 0.67$\pm$0.03 & T6 & T5.6 \\
\nodata (XS) & 0.43$\pm$0.02 & 0.40$\pm$0.02 & 0.13$\pm$0.15 & 0.15$\pm$0.02 & 0.29$\pm$0.02 & 0.50$\pm$0.09 & 0.54$\pm$0.04 & 0.19$\pm$0.03 & 0.08$\pm$0.01 & 0.64$\pm$0.02 & T6 & T5.5 \\
\nodata (NIRES) & 0.44$\pm$0.01 & 0.40$\pm$0.01 & 0.25$\pm$0.01 & 0.15$\pm$0.01 & 0.28$\pm$0.01 & 0.49$\pm$0.01 & 0.45$\pm$0.01 & 0.25$\pm$0.01 & 0.10$\pm$0.01 & 0.61$\pm$0.01 & T6 & T5.5 \\
G~204$-$39B & 0.41$\pm$0.01 & 0.25$\pm$0.01 & 0.10$\pm$0.01 & 0.10$\pm$0.01 & 0.24$\pm$0.01 & 0.40$\pm$0.01 & 0.41$\pm$0.01 & 0.40$\pm$0.01 & 0.20$\pm$0.01 & 0.44$\pm$0.01 & T6 & T6.4 \\
Wolf~1130C & 0.82$\pm$0.03 & 0.42$\pm$0.01 &  \nodata\tablenotemark{b} & 0.01$\pm$0.01 & 0.39$\pm$0.01 &  \nodata\tablenotemark{b} & 0.74$\pm$0.11 &  \nodata\tablenotemark{b} &  \nodata\tablenotemark{b} & 0.75$\pm$0.01 & T6 & T5.4 \\
HN~PegB & 0.63$\pm$0.01 & 0.84$\pm$0.01 & 0.53$\pm$0.01 & 0.38$\pm$0.01 & 0.43$\pm$0.01 & 0.54$\pm$0.01 & 0.55$\pm$0.01 & 0.46$\pm$0.01 & 0.31$\pm$0.01 & 0.98$\pm$0.01 & T3 & T2.4 \\
{Wolf~940B} & 0.12$\pm$0.02 & 0.08$\pm$0.03 & $-$0.03$\pm$0.09 & 0.03$\pm$0.02 & 0.14$\pm$0.04 & 0.29$\pm$0.08 &  \nodata\tablenotemark{b} & 0.24$\pm$0.01 & 0.13$\pm$0.01 & 0.15$\pm$0.06 & T8 & T8.6 \\
\hline
\multicolumn{12}{c}{Subdwarf Candidates} \\
\hline
J0045+7958 & 0.82$\pm$0.09 & 1.12$\pm$0.06 & 0.90$\pm$0.05 & 0.67$\pm$0.11 & 0.60$\pm$0.06 & 0.62$\pm$0.04 & 0.57$\pm$0.17 & 0.63$\pm$0.03 & 0.50$\pm$0.04 & 1.02$\pm$0.05 & T0 & L8.6: \\
J0055+5947 & 0.35$\pm$0.02 & 0.27$\pm$0.02 & 0.16$\pm$0.05 & 0.11$\pm$0.03 & 0.25$\pm$0.04 & 0.36$\pm$0.08 & 0.60$\pm$0.06 & 0.23$\pm$0.01 & 0.11$\pm$0.01 & 0.46$\pm$0.03 & T6 & T6.4 \\
J0140+0150 (SpeX) & 0.59$\pm$0.02 & 0.58$\pm$0.03 & 0.27$\pm$0.05 & 0.28$\pm$0.02 & 0.34$\pm$0.03 & 0.47$\pm$0.05 & 0.45$\pm$0.02 & 0.44$\pm$0.02 & 0.25$\pm$0.01 & 0.82$\pm$0.04 & T4 & T4.2 \\
\nodata (NIRES) & 0.59$\pm$0.01 & 0.56$\pm$0.01 & 0.26$\pm$0.01 & 0.24$\pm$0.01 & 0.32$\pm$0.01 & 0.47$\pm$0.01 & 0.44$\pm$0.01 & 0.44$\pm$0.01 & 0.24$\pm$0.01 & 0.78$\pm$0.01 & T4 & T4.5 \\
J0411+4714 & 0.27$\pm$0.01 & 0.16$\pm$0.01 & 0.10$\pm$0.01 & 0.08$\pm$0.01 & 0.21$\pm$0.01 & 0.46$\pm$0.01 & 0.46$\pm$0.01 & 0.24$\pm$0.01 & 0.10$\pm$0.01 & 0.33$\pm$0.01 & T7 & T7.4 \\
J0429+3201 & 0.85$\pm$0.02 & 0.96$\pm$0.01 & 0.77$\pm$0.01 & 0.56$\pm$0.01 & 0.55$\pm$0.01 & 0.59$\pm$0.01 & 0.90$\pm$0.03 & 0.56$\pm$0.01 & 0.45$\pm$0.01 & 1.03$\pm$0.01 & T1 & T1.1 \\
J0433+1009 & 0.22$\pm$0.01 & 0.16$\pm$0.01 & 0.10$\pm$0.02 & 0.09$\pm$0.01 & 0.22$\pm$0.01 & 0.34$\pm$0.02 & 0.40$\pm$0.03 & 0.34$\pm$0.01 & 0.19$\pm$0.01 & 0.31$\pm$0.02 & T8 & T7.3 \\
J0623+0715 & 0.82$\pm$0.02 & 0.64$\pm$0.01 & 0.58$\pm$0.01 & 0.50$\pm$0.01 & 0.60$\pm$0.02 & 0.77$\pm$0.02 & 0.72$\pm$0.03 & 0.34$\pm$0.01 & 0.25$\pm$0.01 & 0.89$\pm$0.01 & T3 & T3.4 \\
J0659+1615 & 0.87$\pm$0.01 & 0.95$\pm$0.01 & 0.76$\pm$0.01 & 0.62$\pm$0.01 & 0.58$\pm$0.01 & 0.61$\pm$0.01 & 0.65$\pm$0.02 & 0.54$\pm$0.01 & 0.44$\pm$0.01 & 1.03$\pm$0.01 & T1 & T1.0 \\
J0758+5711 & 0.35$\pm$0.01 & 0.26$\pm$0.01 & 0.17$\pm$0.01 & 0.12$\pm$0.01 & 0.23$\pm$0.01 & 0.45$\pm$0.02 & 0.56$\pm$0.02 & 0.27$\pm$0.01 & 0.14$\pm$0.01 & 0.47$\pm$0.01 & T7 & T6.4 \\
J0843+2904 & 0.93$\pm$0.03 & 1.07$\pm$0.04 & 0.90$\pm$0.03 & 0.83$\pm$0.04 & 0.72$\pm$0.04 & 0.75$\pm$0.03 & 0.66$\pm$0.03 & 0.74$\pm$0.03 & 0.67$\pm$0.03 & 1.00$\pm$0.03 & L6 & L5.9 \\
J0845$-$3305 & 0.38$\pm$0.01 & 0.27$\pm$0.01 & 0.22$\pm$0.02 & 0.12$\pm$0.01 & 0.28$\pm$0.01 & 0.42$\pm$0.01 & 0.46$\pm$0.01 & 0.24$\pm$0.01 & 0.12$\pm$0.01 & 0.46$\pm$0.01 & T7 & T6.2 \\
J0911+2146 & 0.19$\pm$0.01 & 0.13$\pm$0.01 & \nodata\tablenotemark{b} & 0.04$\pm$0.01 & 0.17$\pm$0.01 & 0.56$\pm$0.05 & 0.37$\pm$0.02 & 0.19$\pm$0.01 & 0.10$\pm$0.01 & 0.29$\pm$0.02 & T8 & T8.2 \\
J1110$-$1747 & 0.20$\pm$0.01 & 0.10$\pm$0.01 & 0.04$\pm$0.01 & 0.05$\pm$0.01 & 0.17$\pm$0.01 & 0.32$\pm$0.01 & 0.46$\pm$0.01 & 0.32$\pm$0.01 & 0.19$\pm$0.01 & 0.22$\pm$0.01 & T8 & T8.2 \\
J1130+3139 & 0.45$\pm$0.01 & 0.38$\pm$0.01 & 0.26$\pm$0.01 & 0.14$\pm$0.01 & 0.26$\pm$0.01 & 0.47$\pm$0.02 & 0.58$\pm$0.01 & 0.27$\pm$0.01 & 0.11$\pm$0.01 & 0.63$\pm$0.01 & T6 & T5.6 \\
{J1138+7212} & 0.31$\pm$0.02 & 0.23$\pm$0.03 & 0.18$\pm$0.08 & 0.13$\pm$0.03 & 0.26$\pm$0.04 & 0.58$\pm$0.10 & 0.56$\pm$0.04 & 0.21$\pm$0.01 & 0.09$\pm$0.01 & 0.40$\pm$0.05 & T7 & T6.5 \\
J1204$-$2359 & 0.31$\pm$0.01 & 0.24$\pm$0.01 & 0.10$\pm$0.02 & 0.09$\pm$0.01 & 0.21$\pm$0.02 & 0.33$\pm$0.02 & 0.50$\pm$0.03 & 0.25$\pm$0.01 & 0.13$\pm$0.01 & 0.42$\pm$0.02 & T7 & T6.8 \\
J1304+2819 & 0.92$\pm$0.02 & 1.02$\pm$0.01 & 0.90$\pm$0.01 & 0.75$\pm$0.02 & 0.68$\pm$0.01 & 0.69$\pm$0.01 & 0.56$\pm$0.02 & 0.63$\pm$0.01 & 0.44$\pm$0.01 & 0.98$\pm$0.01 & L9 & L7.4 \\
J1308$-$0321 & 0.22$\pm$0.01 & 0.12$\pm$0.01 & 0.12$\pm$0.01 & 0.03$\pm$0.01 & 0.16$\pm$0.01 & 0.36$\pm$0.01 & 0.50$\pm$0.01 & 0.24$\pm$0.01 & 0.11$\pm$0.01 & 0.25$\pm$0.01 & T8 & T7.9 \\
J1401+4325 & 0.42$\pm$0.02 & 0.36$\pm$0.01 & 0.27$\pm$0.03 & 0.09$\pm$0.01 & 0.27$\pm$0.02 & 0.62$\pm$0.04 & 0.38$\pm$0.02 & 0.29$\pm$0.01 & 0.10$\pm$0.01 & 0.60$\pm$0.02 & T7 & T5.6 \\
J1458+1734 & 0.21$\pm$0.01 & 0.10$\pm$0.01 & 0.05$\pm$0.01 & 0.01$\pm$0.01 & 0.18$\pm$0.01 & 0.36$\pm$0.01 & 0.40$\pm$0.02 & 0.30$\pm$0.01 & 0.15$\pm$0.01 & 0.23$\pm$0.01 & T8 & T8.2 \\
{J1515$-$2157} & 0.43$\pm$0.01 & 0.28$\pm$0.01 & 0.23$\pm$0.02 & 0.12$\pm$0.01 & 0.18$\pm$0.01 & 0.39$\pm$0.02 & 0.51$\pm$0.03 & 0.23$\pm$0.01 & 0.10$\pm$0.01 & 0.50$\pm$0.02 & T6 & T6.1 \\
J1524$-$2620 & 0.87$\pm$0.02 & 0.94$\pm$0.01 & 0.77$\pm$0.01 & 0.67$\pm$0.02 & 0.72$\pm$0.02 & 0.72$\pm$0.01 & 0.48$\pm$0.03 & 0.48$\pm$0.01 & 0.36$\pm$0.01 & 1.01$\pm$0.01 & L9 & L9.3 \\
J1710+4537 & 0.40$\pm$0.01 & 0.31$\pm$0.01 & 0.26$\pm$0.02 & 0.14$\pm$0.01 & 0.26$\pm$0.01 & 0.44$\pm$0.02 & 0.41$\pm$0.02 & 0.30$\pm$0.01 & 0.15$\pm$0.01 & 0.51$\pm$0.01 & T6 & T5.9 \\
J1801+4717 & 0.44$\pm$0.01 & 0.43$\pm$0.01 & 0.28$\pm$0.03 & 0.18$\pm$0.01 & 0.33$\pm$0.02 & 0.37$\pm$0.03 & 0.65$\pm$0.03 & 0.34$\pm$0.01 & 0.13$\pm$0.01 & 0.70$\pm$0.02 & T5 & T5.1 \\
J2013$-$0326 & 0.40$\pm$0.01 & 0.34$\pm$0.01 & 0.22$\pm$0.02 & 0.13$\pm$0.01 & 0.30$\pm$0.01 & 0.60$\pm$0.03 & 0.58$\pm$0.02 & 0.25$\pm$0.01 & 0.12$\pm$0.01 & 0.57$\pm$0.02 & T6 & T5.8 \\
J2021+1524 & 0.86$\pm$0.01 & 1.04$\pm$0.01 & 0.91$\pm$0.01 & 0.62$\pm$0.01 & 0.56$\pm$0.01 & 0.59$\pm$0.01 & 0.53$\pm$0.01 & 0.56$\pm$0.01 & 0.39$\pm$0.01 & 1.03$\pm$0.01 & T0 & L9.8 \\
J2112$-$0529 & 0.82$\pm$0.01 & 0.88$\pm$0.01 & 0.77$\pm$0.01 & 0.61$\pm$0.01 & 0.60$\pm$0.01 & 0.68$\pm$0.01 & 0.58$\pm$0.02 & 0.42$\pm$0.01 & 0.25$\pm$0.01 & 1.01$\pm$0.01 & T1 & T1.4 \\
J2112+3030 & 0.78$\pm$0.01 & 0.74$\pm$0.01 & 0.51$\pm$0.01 & 0.53$\pm$0.01 & 0.50$\pm$0.01 & 0.61$\pm$0.01 & 0.93$\pm$0.03 & 0.39$\pm$0.01 & 0.28$\pm$0.01 & 0.93$\pm$0.01 & T3 & T2.6 \\
J2218+1146 & 0.37$\pm$0.01 & 0.24$\pm$0.01 & 0.14$\pm$0.02 & 0.11$\pm$0.01 & 0.30$\pm$0.01 & 0.51$\pm$0.03 & 0.41$\pm$0.02 & 0.26$\pm$0.01 & 0.10$\pm$0.01 & 0.46$\pm$0.02 & T7 & T6.4 \\
J2251$-$0740 & 0.28$\pm$0.01 & 0.19$\pm$0.01 & 0.14$\pm$0.02 & 0.06$\pm$0.01 & 0.16$\pm$0.01 & 0.34$\pm$0.03 & 0.69$\pm$0.03 & 0.26$\pm$0.01 & 0.16$\pm$0.01 & 0.39$\pm$0.02 & T7 & T7.1 \\
\hline
\multicolumn{12}{c}{Metal-poor Comparison Sources} \\
\hline
J0004$-$1336 & 0.76$\pm$0.02 & 0.95$\pm$0.03 & 0.85$\pm$0.05 & 0.57$\pm$0.02 & 0.59$\pm$0.04 & 0.63$\pm$0.04 & 0.55$\pm$0.02 & 0.44$\pm$0.02 & 0.26$\pm$0.01 & 0.96$\pm$0.03 & T1 & T0.6 \\
J0004$-$2604 & 0.70$\pm$0.01 & 0.75$\pm$0.01 & 0.57$\pm$0.02 & 0.47$\pm$0.01 & 0.52$\pm$0.02 & 0.59$\pm$0.02 & 0.57$\pm$0.01 & 0.41$\pm$0.01 & 0.27$\pm$0.01 & 0.90$\pm$0.02 & T3 & T2.5 \\
{J0013+0634} & 0.26$\pm$0.02 & 0.32$\pm$0.04 & 0.21$\pm$0.09 & 0.09$\pm$0.05 & 0.17$\pm$0.02 & \nodata\tablenotemark{b} & 0.50$\pm$0.04 & 0.09$\pm$0.01 & 0.04$\pm$0.01 & 0.38$\pm$0.06 & T7 & T6.2 \\
J0021+1552 & 0.67$\pm$0.04 & 0.57$\pm$0.03 & 0.26$\pm$0.17 & 0.39$\pm$0.05 & 0.46$\pm$0.04 & 0.67$\pm$0.14 & 0.52$\pm$0.11 & 0.27$\pm$0.04 & 0.17$\pm$0.03 & 0.83$\pm$0.05 & T4 & T4.0 \\
J0057+2013 & 0.84$\pm$0.01 & 1.02$\pm$0.01 & 0.95$\pm$0.02 & 0.66$\pm$0.01 & 0.59$\pm$0.01 & 0.70$\pm$0.01 & 0.57$\pm$0.01 & 0.53$\pm$0.01 & 0.38$\pm$0.01 & 0.97$\pm$0.01 & T0 & L8.7 \\
J0301$-$2319 & 0.82$\pm$0.02 & 0.98$\pm$0.02 & 0.73$\pm$0.02 & 0.67$\pm$0.02 & 0.67$\pm$0.02 & 0.63$\pm$0.02 & 0.74$\pm$0.02 & 0.49$\pm$0.01 & 0.33$\pm$0.01 & 1.01$\pm$0.02 & T1 & L9.6 \\
J0309$-$5016 & 0.20$\pm$0.01 & 0.16$\pm$0.01 & 0.16$\pm$0.08 & 0.05$\pm$0.01 & 0.23$\pm$0.01 & 0.51$\pm$0.05 & 0.56$\pm$0.02 & 0.14$\pm$0.01 & 0.06$\pm$0.01 & 0.30$\pm$0.02 & T7 & T7.7 \\
J0348$-$5620 & 0.54$\pm$0.02 & 0.59$\pm$0.02 & 0.30$\pm$0.04 & 0.39$\pm$0.02 & 0.47$\pm$0.02 & 0.61$\pm$0.04 & 0.50$\pm$0.02 & 0.34$\pm$0.02 & 0.20$\pm$0.01 & 0.79$\pm$0.03 & T4 & T4.1 \\
J0414$-$5854 & 0.90$\pm$0.06 & 0.58$\pm$0.07 & \nodata\tablenotemark{b} & 0.22$\pm$0.07 & 0.81$\pm$0.09 & \nodata\tablenotemark{b} & 0.91$\pm$0.09 & 0.04$\pm$0.04 & 0.02$\pm$0.02 & 0.89$\pm$0.09 & T4 & T3.9 \\
J0532+8246 & 0.85$\pm$0.01 & 0.83$\pm$0.01 & 0.81$\pm$0.01 & 0.83$\pm$0.01 & 0.86$\pm$0.01 & 0.88$\pm$0.01 & 0.56$\pm$0.01 & 0.35$\pm$0.01 & 0.17$\pm$0.01 & 1.00$\pm$0.01 & L8 & L4.6 \\
J0616$-$6407 & 0.86$\pm$0.03 & 0.82$\pm$0.02 & 0.95$\pm$0.08 & 0.89$\pm$0.03 & 0.92$\pm$0.03 & 0.84$\pm$0.06 & 0.59$\pm$0.03 & 0.31$\pm$0.02 & 0.15$\pm$0.01 & 0.97$\pm$0.02 & T0 & T1.1 \\
J0645$-$6646 & 0.85$\pm$0.01 & 1.01$\pm$0.01 & 0.87$\pm$0.02 & 0.68$\pm$0.01 & 0.63$\pm$0.01 & 0.66$\pm$0.01 & 0.52$\pm$0.02 & 0.48$\pm$0.01 & 0.34$\pm$0.01 & 1.03$\pm$0.01 & T0 & L9.0 \\
J0850$-$0221 & 0.89$\pm$0.01 & 1.06$\pm$0.01 & 0.95$\pm$0.01 & 0.77$\pm$0.01 & 0.70$\pm$0.01 & 0.71$\pm$0.01 & 0.48$\pm$0.01 & 0.55$\pm$0.01 & 0.40$\pm$0.01 & 0.98$\pm$0.01 & L6 & L6.5 \\
{J0833+0052} & 0.09$\pm$0.02 & 0.12$\pm$0.03 & \nodata\tablenotemark{b} & $-$0.04$\pm$0.04 & 0.07$\pm$0.04 & 0.29$\pm$0.09 & 0.51$\pm$0.05 & 0.17$\pm$0.02 & 0.10$\pm$0.01 & 0.30$\pm$0.05 &  T8 & \nodata \\
J0937+2931 & 0.43$\pm$0.01 & 0.31$\pm$0.01 & 0.18$\pm$0.01 & 0.15$\pm$0.01 & 0.31$\pm$0.01 & 0.53$\pm$0.02 & 0.53$\pm$0.01 & 0.17$\pm$0.01 & 0.07$\pm$0.01 & 0.53$\pm$0.01 & T6 & T5.8 \\
J0939$-$2448 & 0.30$\pm$0.01 & 0.14$\pm$0.01 & 0.10$\pm$0.04 & 0.03$\pm$0.01 & 0.17$\pm$0.01 & 0.52$\pm$0.05 & 0.51$\pm$0.02 & 0.13$\pm$0.01 & 0.06$\pm$0.01 & 0.33$\pm$0.02 & T8 & T7.7 \\
J0953$-$0943 & 0.50$\pm$0.02 & 0.37$\pm$0.03 & 0.21$\pm$0.10 & 0.21$\pm$0.02 & $-$0.18$\pm$0.05 & 0.58$\pm$0.13 & 0.45$\pm$0.02 & 0.25$\pm$0.03 & 0.10$\pm$0.01 & 0.50$\pm$0.06 & T6 & T5.1 \\
J1019$-$3911 & 0.66$\pm$0.01 & 0.55$\pm$0.01 & 0.29$\pm$0.01 & 0.44$\pm$0.01 & 0.40$\pm$0.01 & 0.50$\pm$0.01 & 0.48$\pm$0.01 & 0.36$\pm$0.01 & 0.20$\pm$0.01 & 0.77$\pm$0.01 & T4 & T4.0 \\
J1035$-$0711 & 0.89$\pm$0.02 & 1.07$\pm$0.02 & 0.94$\pm$0.02 & 0.72$\pm$0.01 & 0.63$\pm$0.02 & 0.66$\pm$0.02 & 0.58$\pm$0.01 & 0.56$\pm$0.01 & 0.38$\pm$0.01 & 1.04$\pm$0.02 & L9 & L7.7 \\
J1055+5443 & 0.05$\pm$0.01 & 0.06$\pm$0.01 & 0.05$\pm$0.03 & 0.02$\pm$0.02 & 0.11$\pm$0.02 & 0.27$\pm$0.03 & 0.79$\pm$0.05 & 0.31$\pm$0.01 & 0.23$\pm$0.01 & 0.10$\pm$0.03 & T9 & \nodata \\
J1130$-$1158 & 0.49$\pm$0.03 & 0.33$\pm$0.03 & 0.69$\pm$0.46 & 0.21$\pm$0.06 & 0.38$\pm$0.05 & 1.11$\pm$0.27 & 0.49$\pm$0.05 & 0.13$\pm$0.02 & 0.06$\pm$0.01 & 0.62$\pm$0.06 & T6 & T5.4 \\
J1158+0435 & 0.86$\pm$0.01 & 1.04$\pm$0.01 & 0.89$\pm$0.01 & 0.71$\pm$0.01 & 0.64$\pm$0.01 & 0.69$\pm$0.01 & 0.47$\pm$0.01 & 0.50$\pm$0.01 & 0.32$\pm$0.01 & 1.01$\pm$0.01 & L7 & L8.0 \\
J1307+1510 & 0.79$\pm$0.05 & 1.03$\pm$0.04 & 0.95$\pm$0.12 & 0.66$\pm$0.06 & 0.58$\pm$0.03 & 0.65$\pm$0.06 & 0.49$\pm$0.07 & 0.52$\pm$0.04 & 0.35$\pm$0.03 & 1.04$\pm$0.03 & T1 & T1.1 \\
J1316+0755 & 0.40$\pm$0.02 & 0.35$\pm$0.03 & \nodata\tablenotemark{b} & 0.13$\pm$0.02 & 0.26$\pm$0.04 & \nodata\tablenotemark{b} & 0.69$\pm$0.05 & 0.10$\pm$0.01 & 0.05$\pm$0.01 & 0.51$\pm$0.05 & T7 & T5.7 \\
J1338$-$0229 & 0.86$\pm$0.06 & 1.06$\pm$0.08 & 0.83$\pm$0.07 & 0.68$\pm$0.05 & 0.62$\pm$0.05 & 0.70$\pm$0.07 & 0.44$\pm$0.04 & 0.43$\pm$0.03 & 0.28$\pm$0.02 & 1.12$\pm$0.07 & T0 & L9.2: \\
J1416+1348B & 0.30$\pm$0.01 & 0.21$\pm$0.01 & 0.00$\pm$0.18 & 0.04$\pm$0.01 & 0.17$\pm$0.01 & 0.35$\pm$0.15 & 0.60$\pm$0.01 & 0.07$\pm$0.01 & 0.04$\pm$0.01 & 0.42$\pm$0.02 & T7 & T7.0 \\
J1553+6934 & 0.86$\pm$0.03 & 0.63$\pm$0.02 & 0.65$\pm$0.04 & 0.42$\pm$0.03 & 0.66$\pm$0.03 & 0.86$\pm$0.07 & 0.56$\pm$0.05 & 0.27$\pm$0.01 & 0.16$\pm$0.01 & 0.89$\pm$0.02 & T4 & T3.6 \\
J1810$-$1010 & 0.98$\pm$0.05 & 0.75$\pm$0.05 & \nodata\tablenotemark{b} & 0.53$\pm$0.02 & 1.06$\pm$0.07 & \nodata\tablenotemark{b} & 0.74$\pm$0.01 & 0.37$\pm$0.05 & 0.32$\pm$0.04 & 0.95$\pm$0.05 & T3 & T2.2 \\
J2105$-$6235 & 0.80$\pm$0.01 & 0.85$\pm$0.01 &  \nodata\tablenotemark{b} & 0.55$\pm$0.01 & 0.60$\pm$0.01 &  \nodata\tablenotemark{b} & 0.53$\pm$0.01 &  \nodata\tablenotemark{b} &  \nodata\tablenotemark{b} & 0.98$\pm$0.01 & T2 & T1.8 \\
\enddata 
\tablecomments{Spectral indices are defined as flux ratios, computed as the integral of the spectrum over the defined wavelength ranges in $\mu$m. For the [H-dip] index, the denominator is the median of the fluxes computed in the two specified ranges. Further details are provided in \citet{2006ApJ...639.1095B,2006ApJ...637.1067B} and \citet{2019ApJ...883..205B}.}
\tablenotetext{a}{Spectral type as inferred from direct comparison to spectral standards (``Std'') and indices (``Ind'').}
\tablenotetext{b}{{Missing spectral data or skewed by excess noise.}}
\end{deluxetable} 
\end{longrotatetable}

%Focusing on classification of our candidates and benchmark sources, the index types generally given similar classifications to the template fits

\subsection{K~I Equivalent Widths}

The moderate resolution of a subset of our spectra facilitates measurement of line equivalent widths (EWs) of the K~I doublets at 1.1692/1.1778~$\mu$m and 1.2436/1.2526~$\mu$m (Figure~\ref{fig:spectra-all}). We measured these lines by integrating the unsmoothed spectral flux density relative to a linear fit of the local continuum over $\pm$0.02~$\mu$m windows outside the line regions, and sampled both flux uncertainties and small variations in the integration window through Monte Carlo methods. 
Table~\ref{tab:ews} lists the measured EWs.
In cases where measurements were less than three times their uncertainties, we report 3$\sigma$ upper limits.

\startlongtable
\begin{deluxetable}{lllcccc}
\tablecaption{Equivalent Widths \label{tab:ews}} 
\tabletypesize{\scriptsize} 
\tablehead{ 
\colhead{Name} & 
\colhead{SpT\tablenotemark{a}} & 
\colhead{Instrument} & 
\colhead{1.1692~$\micron$}  & 
\colhead{1.1778~$\micron$}  &
\colhead{1.2436~$\micron$}  & 
\colhead{1.2526~$\micron$}  \\
 & & &
\colhead{({\AA})} & 
\colhead{({\AA})} &  
\colhead{({\AA})} & 
\colhead{({\AA})}   
} 
\startdata 
\hline
\multicolumn{7}{c}{Metallicity Benchmarks} \\
\hline
LHS~6176B & d/sdT7.5 & NIRES & 16.0$\pm$2.4 & $<$ 9 & $<$ 1.0 & $<$ 1.2 \\
{HD~126053B} & sdT7.5 & GNIRS & $<$ 63 & $<$ 93 & $<$ 2.5 & $<$ 2.0  \\
GJ~576B & d/sdT5.5 & XS & $<$ 7 & 13.3$\pm$2.2 & $<$ 1.3 & $<$ 1.5 \\
\nodata & \nodata & NIRES & 14.7$\pm$2.2 & 10.2$\pm$1.5 & 1.6$\pm$0.2 & 1.8$\pm$0.2 \\
Wolf~1130C & (e)sdT6: & Keck & $<$ 19 & $<$ 14 & $<$ 1.7 & $<$ 1.6 \\
\hline
\multicolumn{7}{c}{Subdwarf Candidates} \\
\hline
J0045+7958 & d/sdL9 & TSpec & $<$ 25 & 6.3$\pm$1.1 & 6.9$\pm$1.2 & 7.8$\pm$1.0 \\
J0055+5947 & d/sdT6.5 & NIRES & $<$ 35 & $<$ 26 & $<$ 2.1 & $<$ 1.7 \\
J0140+0150 & rT4.5 & NIRES & 12.8$\pm$1.4 & 13.0$\pm$1.5 & 7.1$\pm$0.5 & 8.9$\pm$0.5 \\
J0411+4714 & d/sdT7.5 & NIRES & 14.8$\pm$2.5 & $<$ 8 & $<$ 1.0 & $<$ 1.1  \\
J0429+3201 & T1 & NIRES & 8.3$\pm$0.9 & 8.5$\pm$0.8 & $<$ 1.7 & 3.6$\pm$0.5 \\
J0433+1009 & T8 & NIRES & $<$ 41 & $<$ 65 & $<$ 1.6 & 3.3$\pm$0.4 \\
J0623+0715 & sdT3 & NIRES & 6.7$\pm$1.2 & 5.5$\pm$0.8 & $<$ 1.1 & $<$ 1.2 \\
J0659+1615 & T1 & NIRES & 5.4$\pm$0.6 & 7.3$\pm$0.5 & 2.2$\pm$0.3 & 4.6$\pm$0.3  \\
J0758+5711 & T6.5 & NIRES & 12.4$\pm$1.9 & 7.7$\pm$1.5 & 2.0$\pm$0.3 & 2.1$\pm$0.3 \\
J1110$-$1747 & rT8 & NIRES & $<$ 12 & $<$ 22 & $<$ 1.0 & $<$ 1.7 \\
J1130+3139 & d/sdT5.5 & NIRES & 13.9$\pm$2.0 & 8.9$\pm$1.5 & 2.0$\pm$0.3 & $<$ 0.8 \\
{J1138+7212} & d/sdT7 & GNIRS & $<$ 67 & $<$ 50 & $<$ 1.3 & $<$ 1.2  \\
J1204$-$2359 & d/sdT7 & NIRES & 16$\pm$3 & 15$\pm$3 & 3.2$\pm$0.4 & $<$ 1.6 \\
J1304+2819 & d/sdL9: & NIRES & 4.8$\pm$0.6 & 9.5$\pm$0.6 & 6.5$\pm$0.6 & 5.3$\pm$0.7 \\
J1308$-$0321 & d/sdT8 & NIRES & $<$ 51 & \nodata\tablenotemark{b} & $<$ 1.1 & $<$ 1.8 \\
J1401+4325 & d/sdT5.5 & NIRES & $<$ 18 & $<$ 9 & $<$ 1.6 & $<$ 1.8 \\
J1458+1734 & T8 & NIRES & $<$ 47 & \nodata\tablenotemark{b} & $<$ 1.4 & $<$ 1.5 \\
J1524$-$2620 & sdT0 & NIRES & 8.9$\pm$1.1 & 9.4$\pm$0.9 & 5.0$\pm$0.7 & 3.7$\pm$0.7 \\
J1710+4537 & T6 & NIRES & 14$\pm$3 & 11.1$\pm$1.9 & $<$ 1.0 & $<$ 1.0 \\
J1801+4717 & d/sdT5 & NIRES & 17$\pm$3 & 9.8$\pm$1.7 & $<$ 1.4 & $<$ 1.7 \\
J2013$-$0326 & d/sdT6 & NIRES & $<$ 10 & 12.6$\pm$2.3 & $<$ 1.2 & $<$ 1.1 \\
J2021+1524 & d/sdL9 & NIRES & 9.7$\pm$0.7 & 12.4$\pm$0.7 & 6.3$\pm$0.6 & 8.0$\pm$0.5 \\
J2112$-$0529 & sdT1 & NIRES & 7.9$\pm$0.8 & 10.4$\pm$0.6 & 5.3$\pm$0.5 & 5.9$\pm$0.4 \\
J2112+3030 & d/sdT2.5 & NIRES & 5.8$\pm$0.9 & 7.0$\pm$0.8 & $<$ 1.1 & 3.0$\pm$0.4 \\
J2218+1146 & d/sdT6.5 & NIRES & $<$ 12 & $<$ 8 & 3.2$\pm$0.4 & $<$ 1.4 \\
J2251$-$0740 & d/sdT7 & NIRES & $<$ 47 & \nodata\tablenotemark{b} & $<$ 1.7 & $<$ 1.8 \\
\hline
\multicolumn{7}{c}{Metal-poor Comparison Sources} \\
\hline
J0021+1552 & d/sdT4 & XS & $<$ 76 & $<$ 6 & $<$ 2.3 & $<$ 1.9 \\
J0532+8246 & esdL8: & NIRES & 10.1$\pm$0.7 & 13.4$\pm$0.7 & 7.0$\pm$0.7 & 6.3$\pm$0.4 \\
J0616$-$6407 & esdT0: & XS & 7.3$\pm$0.7 & 10.9$\pm$0.8 & 7.8$\pm$0.8 & 7.2$\pm$0.4  \\
J0645$-$6646 & d/sdT0 & XS & 10.6$\pm$0.6 & 12.2$\pm$0.5 & 6.8$\pm$0.5 & 8.3$\pm$0.3  \\
{J0833+0052} & d/sdT9 & GNIRS & $<$ 54 & $<$ 65 & $<$ 8 & $<$ 5  \\
{J1019-3911} & T4 & ARCoIRIS2 & 9.0$\pm$1.1 & 6.3$\pm$0.9 & 3.0$\pm$0.4 & $<$ 0.9  \\
J1055+5443 & Y0 & NIRES & \nodata\tablenotemark{b} & \nodata\tablenotemark{b} & $<$ 4 & $<$ 3  \\
J1130$-$1158 & sdT5.5 & ARC & $<$ 57 & $<$ 12 & $<$ 3 & $<$ 3 \\
J1307+1510 & T1 & XS & $<$ 9 & $<$ 10 & 8.1$\pm$1.0 & 8.7$\pm$1.0  \\
{J1316+0755} & sdT6.5 & GNIRS & $<$ 62 & $<$ 11 & $<$ 2.5 & $<$ 2.5  \\
J1553+6933 & sdT4 & NIRES & $<$ 5 & $<$ 4 & $<$ 4 & $<$ 3 \\
J1810$-$1010 & esdT3: & TSpec & $<$ 5 & $<$ 4 & $<$ 3 & $<$ 4 \\
\enddata 
\tablenotetext{a}{{Final spectral type as discussed in Section~\ref{sec:classify}.}}
\tablenotetext{b}{Affected by excess noise in local continuum.}
\end{deluxetable} 
%\endlongtable

\begin{figure}[ht!]
\plottwo{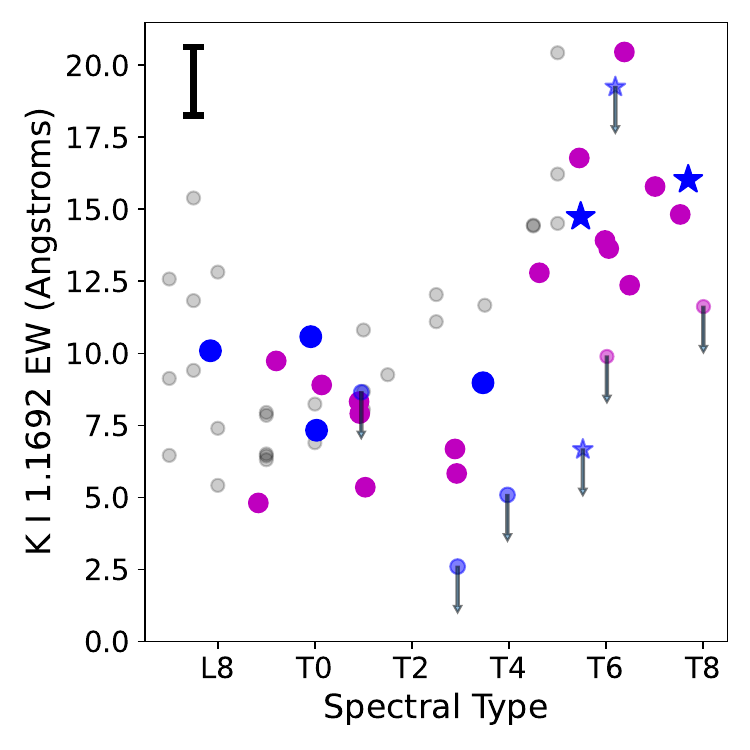}{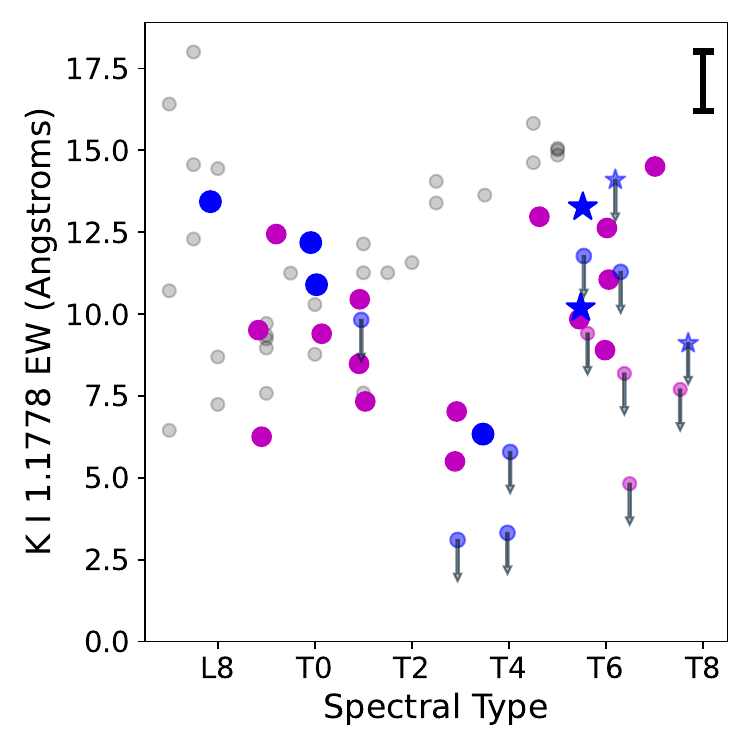}
\plottwo{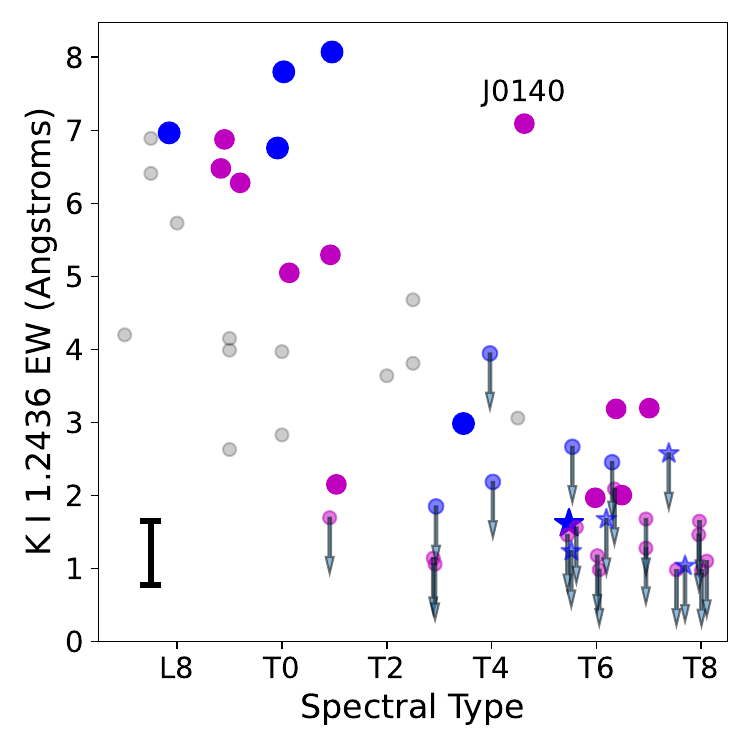}{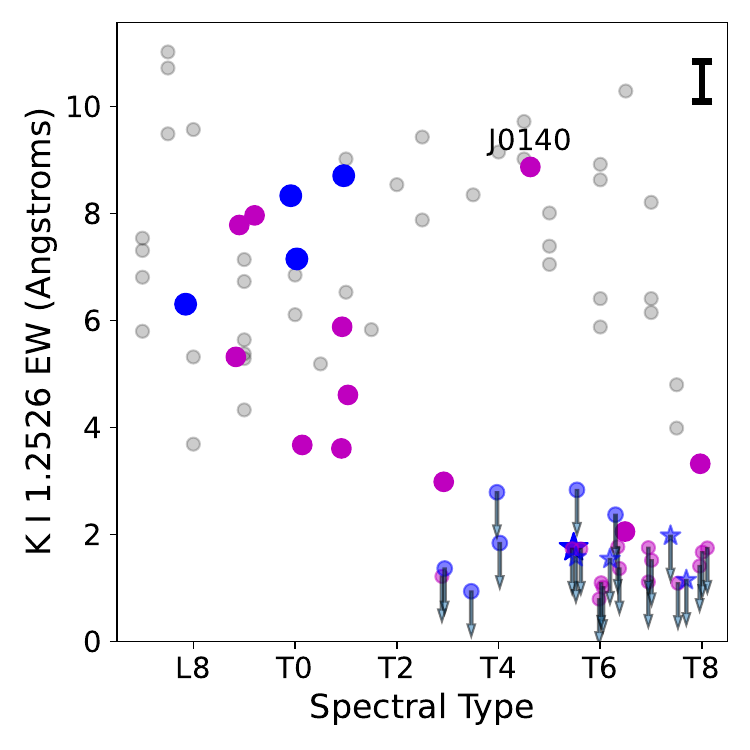}
\caption{Equivalent width measurements of the 1.1692~$\mu$m (top left), 1.1178~$\mu$m (top right), 1.2436~$\mu$m (bottom left) and 1.2526~$\mu$m (bottom right) K~I lines based on medium-resolution data in our spectral sample.
Subdwarf candidates are indicated by magenta circles,
metal-poor comparison sources by blue circles, and
benchmark companions by blue stars. 
Smaller symbols with downward arrows indicate 3$\sigma$ upper limits.
A comparison sample of measurements for local L and T dwarfs with significant detections reported in \citet{2017ApJ...838...73M} are indicated as grey circles.
The typical measurement uncertainties for sources with significant detections are indicated by the black error bar in the corners of each panel.
Each source is given a small random offset in spectral type to help differentiate overlapping data points.
{The unusually red high velocity T dwarf J0140+0150 is specifically labeled in the bottom two panels.}
\label{fig:ews}}
\end{figure}

In Figure~\ref{fig:ews}, we compare these measurements to a compilation of local L and T dwarf spectra from \citet{2017ApJ...838...73M}, which is based on {\ldl} $\approx$ 2000 data from the Keck Near-InfraRed SPECtrometer (NIRSPEC; \citealt{1998SPIE.3354..566M}). 
The dwarf data follow previously-identified trends for K~I absorption, 
with 1.1692~$\mu$m and 1.1778~$\mu$m EWs rising from late-L to late-T (largely driven by the drop in continuum as H$_2$O absorption increases), 1.2436~$\mu$m EWs falling over the same range, and 1.2526~$\mu$m EWs rising from late-L to mid-T then dropping to zero for the latest T dwarfs. 
In contrast, many of the previously-observed T subdwarfs and metal-poor T benchmarks, and several of our candidates, are either well below these trends or lack detectable K~I absorption to limits of 1--2~{\AA}. This is particularly clear for the 1.2526~$\mu$m line, which sits near the $J$-band flux peak and shows the greatest dichotomy between dwarfs and subdwarfs for types T1--T7.
One notable exception in our candidate sample is J0140+0150, which has exceptionally strong absorption at 1.2436~$\mu$m for its T4.5 classification.

\subsection{Spectral Model Fits\label{sec:modelfit}}

\begin{deluxetable}{lcccccll}
\tablecaption{Atmosphere Model Parameters \label{tab:models}} 
\tabletypesize{\scriptsize} 
\tablehead{ 
\colhead{Model} & 
\colhead{{\teff}} & 
\colhead{{\logg}} & 
\colhead{[M/H]} & 
\colhead{C/O} & 
\colhead{{\kzz}} & 
\colhead{Other Parameters} & 
\colhead{Ref} \\  
 & (K) & (cm/s$^2$) & (dex) & & (cm$^2$/s) & \\
} 
\startdata 
LOWZ & 500--1600 & 4.5--5.25 & $-$2.5 to +1.0 & 0.10--0.85 & $-$1 to +10 &    & [1] \\
B06 & {700--2200} & {4.5--5.5} & {$-$0.5 to +0.5} & \nodata & \nodata  & Cloud model f100  & [2] \\
Settl &  500--2200 & 4.5--5.5 & {$-$0.5 to +0.5} & \nodata  & \nodata &  [$\alpha$/Fe] = {0 to +0.2}  & [3] \\
Dusty & 1000--2200 & {4.5--5.5} & {$-$2.5 to +0.0}  & \nodata & \nodata & [$\alpha$/Fe] = 0 to +0.6   & [3] \\
ATMO\tablenotemark{a} & 600--2000 & 4.5--5.5 & 0.0 &  \nodata &  4--8, CE\tablenotemark{b} &  $\log_{10}P_{min}$ = $-$8--0; $\gamma$ = 1.0--1.1  & [4] \\
Sonora & 500--2200 & {4.5--5.5} & $-$0.5 to +0.5 & 0.22--0.83 & \nodata &   & [5] \\
ElfOwl & 525--2200 & {4.5--5.5} & $-$1.0 to +1.0 & 0.22--0.83 & 2--8 &  & [6] \\
{SAND} & {700--2200} & {4.5--6.0} & {$-$2.4 to +0.3} &\nodata & \nodata &  {[$\alpha$/Fe] = 0 to +0.4} & [7] \\
\enddata 
\tablerefs{
[1] \citet{2021ApJ...915..120M};
[2] \citet{2006ApJ...640.1063B};
[3] \citet{2012RSPTA.370.2765A};
[4] \citet{2020AAP..637A..38P};
[5] \citet{2021ApJ...920...85M};
[6] \citet{2024ApJ...963...73M};
{[7] \citet{Alvarado_2024}}.
}
%\tablenotetext{a}{B06 models are constrained to cloud model f100; see \citet{2006ApJ...640.1063B}.}
\tablenotetext{a}{ATMO models are further constrained to the K~I broadening profile from \citet{2016AandA...589A..21A} and the condensate rainout prescription described by \citet{2019MNRAS.482.4503G}; see \citet{2020AAP..637A..38P}.}
\tablenotetext{b}{{Chemical Equilibrium (no mixing).}}
\end{deluxetable} 

To obtain initial estimates of the physical properties of our sources, we compared low-resolution versions of our spectra 
to {eight} sets of spectral models from
\citet[][hereafter B06]{2006ApJ...640.1063B},
\citet[][hereafter Settl and Dusty]{2012RSPTA.370.2765A},
%\citet[][hereafter Dusty]{BTDUSTY2016},
\citet[][hereafter ATMO]{2020AAP..637A..38P},
\citet[][hereafter Sonora]{2021ApJ...920...85M},
\citet[][hereafter LOWZ]{2021ApJ...915..120M}, 
\citet[][hereafter ElfOwl]{2024ApJ...963...73M}, and
{\citet[][hereafter SAND]{Alvarado_2024}.}
The parameters and parameter ranges explored for these models are summarized in Table~\ref{tab:models}.
These models were chosen as they span both low temperatures and solar and subsolar metallicities, with the LOWZ models currently providing the best metallicity coverage for T subdwarf atmospheres. 
%(cf.\ \citealt{2021ApJ...915..120M,lodieu2022,2024ApJ...962..177B}).
%For the ATMO models, we fit to models using the \citet{2005AandA...440.1195A} line-broadening profile (broad = A) and condensate cloud formation (cloud = LC). 
The observed ($O[\lambda]$) and model ($M[\lambda]$) spectra were resampled onto a common wavelength scale with a constant resolution of {\ldl} = 150 over the range 0.9--2.45~$\mu$m using a median filter,\footnote{Smoothed uncertainties were computed as median(\{$\sigma$\})/$\sqrt{N-1}$ for the $N$ uncertainty values \{$\sigma$\} within the sampling region.}
and noisy data at the edge of the observed spectral range and in the 1.35--1.45~$\micron$ and 1.8--1.95~$\micron$ telluric bands were masked from the fit. 
To identify the best-fitting model, we used a $\chi^2$ goodness-of-fit statistic,
\begin{equation}
    \chi^2 = \sum_{i=1}^N\frac{(O[\lambda_i]-{\beta}M[\lambda_i])^2}{\sigma[\lambda_i]^2}
    \label{eqn:chi1}
\end{equation}
where $N$ is the number of unmasked flux values,
$\sigma[\lambda_i]$ is the observed uncertainty spectrum, and the optimal scale factor $\beta$ is computed as
\begin{equation}\label{eqn:chiscale}
    \beta = \frac{\sum_{i=1}^N\frac{O[\lambda_i]M[\lambda_i]}{\sigma[\lambda_i]^2}}{\sum_{i=1}^N\frac{M[\lambda_i]^2}{\sigma[\lambda_i]^2}}
\end{equation}
(cf.\ \citealt{2008ApJ...678.1372C}).
We report here the reduced chi-square, $\chi^2_r \equiv \chi^2/(N-1)$, as a normalized quality of fit.
%We also computed a normalized standard deviation to assess the mean divergence between model and data:
%\begin{equation}
%    \hat\sigma^2 = \frac{1}{N-1}\frac{\sum_{i=1}^N(O[\lambda_i]-{\alpha}M[\lambda_i])^2}{{\rm MAX}(O[\lambda_i])^2}
%\end{equation}
%where MAX is the maximum flux density in the resampled spectrum. 
For sources with measured parallaxes, we scaled the observed spectra to absolute fluxes using their $M_J$ magnitudes, which
%. As the models are computed in surface fluxes, this scaling 
allows us to estimate the source radius from the scale factor
\begin{equation}
    R_\beta = 10\sqrt\beta~{\rm pc} = 2.256\times10^{-9}\sqrt\beta~{\rm R_\odot}.
\end{equation}
All other spectra are scaled to apparent $J$ magnitudes.
No attempt was made to adjust for relative velocity shifts between the observed and model spectra, which are expected to be minimal at this low resolution.

For each model set, an initial fit was made to the individual grid models, and the best-fit (lowest $\chi^2_r$) model was used as an initial estimate of the atmosphere parameters. We employed a simple Metropolis-Hastings Markov Chain Monte Carlo (MCMC) algorithm \citep{1953JChPh..21.1087M,HASTINGS01041970} to explore the proximate parameter space. 
We used a single chain of 2,000 steps for each spectrum and model set comparison, varying the continuous variables of effective temperature ({\teff}), surface gravity ({\logg}), and solar-scaled metallicity ([M/H]) for all models, as well as 
carbon/oxygen abundance ratio (C/O) for LOWZ, Sonora, and ElfOwl models; 
alpha element enrichment ([$\alpha$/Fe]) for Settl, Dusty, {and SAND} models;
and adiabatic coefficient $\gamma = \left(1-\partial\log{T}/\partial\log{P}\right)^{-1}$ \citep{2019ApJ...876..144T} and minimum cloud deck pressure $P_{min}$ for ATMO models.
We also varied the disequilibrium mixing diffusion rate {\kzz} for the
LOWZ, ATMO, and ElfOwl models as a discrete value.
Models were linearly interpolated in logarithmic flux units on a logarithmic parameter grid (i.e,. {\teff} $\Rightarrow$ $\log${\teff}).
In each iteration, we drew new parameters for the continuous variables using a normal distribution centered on the last parameter values in the chain with pre-defined standard deviations.\footnote{These standard deviations were $\sigma_{Teff}$ = 30~K, $\sigma_{\log{g}}$ = 0.1 dex, $\sigma_{[M/H]}$ = 0.05 dex, $\sigma_{C/O}$ = 0.05 dex, $\sigma_{[\alpha/Fe]}$ = 0.05 dex, and $\sigma_\gamma$ = 0.01.} 
The discrete variable {\kzz} was randomly drawn from the available values.
Sequential fits ($i\Rightarrow{i+1}$) were compared using the criterion
\begin{equation}\label{eqn:mcmc}
    \frac{\chi^2(i+1)-\chi^2(i)}{\textrm{MIN}[\chi^2]} < \mathcal{U}(0,0.5)
\end{equation}
where MIN is the minimum of all $\chi^2$ values in the chain and $\mathcal{U}(0,0.5)$ is a number drawn from a uniform distribution between 0 and 0.5. 
If the new fit satisfied this criterion, these parameters were added to the chain; otherwise the previous parameters were added.
We also enforced a limit on $\chi^2$ values exceeding 2$\times$MIN[$\chi^2$], at which point the chain reverted to the minimum $\chi^2$ parameter set. We verified that all fits converged 
using the convergence diagnostic defined by \citet{geweke1992}, requiring that the means of the variable parameters in the first 10\% and last 50\% of each chain differ by less than three times their combined variance.
We also visually confirmed convergence, and examined each fit to ensure it provided an accurate reproduction of the data.

\startlongtable
\begin{deluxetable}{lllcccccccl}
\tablecaption{Spectral Model Fit Parameters \label{tab:modelfit}} 
\tabletypesize{\tiny} 
\tablehead{ 
%  & & & \multicolumn{5}{c}{Best Fit Model} \\
% \cline{4-10}
\colhead{Name} & 
\colhead{SpT} & 
\colhead{Instrument} & 
%\colhead{Model\tablenotemark{a}}  & 
\colhead{{\teff}}  & 
\colhead{{\logg}} & 
\colhead{[M/H]} & 
\colhead{C/O} & 
\colhead{$\log{k_{zz}}$} & 
\colhead{R} &
\colhead{MIN} &  
\colhead{Model} \\ 
 & & &
\colhead{(K)} & 
\colhead{(cm/s$^2$)} & 
\colhead{(dex)} & &
\colhead{(cm$^2$/s)} & 
\colhead{(R$_\odot$)} &  
\colhead{$\chi^2_r$} &    
} 
\startdata 
\hline
\multicolumn{11}{c}{Metallicity Benchmarks} \\
\hline
HD~3651B &  T8 &  SpeX & 856$^{+23}_{-20}$ & 4.56$^{+0.17}_{-0.06}$ & 0.01$^{+0.12}_{-0.10}$ & 0.48$^{+0.11}_{-0.05}$ & $-$1 & 0.070$^{+0.004}_{-0.004}$ & 21 & LOWZ \\ 
\nodata & \nodata &  \nodata & 834$^{+20}_{-31}$ & 4.53$^{+0.15}_{-0.03}$ & 0.00$^{+0.11}_{-0.05}$ & 0.67$^{+0.19}_{-0.04}$ & 4 & 0.077$^{+0.006}_{-0.004}$ & 15 & ElfOwl \\ 
HD~65486B &  d/sdT4.5 &  SpeX & 993$^{+44}_{-47}$ & 4.98$^{+0.20}_{-0.24}$ & $-$0.53$^{+0.13}_{-0.17}$ & 0.38$^{+0.15}_{-0.10}$ & $-$1 & 0.099$^{+0.013}_{-0.010}$ & 5 & LOWZ \\ 
LHS~6176B &  d/sdT7.5 &  NIRES & 846$^{+19}_{-6}$ & 4.52$^{+0.10}_{-0.01}$ & $-$0.21$^{+0.05}_{-0.03}$ & 0.49$^{+0.04}_{-0.02}$ & $-$1 & 0.061$^{+0.001}_{-0.004}$ & 15 & LOWZ \\ 
{Ross~458C} &  rT8.5 &  FIRE & 790$^{+63}_{-10}$ & 4.61$^{+0.13}_{-0.04}$ & 0.95$^{+0.05}_{-0.08}$ & 0.38$^{+0.06}_{-0.00}$ & $-$1 & 0.080$^{+0.000}_{-0.016}$ & 319 & LOWZ \\ 
\nodata & \nodata &  \nodata & 812$^{+38}_{-28}$ & 4.61$^{+0.18}_{-0.09}$ & 0.78$^{+0.09}_{-0.21}$ & 1.11$^{+0.08}_{-0.15}$ & 2 & 0.075$^{+0.005}_{-0.007}$ & 298 & ElfOwl \\ 
{HD~126053B} &  sdT7.5 &  GNIRS & 654$^{+52}_{-39}$ & 4.98$^{+0.27}_{-0.36}$ & $-$0.35$^{+0.11}_{-0.15}$ & 0.56$^{+0.16}_{-0.08}$ & 2 & 0.076$^{+0.016}_{-0.015}$ & 40 & LOWZ \\
\enddata 
\tablecomments{Model parameters are listed for best-fit LOWZ models by default \citep{2021ApJ...915..120M}, 
Sources that reached the LOWZ {\teff} parameter limit or had a significantly better fits ($|\Delta\chi^2_r|$ $>$ 5) to an alternate model are also listed with the alternative model parameters. Model names are 
given in shorthand notation as 
B06 \citep[assuming cloud model f100]{2006ApJ...640.1063B},
Settl and Dusty \citep{2012RSPTA.370.2765A},
%Dusty \citep{BTDUSTY2016},
ATMO \citep{2020AAP..637A..38P}, 
Sonora \citep{2021ApJ...920...85M},
ElfOwl \citep{2024ApJ...963...73M}, or
{or SAND \citep{Alvarado_2024}.}
Instrument names are given in shorthand notation as ARC = SOAR/ARCoIRIS, TSpec = APO/TripleSpec or Palomar/TripleSpec, and XS = VLT/X-Shooter.
Multiple instruments contribute to spectra used for Wolf~1130C (Keck/NIRSPEC and Keck/MOSFIRE), 
J0532+8246 (Keck/LRIS and Keck/NIRES),
and J1810$-$1010 (GTC/OSIRIS, GTC/EMIR, and Palomar/TripleSpec)
}
\tablenotetext{a}{Reaches {\teff} parameter limit for given model set.}
%\tablenotetext{b}{Adiabatic coefficient $\gamma$ for ATMO models.}
\tablenotetext{b}{Alpha element enrichment [$\alpha$/Fe] relative to the Sun for Settl and Dusty models.}
%\tablenotetext{d}{Chemical equilibrium (no vertical mixing).}
\end{deluxetable}

\begin{figure}[ht!]
\centering
\includegraphics[width=0.45\textwidth]{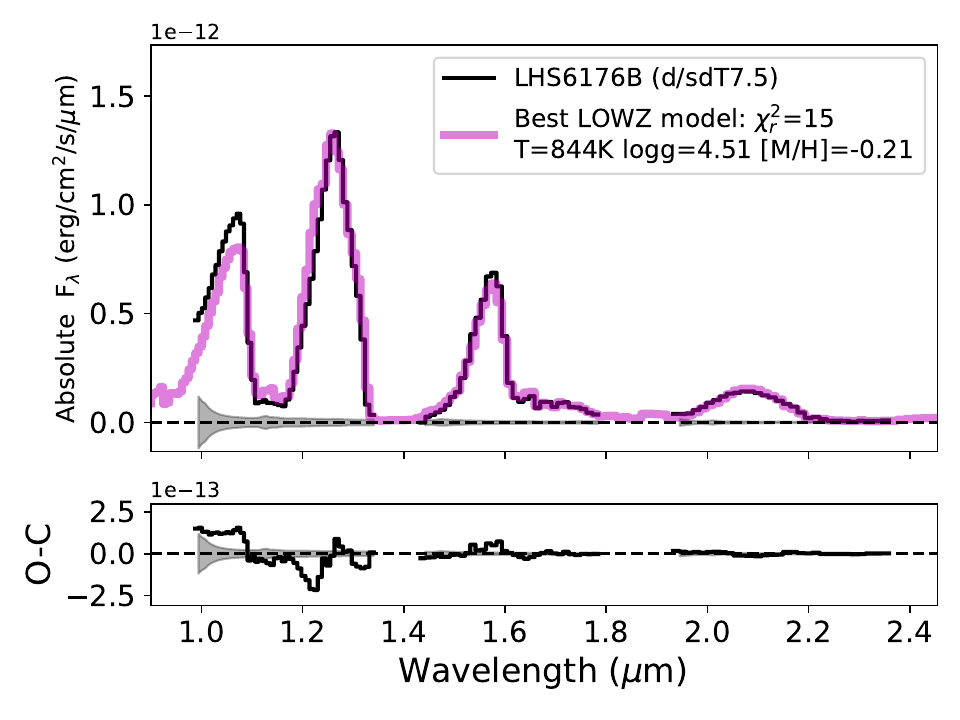}
\includegraphics[width=0.45\textwidth]{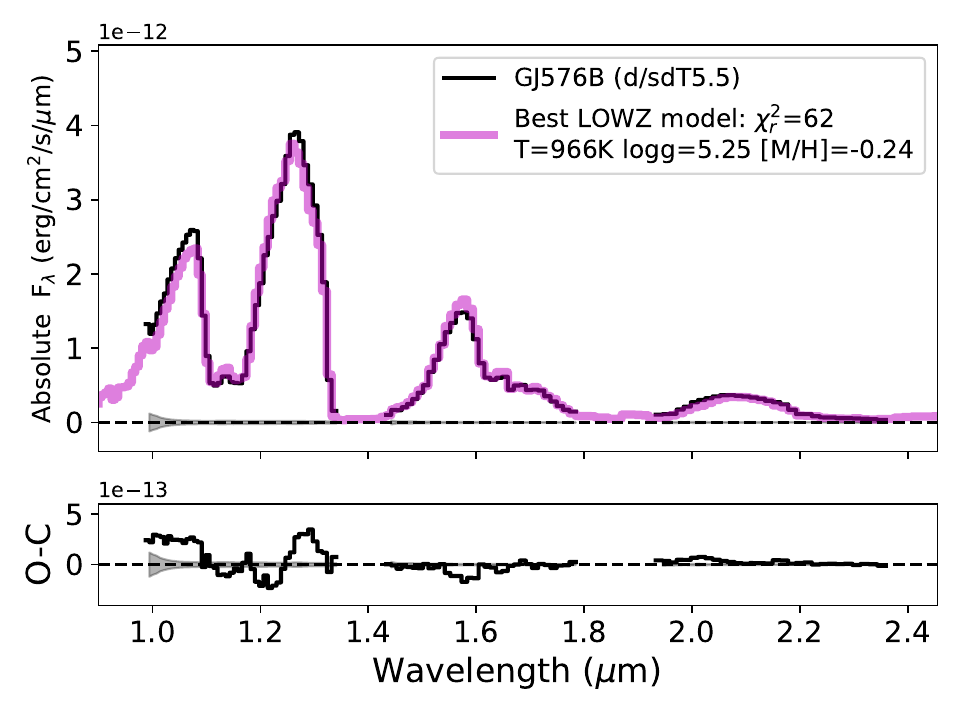} \\
\includegraphics[width=0.45\textwidth]{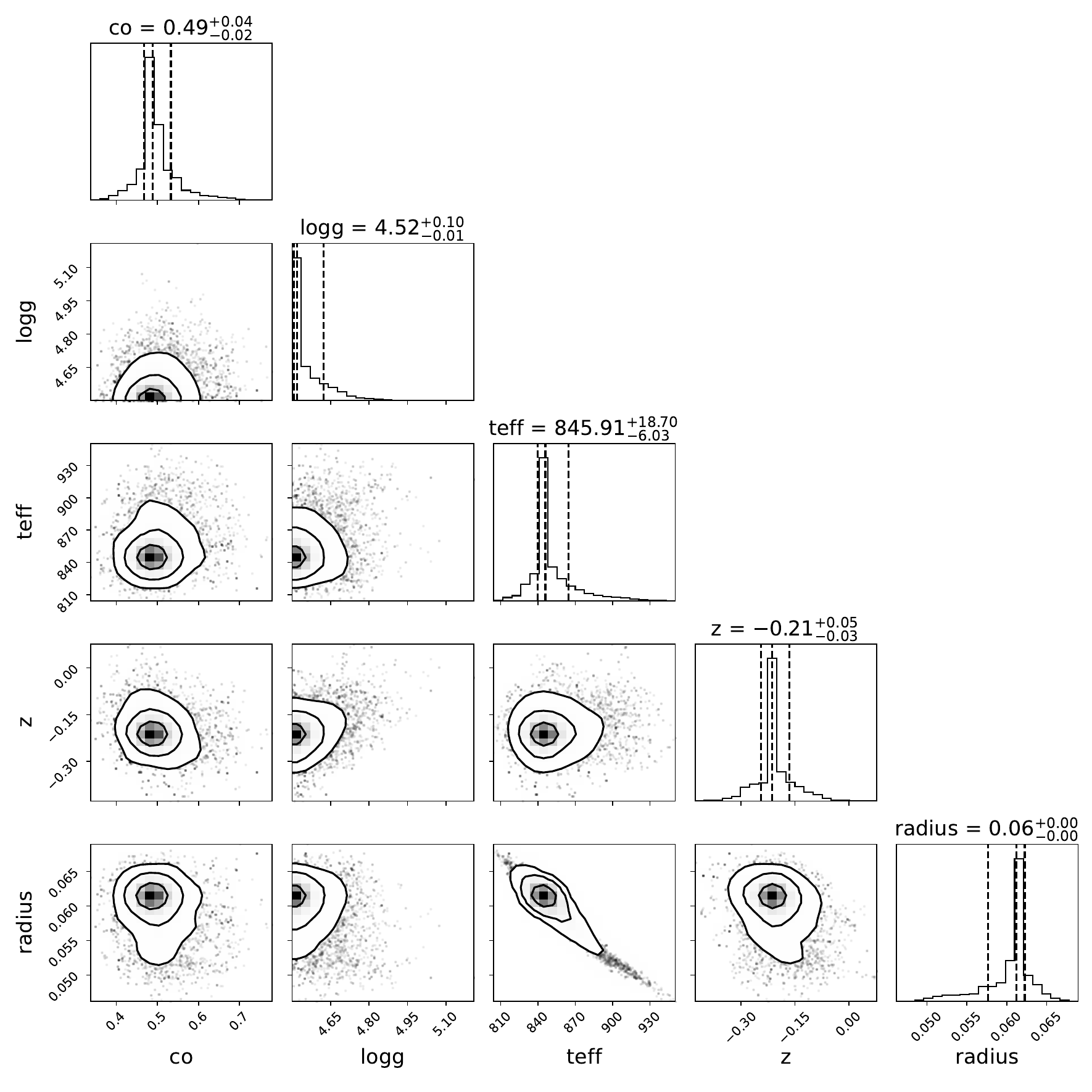}
\includegraphics[width=0.45\textwidth]{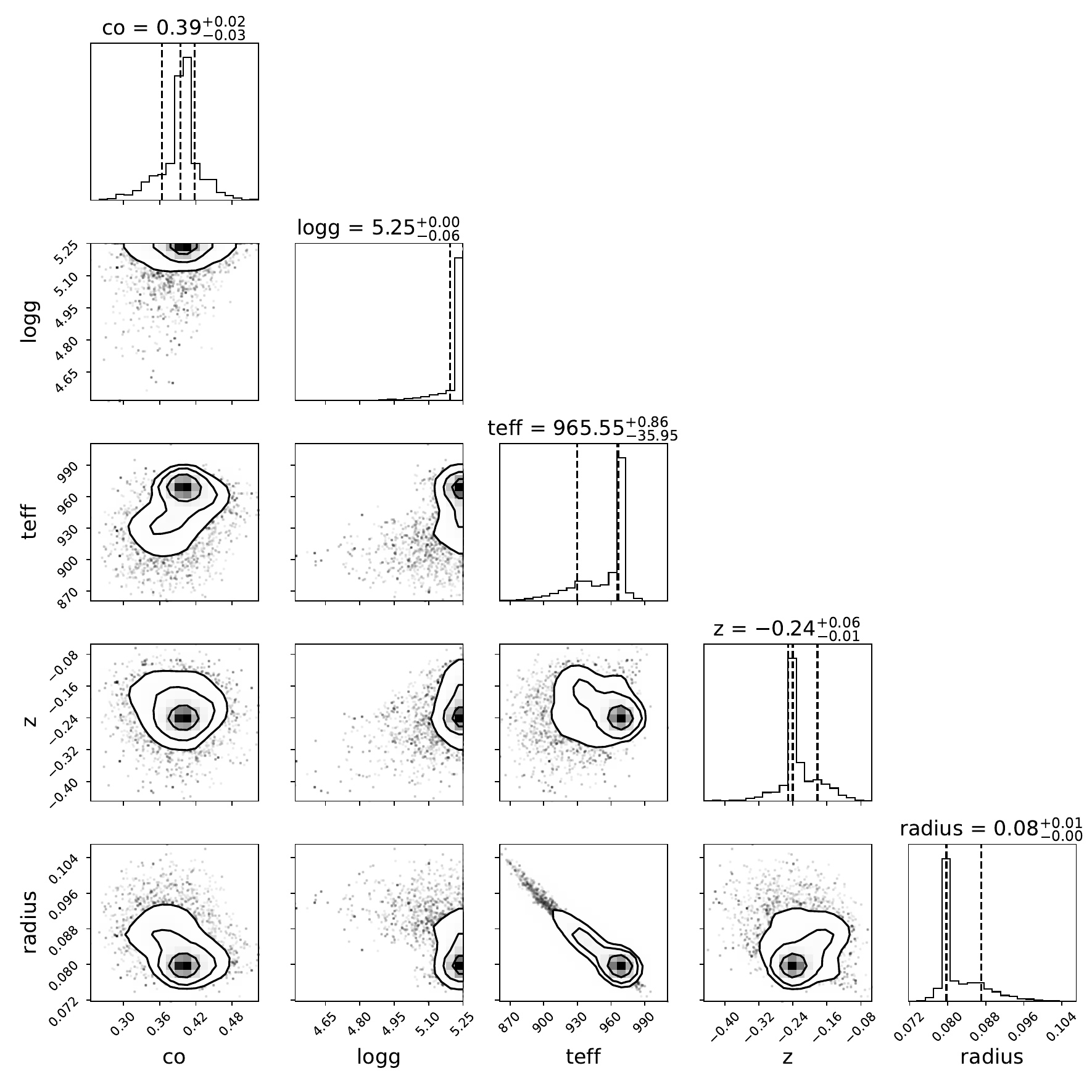}
\caption{
Best-fit LOWZ models and model parameter distributions for {Keck/NIRES data for} the metallicity benchmarks LHS~6176B (left) and GJ~576B (right).
The top panels compare the observed spectra in absolute flux density (black lines) to the best-fit model (magenta line).
The bottom panels show parameter distributions for {\teff}, {\logg}, [M/H], C/O, and radius.
Plots along the diagonal axis show the marginalized posterior distributions for each parameter, with 16\%, 50\%, and 84\% quantiles indicated as vertical dashed lines (see Table~\ref{tab:modelfit}). The remaining contour plots display two-dimensional distributions of parameter pairs in the posterior solutions, highlighting parameter correlations.
Note that parameter distributions are weighted by their $\chi^2$ value following Equation~\ref{eqn:weight}.
\label{fig:mcmcfit1}}
\end{figure}

\begin{figure}[ht!]
\centering
\includegraphics[width=0.45\textwidth]{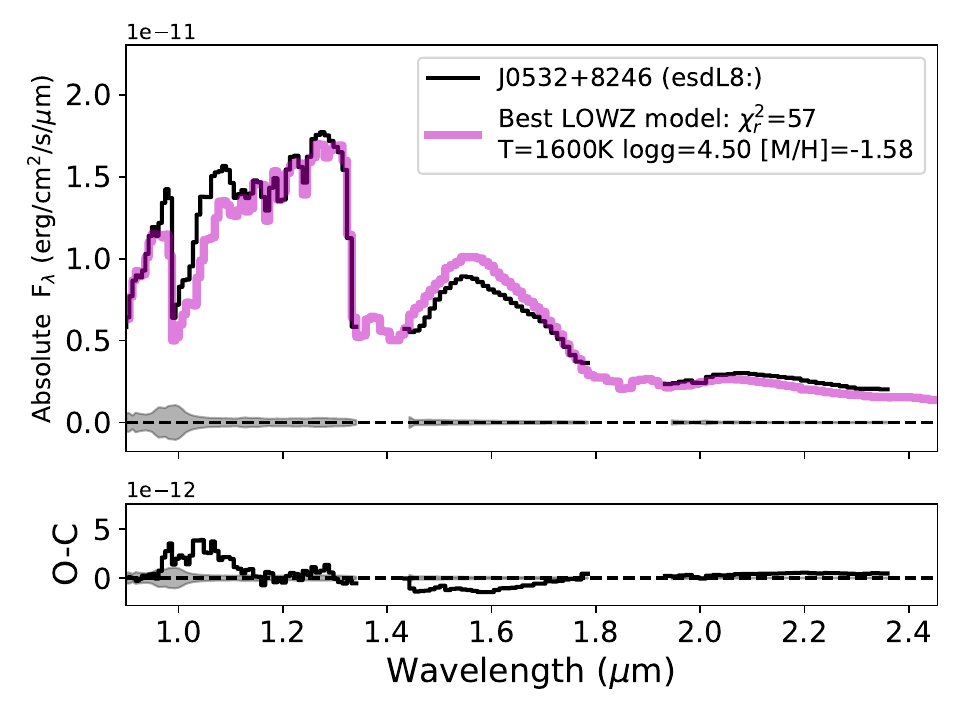}
\includegraphics[width=0.45\textwidth]{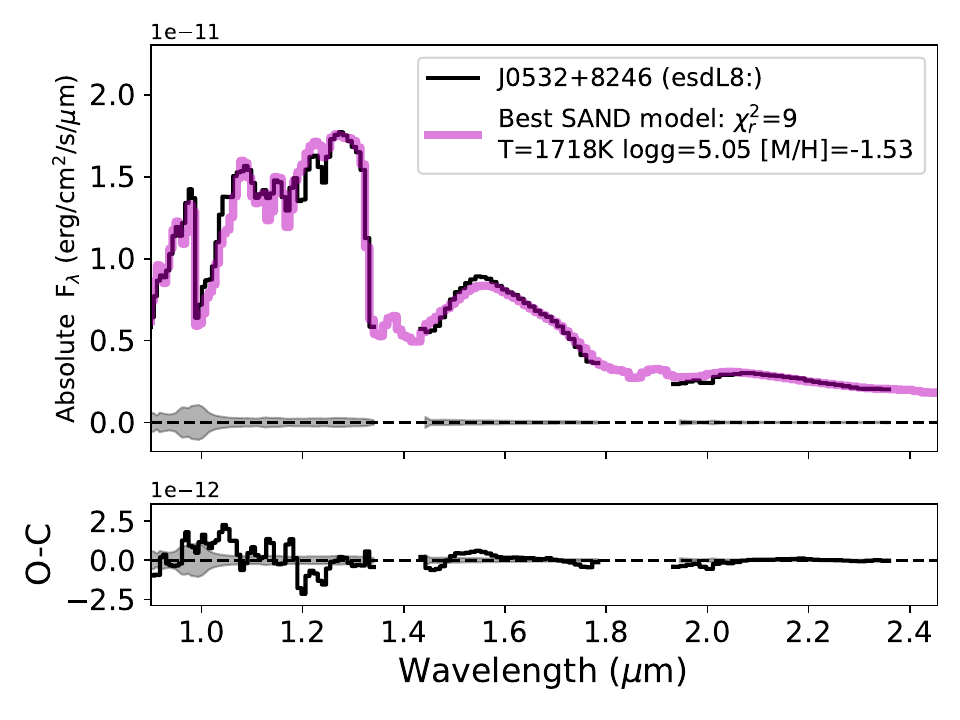} \\
\includegraphics[width=0.45\textwidth]{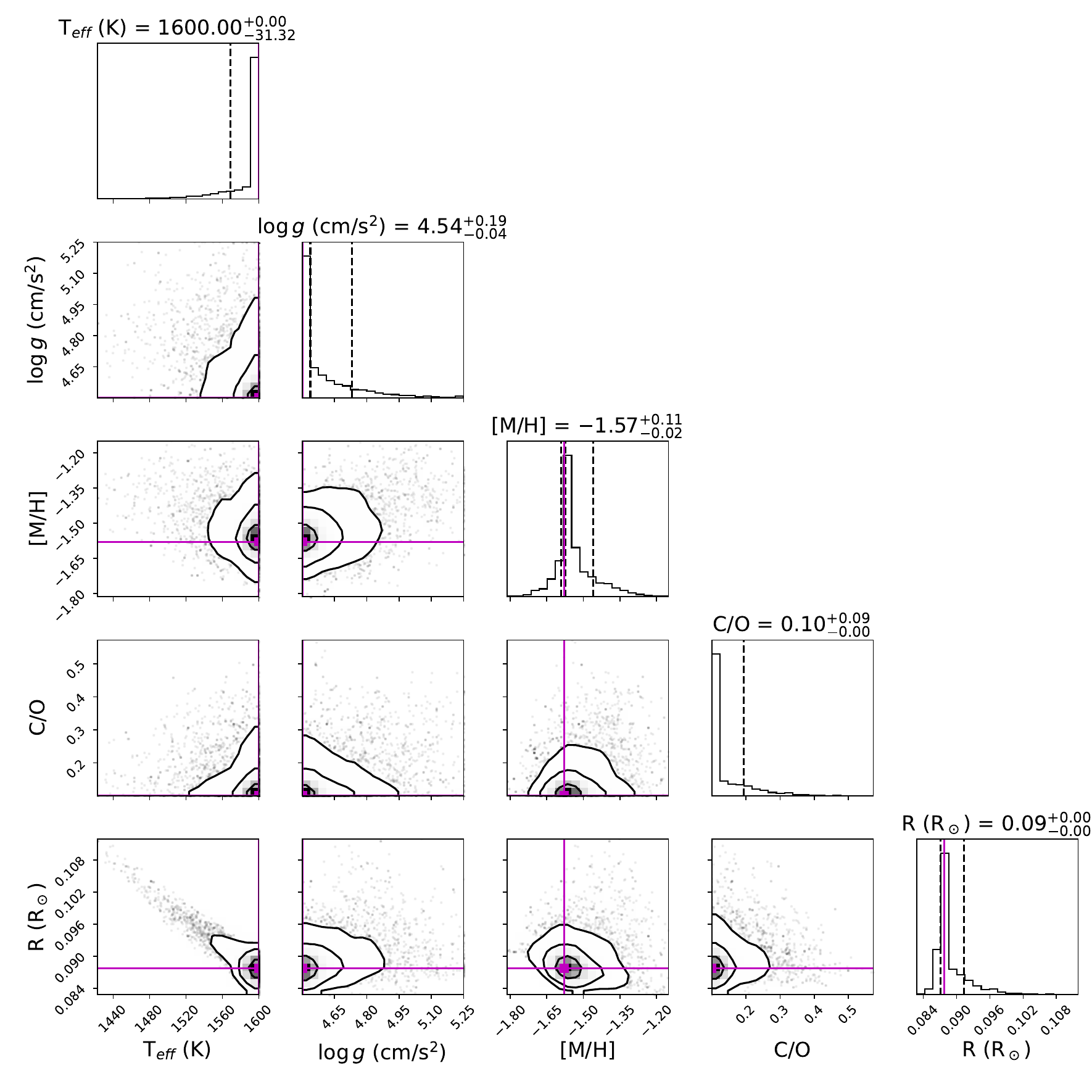}
\includegraphics[width=0.45\textwidth]{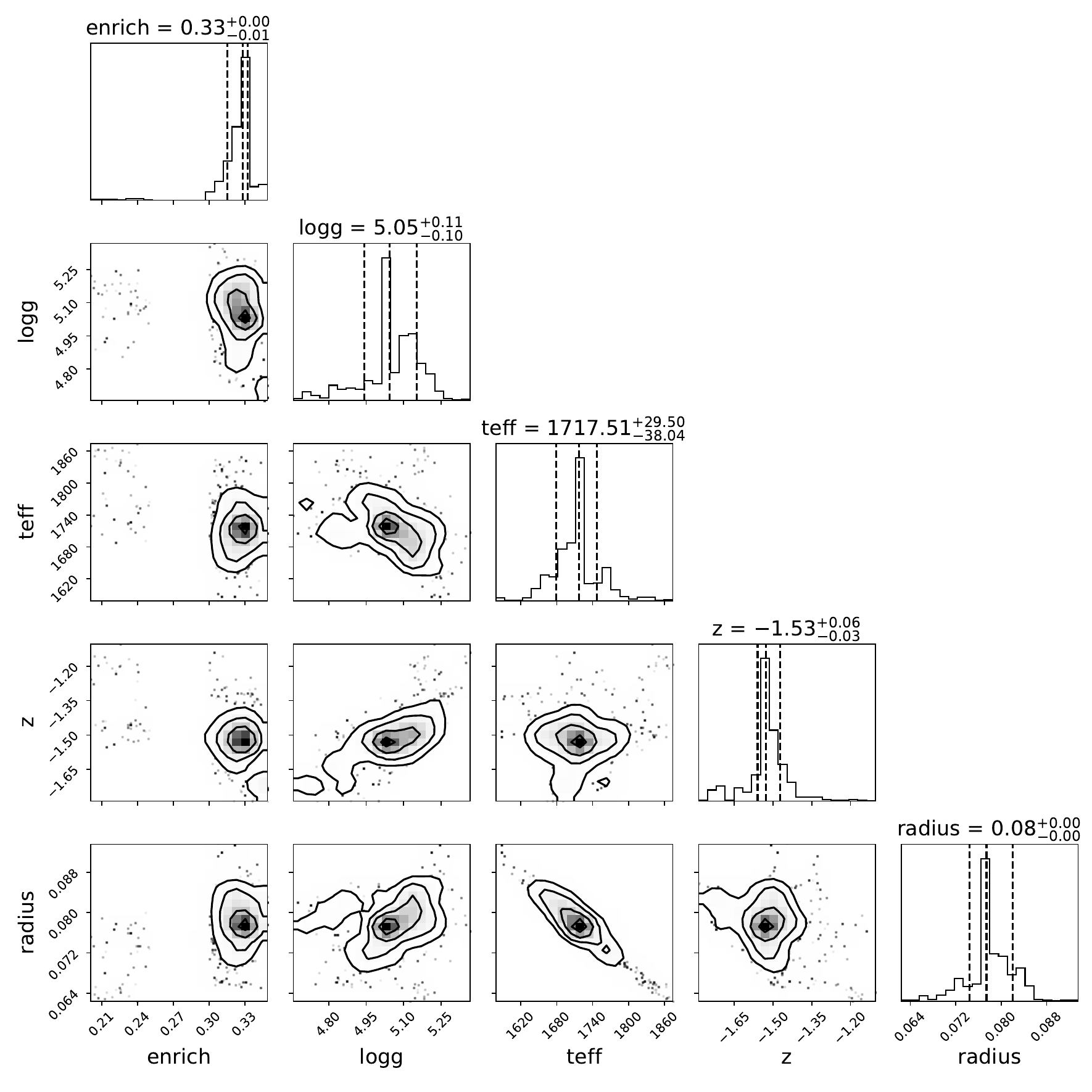}
\caption{
Same as Figure~\ref{fig:mcmcfit1} comparing LOWZ (left) and {SAND} (right) model fits for J0532+8246. 
The former converges to the {\teff} upper limit of the LOWZ grid, while the latter provides a significantly better fit
at a higher {\teff}, {higher surface gravity, and comparable metallicity}. 
\label{fig:mcmcfit2}}
\end{figure}

\begin{figure}[ht!]
\centering
\includegraphics[width=0.45\textwidth]{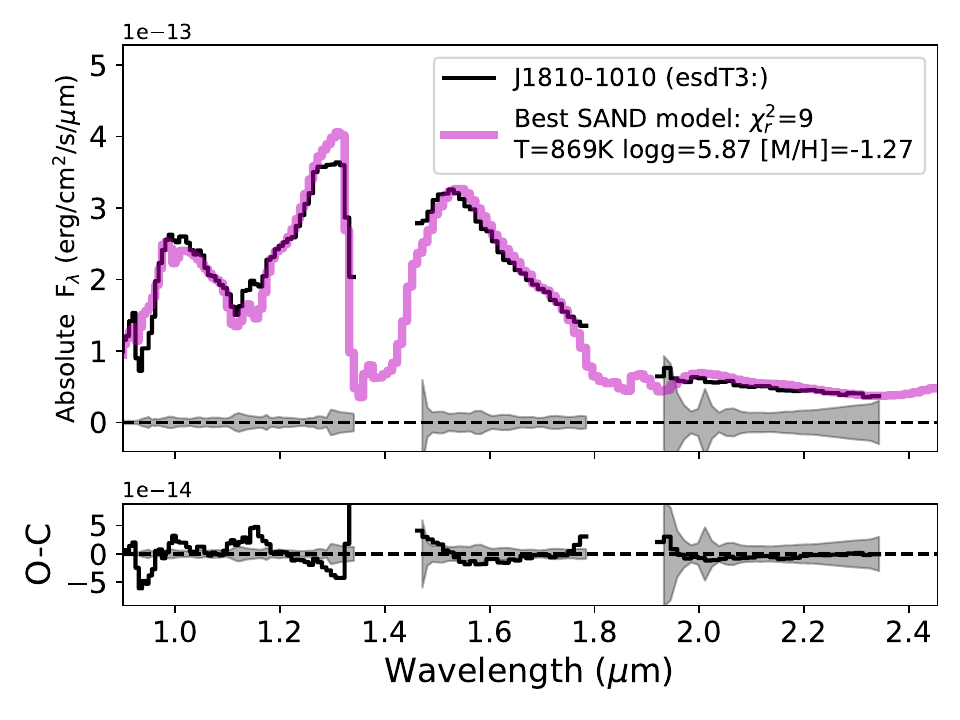} 
\includegraphics[width=0.45\textwidth]{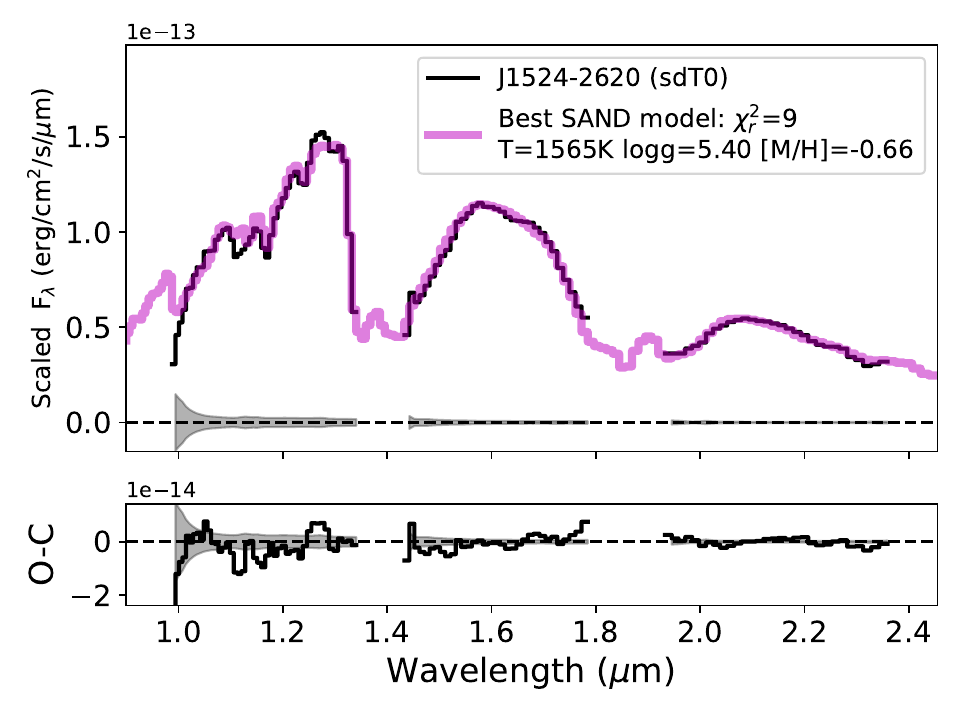}  \\
\includegraphics[width=0.45\textwidth]{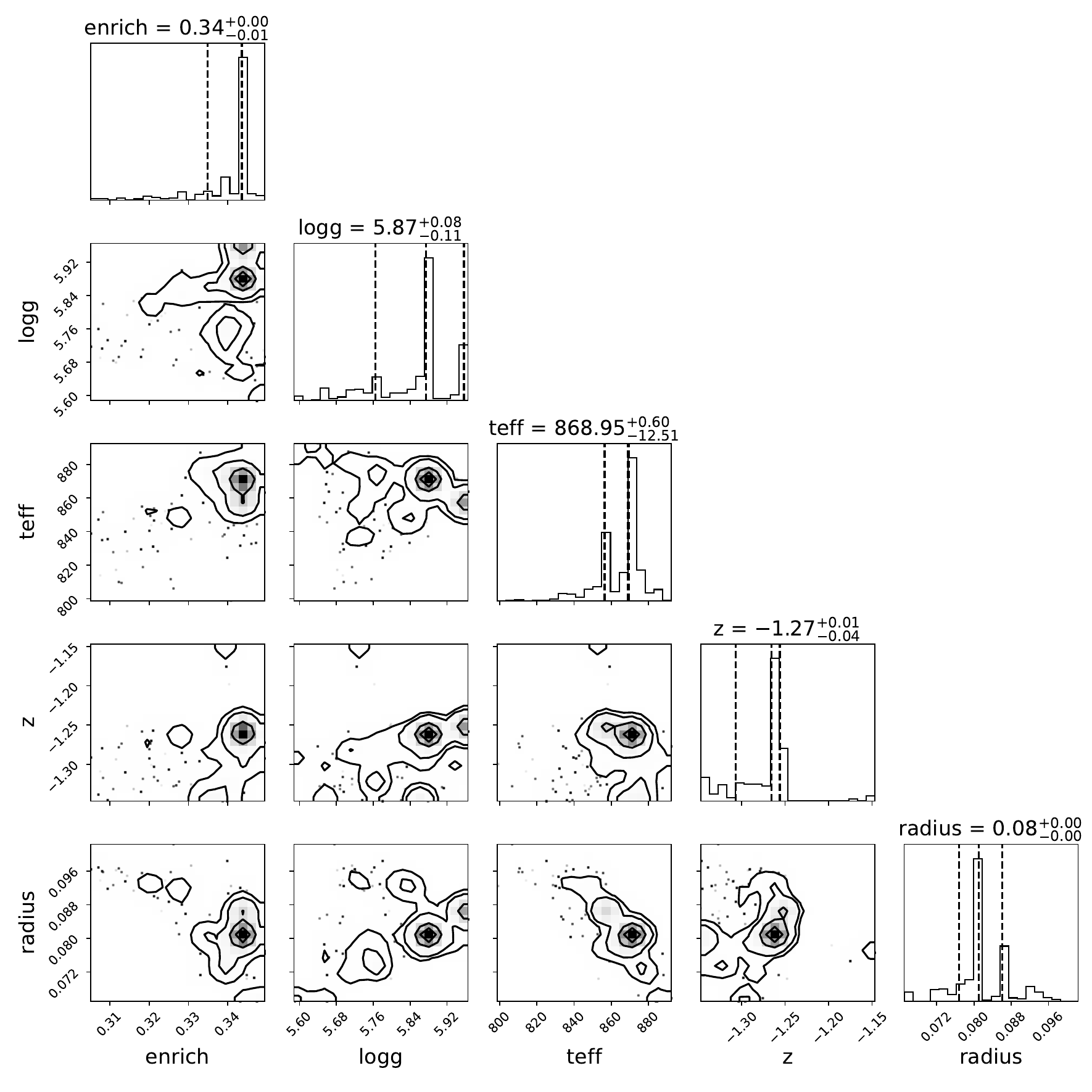}
\includegraphics[width=0.45\textwidth]{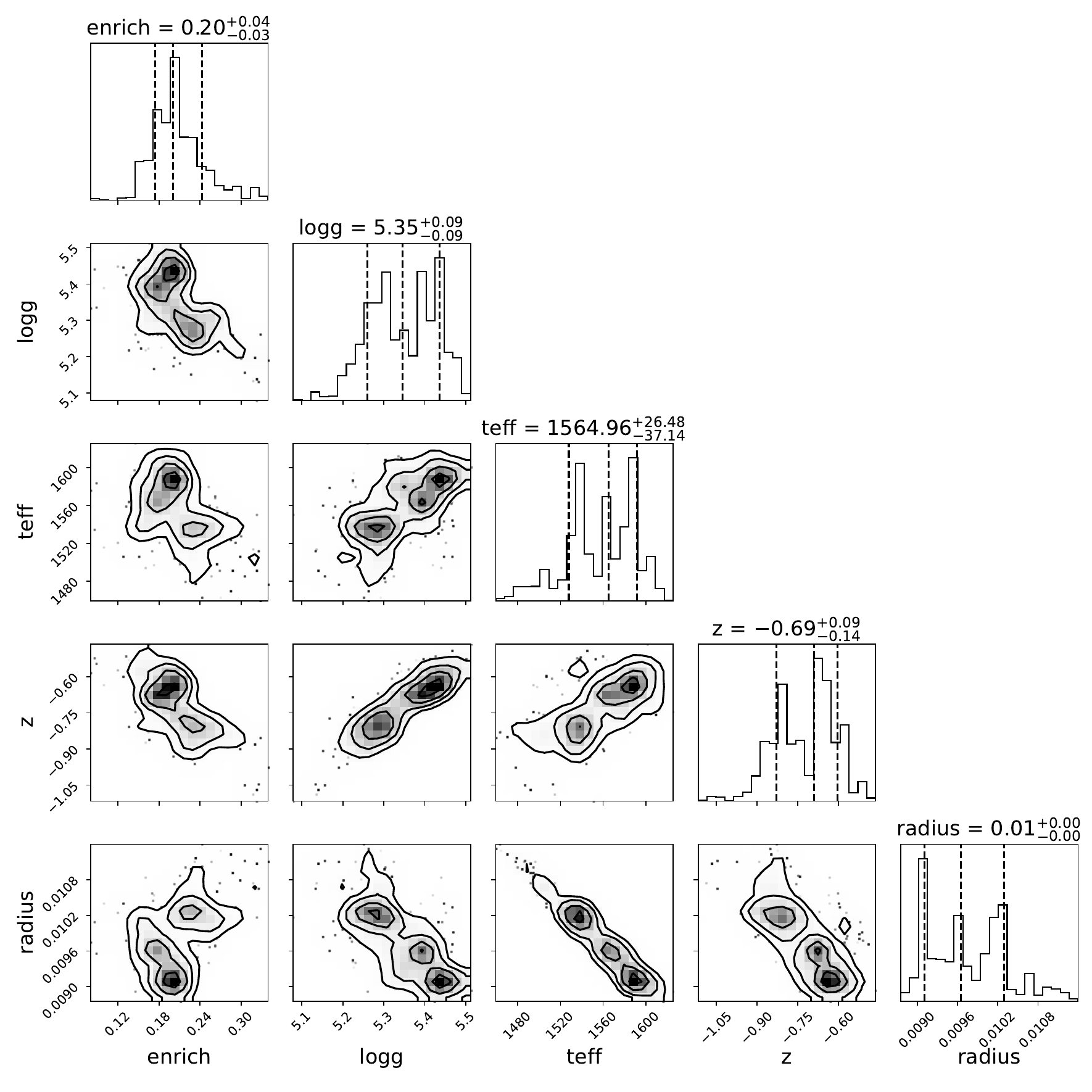}
\caption{
{Same as Figure~\ref{fig:mcmcfit1}, showing best-fit SAND model and parameter distributions for 
the extreme T subdwarf J1810$-$0010 (left) and the 
T subdwarf candidate J1524$-$2620 (right). }
\label{fig:mcmcfit3}}
\end{figure}

\begin{figure}[ht!]
\centering
\includegraphics[width=0.45\textwidth]{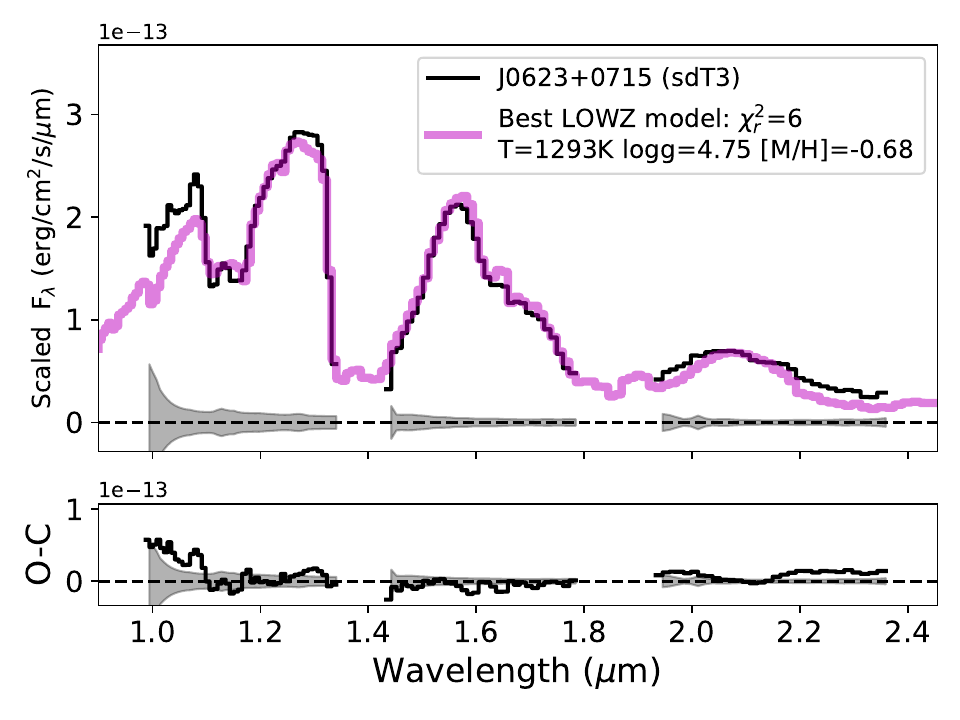}
\includegraphics[width=0.45\textwidth]{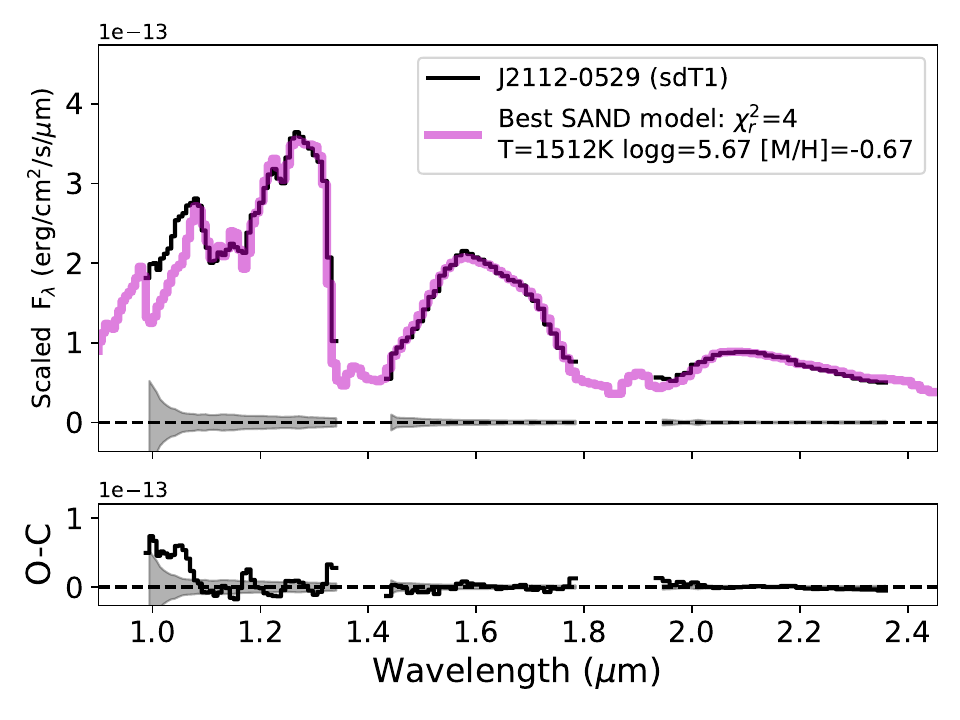} \\
\includegraphics[width=0.45\textwidth]{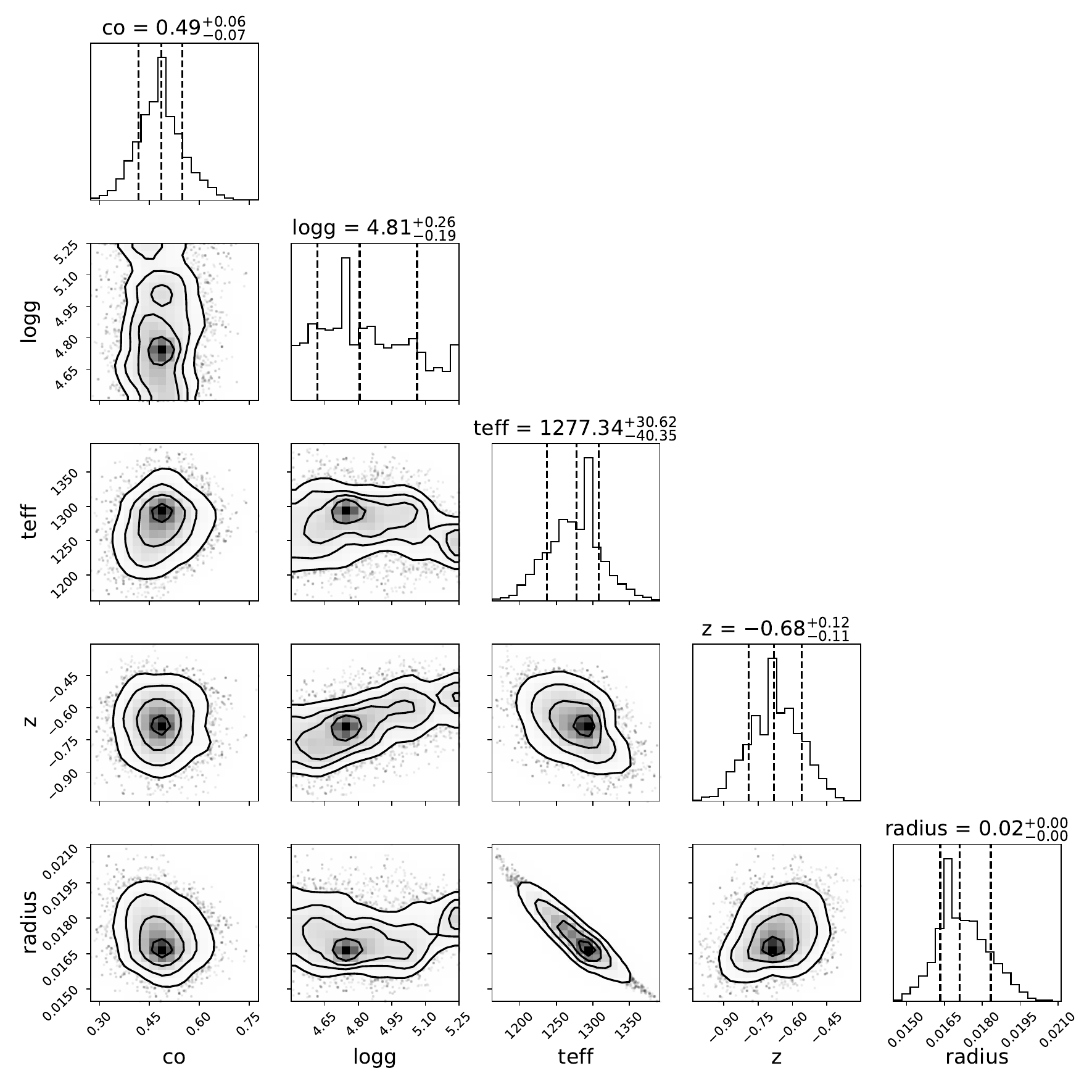}
\includegraphics[width=0.45\textwidth]{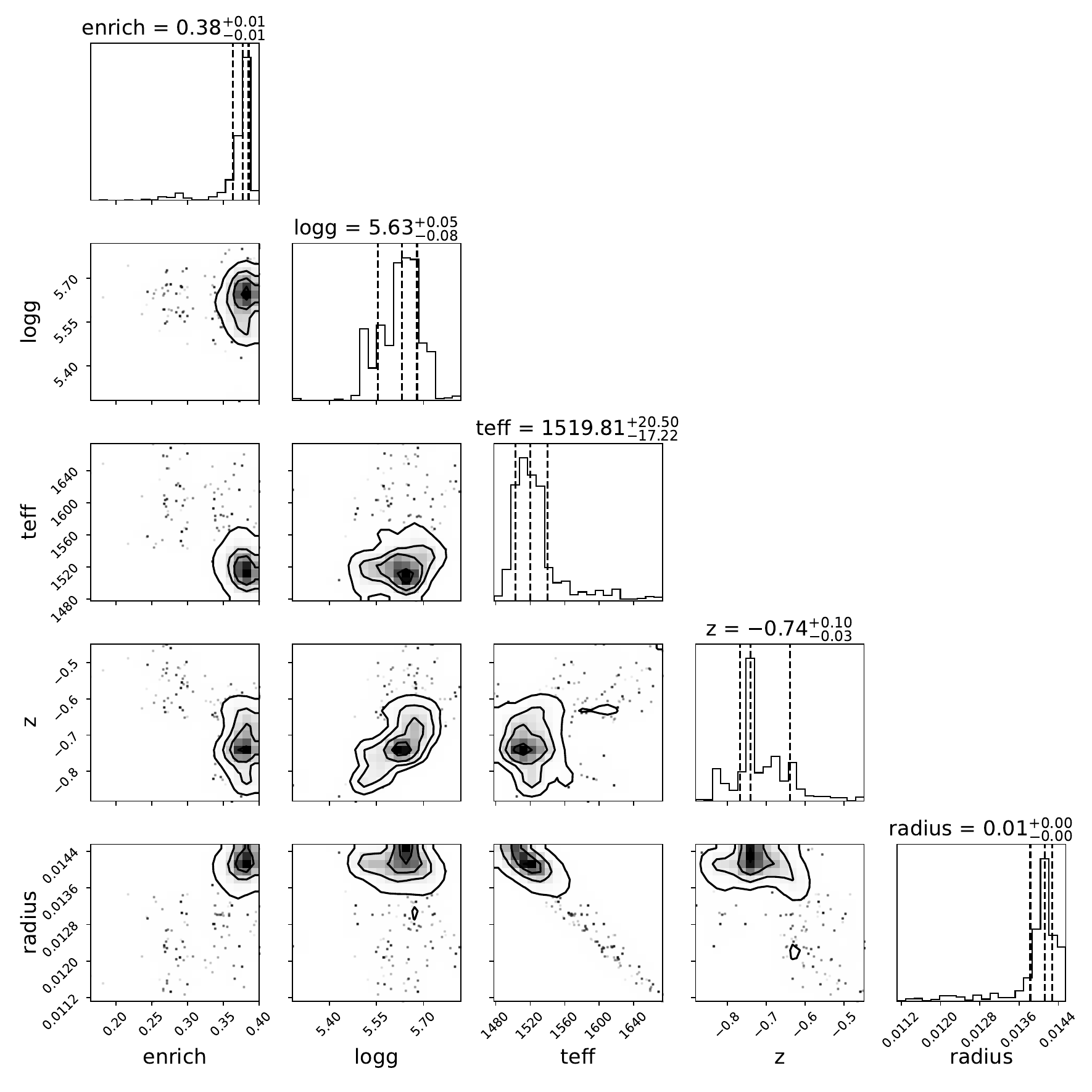}
\caption{
{Same as Figure~\ref{fig:mcmcfit1}, showing best-fit models to the spectra of the T subdwarf candidates J0623+0715 (left, compared to LOWZ models) and J2112$-$0529 (right, compared to SAND models).}
\label{fig:mcmcfit4}}
\end{figure}

%In most cases, the LOWZ model provided the best overall fit, and we use these as our default model parameters. However, in a few cases significant improvement in fit ($\Delta\chi^2_r > 5$) the parameters from the overall best-fit model from all model sets are also reported. 

Table~\ref{tab:modelfit} lists the best-fit model parameters for each of the spectra in our sample, while Figures~\ref{fig:mcmcfit1} through {\ref{fig:mcmcfit4}} display model fits for representative sources.
We find excellent agreement with the LOWZ models for most of our spectra, with temperatures appropriate for their spectral classifications.
%and near-solar or subsolar metallicities.
In particular, we find good agreement between the metallicities inferred for our benchmark companions and the metallicities reported for their primaries {(Figure~\ref{fig:mcmcfit1})}.
Nearly all of the sources with measured parallaxes also have inferred radii consistent with evolved brown dwarfs, with a few cases where radii are either too large (e.g., J0055+5947) or too small (e.g., J1316+0755 and J1810$-$1010), likely reflecting inferred temperatures that are too low or too high, respectively.
Fits to three sources (GJ~576B, J0414$-$5854, and J1316+0755) yield unusually small C/O ratios, which likely offset low temperatures since low C/O reduces the relative strength of CH$_4$ absorption.
%the subdwarf candidate J0055+5847 with $R \approx 0.129R_\odot$, suggesting this source is warmer that the best-fit LOWZ model ({\teff} = 650~K, {\logg} = 5.25,  [M/H] = 0);
%the metal-poor sdT6.5 J1316+0755 with $R \approx 0.027R_\odot$, suggesting this source is cooler that the best-fit LOWZ model ({\teff} = 700~K, {\logg} = 5.0, [M/H] = $-$0.5);
%and the ``extreme'' T subdwarf J1810$-$1010 with $R \approx 0.018R_\odot$, which as discussed above likely has a {\teff} closer to 800~K \citep{lodieu2022}.
%and the T8 J0929$-$2448 whose large inferred radius $R \approx 0.12R_\odot$ is consistent with prior evidence that this is an unresolved binary \citep{2008ApJ...689L..53B,2009ApJ...695.1517L}.
The poorest fits for the LOWZ models occur for the late-L and early-T dwarfs in our sample,
a consequence of the {\teff} = 1600~K maximum for this model set. 

{For sources that hit the parameter limits for the LOWZ models
or have better fits among the other seven model sets ($|\Delta\chi^2_r|$ $>$ 5), 
we also report in Table~\ref{tab:modelfit} the parameters inferred for the best fits among all models.
For the highest-temperature sources, these are typically from the SAND models which yield} higher temperatures outside the LOWZ parameter range, {and generally higher surface gravities and lower metallicities.
An illustrative case is the extreme L subdwarf J0532+8246 (Figure~\ref{fig:mcmcfit2}) for which the SAND models provide an excellent fit to strong FeH absorption at 1~$\mu$m and the smooth near-infrared continuum, while inferring a radius (R = 0.077$^{+0.005}_{-0.003}$~R$_\odot$) in line with theoretical models. 
On the other, other SAND model fits to sources with parallax measurements (e.g., J1158+0435) yield radii too small to be physical, suggesting that the inferred temperatures are too high in these cases. 
SAND models also provide improved fits for some of the lower-temperature metal-poor sources, including Wolf~1130C for which the inferred metallicity ([M/H] = $-$0.65$^{+0.10}_{-0.07}$) is consistent with the metallicity of the system's M dwarf primary ([Fe/H] = $-$0.70$\pm$0.12; \citealt{2018ApJ...854..145M});
%despite being at the lower temperature limit of the model. Another case of note is 
and the extreme T subdwarf J1810$-$1010 (Figure~\ref{fig:mcmcfit3}), for which the inferred temperature ({\teff} = 869$^{+5}_{-13}$~K) and radius (0.081$^{+0.005}_{-0.04}$~R$_\odot$) are consistent with luminosity-based estimates from \citet{lodieu2022}. The better agreement in near-infrared spectral morphology and inferred parameters for these sources suggest that the SAND models are well-suited for metal-poor L and T subdwarfs. 
Other low-temperature T dwarfs with solar/near-solar metallicities are better fit by the} 
% for these alternate models;
% although for one important case, J1810$-$1010, the SAND models converges on a far more realistic radius (see $\S$\ref{sec:J1810}).}
% For the coldest sources, the best alternative models are from 
ElfOwl models, which have
up-to-date opacities for key molecules such as CH$_4$
{and thus better reproduce the strong absorption features present in late-type T dwarf spectra. Nevertheless,} these fits 
%We also report separation model parameters for a handful of sources for which
%altnerate models provide a significantly better fit ($|\Delta\chi^2_r|$ $>$ 5), which 
yield similar atmosphere parameters as LOWZ ($|\Delta${\teff}$| \lesssim$ 100~K, $|\Delta\log{g}| \lesssim$ 0.5, $|\Delta{\rm [M/H]}| \lesssim$ 0.2).
% model  Many of the early T and late L dwarfs in our sample show considerably worse fits to LOWZ models, matching better to B06, Dusty, ATMO, or Sonora models (Figure~\ref{fig:modelfit2}). The failure of the LOWZ models to fit at higher temperatures is likely due to the role of condensate cloud formation in the atmospheres of L/T transition objects \citep{1996AandA...305L...1T,2001ApJ...556..872A},
% which is included in the Dusty, ATMO, and B06 models.
% %or alternatively a change in the adiabatic mixing due to double diffusion convective instabilities
% %\citep{2015ApJ...804L..17T,2016ApJ...817L..19T}.
% In most cases, the alternative models give similar parameters 
% ($|\Delta${\teff}$| \lesssim$ 100~K, $|\Delta\log{g}| \lesssim$ 0.5, $|\Delta{\rm [M/H]}| \lesssim$ 0.5), 
% although alternative fits to the L subdwarfs J0532+8246 and J1158+0435 are 600~K an 400~K hotter than the LOWZ fit,
% and yield significantly smaller (and likely aphysical) radii. For these metal-poor late-L subdwarfs, and the esdT J1810$-$1010,
% current atmosphere models do not yet yield self-consistent fits to absolute fluxes.

{Turning to} our subdwarf candidates, several show evidence of modest subsolar metallicity ($-$0.2 $\gtrsim$ [M/H] $\gtrsim$ $-$0.5) based on the model fits, with {three} sources---J0623+0715, {J1524$-$2620, and J2112$-$0529}---being the most metal-poor {([M/H] $\approx$ $-$0.7)}.
%{based on LOWZ model fits. Additionally, LOWZ fits to
%J1524$-$2620 and J2112$-$0529 yield subsolar metallicities but at the limit of the {\teff} parameter range; improved fits with the B06 and ElfOwl models, respectively, yield modest metal deficiencies.}
Remarkably, {the} candidate J0140+0150 and the previously reported J1055+5443 \citep{2023ApJ...958...94R} {are found} to be significantly metal-rich ([M/H] $\gtrsim$ +0.4) {despite their selection as high velocity objects}. These sources are discussed in further detail in Section~\ref{sec:individual}.

\subsection{Radial Velocities}\label{sec:rv}

The moderate resolution of our Keck/NIRES data facilitates measurement of the radial velocities of our targets via the forest of H$_2$O, CH$_4$, and (for late-L and early T dwarfs) CO molecular features that are marginally resolved in the data. Our approach is based on the forward modeling technique using pre-telluric-corrected spectra, as described in \citet{2015AJ....149..104B} and \citet{2021ApJS..257...45H}.
%and recently applied to Keck/NIRES observations of the red L dwarf CWISE~J0506+0738 \citep{2023ApJ...943L..16S}. 
We performed separate fits in $Y$-band (1.11--1.16~$\mu$m), $J$-band (1.26--1.31~$\mu$m), $H$-band (1.52--1.59~$\mu$m), and $K$-band (2.03--2.23~$\mu$m), regions that encompass strong molecular features in both source and telluric spectra.
We also {conducted additional} fits in the 1.235--1.285~$\mu$m, 1.95--2.40~$\mu$m, and 2.26--2.40~$\mu$m regions for our earliest-type sources, which encompass K~I and CO absorption.
Our spectral data model was defined as
\begin{equation}
    D[\lambda] = \left(C[\lambda]{\times}M[\lambda^\prime]{\times}T^{\alpha}[\lambda]+\delta_F\right)\otimes\kappa_G(v_{broad})
\end{equation}
where $M[\lambda^\prime]$ is a high-resolution ({\ldl} = 50,000) BT-Settl atmosphere model \citep{2012RSPTA.370.2765A},
%and Sonora-Bobcat grids \citep{2021ApJ...920...85M}, 
with {\logg} = 5.0 (cgs) and {\teff} based on the fitting values in Table~\ref{tab:modelfit}.
This model is
evaluated at wavelengths $\lambda^\prime = \lambda\left(1+\frac{v_{bary}+RV}{c}\right)$ where $v_{bary}$ is the barycentric velocity correction for the source at the time of observation, $RV$ is the heliocentric velocity constrained to be within $\pm$200~km/s, and $c$ is the speed of light. 
$T[\lambda]$ is an empirical telluric absorption template from \citet{1991aass.book.....L} scaled by an exponent $\alpha$ to account for varying airmass. 
$C[\lambda]$ is a fifth-order polynomial continuum correction to the ratio of smoothed versions of the data and the combination $M[\lambda^\prime]{\times}T^{\alpha}[\lambda]$. $\delta_F$ is a constant flux offset. 
$\kappa_G(v_{broad})$ is a Gaussian instrumental broadening profile of width $v_{broad} = c\frac{\Delta\lambda_I}{\langle\lambda\rangle}$, 
where $\langle\lambda\rangle$ is the average wavelength over the fitting range
and $\Delta\lambda_I$ is a free parameter.
As $v_{broad}$ is typically of order 100~km/s, we ignore stellar rotational broadening in this analysis. We also fit for a constant wavelength offset $\delta_{\lambda} = \langle\lambda-\lambda_0\rangle$ between the wavelength grid of the data $\lambda_0$ and that of the data model to account for slight offsets (up to $\pm$50~{\AA}) in the original wavelength calibration. 
These {adjustments} significantly improve our RV precisions and motivate our use of pre-telluric-corrected spectra (cf.~\citealt{2021ApJS..257...45H}).  

%% STOPPED HERE

The data model is defined by five free parameters (RV, $\alpha$, $\Delta\lambda_I$, $\delta_F$, $\delta_\lambda$) and six continuum coefficients that are determined at each fitting step. 
We used a similar reduced $\chi^2$ statistic as Eqn.~\ref{eqn:chi1} to assess the quality of the fit,
\begin{equation}
    \chi^2_r = \frac{1}{DOF}\sum_i^N\frac{(O[\lambda_i]-D[\lambda_i])^2}{\sigma^2[\lambda_i]}.
    \label{eqn:chi2}
\end{equation} 
{where $N$ is the number of unmasked data points and}
$DOF = N-11$. Note that the scaling factor $\beta$ in Eqns.~\ref{eqn:chi1} and~\ref{eqn:chiscale}
is incorporated into the continuum correction $C[\lambda]$.
The five free parameters were first optimized using the Nelder-Mead algorithm \citep{nelder65,gao2012} as implemented in the \texttt{scipy} optimize package \citep{2020SciPy-NMeth}. These parameters were then used to seed an MCMC model with
%simple Metropolis-Hastings Markov chain Monte Carlo (MCMC) algorithm \citep{1953JChPh..21.1087M,HASTINGS01041970} to explore the proximate parameter space. We used 
a single chain of 5,000 steps for each source and wavelength range, with verification of convergence 
using the \citet{geweke1992} diagnostic for the RV values and visual confirmation of the chain.
We also examined each fit to ensure the presence of distinct spectral structure from both stellar and telluric absorption, and rejected those fits {lacking sufficient} structure or are in very low signal-to-noise regions.

\begin{figure}
    \centering
    \includegraphics[width=0.45\textwidth]{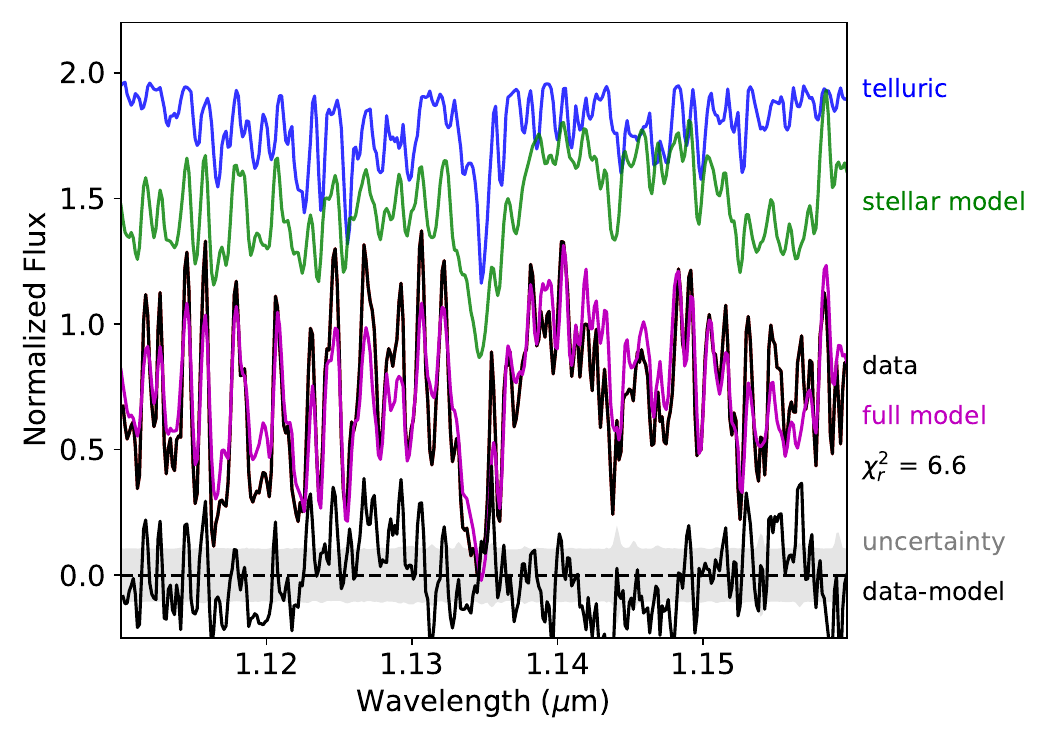}
    \includegraphics[width=0.45\textwidth]{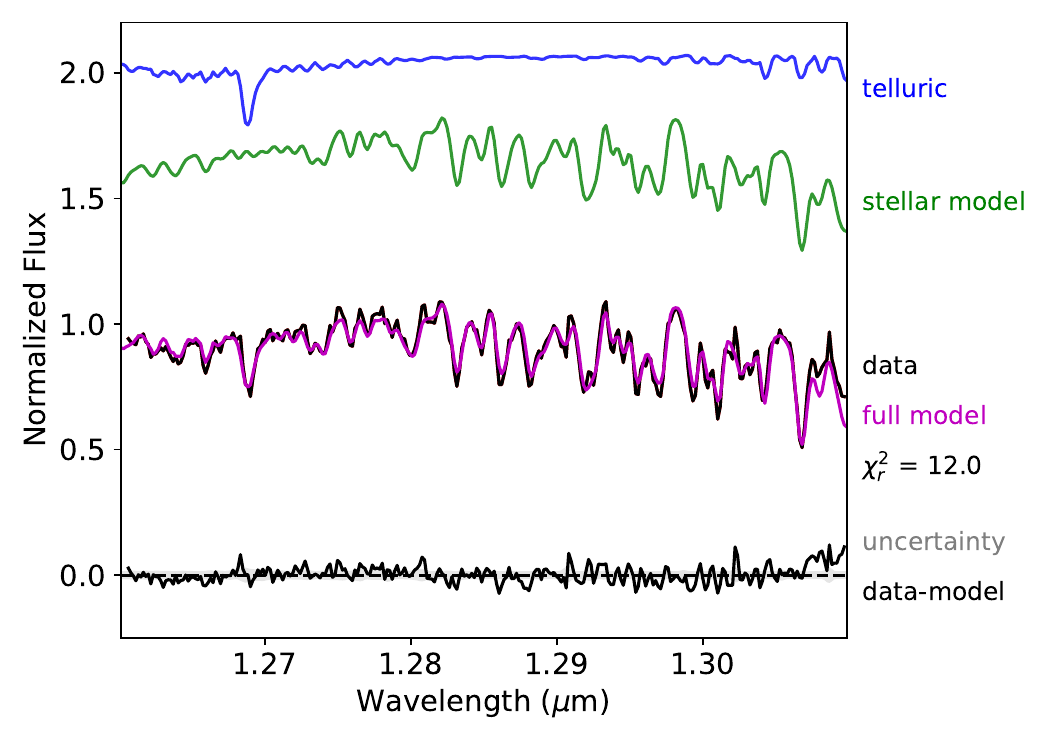} \\
    \includegraphics[width=0.45\textwidth]{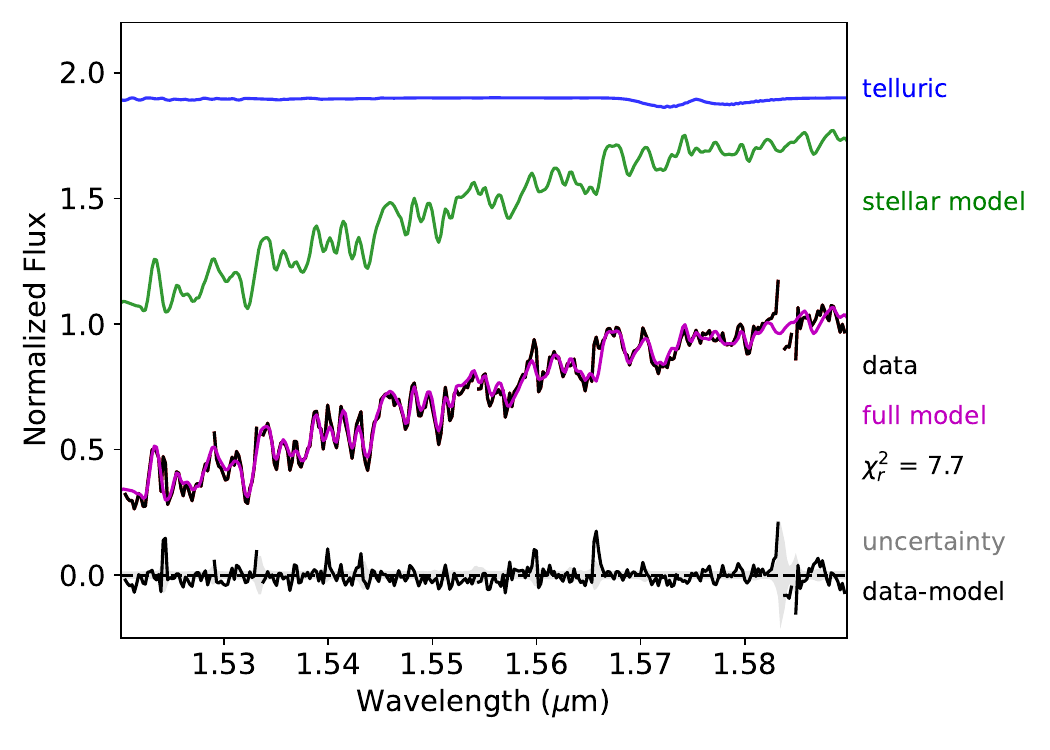}
    \includegraphics[width=0.45\textwidth]{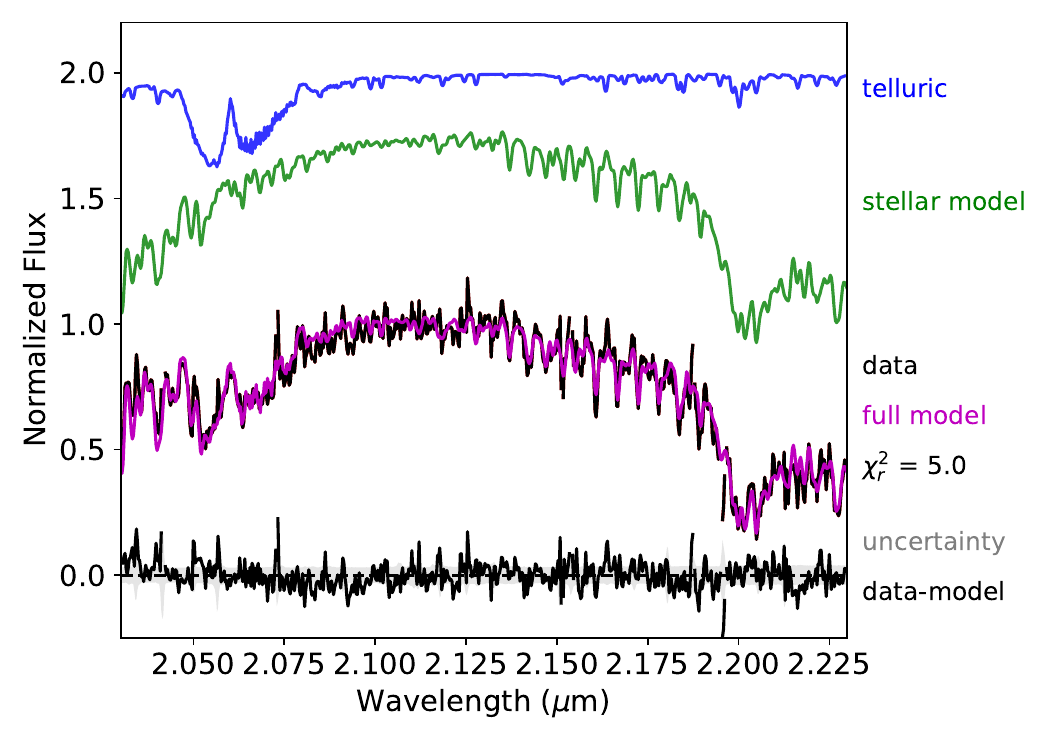} \\
    \caption{Forward model fits to $Y-$ (top left), $J-$ (top right), $H-$ (bottom left), and $K-$band (bottom right) spectra of GJ~576B. Each panel displays from top to bottom: the best-fit telluric model (blue), the best-fit stellar model (green), the observed data (black) compared to the full data model (magenta), and the difference of these (black) compared to the data uncertainty (grey band, $\pm$1$\sigma$).
    The reduced $\chi^2_r$ (Eqn.~\ref{eqn:chi2}) is listed along the right axis.
%    (Bottom panels): MCMC parameter corner plots for the $J-$ (left), $H-$ (middle), and $K-$band (right) fits. Plots along the diagonal axis show the marginalized posterior distributions for each parameter, with 16\%, 50\%, and 84\% quantiles indicated by dashed lines. The remaining contour plots display two-dimensional distributions of parameter pairs in the posterior solutions, highlighting parameter correlations.  Parameter distributions are weighted using the statistic defined in Eqn.~\ref{eqn:weight}.
}
    \label{fig:rvexample}
\end{figure}

Figure~\ref{fig:rvexample} illustrates fits for GJ~576B, demonstrating how the combination of telluric and  stellar absorption are well reproduced by the forward model. For our highest signal-to-noise data, these fits provide robust constraints on RV ($\sigma_{RV} \approx$ 5~km/s, equivalent to roughly 0.2~pixels at native resolution) and consistency between the four fit regions. For lower signal-to-noise data, RV errors increase to tens of km/s.
{Variations in the RV fit value across the spectral regions} are generally consistent with the measurement uncertainties.
%, with exceptions for our highest S/N data.
%(GJ~576B and J0532+8246)

Table~\ref{tab:rvfits} summarizes the individual fit RVs, with median values and uncertainties
computed as weighted 16\%, 50\%, and 84\% quantiles,
using the weighting factor
\begin{equation} \label{eqn:weight}
    w_i = \frac{DOF}{DOF+\chi^2_i-{\rm MIN}(\chi^2)}.
\end{equation}
%where $DOF = N-5$ and MIN$(\chi^2)$ is the minimum chi-square for all fits.
%For those cases in which the Geweke convergence test failed, or the source spectrum contained minimal structure or excessive noise, we excluded the RV for that region in our analysis.
%remaining fits, we find formal consistency across bands for every source, 
%albeit with a wide range of typically asymmetric uncertainties. 
For sources with at least two viable RV measurements, we computed a weighted mean RV and uncertainty by adopting the larger of the lower and upper RV error estimates, $\sigma_{RV}$, as a weighting factor:
% \begin{equation}
%     \langle{RV}\rangle = \frac{\sum_i^M\frac{RV_i}{\sigma^2_{RV,i}}}{\sum_i^M\frac{1}{\sigma^2_{RV,i}}}
% \end{equation}
\begin{equation}
    \langle{RV}\rangle = \frac{\sum_i^M{RV_i}/{\sigma^2_{RV,i}}}{\sum_i^M{1}/{\sigma^2_{RV,i}}}
\end{equation}
\begin{equation} \label{eqn:rvcombunc}
    \sigma^2_{\langle{RV}\rangle} = \frac{1+\sum_i^M{(RV_i-\langle{RV}\rangle)^2}/{\sigma^2_{RV,i}}}{\sum_i^M{1}/{\sigma^2_{RV,i}}}.
%    +\frac{1}{\sum_i^M{1}/{\sigma^2_{RV,i}}}
\end{equation}
The sums are over the $M$ = 2--4 combined measurements, and the form of Eqn.~\ref{eqn:rvcombunc} accounts for the size of the individual measurement uncertainties.
The final RVs are reported in Table~\ref{tab:rvfits}.

\begin{deluxetable}{lccccccccccl}
\tablecaption{Radial Velocity Measurements (Keck/NIRES) \label{tab:rvfits}} 
\tabletypesize{\scriptsize} 
\tablehead{ 
\multicolumn{1}{c}{} & 
\multicolumn{2}{c}{Y} & 
\multicolumn{2}{c}{J} & 
\multicolumn{2}{c}{H} & 
\multicolumn{2}{c}{K} & 
\multicolumn{1}{c}{Adopted} & 
\multicolumn{1}{c}{Prior} & 
\\
\cline{2-3} \cline{4-5} \cline{6-7} \cline{8-9}
\multicolumn{1}{c}{Name} & 
\multicolumn{1}{c}{RV} & 
\multicolumn{1}{c}{$\chi^2_r$}  & 
\multicolumn{1}{c}{RV} & 
\multicolumn{1}{c}{$\chi^2_r$}  & 
\multicolumn{1}{c}{RV} & 
\multicolumn{1}{c}{$\chi^2_r$}  & 
\multicolumn{1}{c}{RV} & 
\multicolumn{1}{c}{$\chi^2_r$}  & 
\multicolumn{1}{c}{RV} & 
\multicolumn{1}{c}{RV} & 
\colhead{Ref} \\
\multicolumn{1}{c}{} & 
\multicolumn{1}{c}{(km/s)} & 
\multicolumn{1}{c}{} & 
\multicolumn{1}{c}{(km/s)} & 
\multicolumn{1}{c}{} & 
\multicolumn{1}{c}{(km/s)} & 
\multicolumn{1}{c}{} & 
\multicolumn{1}{c}{(km/s)} & 
\multicolumn{1}{c}{} & 
\multicolumn{1}{c}{(km/s)} & 
\multicolumn{1}{c}{(km/s)} 
} 
\startdata 
\hline
\multicolumn{12}{c}{Metallicity Benchmarks} \\
\hline
LHS~6176B & \nodata & \nodata & 10$^{+4}_{-4}$ & 4 & $-$9$^{+15}_{-13}$ & 5 & $-$2$^{+11}_{-12}$ & 2 &  8$\pm$7 & 11.7$\pm$2.1\tablenotemark{c} & [1] \\ 
GJ~576B & $-$32$^{+8}_{-10}$ & 7 & $-$23$^{+5}_{-5}$ & 12 & $-$24$^{+6}_{-5}$ & 7 & $-$25$^{+9}_{-12}$ & 5 & ($-$25$\pm$4)\tablenotemark{d} & $-$84.79$\pm$0.12 & [2]  \\ 
\hline
\multicolumn{12}{c}{Subdwarf Candidates} \\
\hline
J0055+5947 & \nodata & \nodata & 38$^{+20}_{-34}$ & 1 & 30$^{+45}_{-31}$ & 1 & 27$^{+60}_{-36}$ & 4 & 34$\pm$25 & \nodata & \\ 
J0140+0150 &  84$^{+24}_{-23}$ & 2 & 96$^{+12}_{-11}$ & 2 & 90$^{+9}_{-9}$ & 2 & 86$^{+15}_{-13}$ & 4 & 91$\pm$7 & \nodata & \\ 
J0411+4714 & \nodata & \nodata & 37$^{+5}_{-5}$ & 137 & 17$^{+12}_{-11}$ & 175 & 40$^{+14}_{-20}$ & 118 & 34$\pm$8 & \nodata & \\
J0429+3201 & \nodata & \nodata & $-$117$^{+29}_{-21}$\tablenotemark{a} & 2  & \nodata & \nodata & $-$63$^{+9}_{-11}$\tablenotemark{b} & 6 & ($-$70$\pm$21)\tablenotemark{d} & \nodata & \\ 
J0433+1009 & \nodata & \nodata & 21$^{+8}_{-8}$ & 2 & 15$^{+18}_{-30}$ & 2 & 21$^{+20}_{-19}$ & 2 & 21$\pm$7 & \nodata & \\ 
J0623+0715 & \nodata & \nodata & $-$14$^{+28}_{-21}$ & 2 & 18$^{+34}_{-23}$ & 2 & $-$2$^{+27}_{-24}$\tablenotemark{b} & 3 & $-$1$\pm$27 & \nodata & \\ 
J0659+1615 & $-$41$^{+13}_{-13}$ & 1  & $-$46$^{+51}_{-25}$ & 2 & $-$29$^{+10}_{-12}$ & 2 & $-$38$^{+12}_{-11}$\tablenotemark{c} & 7 & $-$36$\pm$9 & \nodata \\
J0758+5711 & \nodata & \nodata & 81$^{+8}_{-7}$ & 3 & 75$^{+12}_{-13}$ & 2 & 72$^{+17}_{-19}$ & 2 & 79$\pm$7 & \nodata & \\ 
J1110$-$1747 & \nodata & \nodata & $-$46$^{+4}_{-4}$ & 3 & $-$55$^{+9}_{-10}$ & 5 & $-$44$^{+10}_{-9}$ & 5 & $-$47$\pm$5 & \nodata & \\ 
J1130+3139 & 64$^{+13}_{-11}$ & 3 & 76$^{+9}_{-9}$ & 3 & 76$^{+13}_{-10}$ & 2 & 68$^{+18}_{-17}$ & 2 & 72$\pm$8 & \nodata & \\ 
J1204$-$2359 & \nodata & \nodata & $-$41$^{+10}_{-10}$ & 2 & $-$76$^{+25}_{-17}$ & 3 & $-$57$^{+29}_{-20}$ & 2 & $-$47$\pm$15 & \nodata & \\ 
J1304+2819 & 45$^{+20}_{-37}$ & 1 & 35$^{+31}_{-21}$ & 2 & \nodata & \nodata &  51$^{+8}_{-8}$\tablenotemark{c} & 6 & 50$\pm$9 & \nodata & \\ 
J1308$-$0321 & \nodata & \nodata & $-$40$^{+4}_{-4}$ & 3 & $-$52$^{+12}_{-12}$ & 3 & $-$51$^{+11}_{-10}$ & 2 & $-$42$\pm$6 & \nodata & \\ 
J1401+4325 & \nodata & \nodata & $-$68$^{+25}_{-21}$ & 3 & $-$79$^{+38}_{-36}$ & 2  & \nodata & \nodata & $-$71$\pm$21 & \nodata & \\ 
J1458+1734 & \nodata & \nodata & 19$^{+6}_{-6}$ & 4 & 16$^{+12}_{-13}$ & 3 & 24$^{+14}_{-13}$ & 4 & 19$\pm$6 & \nodata & \\ 
J1524$-$2620 & 34$^{+42}_{-26}$ & 1 & \nodata & \nodata & \nodata & \nodata& 28$^{+19}_{-17}$ & 4 & 29$\pm$17 & \nodata & \\ 
J1710+4537 & $-$1$^{+16}_{-15}$ & 3 & $-$2$^{+9}_{-8}$ & 2 & $-$13$^{+27}_{-23}$ & 3 & \nodata & \nodata & $-$3$\pm$8 & \nodata & \\ 
J1801+4717 & \nodata & \nodata & $-$27$^{+21}_{-22}$ & 2 & $-$33$^{+19}_{-26}$ & 2 & $-$22$^{+91}_{-16}$ & 4  & $-$29$\pm$17 & \nodata & \\ 
J2013$-$0326 & \nodata & \nodata & $-$20$^{+10}_{-11}$ & 2 & $-$22$^{+26}_{-28}$ & 2 & 23$^{+33}_{-49}$  & 3 & $-$18$\pm$13 & \nodata & \\ 
J2021+1524 & 11$^{+22}_{-10}$ & 1  & \nodata & \nodata & 10$^{+15}_{-17}$ & 2 & $-$7$^{+5}_{-5}$\tablenotemark{c} & 5 & $-$5$\pm$7 & \nodata & \\ 
J2112$-$0529 & $-$42$^{+24}_{-19}$ & 2 & $-$24$^{+46}_{-36}$\tablenotemark{a} & 2 &  \nodata & \nodata & $-$37$^{+10}_{-17}$\tablenotemark{b} & 4 & $-$37$\pm$14 & \nodata & \\ 
J2112+3030 & \nodata & \nodata & \nodata & \nodata &  \nodata & \nodata & $-$125$^{+62}_{-40}$ & 4 & $-$125$\pm$62 & \nodata & \\ 
J2218+1146 & $-$23$^{+16}_{-23}$ & 5 & $-$22$^{+8}_{-9}$ & 3 & $-$15$^{+22}_{-47}$ & 3 & $-$43$^{+52}_{-33}$ & 2  & $-$22$\pm$9 & \nodata & \\ 
J2251$-$0740 & \nodata & \nodata & $-$133$^{+16}_{-17}$ & 4 & \nodata & \nodata & $-$197$^{+49}_{-26}$ & 3 & $-$140$\pm$25 & \nodata & \\ 
\hline
\multicolumn{12}{c}{Comparison Sources} \\
\hline
J0532+8246 & $-$155$^{+42}_{-36}$ & 99 & $-$205$^{+22}_{-35}$\tablenotemark{a} & 56 & \nodata & \nodata & $-$128$^{+31}_{-21}$\tablenotemark{b} & 42 & $-$160$\pm$39  & $-$172$\pm$1 & [3] \\ 
J1055+5443 & \nodata & \nodata & 36$^{+11}_{-13}$ & 4 & 3$^{+33}_{-27}$ & 3 & 75$^{+19}_{-27}$ & 4 & 39$\pm$22 & \nodata & \\ 
J1553+6933 & \nodata & \nodata & \nodata & \nodata & \nodata & \nodata & 110$^{+46}_{-90}$ & 5  & 110$\pm$90 & \nodata & \\ 
\enddata 
\tablerefs{
[1] \citet{2020AJ....160...83S};
[2] \citet{2018AandA...619A..81H}; 
[3] \citet{2006AJ....131.1806R}.
}
\tablenotetext{a}{Based on fits to alternative spectral range 1.235--1.280~$\mu$m.}
\tablenotetext{b}{Based on fits to alternative spectral range 1.95--2.40~$\mu$m.}
\tablenotetext{c}{Based on fits to alternative spectral range 2.26-2.40~$\mu$m.}
\tablenotetext{d}{Due to unusually good seeing at the time of observation, offsets of tens of km/s are possible; value should be treated with skepticism and is reported here only for completeness.}
\end{deluxetable} 

For two of the three sources with previously published RVs, LHS~6176B (from its primary) and J0532+82346, we find excellent agreement with our Keck/NIRES RV measurements. For GJ~576B, we measure a precise RV = $-$25$\pm$4~km/s that is nevertheless highly discrepant with its primary RV of $-$84.79$\pm$0.12~km/s \citep{2018AandA...619A..81H}. As the latter value is validated by Gaia DR3 measurements ($-$83.9$\pm$0.2~km/s; \citealt{2021AandA...650C...3G}), this deviation cannot be explained by unresolved multiplicity in the primary, and is of too large of a magnitude to be explained by unresolved multiplicity of the brown dwarf companion.\footnote{The shortest-period very-low-mass binary currently known, LP~413-53AB with $P$ = 17~hr, has an RV semi-amplitude of 24~km/s \citep{2023ApJ...945L...6H}.}
Instead, this discrepancy is likely due to an observational bias. 
Both LHS~6176B and J0429+3201 were observed in conditions of excellent seeing ($\lesssim$0$\farcs$4), smaller than the fixed slit width of NIRES. A slight offset of the source from the slit center by a single pixel (0$\farcs$15, approximately 1/3 of a resolution element or $\sim$40~km/s) would shift the dispersed spectrum {by an} amount sufficient to explain this {discrepancy}. 
Unfortunately, this systematic bias is uncorrectable with the present data.
We therefore exclude the RV of this source and that of J0429+3201 from our analysis.
%We are unable to account for this discrepancy.

\begin{longrotatetable}
\begin{deluxetable}{lcccccccccccl}
\tablecaption{Sample Kinematics \label{tab:kinematics}} 
\tabletypesize{\scriptsize} 
\tablehead{ 
\colhead{Name} & 
\colhead{d\tablenotemark{a}} & 
\colhead{$\mu_\alpha$}  & 
\colhead{$\mu_\delta$}  & 
\colhead{$V_{tan}$}  & 
\colhead{RV} & 
\colhead{U} & 
\colhead{V} & 
\colhead{W} & 
\multicolumn{3}{c}{Probabilities} & 
\colhead{Ref} \\
\cline{10-12}
 & 
\multicolumn{1}{c}{(pc)} & 
\multicolumn{1}{c}{(mas/yr)} & 
\multicolumn{1}{c}{(mas/yr)} & 
\multicolumn{1}{c}{(km/s)} & 
\multicolumn{1}{c}{(km/s)} & 
\multicolumn{1}{c}{(km/s)} & 
\multicolumn{1}{c}{(km/s)} & 
\multicolumn{1}{c}{(km/s)} &
\colhead{thin}  & 
\colhead{thick}  & 
\colhead{halo}  & 
\\
} 
\startdata 
\hline
\multicolumn{13}{c}{Metallicity Benchmarks} \\
\hline
HD~3651B\tablenotemark{b} & 11.108$\pm$0.006 & $-$461.95$\pm$0.07 & $-$369.62$\pm$0.03 & 31.150$\pm$0.017 & $-$32.898$\pm$0.005 & 51.149$\pm$0.015 & $-$7.205$\pm$0.004 & 15.664$\pm$0.008 & 99\% & 1\% & 0\%  & [1],[2] \\
{HD~65486B}\tablenotemark{b} & 18.465$\pm$0.006 & 362.409$\pm$0.012 & $-$245.786$\pm$0.016 & 38.326$\pm$0.013 & $-$8.19$\pm$0.12 & 46.17$\pm$0.05 & 4.24$\pm$0.10 & 22.816$\pm$0.006 & 99\% & 1\% & 0\%  & \\
LHS~6176B\tablenotemark{b} & 19.607$\pm$0.010 & 234.47$\pm$0.03 & $-$360.66$\pm$0.02 & 39.98$\pm$0.02 & 12$\pm$2 & 38.7$\pm$0.9 & $-$18.3$\pm$1.3 & 13.7$\pm$1.3 & 99\% & 1\% & 0\%  & [1],[3] \\
{Ross~458C}\tablenotemark{b} & 11.507$\pm$0.015 & $-$628.72$\pm$0.18 & $-$33.47$\pm$0.13 & 34.34$\pm$0.05 & $-$12.3$\pm$0.3 & $-$18.46$\pm$0.07 & $-$6.02$\pm$0.07 & $-$3.9$\pm$0.3 & 99\% & 1\% & 0\%  & \\
{HD~126053B}\tablenotemark{b} & 17.461$\pm$0.011 & 2243.53$\pm$0.05 & $-$478.28$\pm$0.03 & 43.69$\pm$0.03 & $-$19.21$\pm$0.14 & 32.78$\pm$0.07 & $-$3.052$\pm$0.017 & $-$32.43$\pm$0.12 & 93\% & 7\% & 0\%  & \\
GJ~570D\tablenotemark{b} & 5.886$\pm$0.002 & 1031.47$\pm$0.07 & $-$1724.62$\pm$0.06 & 56.04$\pm$0.02 & 26.987$\pm$0.002 & 59.36$\pm$0.01 & $-$9.792$\pm$0.006 & $-$25.244$\pm$0.019 & 93\% & 7\% & 0\%  & [1],[2] \\
GJ~576B\tablenotemark{b} & 19.046$\pm$0.009 & $-$607.63$\pm$0.02 & $-$506.51$\pm$0.02 & 71.41$\pm$0.03 & $-$84.79$\pm$0.12 & $-$47.25$\pm$0.07 & $-$62.79$\pm$0.03 & $-$49.81$\pm$0.09 & 0\% & 99\% & 0\%  & [1],[4] \\
G~204$-$39B\tablenotemark{b} & 13.987$\pm$0.003 & $-$15.07$\pm$0.02 & 578.14$\pm$0.02 & 38.343$\pm$0.008 & $-$31.150$\pm$0.002 & $-$33.704$\pm$0.008 & $-$7.164$\pm$0.002 & $-$0.288$\pm$0.002 & 99\% & 1\% & 0\%  & [1],[5] \\
Wolf~1130C\tablenotemark{b} & 16.585$\pm$0.007 & $-$1159.52$\pm$0.04 & $-$904.01$\pm$0.03 & 115.58$\pm$0.05 & $-$33.2$\pm$1.3\tablenotemark{c} & 117.46$\pm$0.05 & $-$31.7$\pm$1.2 & 42.2$\pm$0.2 & 18\% & 81\% & 0\%  & [1],[6] \\
HN~PegB\tablenotemark{b} & 18.133$\pm$0.011 & 231.11$\pm$0.03 & $-$113.20$\pm$0.03 & 22.12$\pm$0.014 & $-$16.7230$\pm$0.0013 & $-$3.340$\pm$0.006 & $-$8.848$\pm$0.005 & $-$3.510$\pm$0.012 & 99\% & 1\% & 0\%  & [1],[2] \\
{Wolf~940B}\tablenotemark{b} & 12.386$\pm$0.008 & 769.37$\pm$0.05 & $-$505.68$\pm$0.03 & 54.05$\pm$0.03 & $-$28.4$\pm$0.3 & $-$21.27$\pm$0.12 & $-$33.11$\pm$0.18 & $-$17.72$\pm$0.17 & 92\% & 8\% & 0\%  & \\
\hline
\multicolumn{13}{c}{Subdwarf Candidates} \\
\hline
J0055+5947\tablenotemark{b} & 22.84$\pm$0.03 & 408$\pm$25 & $-$24$\pm$25 & 44$\pm$3 & 34$\pm$25 & $-$44$\pm$15 & 16$\pm$23 & 4$\pm$3 & 96\% & 4\% & 0\%  & [1],[7] \\
J0140+0150 & 27$\pm$2 & 971$\pm$13 & $-$90$\pm$19 & 123$\pm$11 & 91$\pm$7 & $-$117$\pm$8 & $-$45$\pm$8 & $-$53$\pm$7 & 0\% & 99\% & 1\%  &[7] \\
J0411+4714 & 22$\pm$2 & 440$\pm$25 & $-$143$\pm$26 & 49$\pm$5 & 34$\pm$8 & $-$40$\pm$8 & $-$13$\pm$6 & 26$\pm$3 & 97\% & 3\% & 0\%  &[7] \\
J0429+3201\tablenotemark{d} & 35$\pm$3 & 30$\pm$31 & $-$458$\pm$31 & 75$\pm$8 & ($-$70$\pm$21) & (75$\pm$21) & ($-$62$\pm$8) & ($-$25$\pm$7) & 15\% & 85\% & 0\%  & [7] \\
J0433+1009 & 29$\pm$3 & 174$\pm$34 & $-$384$\pm$36 & 58$\pm$7 & 21$\pm$7 & 3$\pm$6 & $-$46$\pm$7 & $-$12$\pm$6 & 83\% & 17\% & 0\%  &[7] \\
J0623+0715 & 25$\pm$2 & 7$\pm$32 & $-$980$\pm$37 & 117$\pm$11 & $-$1$\pm$27 & 56$\pm$22 & $-$82$\pm$12 & $-$46$\pm$6 & 0\% & 99\% & 1\%  &[7] \\
J0659+1615 & 31$\pm$3 & $-$115$\pm$8 & $-$445$\pm$8 & 68$\pm$6 & $-$36$\pm$9 & 55$\pm$9 & $-$27$\pm$5 & $-$42$\pm$4 & 43\% & 57\% & 0\%  & [7] \\
J0758+5711 & 27$\pm$2 & $-$120$\pm$31 & $-$342$\pm$33 & 46$\pm$6 & 79$\pm$7 & $-$75$\pm$6 & $-$3$\pm$5 & 34$\pm$5 & 84\% & 16\% & 0\%  &[7] \\
J1110$-$1747 & 26$\pm$2 & 24$\pm$28 & $-$395$\pm$27 & 49$\pm$5 & $-$47$\pm$5 & 37$\pm$4 & 23$\pm$5 & $-$55$\pm$5 & 28\% & 71\% & 0\%  &[7] \\
J1130+3139 & 32$\pm$3 & $-$1185$\pm$35 & $-$1476$\pm$38 & 284$\pm$25 & 72$\pm$8 & $-$82$\pm$8 & $-$265$\pm$24 & 32$\pm$8 & 0\% & 0\% & 100\%  & [7] \\
J1204$-$2359 & 31$\pm$3 & $-$263$\pm$37 & $-$473$\pm$38 & 79$\pm$9 & $-$47$\pm$15 & $-$6$\pm$6 & $-$4$\pm$11 & $-$82$\pm$10 & 1\% & 98\% & 0\%  &[7] \\
J1304+2819 & 31$\pm$3 & $-$356$\pm$19 & 18$\pm$22 & 52$\pm$5 & 50$\pm$9 & $-$31$\pm$4 & $-$13$\pm$4 & 60$\pm$9 & 49\% & 51\% & 0\%  & [7] \\
J1308$-$0321 & 21$\pm$2 & $-$454$\pm$26 & 189$\pm$29 & 50$\pm$5 & $-$42$\pm$6 & $-$49$\pm$5 & 16$\pm$4 & $-$16$\pm$5 & 96\% & 4\% & 0\%  &[7] \\
J1401+4325 & 34$\pm$3 & $-$172$\pm$33 & $-$659$\pm$34 & 110$\pm$11 & $-$71$\pm$21 & 62$\pm$8 & $-$104$\pm$11 & $-$25$\pm$21 & 0\% & 98\% & 2\%  &[7] \\
J1458+1734 & 32$\pm$3 & $-$476$\pm$20 & 163$\pm$17 & 75$\pm$7 & 19$\pm$6 & $-$37$\pm$6 & $-$14$\pm$4 & 62$\pm$6 & 38\% & 62\% & 0\%  &[7] \\
J1524$-$2620 & 40$\pm$4 & $-$372$\pm$38 & $-$107$\pm$45 & 73$\pm$10 & 29$\pm$17 & 3$\pm$15 & $-$57$\pm$10 & 44$\pm$10 & 35\% & 65\% & 0\%  & [7] \\
J1710+4537 & 36$\pm$3 & $-$404$\pm$36 & 80$\pm$41 & 69$\pm$9 & $-$3$\pm$8 & $-$12$\pm$8 & $-$26$\pm$9 & 61$\pm$10 & 39\% & 61\% & 0\%  & [7] \\
J1801+4717 & 48$\pm$4 & 109$\pm$56 & 478$\pm$63 & 111$\pm$17 & $-$29$\pm$17 & $-$102$\pm$18 & 19$\pm$17 & $-$10$\pm$15 & 56\% & 44\% & 0\%  &[7] \\
J2013$-$0326 & 29$\pm$3 & $-$649$\pm$39 & $-$625$\pm$40 & 124$\pm$12 & $-$18$\pm$13 & 83$\pm$13 & $-$81$\pm$11 & 50$\pm$8 & 1\% & 99\% & 1\%  &[7] \\
J2021+1524 & 22$\pm$2 & $-$547$\pm$3 & $-$441$\pm$3 & 72$\pm$6 & $-$5$\pm$7 & 69$\pm$7 & $-$25$\pm$7 & 31$\pm$3 & 90\% & 10\% & 0\%  & [7] \\
J2112$-$0529 & 30$\pm$3 & $-$556$\pm$24 & $-$748$\pm$25 & 131$\pm$12 & $-$37$\pm$14 & 89$\pm$12 & $-$95$\pm$10 & 42$\pm$8 & 0\% & 99\% & 1\%  & [7] \\
J2112+3030 & 39$\pm$3 & $-$431$\pm$15 & 79$\pm$15 & 80$\pm$8 & $-$125$\pm$62 & 28$\pm$15 & $-$102$\pm$61 & 100$\pm$15 & 0\% & 66\% & 34\%  &[7] \\
J2218+1146 & 27$\pm$2 & 224$\pm$31 & $-$267$\pm$33 & 45$\pm$6 & $-$22$\pm$9 & 1$\pm$4 & $-$31$\pm$8 & $-$16$\pm$8 & 92\% & 8\% & 0\%  &[7] \\
J2251$-$0740 & 29$\pm$3 & 696$\pm$41 & 163$\pm$41 & 97$\pm$10 & $-$140$\pm$25 & $-$117$\pm$10 & $-$66$\pm$11 & 88$\pm$19 & 0\% & 93\% & 7\%  &[7] \\
\hline
\multicolumn{13}{c}{Metal-poor Comparison Sources} \\
\hline
J0532+8246 & 24.6$\pm$0.3 & 2038.8$\pm$0.6 & $-$1663.0$\pm$0.5 & 306$\pm$3 & $-$172$\pm$1 & $-$55$\pm$2 & $-$328$\pm$2 & 66$\pm$2 & 0\% & 0\% & 100\%  & [1],[8] \\
J0937+2931 & 6.12$\pm$0.07 & 973$\pm$6 & $-$1298$\pm$5 & 47.1$\pm$0.5 & $-$4.3$\pm$0.4 & 42.1$\pm$0.4 & $-$21.5$\pm$0.5 & 18.9$\pm$0.4 & 98\% & 2\% & 0\%  & [9],[10] \\
J1055+5443 & 6.9$\pm$0.7 & $-$1519$\pm$2 & $-$223$\pm$2 & 50$\pm$5 & 39$\pm$22 & $-$50$\pm$12 & 1$\pm$6 & 21$\pm$18 & 93\% & 7\% & 0\%  & [11],[7] \\
J1416+1348B\tablenotemark{b} & 9.282$\pm$0.019 & 86.7$\pm$0.3 & 128.0$\pm$0.2 & 6.799$\pm$0.017 & $-$42$\pm$5 & $-$7$\pm$2 & 18.02$\pm$0.12 & $-$31$\pm$5 & 93\% & 7\% & 0\%  & [1],[12] \\
J1553+6933 & 34$\pm$3 & $-$1451$\pm$33 & 1327$\pm$29 & 321$\pm$28 & 110$\pm$90 & $-$299$\pm$31 & $-$29$\pm$73 & 137$\pm$62 & 0\% & 4\% & 96\%  &[7] \\
J1810$-$1010 & 8.9$\pm$0.6 & $-$1027$\pm$4 & $-$264$\pm$4 & 45$\pm$3 & 46$\pm$4 & 62$\pm$4 & $-$2$\pm$3 & 43$\pm$3 & 91\% & 9\% & 0\%  & [13] \\
\enddata 
\tablenotetext{a}{Distance based on parallax measurement of source or primary companion, or estimated from Eqns.\ref{eqn:mj} and~\ref{eqn:mw2}.}
\tablenotetext{b}{Analysis based on kinematics of the primary component of this system.}
\tablenotetext{c}{Center of mass motion of the Wolf~1130AC binary based on orbit fits \citep{2018ApJ...854..145M}.}
\tablenotetext{d}{RV and $UVW$ velocities may be systematically biased by slit-seeing effect and should be treated with caution.}
\tablerefs{
[1] Gaia EDR3 \citep{2021AandA...650C...3G}; 
[2] \citet{2018AandA...616A...7S}; 
[3] \citet{2020AJ....160...83S}; 
[4] \citet{2018AandA...619A..81H};
[5] \citet{2018MNRAS.475.1960F};
[6] \citet{2018ApJ...854..145M};
[7] This paper;
[8] \citet{2006AJ....131.1806R};
[9] \citet{2012ApJ...752...56F};
[10] \citet{2021ApJS..257...45H};
[11] \citet{2021ApJS..253....7K};
[12] \citet{2010AandA...510L...8S};
[13] \citet{lodieu2022}
}
\end{deluxetable} 
\end{longrotatetable}

\begin{figure*}[ht!]
\centering
\plotone{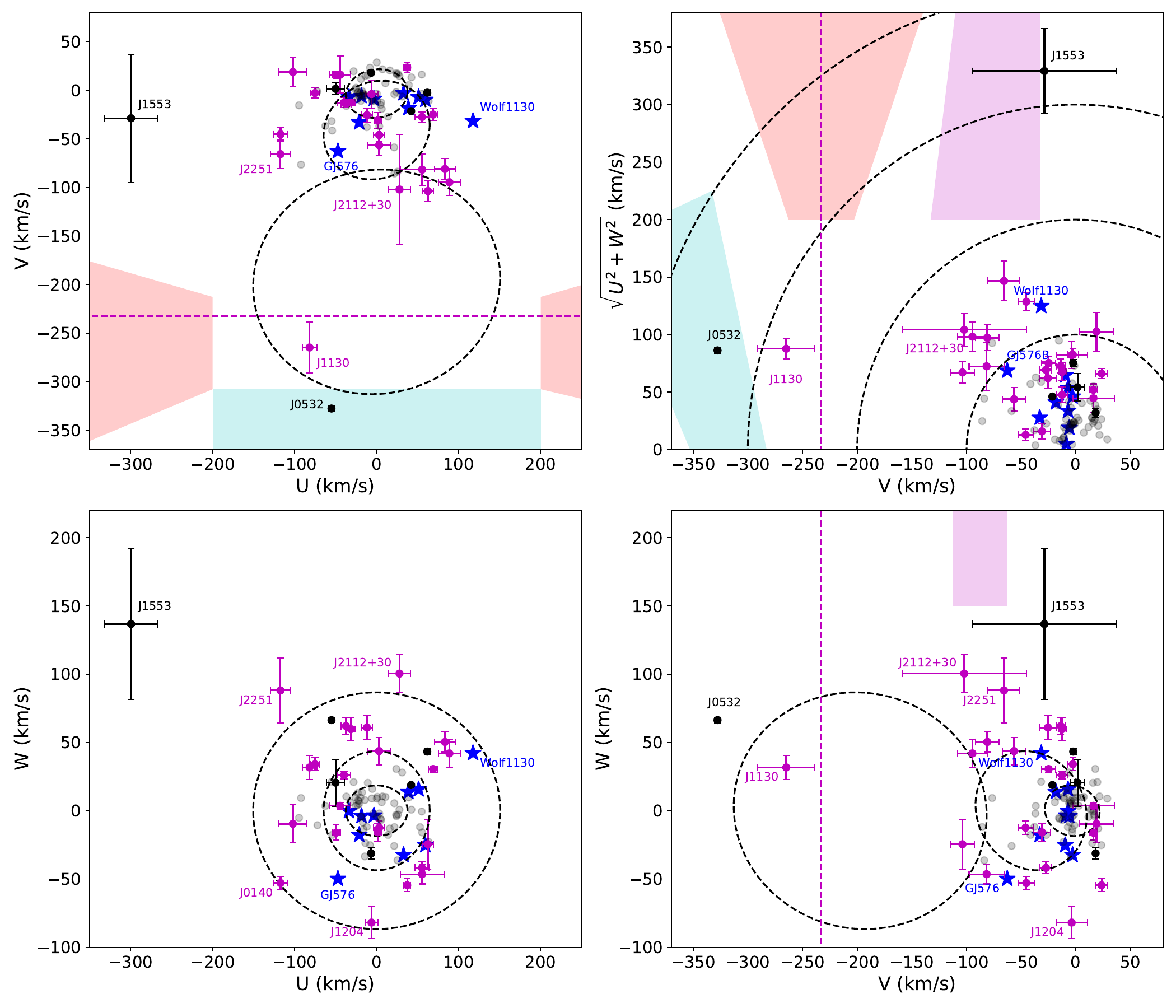}
\caption{
Local Standard of Rest $UVW$ velocities of sources in our Keck/NIRES sample, along with comparison velocities of nearby late-L and T dwarfs from \citet[grey dots]{2021ApJS..257...45H}. Symbol colors segregate subdwarf candidates (magenta), metal-poor comparison sources (black circles), and metallicity benchmarks (blue stars).
In the upper left, lower left, and lower right panels we compare pairs of velocity components, with dashed lines indicating the 1$\sigma$ velocity dispersion spheres for chemically-selected thin disk, thick disk, and halo stars (in increasing size) based on the APOGEE and Gaia DR2 compilation of \citet{2020AJ....160...43A}.
The upper right panel displays a Toomre plot with circles indicating total speeds in steps of 100~km/s.
%relative to the Local Standard of Rest.
In the plots containing the $V$ velocity, we indicate the threshold between prograde and retrograde motion in the Galactic frame of rest (i.e., $L_z$ = 0) by the dashed magenta lines at $V$ = $-$232.8~km/s \citep{2017MNRAS.465...76M}.
We also show schematic regions in the Toomre plot for the Gaia-Enceladus (red), Thamnos (cyan), and Helmi streams (magenta) based on \citet{2019AandA...631L...9K} for velocity pairs where these populations are distinct.
Sources discussed in the text are labeled.
\label{fig:kinematics}}
\end{figure*}

Using the measured proper motions and radial velocities, and parallax or estimated distances, we computed Galactic $U$, $V$, $W$ velocities in the Local Standard of Rest (LSR) assuming a LSR solar velocity of (11.10, 12.24, 7.25)~km/s
\citep{2010MNRAS.403.1829S}.
Uncertainties were propagated by the Monte Carlo method assuming normal distributions.
These values are listed in Table~\ref{tab:kinematics} and plotted in Figure~\ref{fig:kinematics} along with the local late-L and T dwarf kinematic sample of \citet{2021ApJS..257...45H}.
As expected for our kinematic selection, this sample is more widely dispersed than the local dwarf population, with dispersions of $\sigma_U$ = 63~km/s, $\sigma_V$ = 60~km/s, and $\sigma_W$ = 48~km/s, and a mean azimuthal drift velocity of $\langle{V}\rangle$  = $-$17~km/s.
Two candidates, J1130+3139 and J2112+3030, stand out for their large negative $V$ velocities, with J1130+3139 showing retrograde motion with respect to the Galactic rest frame.
The candidates J1204$-$2359, J2112+3030, and J2251$-$0740 show large vertical velocities with $|W|$ $>$ 75~km/s.
Our RV measurement and estimated distance of {the T subdwarf} J1553+6933 places this source in a particularly unusual position in $UVW$ velocity space, with a total LSR speed of $\approx$330~km/s, comparable to J0532+8246 but with distinct $UVW$ component velocities. 
Overall, we find that the majority of our targets lie outside the 1$\sigma$ velocity sphere for thin disk sources defined by \citet{2020AJ....160...43A}.

From the $UVW$ velocities, we computed the kinematic probabilities of membership in the Galactic thin disk ($p_{thin}$), thick disk ($p_{thick}$), and halo ($p_{halo}$) following \citet{2003AandA...410..527B}, assuming $p_{thin}+p_{thick}+p_{halo} = 1$. High probability of thin disk membership does not rule out thick disk or halo membership (cf.\ J1810$-$1010), while high probability of thick disk or halo membership likely rules out thin disk membership. 
Three sources in our sample, the previously identified subdwarfs J0532+8246 and J1553+6933, and the subdwarf candidate J1130+3139, all have high probabilities of halo membership ($>$90\%) due to their extreme LSR velocities; while eight of our subdwarf candidates have thick disk probabilities exceeding 90\%.

\section{Assessment of T Subdwarf Candidates\label{sec:individual}}

\subsection{New T Subdwarfs}

With a thorough assessment of the spectral and kinematic properties of our candidate T subdwarfs, 
%including those with prior spectroscopic observations, 
we qualitatively evaluated
%a case-by-case analysis of each source to assess 
their likelihood of being a metal-poor brown dwarf by considering four criteria: 
(1) the degree of deviation of the near-infrared spectrum from the dwarf standards,
(2) the presence and strength of 1.25~$\mu$m K~I absorption compared to solar-metallicity field dwarfs {for sources with medium-resolution data and spectral types $\leq$T7,}
(3) the metallicity inferred from model fitting, and 
(4) $UVW$ kinematics and the probability of thick disk or halo membership {for sources with RV measurements}.
We treated these criteria in a holistic manner, noting that some criteria vary across spectral type (e.g, metallicity deviations are more pronounced among cloudy late-L and early-T dwarfs) and some may not clarify subdwarf status (e.g., thin disk kinematics does not rule out halo membership). We identified three broad categories for our candidates: likely subdwarfs (sd and esd), ``mild'' subdwarfs (d/sd), and not subdwarfs. Table~\ref{tab:assessment} summarizes our assessments for both candidates and metal-poor comparison sources.

\startlongtable
\begin{deluxetable}{lcccccll}
\tablecaption{Assessment of T Subdwarf Candidates and Comparison Sources \label{tab:assessment}} 
\tabletypesize{\scriptsize} 
\tablehead{ 
 & \multicolumn{4}{c}{Assessment Metrics} &
 & \multicolumn{2}{c}{Spectral Type} \\
\cline{2-5} \cline{7-8}
\colhead{Name} & 
\colhead{deviant?} & 
\colhead{K~I?} & 
\colhead{model?} & 
\colhead{UVW?} & 
% \colhead{\teff} & 
% \colhead{[M/H]} & 
\colhead{$\zeta_T$\tablenotemark{a}} & 
\colhead{Prior} & 
\colhead{Adopted} 
} 
\startdata 
\hline
\multicolumn{8}{c}{Confirmed T Subdwarfs} \\
\hline
J0623+0715  & Y & Y & Y & y & 0.79$\pm$0.09 & \nodata & sdT3 \\
{J1524$-$2620} & y? & y & Y & y & 0.74$\pm$0.09 & \nodata & sdT0 \\
J2112$-$0529 & Y & n & Y & y & 0.79$\pm$0.13 & \nodata & sdT1 \\
\hline
\multicolumn{8}{c}{Mild T Subdwarfs} \\
\hline
{J0045+7958} & y & N & y? & \nodata & 0.99$\pm$0.17 & \nodata & d/sdL9: \\
{J0055+5947} & y? & y & n? & n & 0.78$\pm$0.11 & T8 & d/sdT6.5 \\
J0411+4714 & y & y & y  & n & 0.95$\pm$0.18 & \nodata & d/sdT7.5 \\
J0845$-$3305  & y & \nodata  & y  & \nodata & 0.85$\pm$0.11 & T7 & d/sdT6.5 \\
J0911+2146 & y & \nodata &  y? & \nodata & 0.70$\pm$0.14 & T9 & d/sdT8 \\
J0953$-$0943 & y & \nodata  & y  & \nodata & 0.78$\pm$0.13  & T5.5 & d/sdT6 \\
J1130+3139 & y & Y & y & Y & 0.86$\pm$0.11 & \nodata & d/sdT5.5 \\
{J1138+7212} & Y & \nodata & y & \nodata & 0.76$\pm$0.17 & \nodata & d/sdT7 \\
J1204$-$2359 & y & y & y & y & 0.90$\pm$0.15 & \nodata & d/sdT7 \\
{J1304+2819} & y & n & n & y? & 0.90$\pm$0.13 & \nodata & d/sdL9: \\
J1308$-$0321 & y & \nodata & y & n & 1.08$\pm$0.22 &  \nodata & d/sdT8 \\
J1401+4325 & y? & Y & y & y & 0.87$\pm$0.08 & \nodata & d/sdT5.5 \\
{J1515$-$2157} & y? & n & y? & \nodata & 0.71$\pm$0.11 & \nodata & d/sdT6 \\
{J1801+4717} &  n & Y & y? & n?  & 0.93$\pm$0.08 & \nodata & d/sdT5 \\
J2013$-$0326 & y & Y & y & y & 0.79$\pm$0.08 &  \nodata & d/sdT6 \\
{J2021+1524} &  Y & n & Y & n & 0.94$\pm$0.15 & \nodata & d/sdL9  \\
J2112+3030 & y? & Y & y & Y & 0.96$\pm$0.13 & \nodata & d/sdT2.5 \\
J2218+1146 & y & y & y? & n & 0.85$\pm$0.09 & T7p & d/sdT6.5 \\
J2251$-$0740  & y? & y & n? & y & 0.99$\pm$0.19 & \nodata & d/sdT7 \\
\hline
\multicolumn{8}{c}{Not Subdwarfs} \\
\hline
J0140+0150 & n & N & N & y & 1.11$\pm$0.11 & \nodata & rT4.5 \\
J0429+3201 & n & y & y & \nodata & 1.04$\pm$0.11 & \nodata & T1 \\
J0433+1009 & n & \nodata & n? & n? & 1.12$\pm$0.17 & T8 & T8 \\
J0659+1615 & n & n & y? & y? & 0.97$\pm$0.10 & \nodata & T1 \\
J0758+5711 & n & n  & y & n? & 0.93$\pm$0.12 & \nodata & T6.5 \\
J0843+2904 & n & \nodata  & n? & \nodata & 0.93$\pm$0.11 & \nodata & L6: \\
J1110$-$1747 & n & \nodata & n & y & 1.28$\pm$0.14 & \nodata & rT8 \\
J1458+1734  & n & \nodata & n? & y? & 1.20$\pm$0.17 & \nodata & T8 \\
J1710+4537 &  n & y & n? & y? & 1.00$\pm$0.11 & \nodata & T6 \\
\hline
\multicolumn{8}{c}{Metal-poor Comparison Sources} \\
\hline
{HD~65486B} & n & \nodata & Y & n & 0.95$\pm$0.11 & T4.5 & d/sdT4.5 \\
LHS~6176B &  y & y & y & n &  0.76$\pm$0.12 & T8p & d/sdT7.5 \\
{HD~126053B} & Y & \nodata & Y & n & 0.38$\pm$0.05 & T8p & sdT7.5 \\
GJ~576B &  y & Y & y & y & 0.78$\pm$0.05\tablenotemark{b} & sdT5.5 & d/sdT5.5 \\
Wolf~1130C & Y & Y & y & y & \nodata & T8p & (e)sdT6: \\
J0004$-$1336 & y  & \nodata  & Y  & \nodata  & 0.91$\pm$0.11  & T2p & d/sdT2: \\
J0004$-$2604 & n  & \nodata  & y  & \nodata  & 0.91$\pm$0.06  & sdT2 & T3 \\
{J0013+0634} & Y & \nodata & n & \nodata & 0.33$\pm$0.18 & T8p & sdT7.5 \\
J0021+1552 &  n & Y & y  & \nodata & 0.81$\pm$0.13 & T4p & d/sdT4 \\
J0057+2013 & Y & \nodata & y? & \nodata & \nodata & sdL7 & d/sdL9 \\
J0301$-$2319 & y  & \nodata  & y  & \nodata  & 0.89$\pm$0.11  & sdT1 & d/sdT1 \\
J0309$-$5016 & y  & \nodata  & y  & \nodata  & 0.49$\pm$0.08  & T7p & d/sdT7.5 \\
J0348$-$5620 & y  & \nodata  & n  & \nodata  & 0.89$\pm$0.08  & T3p & d/sdT4 \\
J0414$-$5854 &  Y & \nodata & Y  & \nodata & 0.32$\pm$0.16 & esdT & esdT6: \\
J0532+8246 & Y & n &  Y & Y & 0.58$\pm$0.07 & esdL7 & esdL8: \\
J0616$-$6407  &  Y  & n  &  Y & \nodata & 0.54$\pm$0.06 & esdL6 & esdT0: \\
J0645$-$6646  &  y  & N  &  y? & \nodata & \nodata & sdL8 & d/sdT0 \\
{J0833+0052} & y & \nodata & y? & \nodata & 0.62$\pm$0.13 & T9p & d/sdT9 \\
J0850$-$0221 & y  & \nodata  & y  & \nodata  & \nodata  & sdL7 & d/sdL6 \\
J0937+2931 &  Y  & \nodata  &  Y & n & 0.64$\pm$0.18 & T6p & sdT6 \\
J0939$-$2448 & y & \nodata &  y & \nodata & 0.45$\pm$0.09 & T8p & d/sdT8 \\
J1019$-$3911 & y?  & \nodata  & n  & \nodata  & 1.00$\pm$0.17 & sdT3 & T4 \\
J1035$-$0711 & Y  & \nodata  & y  & \nodata  & \nodata  & sdL7 & d/sdL9 \\
J1055+5443 & N & \nodata & N & n & \nodata & sdT8 & Y0 \\
J1130$-$1158 & Y & Y & Y  & \nodata & 0.45$\pm$0.18 & sdT5.5 & sdT5.5 \\
J1158+0435 &  Y & \nodata  & y  & \nodata & \nodata & sdL7 & d/sdL8 \\
J1307+1510 & n  & N  &  y? & \nodata & \nodata & sdL8 & T1 \\
J1316+0755  & Y & \nodata  &  Y & \nodata & 0.37$\pm$0.17 & sdT6.5 & sdT6.5 \\
J1338$-$0229 & y & \nodata & y & \nodata & \nodata & sdL7 & sdL9 \\
J1416+1348B  & Y  & \nodata  & Y  & n &  0.23$\pm$0.06 & T7.5p & sdT7 \\
J1553+6933 &  Y  & Y & Y  & Y & 0.65$\pm$0.05 & sdT4 & sdT4 \\
J1810$-$1010 &  Y  & Y & Y  & n & 0.49$\pm$0.07 & esdT & esdT3: \\
J2105$-$6235 & y?  & \nodata  & y  & \nodata  & 0.89$\pm$0.05  & sdT1.5 &  d/sdT2: \\
\enddata 
\tablenotetext{a}{Metallicity index from infrared indices; see Section~\ref{sec:zeta} and Table~\ref{tab:zeta}.}
\tablenotetext{b}{Weighted average from multiple spectra.}
\tablecomments{The qualitative assessment metrics are as follows: 
``Y'': strong positive evidence,
``y'': marginal positive evidence,
``n'': marginal negative evidence,
``N'': strong negative evidence, and
``...'': unmeasured or unable to evaluate for this source (e.g., K~I for $\geq$T8 sources {and low resolution spectra}).
}
\end{deluxetable} 

From our candidate sample, we identify J0623+0715, {J1524$-$2620,} and J2112$-$0529 as high probability early-type T subdwarfs, as {all three} exhibit strong deviations from dwarf templates, [M/H] $\lesssim$ $-$0.7 from model fits, and thick disk kinematics. J0623+0715 also lacks detectable K~I absorption, while {J1524$-$2620 and J2112$-$0529 have} weak K~I absorption compared to dwarf L/T transition objects. 
We identify a further {19} candidates as mild subdwarfs, predominantly late-type T dwarfs with $K$-band suppression, modestly subsolar metallicities ([M/H] $\approx$ $-$0.3) from model fits, and a mixture of population assignments. The remaining {9} candidates appear to be normal late-L and T dwarfs, although a few of these sources warrant further scrutiny as discussed below.

We also assessed the subdwarf status of our comparison sample following the same rubric. While the majority {are found to be true subdwarfs}, several sources {appear} to be normal T dwarfs.
We reclassify
J0004$-$2604 and J1019$-$3911, classified sdT2 and sdT3 by \citet{2019AJ....158..182G}, as T3 and T4 dwarfs based on their excellent agreement with the dwarf standards.
We find J1307+1510, classified sdL8 by \citet{2018MNRAS.480.5447Z}, to be well-matched to the T1 dwarf standard with strong K~I absorption.
We also confirm the analysis of \citet{2023ApJ...958...94R} that J1055+5443 is not an sdT8, as photometrically estimated by \citet{2021ApJS..253....7K}, but rather a nearby (6.9$\pm$0.7~pc) Y0 dwarf with unusual colors. Indeed, our spectral model fits suggests this source is highly metal-rich: [M/H] = +0.70$^{+0.21}_{-0.17}$ from LOWZ model fits and [M/H] = +0.78$^{+0.19}_{-0.28}$ from {equivalently matching} ElfOwl model fits. However, both fits predict small radii ($\sim$0.02~R$_\odot$) that are not physical, indicating an overestimated temperature and a likely unreliable metallicity. Our inferred radial velocity of 39$\pm$22~km/s for this source has insufficient precision to confirm or rule out membership in the $\sim$180~Myr Crius 197 association \citep{2022ApJ...939...94M} as proposed by \citet{2023ApJ...958...94R}.

\subsection{The Remarkable Galactic Orbits of J1130+3139 and J1553+6933}

\begin{figure}[tbp]
\centering
\includegraphics[width=0.4\textwidth]{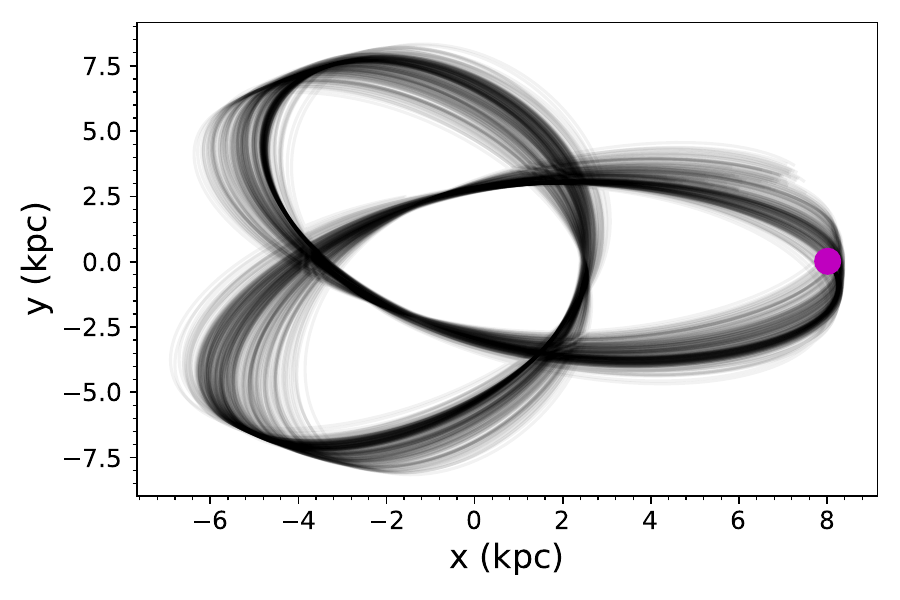}
\includegraphics[width=0.4\textwidth]{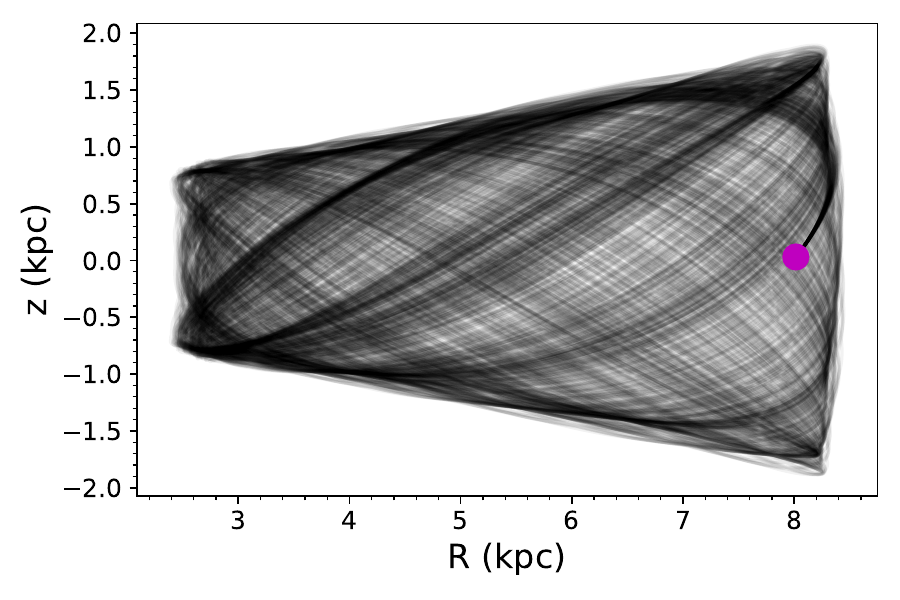} \\
\includegraphics[width=0.4\textwidth]{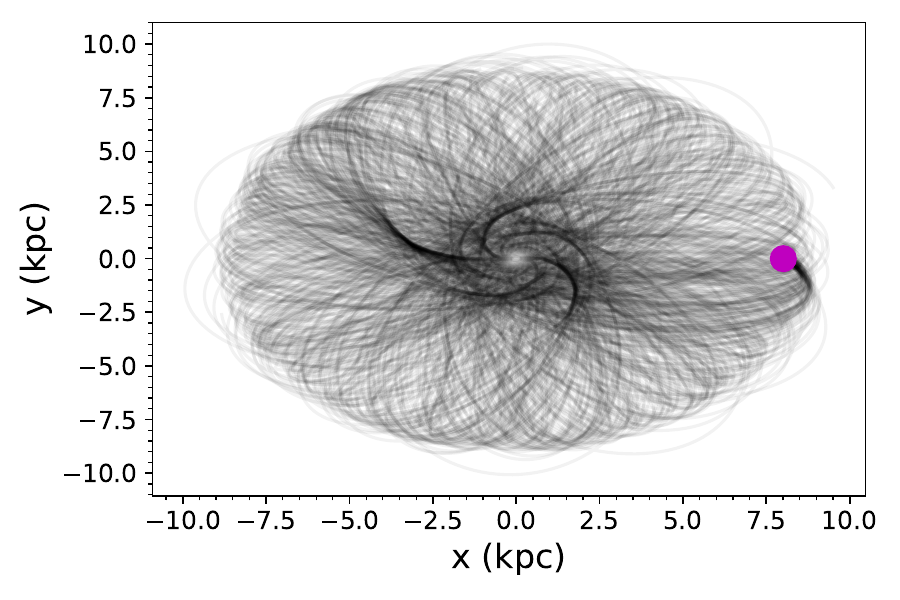}
\includegraphics[width=0.4\textwidth]{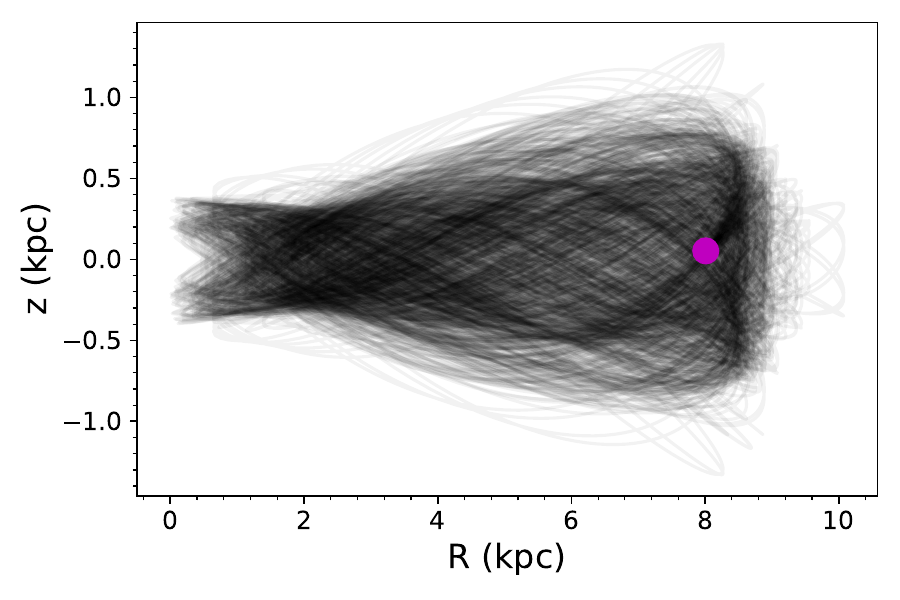} \\
\includegraphics[width=0.4\textwidth]{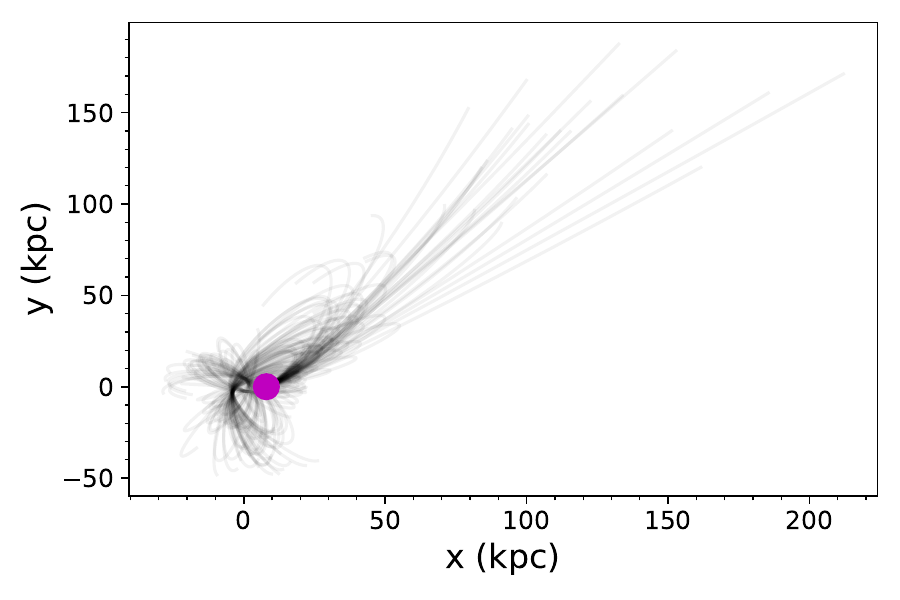}
\includegraphics[width=0.4\textwidth]{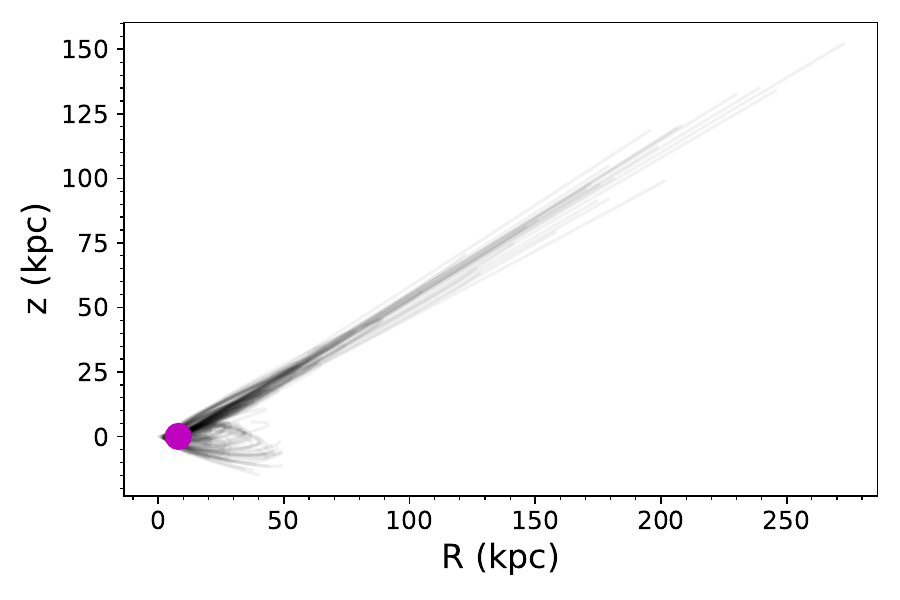} \\
\includegraphics[width=0.4\textwidth]{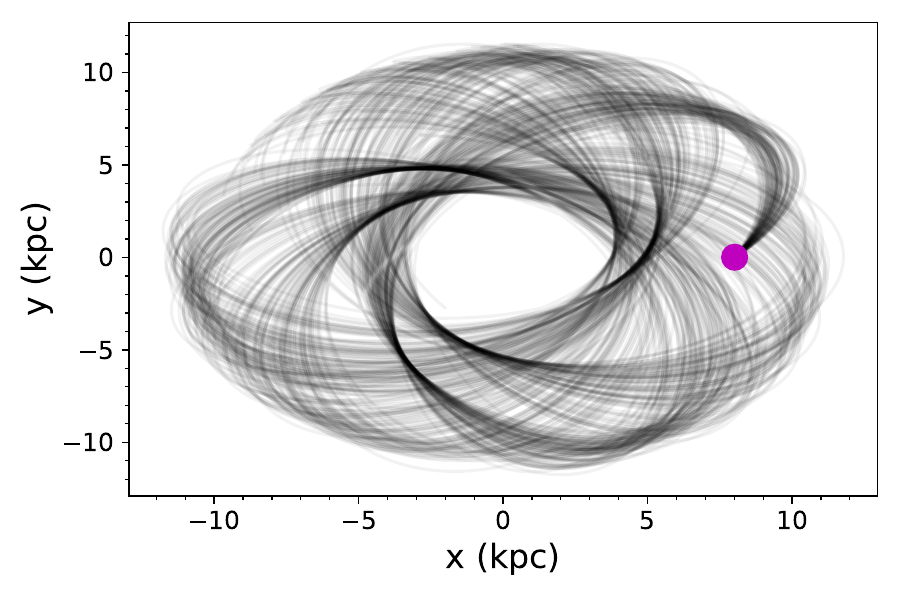}
\includegraphics[width=0.4\textwidth]{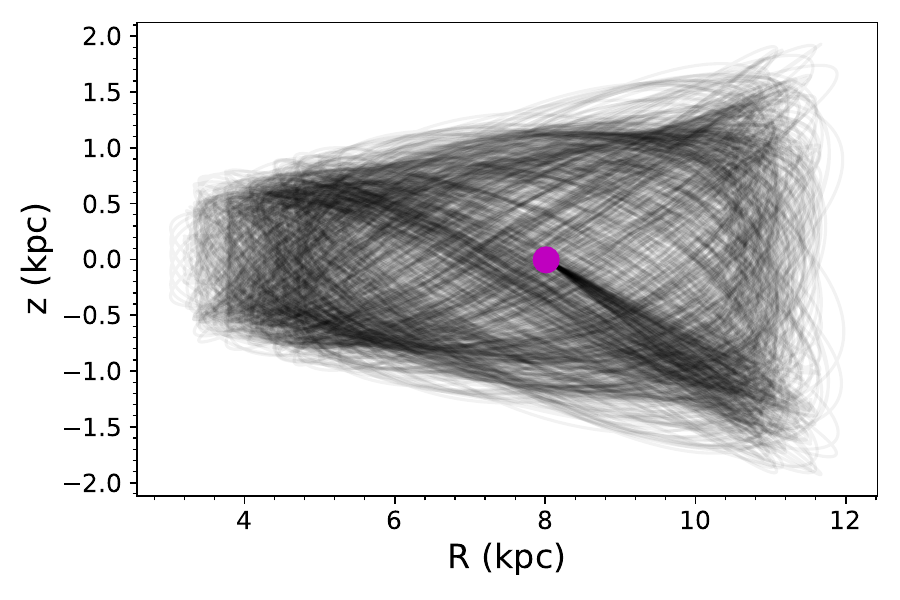} \\
\caption{Galactic orbits for the high velocity sources (from top to bottom): J0532+8246, J1130+3139, J1553+6933, and J0140+0150. Orbits were calculated {forward in time by} 1~Gyr using \texttt{galpy} \citep{2015ApJS..216...29B} with the MWPotential2014 potential. {We assumed} a solar motion from \citet{2010MNRAS.403.1829S}, a local circular velocity of 220~km/s, and R$_\odot$ = 8~kpc. For J1553+6933, we extended the integration to 5~Gyr. The current location of each source is indicated by the magenta circle. We show 100 realizations of each object's orbit by sampling uncertainties in distance, proper motion, and RV.
\label{fig:orbits}}
\end{figure}

Our $UVW$ analysis identified three sources that stand out in their Galactic motion. 
The L subdwarf J0532+8246 has previously been recognized as an object on a retrograde Galactic orbit \citep{2008ApJ...672.1159B}.
The subdwarf candidate J1130+3139 was found to have similar $UV$ component velocities that place it on a nearly radial Galactic orbit. The orbits for both objects are illustrated in Figure~\ref{fig:orbits},
and were computed using the \texttt{galpy} package \citep{2015ApJS..216...29B} assuming the MWPotential2014 potential, solar motion from \citet{2010MNRAS.403.1829S}, a solar radius of 8~kpc, and a local circular velocity of 220~km/s. These figures display 100 orbit realizations sampling uncertainties in distance, proper motion, and radial velocity, all of which dominate over possible model uncertainties for J1130+3139 (e.g., choice of Galactic potential). Both sources are currently near apogal, with J0532+8246 reaching an inner radius of 2.6~kpc and J1130+3139 reaching an inner radius of 0.4--1.8~kpc, well within the bulge. Both sources are also closely aligned in velocity space with the Thamnos streams identified by \citet{2019AandA...631L...9K}, whose members span a broad range of metallicities that encompass those inferred from model fits for these two sources.

The other T subdwarf with unusual kinematics is J1553+6933, previously reported in \citet{2021ApJ...915..120M}.
This source's $UVW$ velocities are distinct from all of the other sources in our sample. 
Its large inferred Galactic radial ($U$) velocity places it on an orbit that reaches R $>$ 25~kpc at apogal, with distribution tails extending as far as 200~kpc from the Galactic center. However, this source also has the most poorly measured RV in our sample, +110$\pm$90~km/s, due to the lack of clear spectral features across the near-infrared region and very low S/N. Nevertheless, with a $V_{tan} >$ 300~km/s (similar to J0532+8246), and a velocity that aligns it with the prograde Helmi streams \citep{1999Natur.402...53H}, this source warrants further kinematic study.

\subsection{Metal-rich Thick-Disk Sources}

Three sources in our sample, J0140+0150, J1110$-$1747, and J1458+1734 show evidence of kinematic membership in the Galactic thick disk population
%, with membership probabilities of 99\%, 73\%, and 62\% based on relatively precise velocities ($\sigma_{UVW}$ $\lesssim$ 10~km/s). However, the 
but have spectra best fit by {solar or} supersolar metallicity models. J0140+0150, with 99\% thick disk membership probability, is found to be the most metal-rich at [M/H] = +0.40$^{+0.11}_{-0.19}$, and
has the strongest K~I lines in the sample (see Table~\ref{tab:ews} and Figure~\ref{fig:ews}).
J1110$-$1747 is one of the few sources to show excess flux at $K$-band compared to its closest-match dwarf standard (Figure~\ref{fig:spectra-all}).
%while the lack of K~I in J1110$-$1747 and J1458+1734 is consistent with other T8 dwarfs.
The nature of these sources suggests that they may be members of the high velocity thick disk, which have high LSR speeds ($v_{tot} > $ 220~km/s) and supersolar metallicities of up to [Fe/H] $\sim$ +0.5 \citep{2020ApJ...903..131Y,2021ApJ...915....9Z}. 
\citet{2017ApJ...845..101B} have demonstrated that these stars likely formed in the inner Milky Way and scattered outward through radial migration. Indeed, the Galactic orbit of J0140+0150 is modestly eccentric (e = 0.45), extending over a radius range of 4--11~kpc (Figure~\ref{fig:orbits}). All three sources have relatively large vertical velocities, $|W| > $ 50~km/s, pushing them to maximum displacements of $\pm$1--2~kpc. While not formally part of the ``high'' velocity tail, these sources demonstrate that thick disk kinematics are not sufficient evidence for subsolar metallicity.

\section{Establishing a T Dwarf Metallicity Classification System\label{sec:classify}}

\subsection{Building a Temperature and Metallicity Grid}

Our comprehensive spectral sample of solar- and subsolar-metallicity T dwarfs
provides the opportunity to establish an empirical classification system
for cold brown dwarfs that encapsulates both temperature and metallicity effects.
We followed prior approaches to qualifying and quantifying metallicity effects in M dwarfs (e.g., \citealt{1997AJ....113..806G,2007ApJ...669.1235L,2019ApJS..240...31Z}) and L dwarfs (e.g., \citealt{2007ApJ...657..494B,2017MNRAS.464.3040Z,2018MNRAS.480.5447Z,2019AJ....158..182G}).
In particular, we aimed to define T subdwarf metallicity classes that roughly align with the metallicities of the existing M and L subdwarf classes.

\citet{1997AJ....113..806G} originally proposed two metallicity classes for metal-poor M subdwarfs with [M/H] $\approx$ $-$1.2 for sdMs and [M/H] $\approx$ $-$2 for edMs.
Subsequent expansion of {this} system into dM, sdM, esdM, and usdM classes by \citet{2007ApJ...669.1235L}, and calibration of metallicity scales using companions of metal-poor FGK primaries and improved spectral models 
have refined these estimates. 
A recent analysis of over 1500 M dwarf and subdwarf optical spectra by \citet{2020AJ....159...30H} with classifications on the \citet{2007ApJ...669.1235L} system yields median metallicities of 
[M/H] = 0.15$\pm$0.27 for dM,
[M/H] = $-$0.75$\pm$0.43 for sdM,
[M/H] = $-$1.42$\pm$0.40 for esdM, and
[M/H] = $-$1.88$\pm$0.18 for usdM
(see also \citealt{2009PASP..121..117W,2019AandA...628A..61L}).
\citet{2017MNRAS.464.3040Z} used model fits to optical and infrared spectra to argue for a similar metallicity scale for their defined L dwarf metallicity classes:
[M/H] $\gtrsim$ $-$0.3 for dL,
$-$1.0 $\lesssim$ [M/H] $\lesssim$ $-$0.3 for sdL,
$-$1.7 $\lesssim$ [M/H] $\lesssim$ $-$1.0 for esdL, and
[M/H] $\lesssim$ $-$1.7 for usdL sources.
Given the range of metallicities inferred from our own model fits of the T dwarfs examined here, we restrict our metallicity grid to dT, sdT, and esdT classes, and introduce a mild subdwarf class d/sdT to 
identify sources with modest subsolar metallicities which are nevertheless distinguishable from solar-metallicity dwarfs (cf.~\citealt{1967MNRAS.136..403C,2007ApJ...657..494B,2009ApJ...706.1114B}).
%recognize the sensitivity of lower-temperature atmospheres to metallicity variations.
%, in particular through reduced cloud opacity and enhanced H$_2$ absorption.
%with {\teff} $\leq$ 3200~K (spectral type $\geq$ M2).
% lodieu: [Fe/H] = −0.5, −1.5, and −2.0 ± 0.5 dex

% and surface gravity effects in late-M and L dwarfs (e.g., \citealt{2013ApJ...772...79A})
% by first defining a set of empirical indices that segregate sources based on metallicity,
% then attempting to calibrate these indices to metallicity values using benchmark sources and atmosphere models.

We adopted a qualitative {empirical} approach to define temperature and metallicity subclasses,
based on principles inherent to the \citet{1943assw.book.....M} stellar classification system. Our system is defined by a grid of standard spectra covering a common wavelength range and resolution (see \citealt{1973ARAandA..11...29M,2005ARAandA..43..195K}).
We resampled our entire spectral sample onto a common wavelength scale spanning 1.0--2.4~$\mu$m at a fixed resolution {\ldl} = 300, and normalized all data at the peak flux between 1.2--1.3~$\mu$m. At this resolution, we are primarily focused on the behavior of broad molecular absorption bands and overall spectral shape, rather than atomic lines. 
To anchor our system to the existing T dwarf classification system \citep{2006ApJ...637.1067B}, we used numerical subtypes that parallel the evolution of molecular features through the dwarf track, notably the 1.6 $\mu$m CH$_4$ and 1.15~$\mu$m H$_2$O+CH$_4$ absorption features,
%; increasingly blue near-infrared spectral energy distributions driven by these band and strengthening H$_2$ absorption; 
and the gradual narrowing of the 1.25~$\mu$m and 1.6~$\mu$m spectral peaks.
Metallicity types were then segregated based on deviations associated with metallicity effects: unusually blue near-infrared spectral energy distributions driven by reduced cloud opacity in late-L and early-T dwarfs, and enhanced H$_2$ in later T dwarfs; 
a ``muting'' of H$_2$O and CH$_4$ bands beyond 1.3~$\mu$m as H$_2$ dominates near-infrared opacities; and a broadening of the 1.1~$\mu$m spectral trough, possibly driven by enhanced pressure-broadening of the 0.77~$\mu$m K~I alkali line wings. 
Mild subdwarfs have spectra similar to their dwarf counterparts but with a slightly enhanced 1.05~$\mu$m peak and suppressed 2.1~$\mu$m peak.
Extreme subdwarfs have highly distinct spectra from their closest dwarf counterpart, which were
assigned by matching the depth of the 1.1~$\mu$m trough. 

\begin{figure*}[ht!]
\centering
\includegraphics[width=0.32\textwidth]{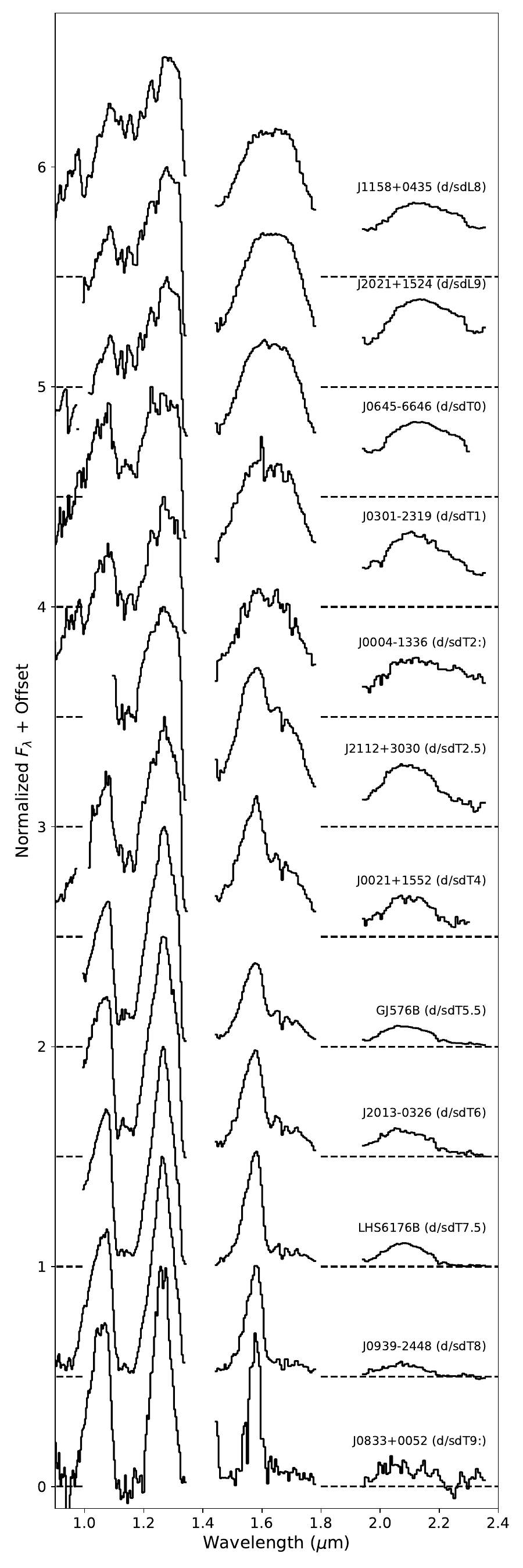}
\raisebox{1in}{\includegraphics[width=0.32\textwidth]{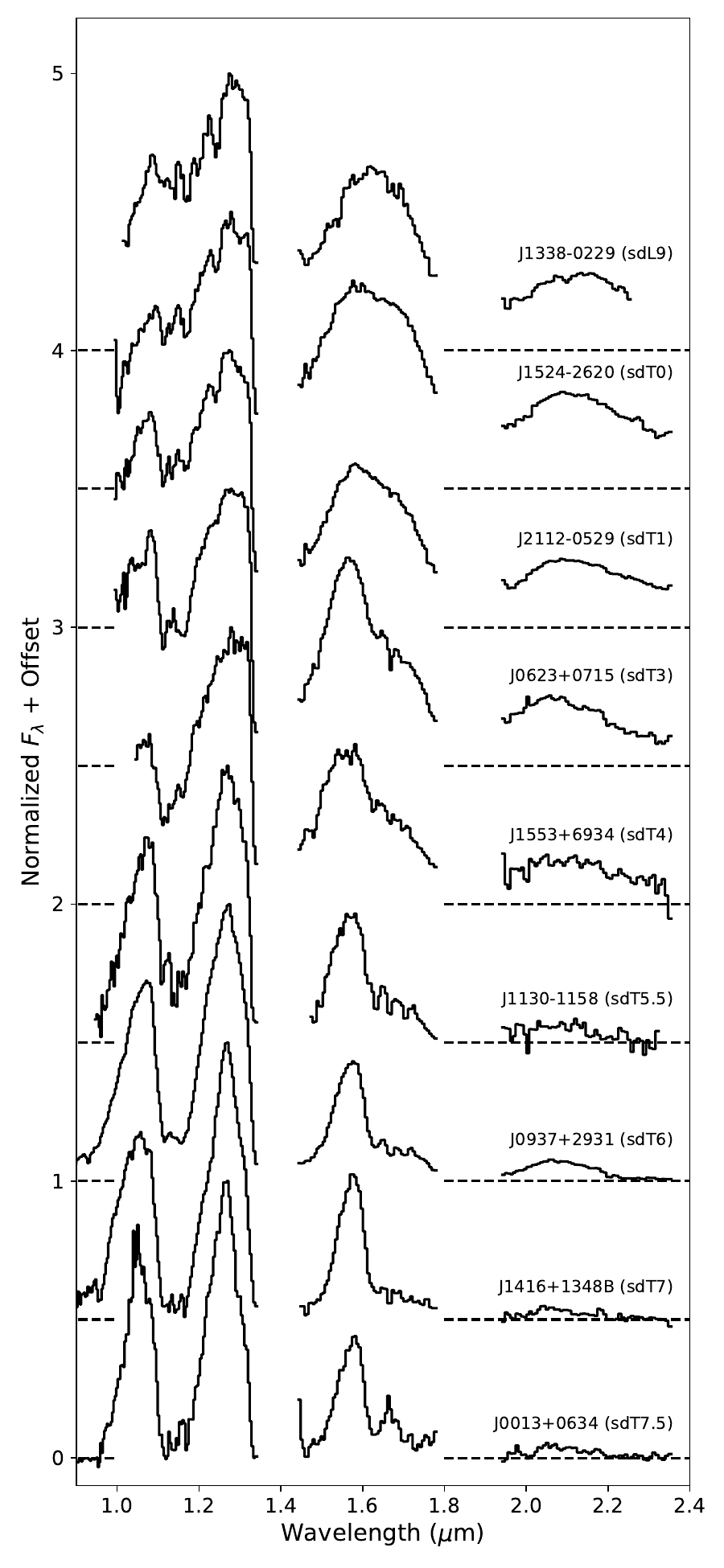}}
\raisebox{1.6in}{\includegraphics[width=0.32\textwidth]{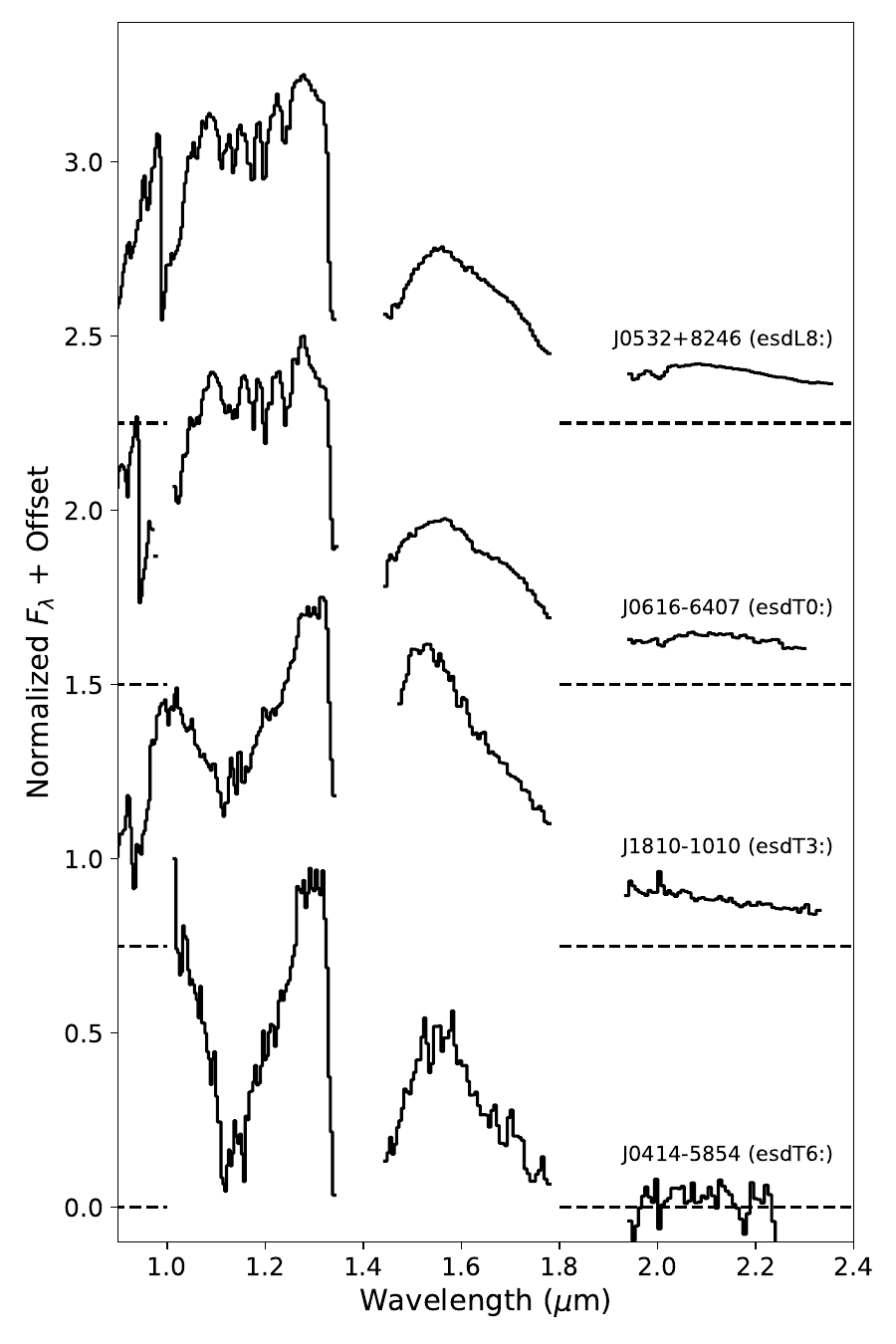}}
\caption{Spectral sequences for mild subdwarfs (d/sd; left), subdwarfs (sd; middle) and extreme subdwarfs (esd; right) for late-L and T dwarfs in our spectral sample. All data are smoothed to an equivalent resolution of {\ldl} = 300, normalized at the 1.25~$\mu$m peak and offset by constants (dashed lines; 0.5 steps for d/sd and sd, 0.75 steps for esd). Standards are identified by name and spectral type.
\label{fig:seq1}}
\end{figure*}

\begin{figure*}[ht!]
\centering
\includegraphics[width=0.32\textwidth]{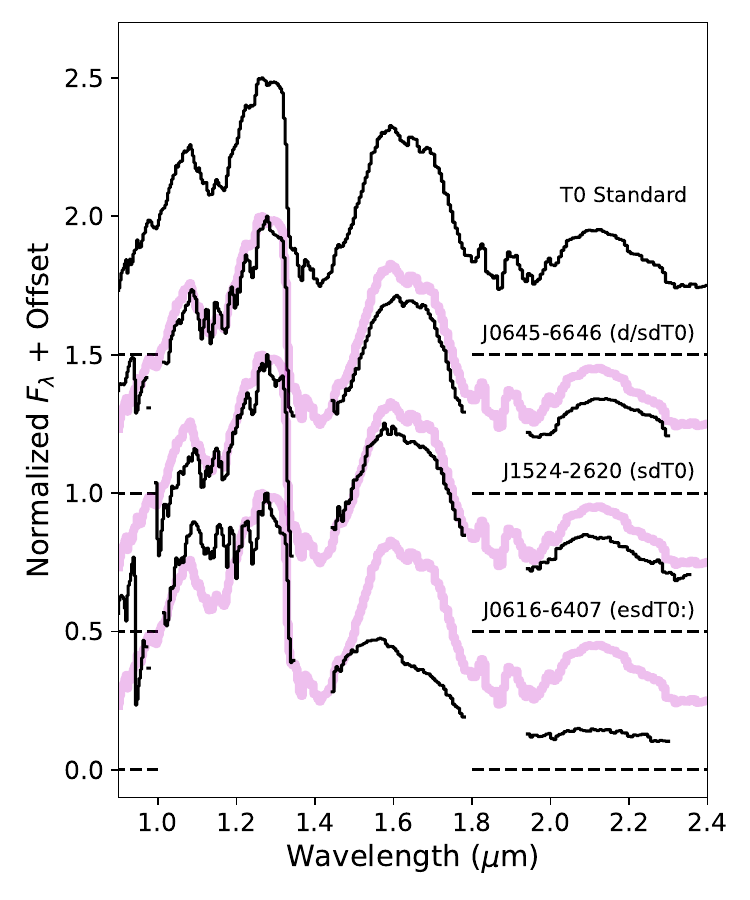}
\includegraphics[width=0.32\textwidth]{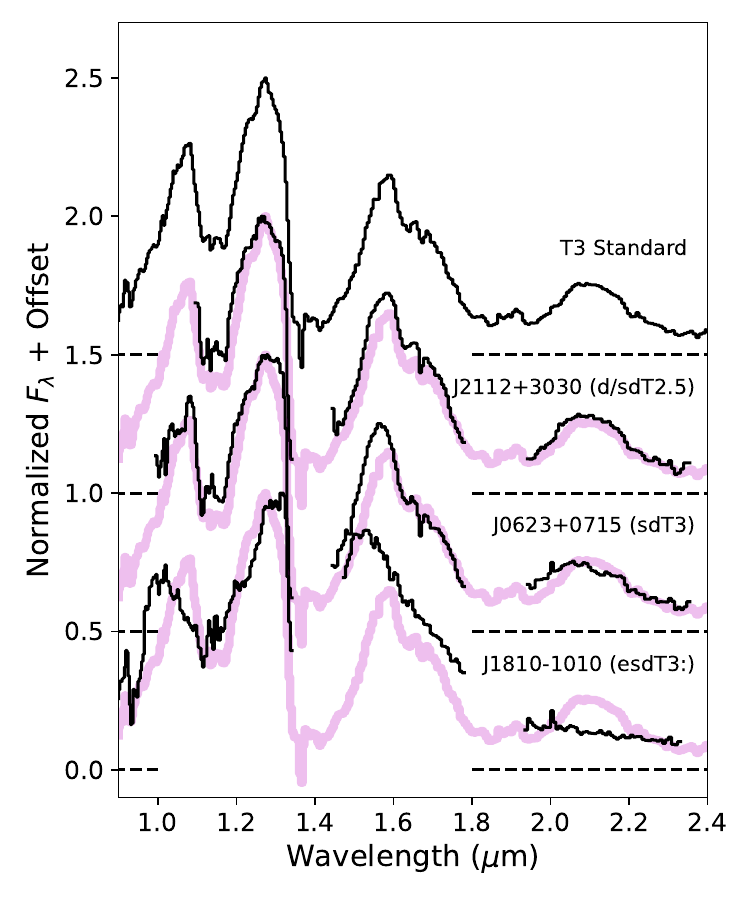}
\includegraphics[width=0.32\textwidth]{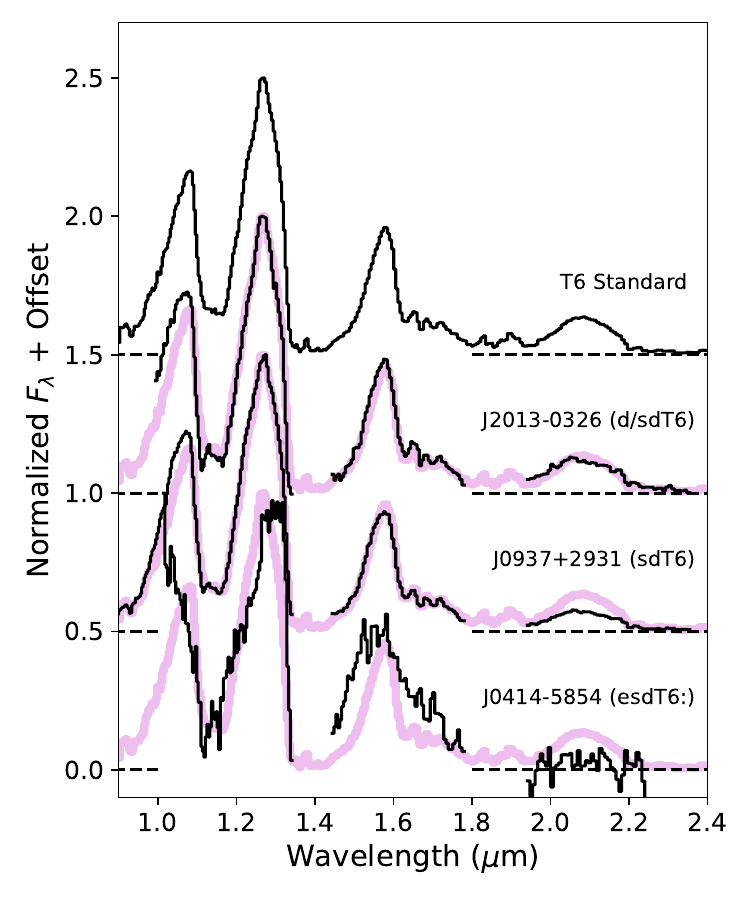}
\caption{Metallicity sequences of spectral type standards for the {T0} (left), T3 (middle), and T6 (right) subclasses, showing from top to bottom dwarf, mild subdwarf (d/sd), subdwarf (sd) and extreme subdwarf (esd) standards. All data are smoothed to an equivalent resolution of {\ldl} = 300, normalized at the 1.25~$\mu$m peak and offset by constants (dashed lines). The dwarf standard is repeated in magenta to compare to the d/sd, sd, and esd standards.
\label{fig:seq2}}
\end{figure*}

We iteratively evaluated these variations by direct comparison between spectra.
We then built a two-dimensional grid of representative standards that spanned our available sample.
Selection of these standards favored
sources with high signal-to-noise spectra and/or accessible declinations when more than one option was available. These standards are listed in Table~\ref{tab:stds} and displayed in sequence in Figure~\ref{fig:seq1}. There are gaps among subclasses {in the sd and esd sequences} where a clear standard was not available. However, as illustrated in Figure~\ref{fig:seq2}, those subtypes with representatives across metallicity classes display continuous variations in spectral features tied to metallicity (e.g., near-infrared slope, breadth of the 1.15~$\mu$m feature), while maintaining agreement with the dwarf standards in spectral features tied to temperature (e.g., strength of the 1.6~$\mu$m CH$_4$ band, depth of the 1.15~$\mu$m {H$_2$O+CH$_4$} feature).
The highly distinct spectra of the extreme subdwarfs J0532+8246, J0616$-$6407, J1810$-$1010, and J0414$-$5854  make their numerical classifications, anchored to the 1.1~$\mu$m trough, highly uncertain and are indicated as such with a ``:'' suffix.

\begin{deluxetable}{cc|cc|cc|cc}
\tablecaption{T Subdwarf Spectral Standards \label{tab:stds}} 
%\tabletypesize{\scriptsize} 
\tablehead{ 
\multicolumn{2}{c|}{Dwarfs (d)\tablenotemark{a}} & 
\multicolumn{2}{c|}{Mild Subdwarfs (d/sd)} & 
\multicolumn{2}{c|}{Subdwarfs (sd)} & 
\multicolumn{2}{c}{Extreme Subdwarfs (esd)} \\
\colhead{SpT} & 
\multicolumn{1}{c|}{Name} & 
\colhead{SpT} & 
\multicolumn{1}{c|}{Name} & 
\colhead{SpT} & 
\multicolumn{1}{c|}{Name} & 
\colhead{SpT} & 
\colhead{Name}
} 
\startdata 
dL8 & J1632+1904  & d/sdL8 & J1158+0435 & sdL8 & \nodata & esdL8 & J0532+8246 \\
dL9 & J0255$-$4700  & d/sdL9 & {J2021+1524} & sdL9 & {J1338$-$0229} & esdL9 & \nodata \\
dT0 & J1207+0244  & d/sdT0 & J0645$-$6646 & sdT0 & {J1524$-$2620} & esdT0 & J0616$-$6407 \\
dT1 & J0837$-$0000  & d/sdT1 & {J0301$-$2319} & sdT1 & J2112$-$0529 & esdT1 & \nodata \\
dT2 & J1254$-$0122  & d/sdT2 & {J0004$-$1336} & sdT2 & \nodata & esdT2 & \nodata \\
dT3 & J1209$-$1004  & d/sdT2.5 & J2112+3030 & sdT3 & J0623+0715 & esdT3 & J1810$-$1010 \\
dT4 & J2254+3123  & d/sdT4 & J0021+1552 & sdT4 & J1553+6933 & esdT4 & \nodata \\
dT5 & J1503+2525  & d/sdT5.5 & {GJ~576B} & sdT5.5 & J1130$-$1158 & esdT5 & \nodata \\
dT6 & J1624+0029  & d/sdT6 & {J2013-0326} & sdT6 & J0937+2931 & esdT6 & J0414$-$5854 \\
dT7 & J0727+1710  & d/sdT7.5 & {LHS~6176B} & sdT7 & J1416+1348B & esdT7 & \nodata \\
dT8 & J0415$-$0935  & d/sdT8 & J0939$-$2448 & sdT7.5 & {J0013+0634} & esdT8 & \nodata \\
dT9 & {J0722$-$0540}  & d/sdT9 & {J0833+0052} & sdT9 & \nodata & esdT9 & \nodata \\
\enddata
\tablenotetext{a}{Dwarf standards from \citet{1999ApJ...519..802K,2010ApJS..190..100K} for L dwarfs and \citet{2006ApJ...637.1067B} for T dwarfs; see these references for full source designations.}
\end{deluxetable} 

We used these standards to determine revised classifications for the remainder of our candidate and comparison sample, listed in Table~\ref{tab:assessment}. Note that this analysis resulted in several significant changes in the classifications of previously identified metal-poor late L and T dwarfs. In addition to reclassifying the previously reported subdwarfs J0004$-$2604, J1019$-$3911, and J1307+1510 as normal early T dwarfs, we reassigned several sources as mild subdwarfs given their relatively weak metallicity signatures. 
We also reclassified the L subdwarfs J0616$-$6407 and J0645$-$6646 as early T subdwarfs based on the presence of weak CH$_4$ absorption in their $H$- or $K$-band spectra. For J0616$-$6407, this is a substantial shift from its prior esdL6 classification \citep{2017MNRAS.464.3040Z}, but we assert this reclassification is warranted given the {emergent} CH$_4$ feature at 1.6~$\mu$m present in its spectrum, {a defining trait for T dwarfs \citep{2006ApJ...637.1067B}}. For sources with incomplete 1--2.5~$\mu$m coverage (e.g., Wolf~1130C, J2105$-$6235), we indicate uncertainty in their classifications with a ``:'' suffix.

As with any empirical stellar classification system, the relationship between {\teff} and spectral type can vary with metallicity, as discussed below. Nevertheless, our model fits indicate that {model-fit metallicities of the metallicity class standards} align with the metallicity ranges for M and L subdwarf {classes}, 
%While our standards were chosen empirically based on variations in spectral morphology, their estimated metallicities roughly align with those of the M and L subdwarf classes, 
with uncertainty-weighted means of
[M/H] = {$-$0.36$\pm$0.09} for d/sdT standards,
[M/H] = {$-$0.63$\pm$0.09} for sdT standards, and
[M/H] = {$-$1.32$\pm$0.13} for esdT standards.
We present this initial grid as a starting point, with the expectation of future expansion and/or reorganization as the known T subdwarf population expands.

\subsection{Defining a Metallicity Index}\label{sec:zeta}

\begin{figure*}[ht!]
\centering
\includegraphics[width=0.45\textwidth]{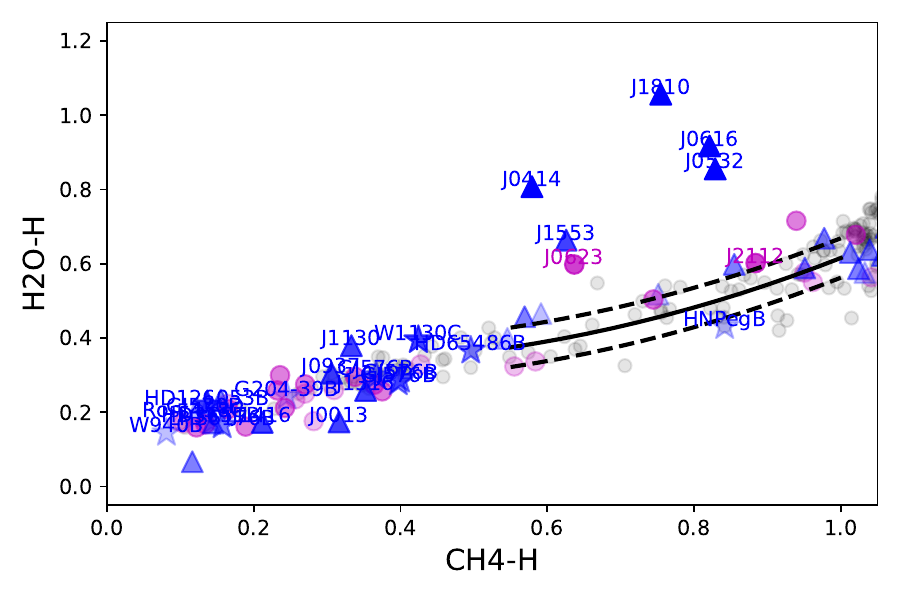} 
\includegraphics[width=0.45\textwidth]{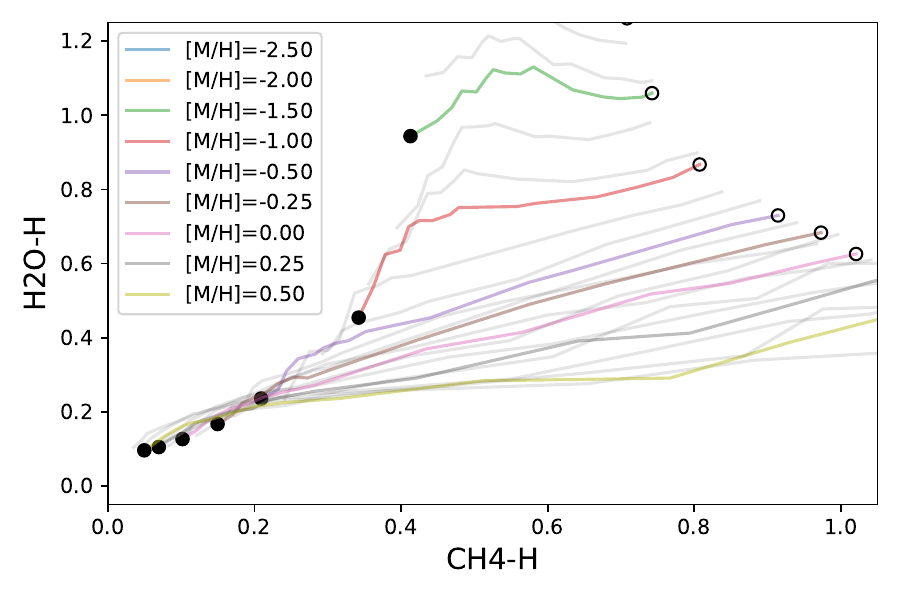}  \\
\includegraphics[width=0.45\textwidth]{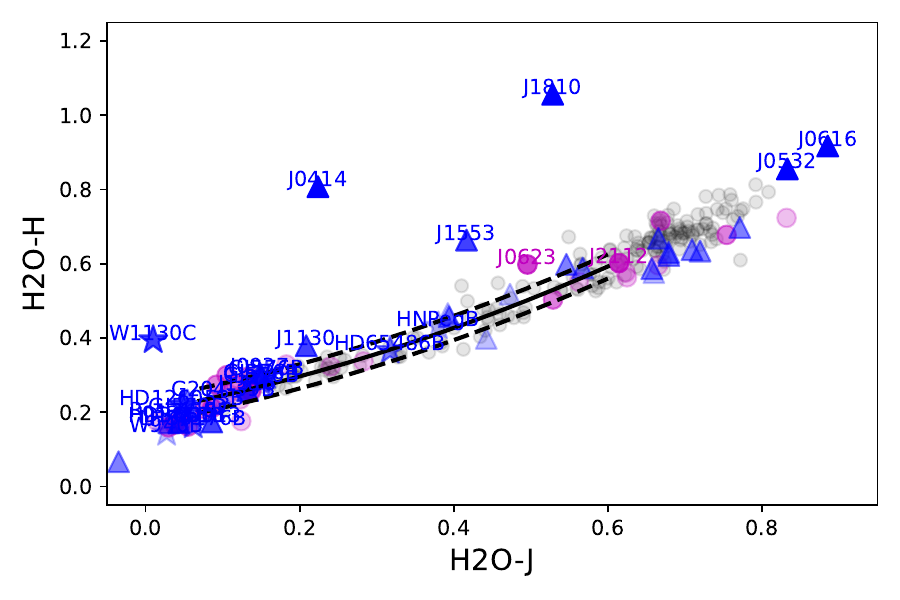}
\includegraphics[width=0.45\textwidth]{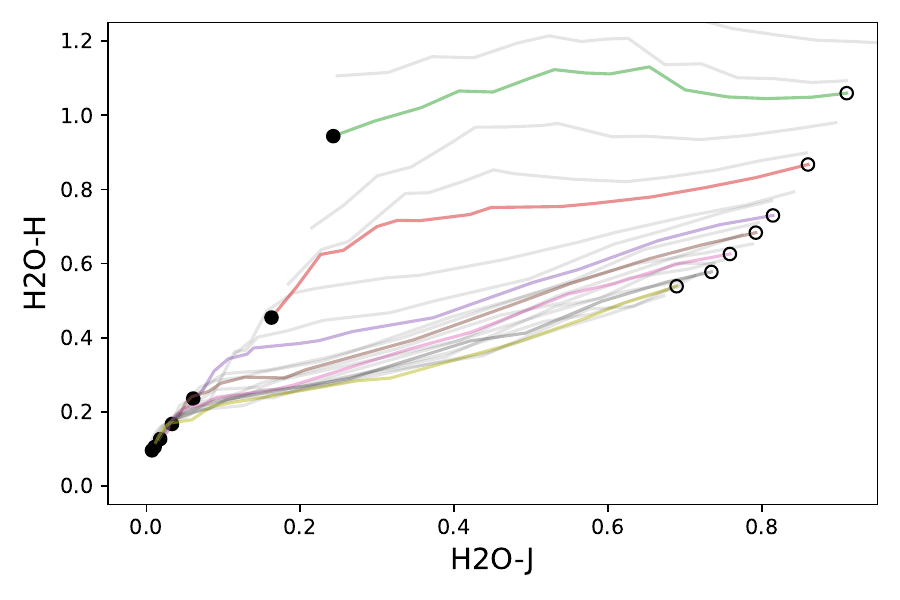} \\
\includegraphics[width=0.45\textwidth]{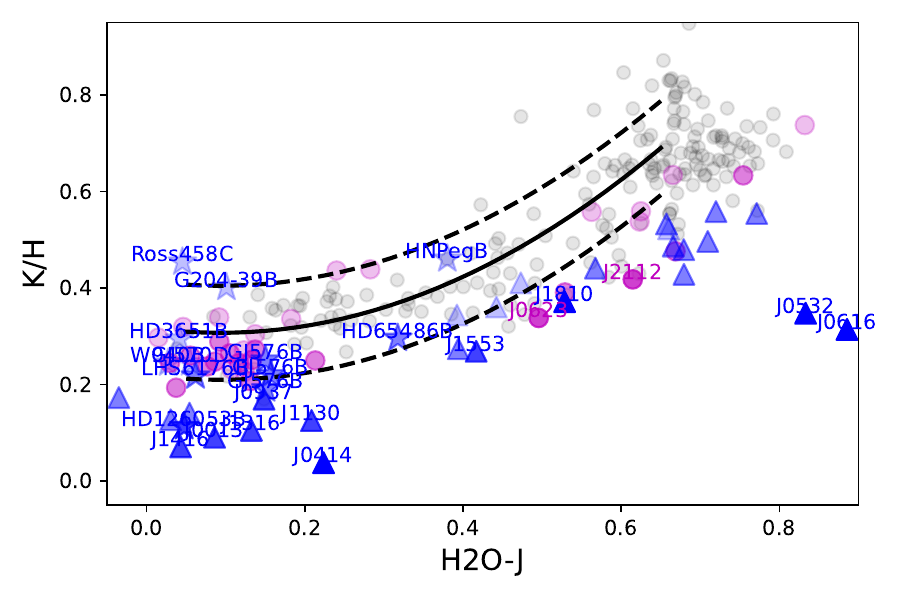}
\includegraphics[width=0.45\textwidth]{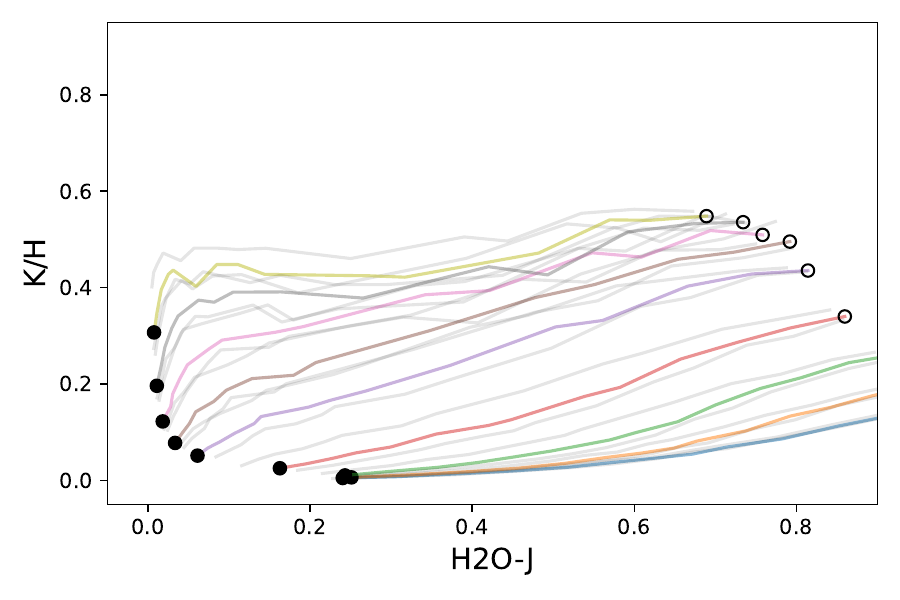} \\
\caption{
(Left panels): Three of the spectral index pairings that segregate metal-poor and solar-metallicity late-L and T dwarfs. 
Previously-classified subdwarfs are indicated by blue triangles, 
T dwarf benchmarks are indicated by blue stars, 
local T dwarfs are indicated by grey circles, 
and observed candidates are indicated by magenta circles.
The transparency of each symbol is based on the metallicity class, with dwarfs being most transparent and extreme subdwarfs being most opaque.
Sources identified as sdT or esdT and all benchmarks are labeled. 
Solid and dashed lines trace the mean and $\pm1\sigma$ standard deviation for second-order polynomial fits to the local T dwarf sample over the ranges specified in Table~\ref{tab:index-relations}.
(Right panels): Same index pairings measured on LOWZ models for
600~K $\leq$ {\teff} $\leq$ 1600~K (filled to open circles)
%4.0 $\leq$ {\logg} $\leq$ 5.25 (cgs)
and $-$2.5 $\leq$ [M/H] $\leq$ +0.5 (colored lines).
The color lines trace {\logg} = 5.0 (cgs), while the semi-transparent grey lines trace 
{\logg} = 4.0 and 5.25.
\label{fig:index1}}
\end{figure*}

\begin{figure*}[ht!]
\centering
\includegraphics[width=0.45\textwidth]{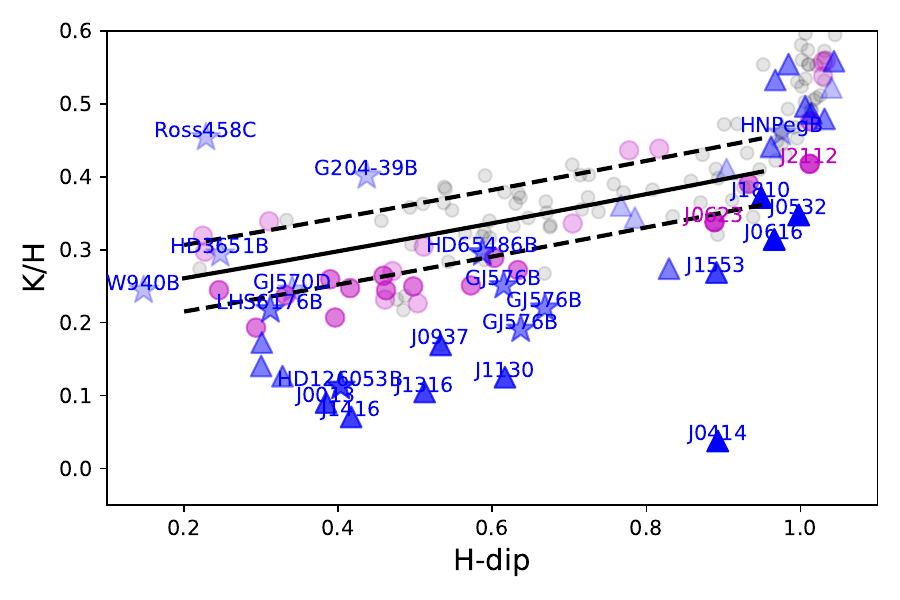}
\includegraphics[width=0.45\textwidth]{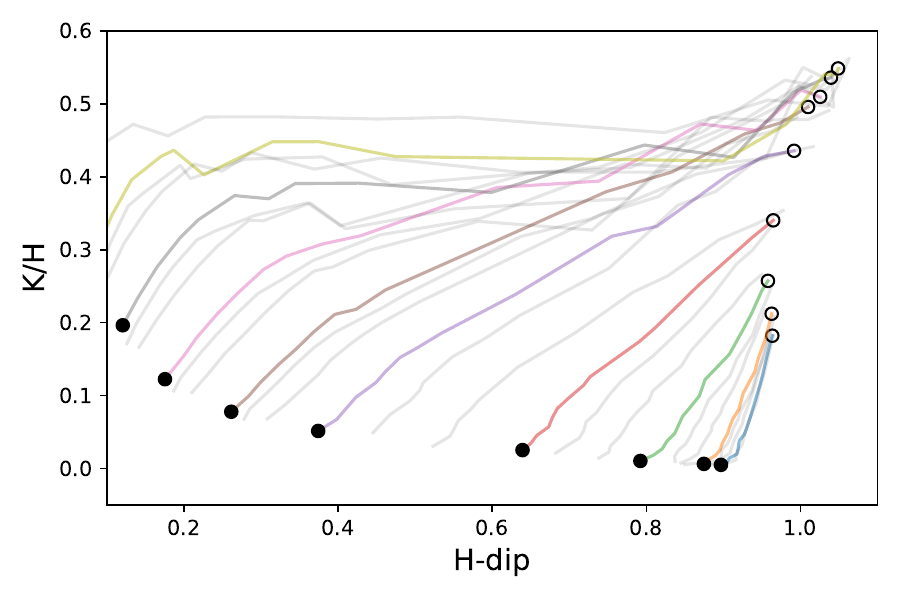} \\
\includegraphics[width=0.45\textwidth]{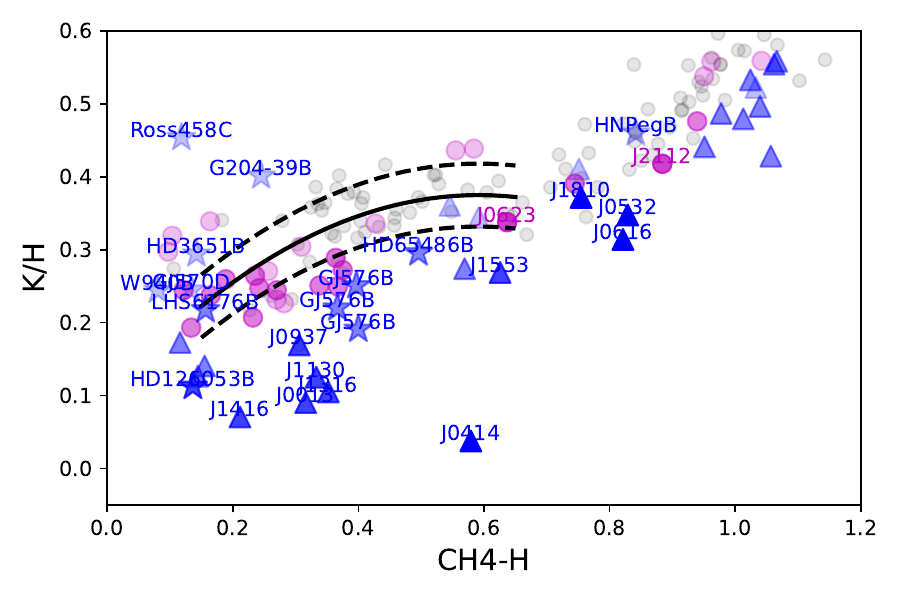}
\includegraphics[width=0.45\textwidth]{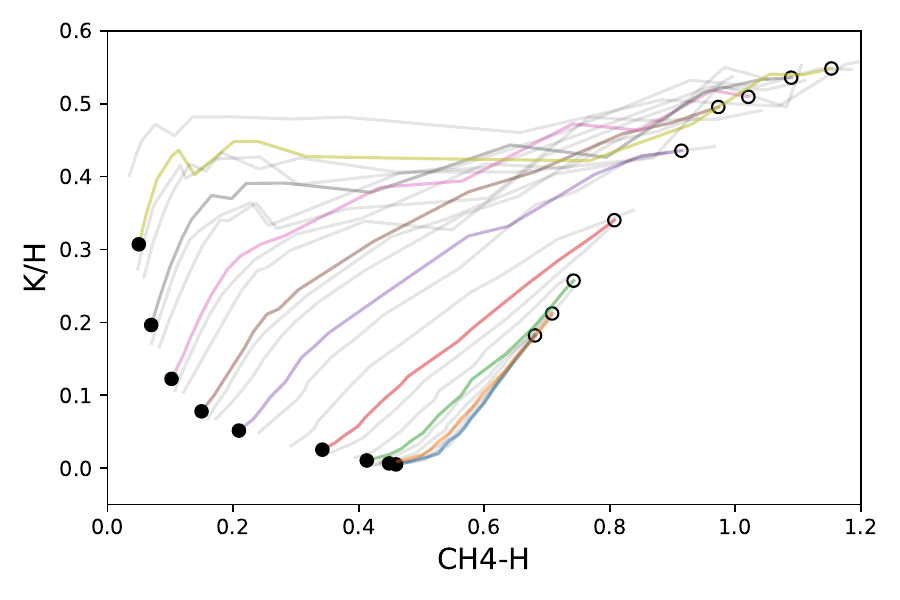} \\
\includegraphics[width=0.45\textwidth]{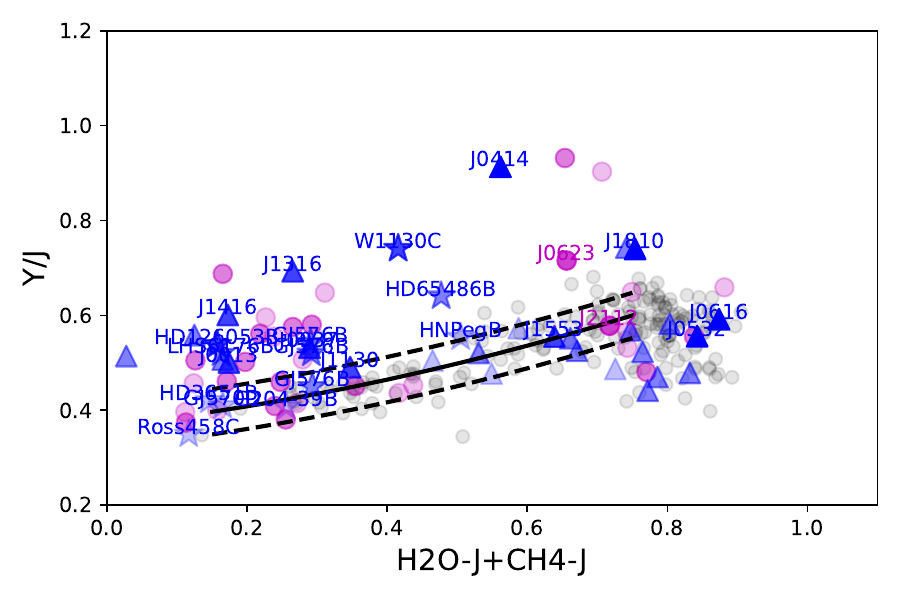}
\includegraphics[width=0.45\textwidth]{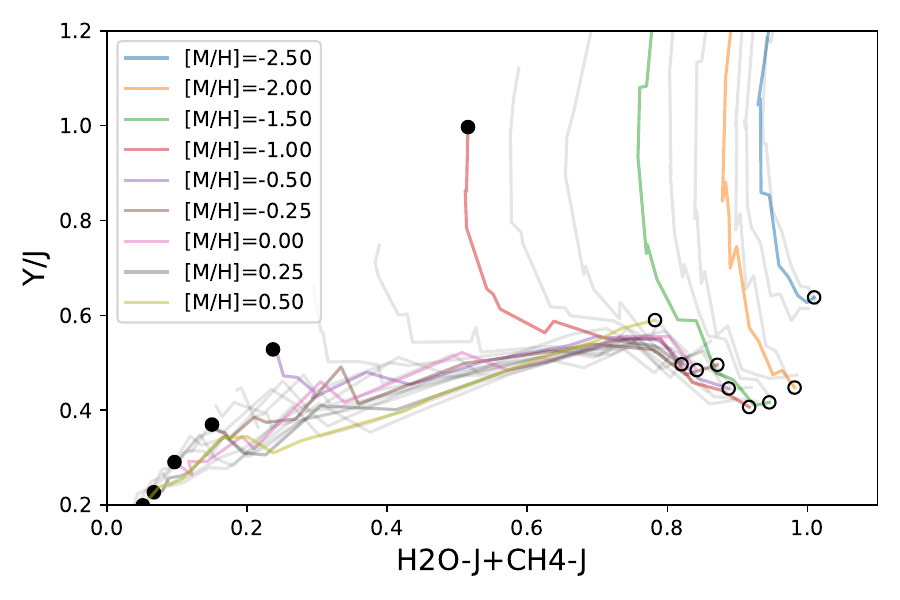}
\caption{Same as Figure~\ref{fig:index1} for three additional index pairings.
\label{fig:index2}}
\end{figure*}

While direct comparison to standards is the optimal method for determining  metallicity classifications, metallicity-sensitive spectral indices can be used as estimators of the {quantified} metallicity of a source.
For M subdwarfs, metallicity-sensitive spectral features
in the optical (e.g., CaH/TiO; \citealt{2006PASP..118..218W,2007ApJ...669.1235L,2009PASP..121..117W,2012AJ....143...67D,2014AandA...568A.121N}) or infrared (e.g., Na/H$_2$O/CO/FeH; \citealt{2012ApJ...748...93R,2013AJ....145...52M,2012ApJ...747L..38T,2014AJ....147...20N}) are
calibrated with sources of known metallicity, typically companions to FGK dwarfs and resolved binaries \citep{2013AJ....145...52M,2018MNRAS.479.1332M,2024MNRAS.527.11866}.
%the zeta-index ($\zeta$), which compares the relative strength of TiO and CaH absorption features in the 6800--7150~{\AA} optical band to a sample of kinematically-selected thin disk stars \citep{2007ApJ...669.1235L}. 
% Comparable metallicty indices have been defined at both optical \citep{2006PASP..118..218W} and infrared wavelengths \citep{2012ApJ...748...93R,2013AJ....145...52M,2012ApJ...747L..38T}
% By calibrating this index with M dwarfs of known metallicities, primarily companions to FGK dwarfs and resolved binaries \citep{2013AJ....145...52M,2018MNRAS.479.1332M,2024MNRAS.527.11866}, or models \citep{}, empirical relations have been established to directly infer metallicities from $\zeta$ values 
We applied the same approach to our T dwarf sample. We first identified spectral indices in the near-infrared that clearly segregate spectroscopically-identified T subdwarfs and low-metallicity benchmark companions from the predominantly solar-metallicity local T dwarf population. 
%From these pairings, we created composite indices that amplified separations, four of which are displayed in Figure~\ref{fig:index}. 
The ten spectral indices reported in Section~\ref{sec:indices} provide 45 unique index pairings, of which six demonstrate consistent separation between dwarfs and subdwarfs. These pairs as listed in Table~\ref{tab:index-relations}
and displayed in Figures~\ref{fig:index1} and~\ref{fig:index2}.
We combined two index pairings, [H2O-J] versus [Y/J] and [CH4-J] versus [Y/J] into a composite index.
%\footnote{The composite index is signified as $\langle$H2O-J,CH4-J$\rangle$ and is equal to either the average of the two indices or one of the values if the other is missing/unmeasureable.}
% or single measured value between the pairs if one is missing.} as these showed nearly identical trends.
% , which  of indices which showed similar behaviors were combined into composite indices
% $\langle$H2O-J,H2O-H$\rangle$ and 
For the three pairings with ordinate indices [H2O-H] and [Y/J], the subdwarfs were found to lie above the dwarf sequence, 
%compared to the local comparison sources and solar-metallicity benchmarks. These offsets are 
consistent with weakened H$_2$O absorption at 1.5~$\mu$m and a brighter $Y$-band peak, both noted previously as metallicity indicators.
The other three pairings with the ordinate index [K/H] have subdwarfs lying below the dwarf sequence, consistent with suppressed $K$-band peaks due to enhanced CIA H$_2$ absorption.
\citet{2006ApJ...637.1067B} had previously proposed use of the [K/J] color index as an indicator of enhanced CIA H$_2$ absorption, but probable errors in the order-to-order flux calibration of the multi-order spectral data in our sample (cf.\ Section~\ref{sec:additional}) results in higher scatter for this color index, hence we excluded it from our metallicity index set.

\begin{deluxetable}{lccccccc}
\tablecaption{Spectral Index Pairings That Segregate Metal-poor Late-L and T Dwarfs \label{tab:index-relations}} 
%\tabletypesize{\scriptsize} 
\tablehead{ 
$\zeta_{T,i}$ & \multicolumn{2}{c}{Abscissa (x)} & \multicolumn{1}{c}{Ordinate (y)} & \multicolumn{4}{c}{Solar Metallicity Sequence Polynomial Fit} \\
\hline
\colhead{} & 
\colhead{Index} & 
\colhead{Range} & 
\colhead{Index} & 
\colhead{$c_2$} & 
\colhead{$c_1$} & 
\colhead{$c_0$} & 
\colhead{$\sigma_y$}  
} 
\startdata 
$\zeta_{T,1}$ & [CH4-H] & 0.55--1.00 & [H2O-H] &  0.52753311 & $-$0.28280498 & 0.37088006 & 0.053 \\ 
$\zeta_{T,2}$ & [H2O-J] & 0.05--0.60 & [H2O-H] &  0.44401774 & 0.38492182 & 0.20196172 & 0.033 \\ 
$\zeta_{T,3}$ & [H2O-J] & 0.05--0.65 & [K/H] &  1.23813044 & $-$0.23225536  & 0.31791332 & 0.097 \\
$\zeta_{T,4}$ & [H-dip] & 0.20--0.90 & [K/H] &  0.01818876 & 0.17387556 & 0.22533516 & 0.046 \\ 
%$\zeta_{T,3}$ & $\langle$[H2O-J],[H2O-H]$\rangle$ & 0.10--0.85 & [K/H] &  0.01818876 & 0.17387556 & 0.22533516 & 0.091 \\
\nodata\tablenotemark{a} & [CH4-H] & 0.10--0.65 & [K/H] &  $-$0.78127813 & 0.92425434  & 0.1014752 & 0.043 \\ 
\nodata\tablenotemark{a} & $\langle$[H2O-J],[CH4-J]$\rangle$ & 0.15--0.75 & [Y/J] &  0.19243998 & 0.16534304 & 0.3675336 & 0.048 \\
\enddata
\tablecomments{
$\langle{SP_1,SP_2}\rangle$ indicates an average of two indices, or the value of one index if the other is unmeasured.
Fit relations for the dwarf standards are defined as $y = \sum_{i=0}^2c_ix^i$ for indices (x,y) over the ranges specified. The quantity $\sigma_y$ measures the root mean square deviation of y-values about this relation. See Figures~\ref{fig:index1} and~\ref{fig:index2} for visualization of these fits.
}
\tablenotetext{a}{Not used as a metallicity index.}
\end{deluxetable}

For each of these pairings, we fit second-order polynomials to the index values of the local T dwarf sample over an abscissa range that had the clearest offset from the T subdwarfs. 
In some cases, the abscissa range encompasses a subset of our sample; e.g., only early-type or only late-type T dwarfs.
% For the index pairings, [H2O-H] versus [CH4-H] and [K/H] versus [CH4-H], we found differing degrees of segregation for low and high values for [CH4-H], corresponding to early-type and late-type T dwarfs, respectively.
Fit parameters and scatter about the best fits are listed in Table~\ref{tab:index-relations} and displayed in Figures~\ref{fig:index1} and~\ref{fig:index2}. 
%We found that several of subdwarfs are several standard deviations away from these fit lines, while a handful of our candidates are $>$1$\sigma$ deviant. 

%The two combined index pairs, [H2O-J+H2O-H] and [H2O-J+CH4-J], 
%We also see good separation by metallicity when comparing the color index [K/H] to [H-dip], [H2O-J], and [H2O-H] indices, and we averaged the last two to form a composite index.
%The [H2O-J] and [H2O-H] pair also separates off metal-poor T subdwarfs.
%Finally, The color ratio [Y/J] shows good separation when compared to [H2O-J] and [CH4-J], and combined these to form another composite index.

To guide our interpretation of these trends, we also show the same spectral indices measured on the LOWZ atmosphere models over the temperature range 600~K $\leq$ {\teff} $\leq$ 1600~K, surface gravities 4.0 $\leq$ {\logg} $\leq$ 5.25 (cgs), metallicities $-$2.5 $\leq$ [M/H] $\leq$ +0.5, C/O = 0.55, and $\log\kappa_{zz}$ = 2. We find that five of these index pairs show good separation of metallicity and temperature and reasonable overlap with our measured values.
The [Y/J] versus $\langle$H2O-J,CH4-J$\rangle$ pair shows the most discrepant values between models and measurements, and poor metallicity separation in the models for [M/H] $\leq$ $-$0.5, so we reject this pair as a viable metallicity diagnostic.
%possibly due to poor modeling of the pressure-broadened K~I doublet at 0.77~$\mu$m or flux calibration issues at the edges of the Keck/NIRES and Magellan/FIRE data. 
Of the remaining indices, [H2O-H] versus [CH4-H], [H2O-H] versus [H2O-J], and [K/H] versus [H2O-J] show reasonably orthogonal metallicity and temperature trends in the models, although there is disagreement between models and measurements for the last index for the late-L and early-T dwarfs, likely due to unincorporated cloud opacity effects. The [K/H] versus [H-dip] and [K/H] versus [CH4-H] pairings show greater degeneracy between metallicity and temperature, while [K/H] versus [H-dip] shows greater dynamic range; we therefore prioritize this index pair over
[K/H] versus [CH4-H].
%We therefore rejected both [Y/J] versus $\langle$H2O-J,CH4-J$\rangle$ and [K/H] versus [H-dip] from our final metallicity set.
%The [K/H] index is influenced by both temperature and metallicity modulations on the strength of CIA H$_2$, resulting in correlated trends. 
%The refine our quantification of metallicity among these sources, we constructed a set of metallicity indices based on the first five pairings shown in Figures~\ref{fig:index1} and~\ref{fig:index2}.
%We follow the approach of \citet{2003AJ....125.1598L} 

Based on these analyses, we defined four metallicity indices, $\zeta_{T,i}$, as the ratio between the metallicity-sensitive index [H2O-H] or [K/H] and their expected values based on the solar-metallicity dwarf sequences:
% \begin{equation}
% \zeta \equiv \frac{\rm [SP_y]}{\rm SP_{y,\odot}([SP_x])}
% \end{equation}
% Here, SP$_x$ and SP$_y$ are the index pairs listed in Table~\ref{tab:index-relations} and ${\rm SP_{y,\odot}([SP_x])}$ is the polynomial fit for solar-metallicity sources for each pair also listed in this Table.
% This definition provides values less than one for subsolar metallicity sources when their index values fall below the solar metallicity sequence. For index pairs where a subsolar metallicity source has an ordinate index value above the solar metallicity trend, we invert this relation. Thus, we have the five metallicity indices:
\begin{equation}
\zeta_{T,1} \equiv \frac{\rm [H2O-H]_\odot([CH4-H])}{\rm [H2O-H]}
\end{equation}
\begin{equation}
\zeta_{T,2} \equiv \frac{\rm [H2O-H]_\odot([H2O-J])}{\rm [H2O-H]}
\end{equation}
\begin{equation}
\zeta_{T,3} \equiv \frac{\rm [K/H]}{\rm [K/H]_\odot([H2O-J])}
\end{equation}
% \begin{equation}
% \zeta_{T,4} \equiv \frac{\rm [K/H]}{\rm [K/H]_\odot([CH4-H])}
% \end{equation}
\begin{equation}
\zeta_{T,4} \equiv \frac{\rm [K/H]}{\rm [K/H]_\odot([H-dip])}.
\end{equation}
Here, the notation ${\rm [SP_y]_\odot([SP_x])}$ is the polynomial fit for the dwarf sequence for the index pairings $[SP_x]$ and $[SP_y]$ listed in Table~\ref{tab:index-relations}. These relationships are defined such that subsolar metallicity objects are expected to have $\zeta_{T,i} \lesssim 1$.

\startlongtable
\begin{deluxetable}{llccccccc}
\tablecaption{Metallicity Indices for the T Subdwarf Sample \label{tab:zeta}} 
%\tabletypesize{\scriptsize} 
\tablehead{ 
 & & & & & & & \colhead{Prior} & \colhead{Model} \\
\colhead{Name} & 
\colhead{SpT} & 
\colhead{$\zeta_{T,1}$} & 
\colhead{$\zeta_{T,2}$} & 
\colhead{$\zeta_{T,3}$} & 
\colhead{$\zeta_{T,4}$} & 
\colhead{$\zeta_{T}$} & 
\colhead{[Fe/H]} & 
\colhead{[M/H]\tablenotemark{a}}  
} 
\startdata 
\hline
\multicolumn{9}{c}{Benchmark Companions} \\
\hline
HD~3651B & T8 &  \nodata &  \nodata &  \nodata & 1.09$\pm$0.20 & 1.09$\pm$0.20 & +0.14 & 0.00 \\
{HD~65486B} & d/sdT4.5 &  \nodata & 1.01$\pm$0.11 & 0.80$\pm$0.29 & 0.89$\pm$0.16 & 0.95$\pm$0.11 & $-$0.28 & $-$0.53 \\
LHS~6176B & d/sdT7.5 &  \nodata &  \nodata & 0.71$\pm$0.34 & 0.77$\pm$0.13 & 0.76$\pm$0.12 & $-$0.30 & $-$0.21  \\
{Ross~458C} & rT8.5 &  \nodata &  \nodata &  \nodata & 1.71$\pm$0.30 & 1.71$\pm$0.30 & +0.25 & +0.95 \\
{HD~126053B} & sdT7.5 &  \nodata &  \nodata &  \nodata & 0.38$\pm$0.08 & 0.38$\pm$0.08 & $-$0.38 & $-$0.35 \\
GJ~570D & T7.5 &  \nodata &  \nodata & \nodata & 0.85$\pm$0.16 & 0.85$\pm$0.16 & +0.05 & $-$0.11 \\
GJ~576B (GNIRS) & d/sdT5.5 &  \nodata & 0.92$\pm$0.12 & 0.71$\pm$0.26 & 0.63$\pm$0.10 & 0.75$\pm$0.16 & $-$0.37 & $-$0.47 \\
\nodata (XS) & d/sdT5.5 &  \nodata & 0.94$\pm$0.14 & 0.61$\pm$0.41 & 0.56$\pm$0.10 & 0.71$\pm$0.21 & $-$0.37 & $-$0.53 \\
\nodata (NIRES) & d/sdT5.5 &  \nodata & 0.96$\pm$0.11 & 0.81$\pm$0.43 & 0.74$\pm$0.09 & 0.83$\pm$0.13 & $-$0.37 & $-$0.21 \\
G~204-39B & T6.5 &  \nodata & 1.01$\pm$0.15 & \nodata & 1.32$\pm$0.21 & 1.11$\pm$0.19 & $-$0.04 & +0.43 \\
% Wolf~1130C & sdT6: &  \nodata &  \nodata &  \nodata &  \nodata &  \nodata & $-$0.64 & $-$0.48 \\
HN~PegB & T2.5 & 1.17$\pm$0.13 & 0.95$\pm$0.07 & 1.13$\pm$0.33 &  \nodata & 1.01$\pm$0.11 & $-$0.08 & $-$0.11 \\
\hline
\multicolumn{9}{c}{Subdwarf Candidates} \\
\hline
% J0045+7958 & L9.5 &  \nodata &  \nodata &  \nodata &  \nodata &  \nodata & \nodata & +0.34 \\
J0055+5947 & d/sdT6.5 &  \nodata & 0.99$\pm$0.21 & 0.75$\pm$0.34 & 0.75$\pm$0.12 & 0.81$\pm$0.15 & \nodata & +0.04 \\
J0140+0150 (SpeX) & rT4.5 & 1.14$\pm$0.22 & 1.03$\pm$0.17 & 1.25$\pm$0.40 & 1.16$\pm$0.16 & 1.12$\pm$0.11 & \nodata & +0.21 \\
\nodata (NIRES) & rT4.5 & 1.16$\pm$0.16 & 0.99$\pm$0.10 & \nodata & 1.17$\pm$0.16 & 1.07$\pm$0.11 & \nodata & +0.40 \\
J0411+4714 & d/sdT7.5 &  \nodata & 1.12$\pm$0.17 & 0.77$\pm$0.32 & 0.83$\pm$0.14 & 0.95$\pm$0.18 & \nodata & $-$0.12 \\
J0429+3201 & T1 & 1.07$\pm$0.10 & 1.01$\pm$0.06 & 0.96$\pm$0.17 &  \nodata & 1.02$\pm$0.06 & \nodata & \nodata \\
J0433+1009 & T8 &  \nodata & 1.08$\pm$0.16 & \nodata & 1.21$\pm$0.40 & 1.10$\pm$0.16 & \nodata & +0.12 \\
J0623+0715 & sdT3 & 0.68$\pm$0.09 & 0.84$\pm$0.06 & 0.67$\pm$0.14 & 0.86$\pm$0.12 & 0.79$\pm$0.09 & \nodata &  $-$0.68\\
J0659+1615 & T1 & 1.00$\pm$0.09 &  \nodata & 0.82$\pm$0.14 &  \nodata & 0.95$\pm$0.11 & \nodata & $-$0.03 \\
J0758+5711 & T6.5 &  \nodata & 1.09$\pm$0.16 & \nodata & 0.87$\pm$0.14 & 0.97$\pm$0.12 & \nodata & $-$0.17 \\
% J0843+2904 & L6 &  \nodata &  \nodata &  \nodata &  \nodata &  \nodata & \nodata &  \\
J0845$-$3305 & d/sdT6.5 &  \nodata & 0.92$\pm$0.11 & 0.79$\pm$0.28 & 0.79$\pm$0.18 & 0.85$\pm$0.11 & \nodata & $-$0.35 \\
J0911+2146 & d/sdT8 &  \nodata &  \nodata &  \nodata & 0.70$\pm$0.17 & 0.70$\pm$0.17 & \nodata & 0.00 \\
J1110$-$1747 & rT8 &  \nodata &  \nodata &  \nodata & 1.20$\pm$0.24 & 1.20$\pm$0.24 & \nodata &  +0.20\\
J1130+3139 & d/sdT5.5 &  \nodata & 1.03$\pm$0.15 & \nodata & 0.79$\pm$0.11 & 0.88$\pm$0.12 &  \nodata & $-$0.17 \\
{J1138+7212} & d/sdT7 &  \nodata & 1.00$\pm$0.24 & 0.67$\pm$0.34 & 0.70$\pm$0.12 & 0.76$\pm$0.17 & \nodata & $-$0.36 \\
J1204$-$2359 & d/sdT7 &  \nodata & 1.13$\pm$0.20 & 0.81$\pm$0.39 & 0.82$\pm$0.15 & 0.91$\pm$0.18 & \nodata & $-$0.15 \\
% J1304+2819 & d/sdT0 &  \nodata &  \nodata &  \nodata &  \nodata &  \nodata & \nodata &  \\
J1308$-$0321 & d/sdT8 &  \nodata &  \nodata &  \nodata & 0.91$\pm$0.18 & 0.91$\pm$0.18 & \nodata & $-$0.11 \\
J1401+4325 & d/sdT5.5 &  \nodata & 0.88$\pm$0.13 & \nodata & 0.86$\pm$0.14 & 0.87$\pm$0.10 & \nodata & $-$0.13 \\
J1458+1734 & T8 &  \nodata &  \nodata &  \nodata & 1.12$\pm$0.18 & 1.12$\pm$0.18 & \nodata & +0.13 \\
{J1515$-$2157} & d/sdT6 &  \nodata & \nodata & 0.73$\pm$0.31 & 0.71$\pm$0.11 & 0.71$\pm$0.11 & \nodata & $-$0.08  \\
J1524$-$2620 & sdT0 & 0.80$\pm$0.08 &  \nodata &  \nodata &  \nodata & 0.80$\pm$0.08 & \nodata &  $-$0.69\\
J1710+4537 & T6 &  \nodata & 1.01$\pm$0.13 & 0.98$\pm$0.40 & 0.95$\pm$0.15 & 0.98$\pm$0.10 & \nodata & +0.02 \\
J1801+4717 & d/sdT5 &  \nodata & 0.87$\pm$0.10 & \nodata & 0.94$\pm$0.12 & 0.91$\pm$0.09 & \nodata & $-$0.09 \\
J2013$-$0326 & d/sdT6 &  \nodata & 0.88$\pm$0.11 & \nodata & 0.76$\pm$0.12 & 0.83$\pm$0.10 & \nodata & $-$0.21 \\
J2021+1524 & d/sdL9 &  \nodata &  \nodata & 0.85$\pm$0.13 &  \nodata & 0.85$\pm$0.13 & \nodata & $-$0.35 \\
J2112$-$0529 & sdT1 & 0.89$\pm$0.08 &  \nodata & 0.65$\pm$0.14 &  \nodata & 0.79$\pm$0.13 & \nodata & $-$0.74 \\
J2112+3030 & d/sdT2.5 & 0.90$\pm$0.11 & 1.05$\pm$0.07 & 0.72$\pm$0.16 &  \nodata & 0.95$\pm$0.14 & \nodata & $-$0.43 \\
J2218+1146 & d/sdT6.5 &  \nodata & 0.83$\pm$0.13 & \nodata & 0.86$\pm$0.15 & 0.84$\pm$0.09 & \nodata & $-$0.10 \\
J2251$-$0740 & d/sdT7 &  \nodata &  \nodata & 0.84$\pm$0.39 & 0.88$\pm$0.17 & 0.88$\pm$0.16 &  \nodata & 0.00  \\
\hline
\multicolumn{9}{c}{Metal-Poor Comparison Sources} \\
\hline
J0004$-$1336 & d/sdT2: & 0.98$\pm$0.11 & 0.96$\pm$0.08 & 0.75$\pm$0.13 &  \nodata & 0.91$\pm$0.11 & \nodata & $-$0.56 \\
J0004$-$2604 & T3 & 0.88$\pm$0.10 & 0.93$\pm$0.07 & 0.85$\pm$0.18 &  \nodata & 0.91$\pm$0.06 & \nodata & $-$0.40 \\
{J0013+0634} & sdT7.5 &  \nodata & 1.37$\pm$0.36 & \nodata & 0.31$\pm$0.06 & 0.32$\pm$0.15 & \nodata & $-$0.61 \\
J0021+1552 & d/sdT4 & 0.83$\pm$0.15 & 0.92$\pm$0.13 & 0.65$\pm$0.22 & 0.72$\pm$0.14 & 0.81$\pm$0.13 & \nodata & $-$0.44 \\
% J0057+2013 & d/sdT0 &  \nodata &  \nodata &  \nodata &  \nodata &  \nodata & \nodata &  \\
J0301$-$2319 & d/sdT1 & 0.89$\pm$0.09 &  \nodata &  \nodata &  \nodata & 0.89$\pm$0.09 & \nodata & $-$0.37 \\
J0309$-$5016 & d/sdT7.5 &  \nodata &  0.96$\pm$0.16 & \nodata & 0.50$\pm$0.09 & 0.63$\pm$0.22\tablenotemark{b} & \nodata & $-$0.36 \\
J0348$-$5620 & d/sdT4 & 0.83$\pm$0.13 & 0.90$\pm$0.09 & 0.82$\pm$0.35 & 0.92$\pm$0.13 & 0.89$\pm$0.08 & \nodata & 0.00 \\
J0414$-$5854 & esdT6: & 0.48$\pm$0.10 & 0.38$\pm$0.08 & 0.12$\pm$0.14 & 0.10$\pm$0.10 & 0.27$\pm$0.16\tablenotemark{b} & \nodata & $-$1.34 \\
J0532+8246 & esdL8: & 0.58$\pm$0.07 &  \nodata &  \nodata &  \nodata & 0.58$\pm$0.07 & {$-$1.6} & $-$1.53 \\
J0616$-$6407 & esdT0: & 0.54$\pm$0.06 &  \nodata &  \nodata &  \nodata & 0.54$\pm$0.06 & {$-$1.6} & $-$1.13 \\
% J0645$-$6646 & d/sdT0 &  \nodata &  \nodata &  \nodata &  \nodata &  \nodata & \nodata &  \\
% J0850$-$0221 & d/sdL6.5 &  \nodata &  \nodata &  \nodata &  \nodata &  \nodata & \nodata &  \\
{J0833+0052} & d/sdT9 &  \nodata &  \nodata &  \nodata & 0.62$\pm$0.12 & 0.62$\pm$0.12 &  \nodata & $-$0.08 \\
J0937+2931 & sdT6 &  \nodata & 0.88$\pm$0.09 & 0.55$\pm$0.33 & 0.52$\pm$0.08 & 0.64$\pm$0.18 & \nodata & $-$0.59 \\
J0939$-$2448 & d/sdT8 &  \nodata &  \nodata &  \nodata & 0.45$\pm$0.10 & 0.45$\pm$0.10 & {$-$0.24} & $-$0.35  \\
J0953$-$0943 & d/sdT6 &  \nodata & \nodata & 0.77$\pm$0.24 & 0.79$\pm$0.16 & 0.78$\pm$0.13 & \nodata & $-$0.43 \\
J1019$-$3911 & T4 &  \nodata & 1.15$\pm$0.15 & 0.79$\pm$0.24 & 0.98$\pm$0.13 & 1.00$\pm$0.17 & \nodata & 0.00 \\
% J1035$-$0711 & d/sdT0 &  \nodata &  \nodata &  \nodata &  \nodata &  \nodata & \nodata &  \\
% J1055+5443 & Y0 &  \nodata &  \nodata &  \nodata &  \nodata &  \nodata &  \nodata & \\
J1130$-$1158 & sdT5.5 &  \nodata & 0.79$\pm$0.15 & 0.39$\pm$0.15 & 0.37$\pm$0.10 & 0.52$\pm$0.19\tablenotemark{b} & \nodata & $-$0.56 \\
% J1158+0435 & d/sdL8 &  \nodata &  \nodata &  \nodata &  \nodata &  \nodata & \nodata &  \\
% J1307+1510 & T1 &  \nodata &  \nodata &  \nodata &  \nodata &  \nodata & \nodata &  \\
J1316+0755 & sdT6.5 &  \nodata & 1.01$\pm$0.23 & 0.34$\pm$0.18 & 0.33$\pm$0.06 & 0.56$\pm$0.32\tablenotemark{b} & \nodata & $-$0.65 \\
% J1338$-$0229 & d/sdT0 &  \nodata &  \nodata &  \nodata &  \nodata &  \nodata & \nodata &  \\
J1416+1348B & sdT7 &  \nodata &  \nodata &  \nodata & 0.23$\pm$0.05 & 0.23$\pm$0.05 & {$-$0.35} &  $-$0.67 \\
J1553+6933 & sdT4 & 0.60$\pm$0.08 & 0.66$\pm$0.07 & 0.62$\pm$0.17 & 0.68$\pm$0.09 & 0.65$\pm$0.05 & \nodata & $-$0.96  \\
J1810$-$1010 & esdT3: & 0.43$\pm$0.06 & 0.50$\pm$0.05 & 0.69$\pm$0.18 &  \nodata & 0.49$\pm$0.07 & {$-$1.5} & $-$1.27 \\
J2105$-$6235 & d/sdT2: & 0.86$\pm$0.09 & 0.91$\pm$0.05 &  \nodata &  \nodata & 0.89$\pm$0.05 & \nodata & $-$0.32 \\
\enddata
\tablenotetext{a}{Based on best-fit spectral model; see Table~\ref{tab:modelfit}.}
\tablenotetext{b}{Individual $\zeta_{T,i}$ values differ by more than the uncertainties; reported $\zeta_T$ value and uncertainty is the unweighted mean and standard deviation of the individual $\zeta_{T,i}$ values.}
\end{deluxetable} 

\begin{figure}[ht!]
\centering
\includegraphics[width=0.45\textwidth]{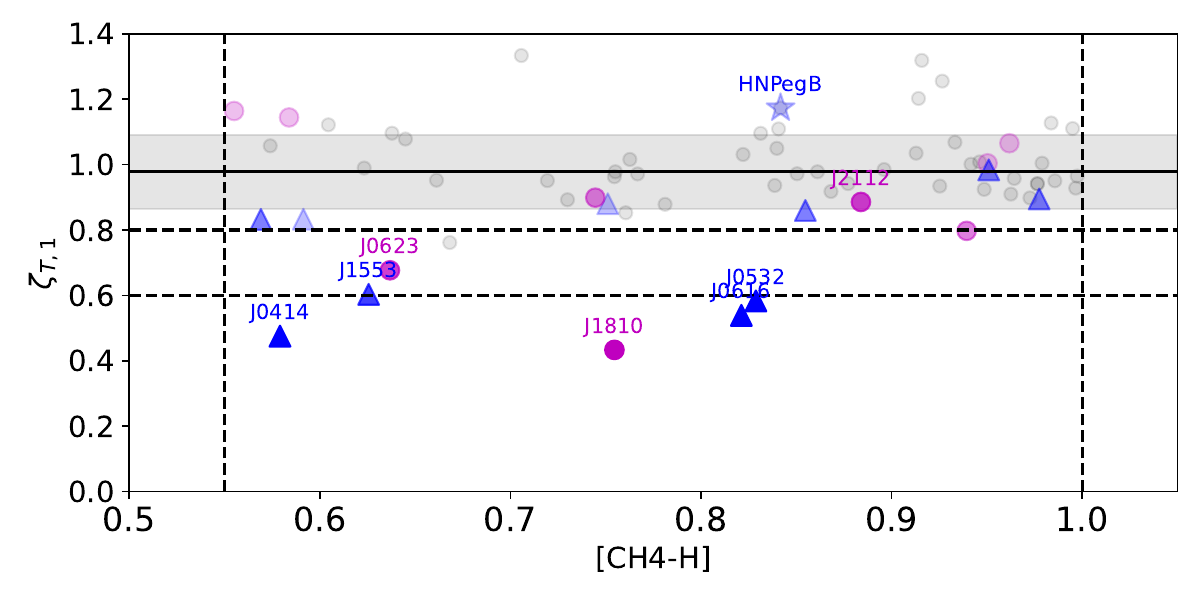} 
\includegraphics[width=0.45\textwidth]{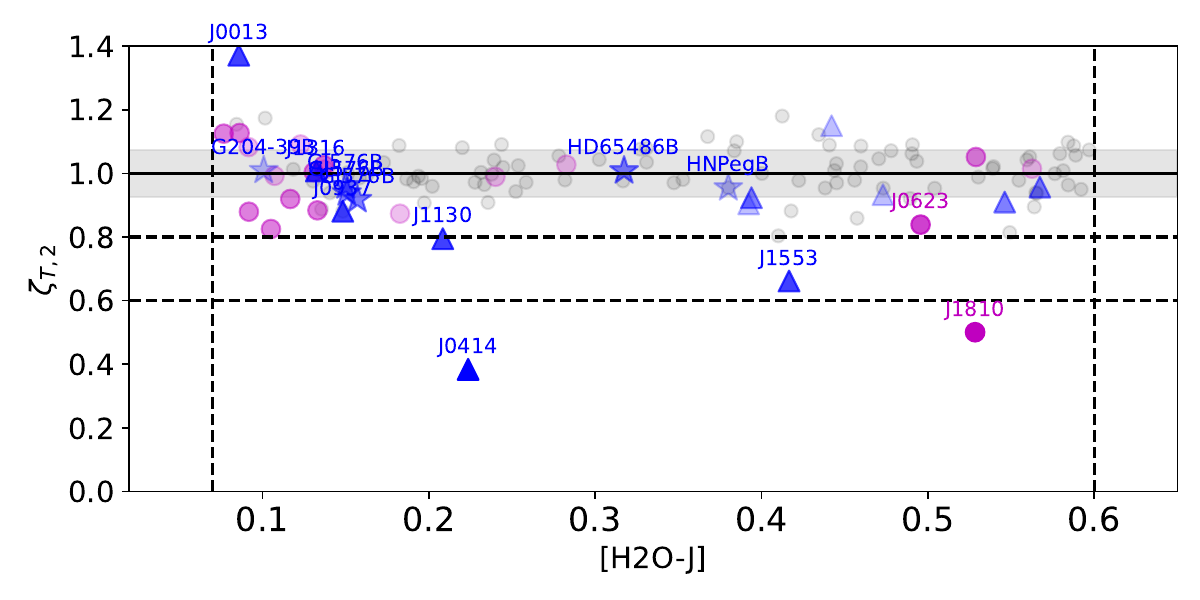} \\
\includegraphics[width=0.45\textwidth]{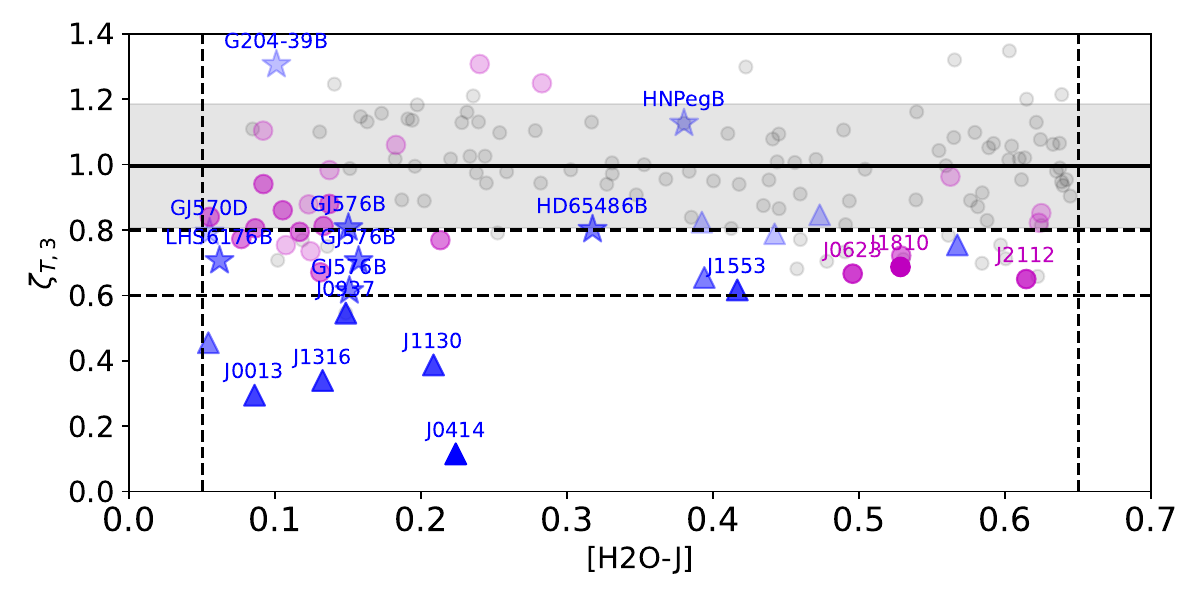}
\includegraphics[width=0.45\textwidth]{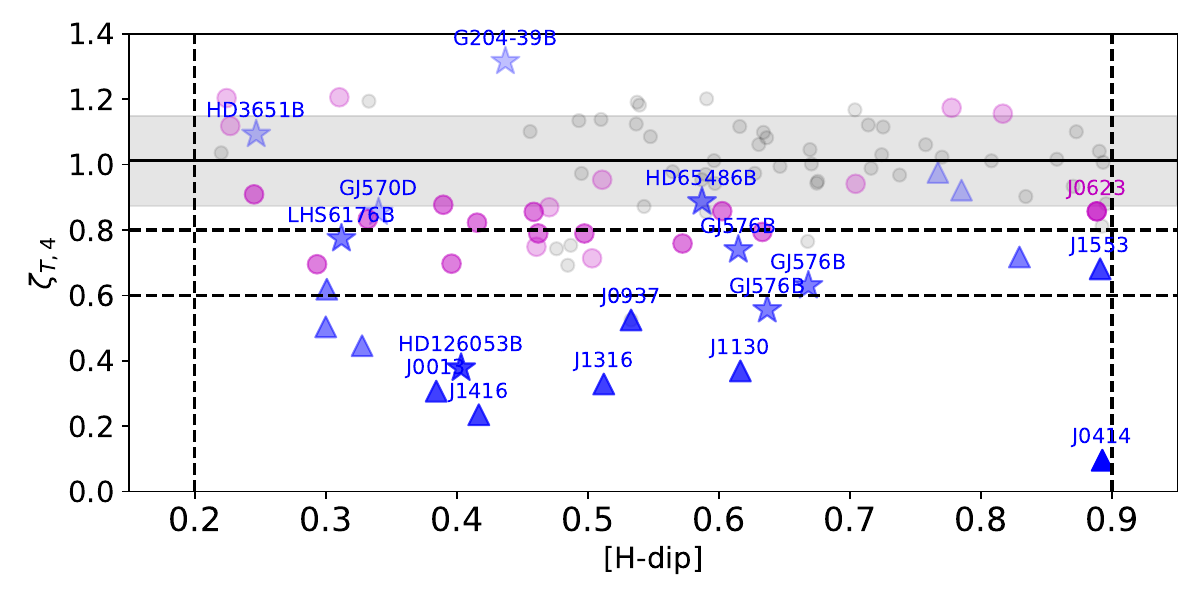} \\
\includegraphics[width=0.8\textwidth]{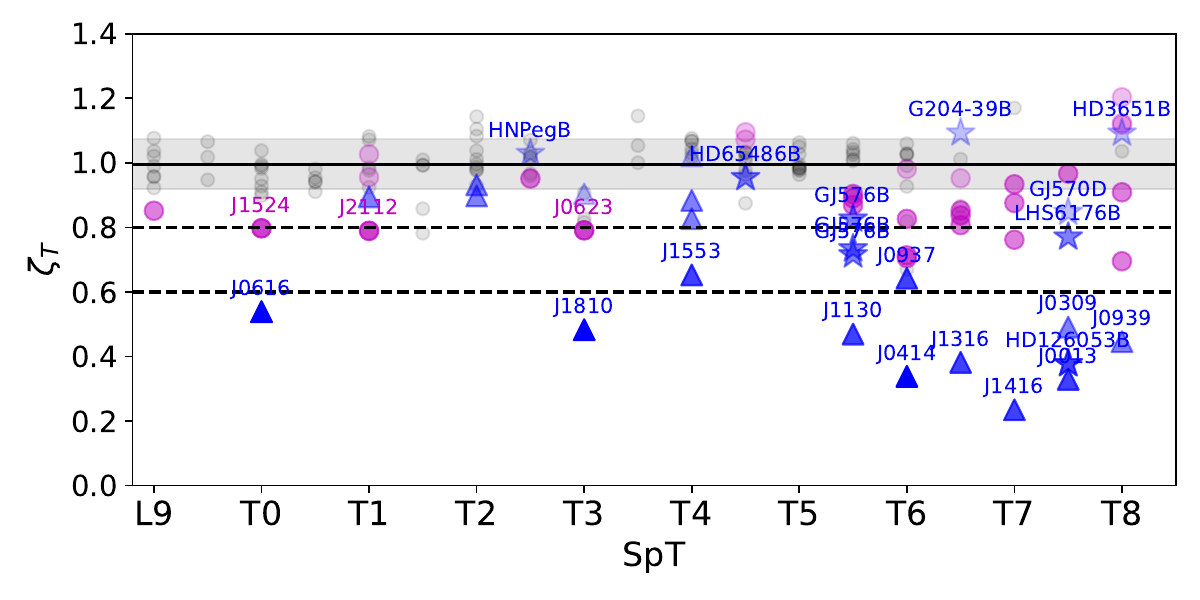}
\caption{Individual metallicity indices $\zeta_{T,i}$ (top four panels) and mean metallicity index $\zeta_T$
(bottom panel) for our spectral sample.
Symbols are the same as Figure~\ref{fig:index1}.
Each panel shows the median (solid line) and $\pm$1$\sigma$ standard deviation (grey band) for local T dwarfs,
with vertical dashed lines in the top four panels indicating the abscissa index range over which the solar-metallicity 
relation is defined (see Table~\ref{tab:index-relations}).
Horizontal dashed lines in each panel indicate threshold values of $\zeta_T$ = 0.8 and 0.6.
sdT and esdT candidates and comparison sources are labeled, as are all benchmark companions.
\label{fig:zeta}}
\end{figure}

Table~\ref{tab:zeta} lists the individual $\zeta_{T,i}$ values for our spectral sample, as well as the combined index $\zeta_T$, while Figure~\ref{fig:zeta} displays the distributions of these indices.
Uncertainties for the individual $\zeta_{T,i}$ values were estimated from the original spectral index uncertainties and the scatter in the dwarf sequence fits using Monte Carlo methods, and are typically 0.1--0.2. 
Note that $\zeta_{T,3}$ shows larger uncertainties due to the greater scatter of its dwarf sequence fit. 
We rejected all values of $\zeta_{T,i}$ with uncertainties greater than 0.4, then computed 
$\zeta_T$ as the uncertainty-weighted mean 
%with an uncertainty that incorporates both individual uncertainties and scatter among the $\zeta_{T,i}$ values 
(Eqn.~\ref{eqn:weight}).
For some sources the scatter among $\zeta_{T,i}$ values exceeded their individual uncertainties (i.e., $\chi^2_r > 1$), in which case we adopted a straight mean with uncertainty equal to the standard deviation of the $\zeta_{T,i}$ values. 
The combined $\zeta_T$ index has a typical uncertainty of 0.1--0.2.
%We find excellent agreement in 
Metallicity index values for sources with multiple spectra (GJ~576B and J0140+0150)
agree within these uncertainties.
Several sources in our sample, including the benchmark companions {Wolf~940B and} Wolf~1130C, fall outside the fit ranges in Table~\ref{tab:index-relations} or had incomplete spectral coverage and thus $\zeta_T$ values were unable to be measured. 
%We find consistent trends for each of the metallicity indices, with local T dwarfs within 10-15\% of $\zeta_{T,i}$ = 1 and classified subdwarfs and extreme subdwarfs below $\zeta_{T,i} \lesssim 0.8$ and $\zeta_{T,i} \lesssim 0.6$, respectively. 

We found excellent alignment between $\zeta_T$ and our metallicity classifications.
Sources with dwarf classifications, including our local T dwarf sample, have $\zeta_{T,i}$ = 1 within 10-15\%,
while subdwarfs and extreme subdwarfs typically having $\zeta_{T,i} \lesssim 0.8$ and $\zeta_{T,i} \lesssim 0.6$, respectively.
{All three of} our new subdwarf discoveries, J0623+0715, {J1524$-$2620,} and J2112$-$0529 have $\zeta_T$ {$\approx$ 0.8}.
On the other hand, there are outliers among our late-type T dwarfs, notably
%the d/sdT8 candidate J0911+2146 with $\zeta_T = 0.70$, 
the d/sdT9 {J0833+0052 with $\zeta_T = 0.62$,} 
the d/sdT8 J0939$-$2448 with $\zeta_T = 0.45$,
and the sdT7 standard J1416+1348B with $\zeta_T = 0.23$,
all smaller than expected for their metallicity classifications.
{In addition, the peculiar red T8.5 Ross 458C, which has an unusually bright $K$-band peak, has a large $\zeta_T = 1.71$.} 
These sources have only the $\zeta_{T,4}$ index measured, which is based on the [K/H] versus [H-dip] index pairing. As noted above, the LOWZ models show a coupling between temperature and metallicity for these index pairs at low temperatures (Figure~\ref{fig:index2}) 
{and, in the case of Ross~458C, may not account for enhanced cloud opacity \citep{2010ApJ...725.1405B,2012ApJ...756..172M}.}
Thus, the {discrepant} $\zeta_T$ values of these sources, and potentially other very late-type T dwarfs, may reflect a {temperature} bias
in this particular metallicity index.
%The esdT3: J1810$-$1010 also stands apart as having a low $\zeta_T$ for its modest 

\begin{figure*}[ht!]
\centering
\includegraphics[width=0.9\textwidth]{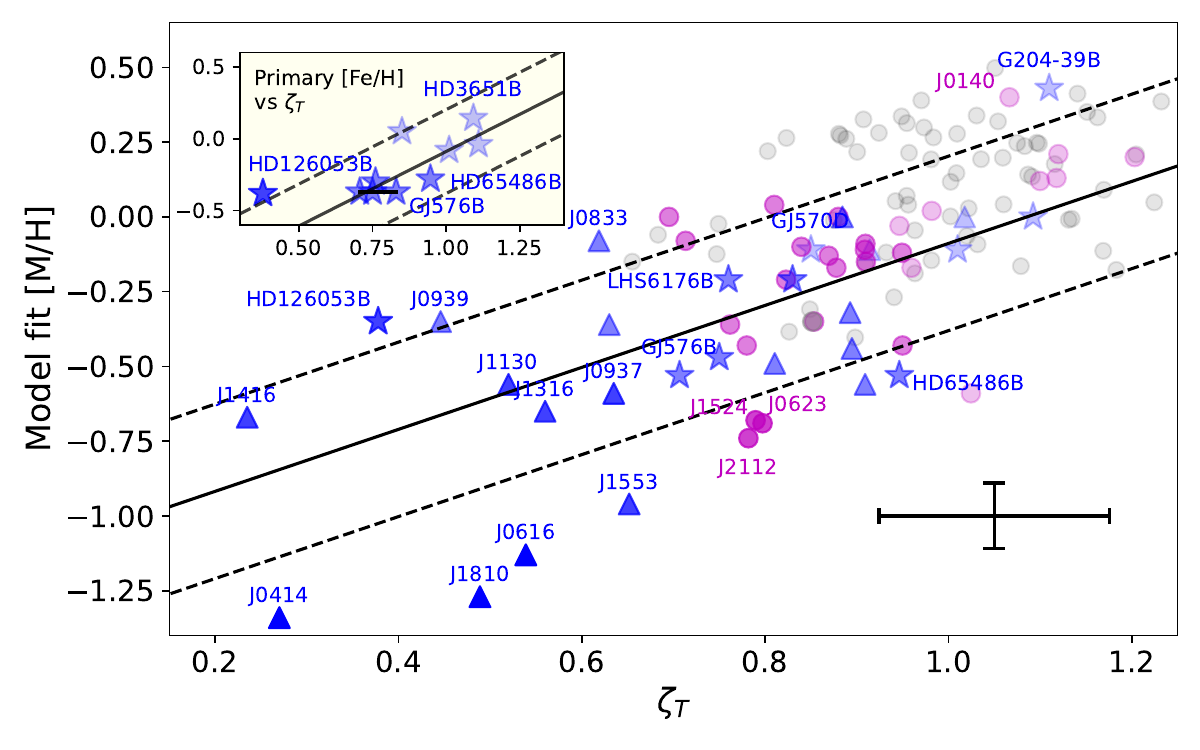}
\caption{(Main panel) Comparison of $\zeta_T$ to total metallicity [M/H] from spectral model fits (Table~\ref{tab:modelfit}). 
Symbols are the same as Figure~\ref{fig:index1}.
The error bar at bottom right indicates the median uncertainties in $\zeta_T$ and [M/H] for the entire sample.
We include values for local T dwarfs (grey dots) based on best-fit LOWZ grid model fits, requiring model \teff\ $<$ 1600~K and robust $\zeta_T$ measurements; an additional Gaussian scatter of 0.05 in $\zeta_T$ and [M/H] has been added to these values for clarity.
The diagonal solid and dashed lines delineate the linear correlation between  $\zeta_T$ and [M/H] and its 1$\sigma$ uncertainty of {0.29}~dex.
Specific sources discussed in the text are labeled.
(Inset panel) Comparison of $\zeta_T$ to iron abundance [Fe/H] for the primaries of our benchmark companions.
HD~3651B, {HD~65486B},{HD~126053B}, and GJ~576B are labeled. 
Note that Wolf~1130C {and Wolf~940B do not appear on this plot as they lack measured $\zeta_T$ values.
The diagonal solid and dashed lines delineate the linear correlation between  $\zeta_T$ and primary [Fe/H] and its 1$\sigma$ uncertainty of 0.12~dex}.
\label{fig:zeta-metallicity}}
\end{figure*}

To further {assess} the reliability of the $\zeta_T$ index, Figure~\ref{fig:zeta-metallicity} compares this index to total metallicity [M/H] based on the best-fit spectral models (Table~\ref{tab:modelfit}), and to the iron abundances [Fe/H] of the primaries of our benchmark companions (Tables~\ref{tab:observations} and~\ref{tab:additional}). The majority of our sample shows a roughly linear trend between $\zeta_T$ and [M/H], with a Pearson r correlation of {0.46$\pm$0.07.
%, that encompasses sources with the most subsolar metallicity (J0414$-$5854 with [M/H] = $-$1.34 and $\zeta_T$ = 0.27) to the most supersolar metallicity sources (G~204-39B with [M/H] = 0.43 and $\zeta_T$ = 1.11). 
The dispersion becomes larger at the lowest metallicities and $\zeta_T$ values, and may reflect the temperature-metallicity coupling for} 
%A deviant spur of very late-type, modestly metal-poor T dwarfs with unusually low values of $\zeta_T$ is visible in this Figure, 
the $\zeta_{T,4}$ index for very late-type T dwarfs.
%The extreme subdwarf J1810$-$1010 also appears to stand apart for the mean trend, likely due to its model-fit metallicity being overestimated for its metallicity classification (see Section~\ref{sec:J1810}). 
Excluding {the L dwarfs in our sample and the unusually red T8.5p dwarf Ross~458C,} we find a roughly linear trend between $\zeta_T$ and [M/H] for {57} sources:  
\begin{equation}
    [M/H] = 1.03611\zeta_T - 1.12491
    \label{eqn:zetamh}
\end{equation}
{spanning 0.24 $\leq$ $\zeta_T$ $\leq$ 1.2 and $-$1.34 $\leq$ [M/H] $\leq$ +0.43, with a metallicity scatter of 0.29~dex.}
%, comparable to the typical inferred metallicity uncertainty. 
We emphasize that this is a {\em model-dependent} correlation, and may be skewed by inaccuracies in the model fitting.
{However, this trend also accurately reproduces the correlation between $\zeta_T$ and [Fe/H] measurements for the primaries of our benchmark companions, with two key exceptions: the aforementioned Ross~458C and the sdT7.5 HD~1260538B, which has a low value of $\zeta_T = 0.38$ for its primary [Fe/H] = $-$0.38. 
Excluding these sources, we find a tight correlation among the 9 remaining benchmarks (Pearson r = 0.52$\pm$0.23) with a similar linear trend}
\begin{equation}
    [Fe/H] = 0.99844\zeta_T - 1.07333
    \label{eqn:zetafeh}
\end{equation}
{for 0.71 $\leq$ $\zeta_T$ $\leq$ 1.11 and $-$0.37 $\leq$ [M/H] $\leq$ +0.14, with a metallicity scatter of 0.12~dex.}
The small number of benchmark companions with both [Fe/H] and $\zeta_T$ measurements, and the limited metallicity range of this sample, warrants some caution in the use of this relationship.   
Nevertheless, these comparisons indicate an overall self-consistency between the metallicity classifications, the $\zeta_T$ index, and the actual metallicities of our sources, whether inferred from spectral model fits or stellar primaries.

\section{Discussion} \label{sec:discussion}

\subsection{Effective Temperature Trends for T Subdwarfs} \label{sec:temperature}

The classification of a star is a proxy for its physical parameters, including temperature, luminosity, surface gravity, and (in the case of metallicity classes) composition. For cool stars and brown dwarfs, there are numerous relations linking dwarf numerical spectral types to luminosities and effective temperatures, the latter either requiring a measured or estimated radius inferred from atmosphere model fitting
\citep{2013AJ....146..161M,2013ApJS..208....9P,2015ApJ...810..158F,2017ApJS..231...15D,2021ApJS..253....7K,2023ApJ...959...63S}.
There are well-documented metallicity offsets in the spectral type/temperature relations of M and L dwarfs, with subdwarfs typically found to be warmer by several 100~K than their equivalently-classified dwarf counterparts \citep{2006ApJ...645.1485B,2007ApJ...657..494B,2014AandA...564A..90R,2017MNRAS.464.3040Z}.
It is thus useful to explore how effective temperature trends vary with metallicity in the T dwarf regime.

\begin{figure}[tbp]
\centering
\plotone{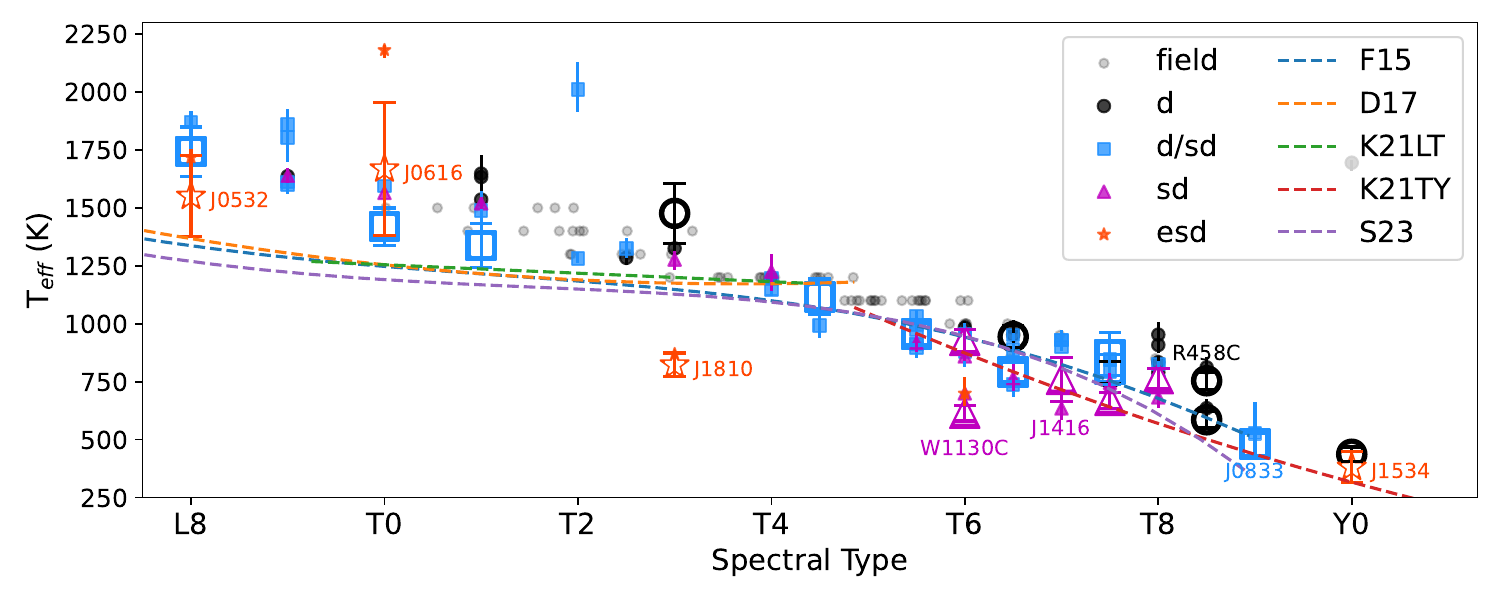}
\caption{Inferred effective temperatures of sources in our spectral sample based on model fits as a function of numerical spectral type.
Symbol shapes and colors encode the metallicity class, with small symbols representing temperatures from spectral model fits (Table~\ref{tab:modelfit}) and large symbols representing bolometric temperatures for sources with measured parallaxes (Table~\ref{tab:teff}).
The light grey circles represent spectral model fit temperatures for our local T dwarf sample.
Five spectral type/temperature relations for solar-metallicity dwarfs are shown for comparison:
% \citet[M2013]{2013AJ....146..161M},
% \citet[P2013]{2013ApJS..208....9P},
\citet[F15]{2015ApJ...810..158F},
\citet[D17]{2017ApJS..231...15D},
\citet[K21LT and K21TY]{2021ApJS..253....7K},
and \citet[S23]{2023ApJ...959...63S}.
%See the main text for comments on the discrepancies between dwarf and subdwarf temperatures for the late-type L dwarfs and early-type T dwarfs.
}
\label{fig:teff}
\end{figure}

Our first approach was to compare effective temperatures from our model fits to assigned spectral classifications, as illustrated in Figure~\ref{fig:teff}. We see the expected decline of {\teff} with later spectral type, although {many} of our sources---including our local comparison sample---are warmer than previously-established dwarf empirical relations. These offsets are particularly large for the late-L and early-T dwarfs. For example, our model fit temperature for {J0616$-$6407, {\teff} = 2181$^{+14}_{-26}$~K, is over 800~K warmer than the equivalent solar-metallicity T0 dwarf ({\teff} $\approx$ 1250~K}; \citealt{2015ApJ...810..158F}). As this source has a measured parallax, we can also deduce that its model-fit temperature is overestimated based on the {non-physically} small radius required to match its absolute fluxes.
%R$_\beta$ = 0.051~R$_\odot$; 
Similar scaling issues are seen for other sources in our sample with trigonometric parallaxes (Table~\ref{tab:modelfit}).

We therefore considered an alternative estimate for {\teff} based on bolometric magnitude
\begin{equation}
    T_B = T_\odot\sqrt{\frac{R_\odot}{R}}10^{(M_{bol,\odot}-M_{bol})}
\end{equation}
where $T_\odot$ = 5772~K, $R_\odot$ = 6.957$\times$10$^{10}$~cm, and $M_{bol,\odot}$ = 4.74 are the temperature, radius and bolometric magnitude of the Sun \citep{mamajek2015iau}. 
Ignoring interstellar absorption, the bolometric magnitude can be inferred from the absolute magnitude in filter $x$ via the bolometric correction (BC)
\begin{equation}
    BC_x = M_{bol} - M_x = M_{bol} - m_x + 5\log_{10}\left(\frac{d}{10~pc}\right).
\end{equation}
While empirical BCs for solar-metallicity T dwarfs exist in the literature
(e.g., \citealt{2010ApJ...722..311L,2015ApJ...810..158F,2023ApJ...959...63S}), we chose to compute BCs from the LOWZ atmosphere models to incorporate metallicity effects. For models computed in surface fluxes,
\begin{equation}
    BC_x = 4.74 - 10\log_{10}\left(\frac{T}{T_\odot}\right) - 5\log_{10}\left(\frac{R_\odot}{10~pc}\right) - m_x
\end{equation}
\citep{2014MNRAS.444..392C}, where $m_x$ is the apparent magnitude computed directly from the {model spectra}. As our available photometry is on the Vega system, we computed spectrophotometric model magnitudes as
\begin{equation}
    m_x = -2.5\log_{10}\left(\frac{\int\lambda{f_\lambda}(\lambda)T_x(\lambda)d\lambda}{\int\lambda{f^V_\lambda}(\lambda)T_x(\lambda)d\lambda}\right)
\end{equation}
using a model spectrum of Vega $f^V_\lambda$ from \citet{1993sssp.book.....K} and filter profiles $T_x$ from the Spanish Virtual Observatory Filter Profile Service\footnote{\url{http://svo2.cab.inta-csic.es/theory/fps/}.} \citep{2012ivoa.rept.1015R,2020sea..confE.182R}.
Appendix~\ref{app:bc} provides a list of bolometric corrections for various infrared filters for the LOWZ models; here, we focus on $J$ and $W2$ photometry for sources with measured parallaxes. As the bolometric corrections are inherently dependent on the physical properties of the atmosphere, we assumed {\logg} = 5.0 (cgs), $\log{k}_{zz}$ = 2 (cgs), C/O = 0.55, and a metallicity based on the classification: [M/H] = 0 for dT, [M/H] = $-$0.25 for d/sdT, [M/H] = $-$0.75 for sdT, and [M/H] = $-$1.5 for esdT. 
We also assumed a theoretical radius of $R_{th}$ = 0.077~R$_\odot$ based on the solar-metallicity evolutionary models of \citet{2001RvMP...73..719B} and \citet{2003AandA...402..701B}, and assumed a 10\% uncertainty on this value.
We iteratively converged to a self-consistent bolometric temperature starting from the model fit temperature, and estimated the uncertainty in $T_B$ as the greater of the combined formal uncertainty from both filters,
\begin{equation}
    \left(\frac{\sigma_{T_B}}{T_B}\right)^2 = \left(\frac{1}{2}\frac{\sigma_{R_{th}}}{R_{th}}\right)^2 + \left(\frac{1}{2}\frac{\sigma_\pi}{\pi}\right)^2 + \left(\frac{0.1}{\ln{10}}\sigma_x\right)^2
\end{equation}
(here $\pi$ is parallax, x $\in$ \{$J$,$W2$\}) and half of the difference between the temperatures inferred from $J$ and $W2$.

\begin{deluxetable}{llcccccccccc}
\tablecaption{Bolometric Temperatures for T {and Y Dwarf Benchmarks and} Subdwarfs\label{tab:teff}} 
\tabletypesize{\tiny} 
\tablehead{ 
 & & \multicolumn{4}{c}{MKO $J$} &\multicolumn{4}{c}{WISE $W2$} & \colhead{Adopted} & \colhead{Model} \\
 \cline{3-6} \cline{7-10}
\colhead{Name} & 
\colhead{SpT} & 
\colhead{$M_J$} & 
\colhead{$BC_J$\tablenotemark{a}} & 
\colhead{$M_{bol}$} & 
\colhead{$T_B$} & 
\colhead{$M_{W2}$} & 
\colhead{$BC_{W2}$\tablenotemark{a}} & 
\colhead{$M_{bol}$} & 
\colhead{$T_B$} & 
\colhead{$T_B$} & 
\colhead{$T_B$} \\
 & & 
\colhead{(mag)} &
\colhead{(mag)} &
\colhead{(mag)} &
\colhead{(K)} &
\colhead{(mag)} &
\colhead{(mag)} &
\colhead{(mag)} &
\colhead{(K)} &
\colhead{(K)} &
\colhead{(K)} \\
} 
\startdata 
\multicolumn{12}{c}{T Dwarf Benchmarks} \\
\cline{1-12}
HD~3651B &  T8 &  16.08$\pm$0.03 & 2.73 & 18.81 & 801$\pm$80 & \nodata\tablenotemark{b} & \nodata & \nodata & \nodata & 801$\pm$80 & 824$^{+20}_{-31}$  \\ 
{HD~65486B} &  d/sdT4.5 &  14.79$\pm$0.08 & 2.29 & 17.08 & 1193$\pm$119 & 12.53$\pm$0.02 & 5.15 & 17.68 & 1039$\pm$104 & 1116$\pm$77 & 993$^{+44}_{-47}$ \\ 
LHS~6176B &  d/sdT7.5 &  16.56$\pm$0.03 & 2.61 & 19.17 & 738$\pm$74 & 13.03$\pm$0.02 & 5.44 & 18.47 & 867$\pm$87 & 802$\pm$64 & 846$^{+19}_{-6}$ \\ 
{Ross~458C} &  rT8.5 &  15.02$\pm$0.05 & 2.73 & 17.75 & 1023$\pm$102 & 13.59$\pm$0.02 & 5.52 & 19.10 & 750$\pm$75 & 887$\pm$137 & 812$^{+38}_{-28}$  \\ 
{HD~126053B} &  sdT7.5 &  17.50$\pm$0.05 & 2.20 & 19.69 & 654$\pm$65 & 13.57$\pm$0.03 & 5.93 & 19.49 & 685$\pm$69 & 670$\pm$34 & 746$^{+36}_{-28}$ \\ 
GJ~570D &  T7.5 &  16.47$\pm$0.05 & 2.67 & 19.15 & 742$\pm$74 & 13.29$\pm$0.01 & 5.34 & 18.63 & 836$\pm$84 & 789$\pm$47 & 837$^{+20}_{-17}$ \\ 
GJ~576B &  d/sdT5.5 &  15.19$\pm$0.02 & 2.71 & 17.90 & 988$\pm$99 & 12.91$\pm$0.02 & 5.32 & 18.23 & 917$\pm$92 & 953$\pm$48 & 995$^{+19}_{-21}$ \\ 
G~204$-$39B &  T6.5 &  15.42$\pm$0.09 & 2.66 & 18.09 & 947$\pm$95 & 13.11$\pm$0.01 & 5.18 & 18.28 & 905$\pm$91 & 926$\pm$46 & 984$^{+25}_{-16}$  \\ 
Wolf~1130C &  (e)sdT6: &  18.54$\pm$0.09 & 1.25 & 19.79 & 640$\pm$64 & 13.77$\pm$0.05 & 5.97 & 19.74 & 648$\pm$65 & 644$\pm$32 & 700$^{+15}_{-0}$\tablenotemark{c} \\ 
{Wolf~940B} &  T8.5 &  17.56$\pm$0.06 & 2.36 & 19.92 & 621$\pm$62 & 14.40$\pm$0.06 & 6.05 & 20.45 & 550$\pm$55 & 586$\pm$35 & 636$^{+32}_{-29}$ \\ 
HN~PegB &  T2.5 &  14.57$\pm$0.03 & 2.19 & 16.75 & 1288$\pm$129 & \nodata\tablenotemark{b} & \nodata & \nodata & \nodata & 1288$\pm$129 & 1284$^{+42}_{-23}$\\ 
\cline{1-12}
\multicolumn{12}{c}{{T and Y Subdwarfs}} \\
\cline{1-12}
%J0055+5947 &  T6.5 &  16.10$\pm$0.05 & 2.73 & 18.83 & 797$\pm$80 & 13.33$\pm$0.03 & 5.48 & 18.81 & 802$\pm$80 & 800$\pm$40 & 738$^{+35}_{-46}$\\ 
%J0004$-$2604 &  T3 &  14.09$\pm$0.34 & 2.20 & 16.29 & 1433$\pm$145 & 11.75$\pm$0.31 & 4.29 & 16.03 & 1520$\pm$153 & 1476$\pm$128 & 1325$^{+23}_{-24}$\\ 
J0301$-$2319 &  d/sdT1 &  14.49$\pm$0.26 & 2.24 & 16.74 & 1292$\pm$130 & 11.95$\pm$0.22 & 4.49 & 16.44 & 1383$\pm$139 & 1338$\pm$96 & 1488$^{+77}_{-73}$\\ 
J0309$-$5016 &  d/sdT7.5 &  16.29$\pm$0.13 & 2.66 & 18.96 & 775$\pm$78 & 12.75$\pm$0.13 & 5.27 & 18.02 & 961$\pm$96 & 868$\pm$93 & 795$^{+55}_{-27}$ \\ 
J0532+8246 &  esdL8: &  13.23$\pm$0.06 & 2.26 & 15.48 & 1725$\pm$173 & 11.73$\pm$0.03 & 4.73 & 16.47 & 1376$\pm$138 & 1550$\pm$175 & 1730$^{+43}_{-46}$\\ 
J0616$-$6407 &  esdT0: &  12.8$\pm$0.7 & 2.3 & 15.10 & 1884$\pm$197 & 11.7$\pm$0.7 & 4.58 & 16.2 & 1452$\pm$152 & 1668$\pm$288 & 2181$^{+14}_{-26}$\tablenotemark{d} \\ 
J0645$-$6646 &  d/sdT0 &  14.18$\pm$0.12 & 2.04 & 16.21 & 1458$\pm$146 & 11.96$\pm$0.12 & 4.49 & 16.45 & 1380$\pm$138 & 1419$\pm$81 & 1594$^{+25}_{-20}$ \\ 
{J0833+0052} &  d/sdT9 &  19.86$\pm$0.15 & 1.62 & 21.49 & 433$\pm$43 & 14.49$\pm$0.12 & 6.13 & 20.63 & 528$\pm$53 & 480$\pm$48 & 528$^{+129}_{-28}$\tablenotemark{d} \\ 
J0850$-$0221 &  d/sdL6.5 &  12.70$\pm$0.11 & 1.96 & 14.66 & 2086$\pm$209 & 10.36$\pm$0.10 & 4.23 & 14.58 & 2121$\pm$212 & 2104$\pm$115 & 1852$^{+48}_{81}$\\ 
J0937+2931 &  sdT6 &  15.71$\pm$0.04 & 2.57 & 18.28 & 906$\pm$91 & 12.74$\pm$0.03 & 5.33 & 18.07 & 952$\pm$95 & 929$\pm$47 & 859$^{+28}_{-15}$\\ 
J0939$-$2448 &  d/sdT8 &  17.13$\pm$0.12 & 2.37 & 19.50 & 683$\pm$68 & 13.10$\pm$0.12 & 5.44 & 18.54 & 853$\pm$85 & 768$\pm$85 & 682$^{+51}_{-40}$\\ 
%J1055+5443 &  Y0 &  19.67$\pm$0.23 & 1.91 & 21.58 & 423$\pm$43 & 15.18$\pm$0.22 & 6.14 & 21.31 & 450$\pm$45 & 437$\pm$31 & 670$^{+105}_{-79}$\tablenotemark{d}\\ 
J1158+0435 &  d/sdL8 &  13.47$\pm$0.16 & 1.96 & 15.43 & 1745$\pm$175 & 11.22$\pm$0.16 & 4.23 & 15.45 & 1739$\pm$174 & 1742$\pm$107 & 1870$^{+40}_{-64}$\\ 
J1416+1348B &  sdT7 &  17.42$\pm$0.02 & 2.20 & 19.62 & 666$\pm$67 & 12.94$\pm$0.04 & 5.59 & 18.54 & 854$\pm$85 & 760$\pm$94 & 635$^{+68}_{-42}$\\ 
J1810$-$1010 &  esdT3: &  17.52$\pm$0.16 & 1.42 & 18.95 & 777$\pm$78 & 12.84$\pm$0.16 & 5.50 & 18.34 & 893$\pm$90 & 835$\pm$58 & 869$^{+5}_{-13}$ \\ 
{J1534$-$1043} &  (e)sdY: &  23.4$\pm$0.3 & $-$0.56\tablenotemark{d} & 22.9 & 314$\pm$32 & 15.09$\pm$0.19 & 6.26\tablenotemark{e} & 21.35 & 447$\pm$45 & 381$\pm$67 & \nodata \\ 
\enddata
\tablenotetext{a}{Bolometric corrections computed from the LOWZ models assuming {\logg} = 5.0 (cgs), $\log{k}_{zz}$ = 2 (cgs), C/O = 0.55, a metallicity based on the source classification ([M/H] = 0 for dwarfs, [M/H] = $-$0.25 for mild subdwarfs, [M/H] = $-$0.75 for subdwarfs, and [M/H] = $-$1.5 for extreme subdwarfs), and the closest temperature to $T_B$ in the model grid. See Appendix~\ref{app:bc} for the full list of bolometric corrections.}
\tablenotetext{b}{The proximity of a bright stellar companion prevents WISE detection of this source.}
\tablenotetext{c}{{Fit {\teff} at limit of model parameters.}}
\tablenotetext{d}{Model fit produces {non-physically} small radius (R $\lesssim$ 0.05~R$_\odot$), suggesting an overestimated {\teff}.}
\tablenotetext{e}{{Using the BC value for {\teff} = 500~K and [M/H] = $-$1.5.}}
\end{deluxetable} 

Table~\ref{tab:teff} lists the resulting bolometric temperatures, which are displayed in Figure~\ref{fig:teff}. These temperatures are {generally} lower than those inferred from the model fits, {in some cases by hundreds of degrees Kelvin (e.g., J0616$-$6407),} in alignment with {the} underestimated radii {from the model fits for these sources}. 
Nevertheless, the {bolometric temperatures for the} late-L and early-T subdwarfs remain warmer than the dwarf sequence. The mid- and late-type T subdwarfs, on the other hand, appear to be cooler than their equivalently-classified dwarf counterparts, notably so for 
J1810$-$1010 ($T_B$ = 790~K compared to 1150~K for a typical T3 dwarf) and
Wolf~1130C ($T_B$ = 645~K compared to 940~K for a typical T6 dwarf).
This apparent reversal in the metallicity-induced temperature offset may be caused by a flattening of the H$_2$O and CH$_4$ features by enhanced H$_2$ absorption, {or reduced absorption from these molecules due to reduced C and O abundances.
Both cases lead} to earlier spectral classifications at a given temperature. 
We also note that this temperature reversal occurs around the L/T transition, where dwarf spectra evolve due
to changes in photospheric cloud properties (or equivalently the adiabatic gradient; \citealt{2016ApJ...817L..19T}), producing a flat temperature/spectral type relation; i.e., large changes in spectral morphology with minimal change in {\teff} \citep{2002ApJ...571L.151B,2012ApJS..201...19D}.
The absence of significant condensate opacity in metal-poor brown dwarf atmospheres could also modify the relationship between temperature and the evolution of spectral morphology.
This {discussion} highlights the non-trivial systematic effects that may be present when inferring subdwarf temperatures from either spectral types or model fitting, and emphasizes the need for additional parallax measurements to infer robust atmospheric properties.

\subsection{Prospects for Growing the Sample of T Subdwarfs} \label{sec:prospects}

To deepen our understanding of the formation, evolution, and atmosphere properties of metal-poor brown dwarfs, it is necessary to expand the known sample of T and Y subdwarfs.
Increasing the local sample (e.g., d $\lesssim$ 20~pc) is a challenge given the sparsity of metal-poor stars in general and metal-poor brown dwarfs in particular, the latter estimated to be as low as 0.1\% \citep{2020ApJ...898...77S}.
%to up to 9\% \citep{lodieu2022} of the local brown dwarf population.
Nevertheless, this study's focus on relatively bright sources ($J < 18.5$) accessible from the northern hemisphere
means that several promising T subdwarf candidates remain to be characterized in the BYW sample alone
(cf.\ \citealt{2022AJ....163...47B,2023AJ....166...57M}).
%Fortunately, given the long-term cooling of thick disk and halo brown dwarfs, T and Y subdwarfs are expected to be relatively more common as compared to warmer late-M and L subdwarfs, enhancing the preference for cold brown dwarfs among the local solar-metallicity population \citep{2021ApJS..253....7K}.

Current and forthcoming deep imaging and spectroscopic surveys are also expected to substantially increase the number of halo brown dwarfs with {photometric}, spectroscopic, and astrometric 
characterization.
In addition to the individual discoveries now being made in deep pointings with JWST to limiting magnitudes of $\approx$30~AB
\citep{2023ApJ...942L..29N,2023ApJ...947L..25G,2023MNRAS.523.4534W,2024ApJ...962..177B,2024ApJ...964...66H,2024MNRAS.529.1067H}, the recently-launched Euclid mission is expected to detect of order 10$^4$ thick disk and 10$^3$ halo T dwarfs in its 15,000~deg$^2$ wide-field survey, reaching $YJH$ $\approx$ 24.5 AB \citep{2011arXiv1110.3193L,2019MNRAS.486.1260Z,2021MNRAS.501..281S}. 
A significant fraction of these sources will also be observed with Euclid's 
slitless spectroscopic mode, yielding 1.25--1.85~$\mu$m spectra at $\lambda/\Delta\lambda$ $\approx$ 250 for potentially hundreds of metal-poor brown dwarfs. 
Starting in the late 2020s, the Nancy Grace Roman Space Telescope's High Latitude Wide Area Imaging Survey, spanning 2,000~deg$^2$ to $YJH$ $\approx$ 27 AB \citep{2021MNRAS.501.2044T}, is expected to detect 5--10$\times$ more sources than Euclid \citep{2019MNRAS.486.1260Z}. This survey is also matched to a spectroscopic survey 
that will acquire 1.0--1.93~$\mu$m spectra at $\lambda/\Delta\lambda$ $\approx$ 400--800,
covering many of the metallicity-sensitive features examined in this study.
These infrared surveys will be complemented by optical photometry from the Vera Rubin Observatory Large Survey of Space and Time (LSST; \citealt{2019ApJ...873..111I}). 
{While this survey will detect few} halo brown dwarfs due to its poor sensitivity at low temperatures (10s to 100s over the Southern sky; \citealt{2019MNRAS.486.1260Z}), it will provide useful color limits and $y$-band dectections for refining candidate samples.

The multi-epoch astrometry of all three surveys will enable identification of candidate nearby halo brown dwarfs through reduced proper motion constraints, candidates which can be spectroscopically confirmed with JWST or planned 30-meter class ground-based facilities.
In short, the limited sample presented here is a small fraction of the halo brown dwarf population expected to be discovered and explored over the next decade, primarily by space-based infrared surveys that continue the legacy of WISE.

\section{Summary} \label{sec:summary}

This work has presented a comprehensive study of the spectroscopic properties of {currently known} metal-poor T dwarfs, augmented with new discoveries
%of T dwarfs of varying metallicities 
from the Backyard Worlds: Planet 9 program. Our key results are as follows:

\begin{itemize}
    \item Using photometric and astrometric data from several wide-field infrared surveys, most notably the CatWISE2020 catalog, and imposing color and reduced proper motion constraints, we identified a sample of 95 high-probability, bright ($J$ $\leq$ 18.5) T subdwarf candidates visible from the northern hemisphere, which includes several previously reported metal-poor brown dwarfs 
    \citep{2020ApJ...898...77S,2020ApJ...899..123M,2021ApJS..253....7K,2024ApJS..271...55K}.
    \item Near-infrared spectroscopic follow-up with Keck/NIRES, IRTF/SpeX, APO/TSpec, {Magellan/FIRE, and Gemini-N/GNIRS} of a subset of these candidates has allowed us to identify {three} new T subdwarfs: {the sdT0 J1524$-$2620, the sdT1 J2112$-$0529, and the sdT3 J0623+0715. We also identify 19} new mildly metal-poor T dwarfs. These identifications are based on metallicity-sensitive spectral features, including a broadened 1.1~$\mu$m absorption band, suppressed 2.1~$\mu$m peak emission, and weak or absent K~I lines at 1.17 and 1.25~$\mu$m.
%    {Eight} late-L and T dwarfs {observed in this study} that do not show significantly subsolar metallicities.
    %, one of which (J2021+1524) is likely an unresolved brown dwarf binary.
    %based on its spectral peculiarities, an assessment that will need to be verified by parallax measurement and/or high resolution imaging. 
    \item Atmosphere model fits to the spectra of candidate and previously reported late-L and T subdwarfs and benchmark companions allowed us to validate the metal-poor nature of our candidates, as well as identify several prior discoveries as normal dwarfs. Comparing to several sets of spectral models, we find that the LOWZ models \citep{2021ApJ...915..120M} typically provide the best fits for {all brown dwarfs with} 800~K $\lesssim$ {\teff} $\lesssim$ 1600~K, {the SAND models \citep{Alvarado_2024} perform better at higher temperatures and low metallicities, and} the ElfOwl models \citep{2024ApJ...963...73M} perform better at lower temperatures {and near-solar metallicities.
    Notably, the SAND models provide self-consistent fits to the spectra of extreme L subdwarfs (e.g., J0532+8246) and T subdwarfs (e.g., J1810$-$1010), indicating that these models are well-suited for metal-poor brown dwarfs.}
%    We show that there are continued issues with self-consistent modeling of the exemplary extreme T subdwarf J1810$-$1010, possibly due to missing opacities in the LOWZ models or errors in the spectral data calibration.
    \item By forward-modeling moderate-resolution Keck/NIRES spectral data, we measured the radial velocities of {27} candidate and comparison late-L and T subdwarfs. The corresponding $UVW$ velocities are more dispersed than the local T dwarfs, and several sources have velocities consistent with the thick disk or halo populations. Two subdwarfs, the esdL8 J0532+8246 and the d/sdT5.5 J1130+3139, have retrograde Galactic orbits potentially consistent with the Thamnos stream; while the sdT4 J1553+6933 has a motion potentially consistent with the Helmi stream.
%    , albeit with significant uncertainty. 
    We also identify three sources with thick disk kinematics that appear to be metal-rich, and may be part of the high-velocity thick disk that originated in the inner Milky Way.
    \item We used our spectral sample to construct a metallicity classification system for T dwarfs anchored to spectral standards. Our scheme encompasses classes of mild subdwarfs (d/sdT), subdwarfs (sdT), and extreme subdwarfs (esdT), with estimated metallicities of [M/H] $\approx$ {$-$0.4}, [M/H] $\approx$ {$-$0.6}, and [M/H] $\approx$ {$-$1.3}, respectively, matching the metallicity scales of M and L subdwarfs. 
    %We use these standards to reclassify our candidates and previously-identified T subdwarfs, several of which are re-categorized as normal T dwarfs.
    We have defined a near-infrared index to quantify metallicity, $\zeta_T$, and demonstrated that this index correlates with the metallicities inferred from spectral model fitting and the iron abundances measured from the primaries of a small sample of benchmark T dwarf companions. 
    \item We examined the relationship between effective temperature and spectral type for brown dwarfs of differing metallicities, and find that late-L and early-T subdwarfs are generally warmer and mid- and late-T subdwarfs generally cooler than equivalently-classified solar-metallicity dwarfs. The latter offset may be due to the enhanced H$_2$ absorption and reduced C and O abundances weakening the molecular bands used for spectral classification, while the switch from warmer to colder may be driven by {a reduction in} condensate opacity in subdwarfs.
\end{itemize}

%that largely drives the evolution of spectra across the L dwarf/T dwarf transition for solar-metallicity dwarfs. 
%    For the handful of sources with parallax measurements, we find that spectral model fits produce excessively warm temperatures for late-L and T subdwarfs, most notably for J1810$-$1010 for which existing atmosphere models fail to provide a consistent fit to the absolute spectral fluxes across the near-infrared band. 

The sample explored here encompasses nearly all of the lowest-temperature, metal-poor brown dwarfs currently known, but it is in no way complete. {In addition to T subdwarf candidates identified at southern declinations and fainter magnitudes by the BYW program, additional yields from JWST, Euclid, LSST and the}
% \citep{2021ApJ...915..120M,2022AJ....163...47B}, and follow-up is continuing for these sources. In addition, the recent start of the Euclid mission \citep{2010SPIE.7731E..1HL,2023arXiv231201903M}, 
% discoveries being made in deep imaging and spectroscopic surveys with JWST \citep{2023ApJ...942L..29N,2023ApJ...947L..25G,2023MNRAS.523.4534W,2024ApJ...964...66H,2024MNRAS.tmp..348H,2024ApJ...962..177B},
% and future deep surveys with the Vera Rubin Observatory and 
Nancy Grace Roman Space Telescope 
{are expected to} dramatically expand our sample of T subdwarfs from tens to thousands.
Adding these discoveries to the known T subdwarf sample will provide a more robust assessment of the diversity of metal-poor brown dwarf atmospheres, including variations in individual elemental abundances that are tied to distinct Milky Way populations.
%such as the thick disk and halo, stellar streams, and globular clusters. 
Both near-infrared spectroscopy and parallax measurements for these sources {will be} crucial, the latter providing an essential constraint on bolometric luminosity to calibrate spectral model fits.
% Cooler and fainter sources, such as the possible Y subdwarf WISE~J1534$-$1043, will require more sensitive spectroscopic follow-up with
% larger ground-based facilities or JWST, the latter providing broader infrared spectral coverage 
% where additional metallicity indicators may be present \citep{2024ApJ...962..177B,2024ApJ...963...73M}. 
Indeed, our study demonstrates that advances in metal-poor atmosphere modeling, {as illustrated by the SAND models, yield more reliable fits and self-consistent radii for absolute flux densities.}
%existing sample of T subdwarf spectra.
%, and to characterize future metal-poor brown dwarf discoveries.
We anticipate that our assessment of the currently-known T subdwarf population and this first approach at {T dwarf} metallicity classification will require revision as future searches, observations, and modeling are undertaken.

\begin{acknowledgments}
{\footnotesize We thank our telescope operators and instrument scientists at Keck Observatory, the NASA Infrared Telescope Facility, and Apache Point Observatory for their assistance in the acquisition of the spectral data reported here.
APO/TSpec data were acquired and reduced by Katelyn Allers.
We thank Ben Burningham, Nicolas Lodieu, Kevin Luhman, Greg Mace, David Pinfield, and Zenghua Zhang for making digital versions of their published data available;
and Michael Cushing for providing modified versions of Spextool avaialble for Keck/NIRES and APO/TSpec reduction.
{We also thank our referee, Zenghua Zhang, for a careful review of our manuscript and constructive feedback that improved the analysis.}
The Backyard Worlds: Planet 9 team acknowledges the many Zooniverse
volunteers who have participated in this project, from
providing feedback during the beta review stage to classifying
flipbooks to contributing to discussions on TALK. 
This material is based upon work supported by the National
Science Foundation under grant No.\ 2009136. 
This research has made use of the Spanish Virtual Observatory (https://svo.cab.inta-csic.es) project funded by MCIN/AEI/10.13039/501100011033/ through grant PID2020-112949GB-I00
Data presented herein were
obtained at the W. M. Keck Observatory, which is operated as
a scientific partnership among the California Institute of
Technology, the University of California, and the National
Aeronautics and Space Administration. The Observatory was
made possible by the generous financial support of the W.M.
Keck Foundation. 
This publication makes use of data products
from the Wide-field Infrared Survey Explorer, which is a joint
project of the University of California, Los Angeles, and the Jet
Propulsion Laboratory/California Institute of Technology, and
NEOWISE, which is a project of the Jet Propulsion
Laboratory/California Institute of Technology. WISE and
NEOWISE are funded by the National Aeronautics and Space
Administration. Part of this research was carried out at the Jet
Propulsion Laboratory, California Institute of Technology,
under a contract with the National Aeronautics and Space
Administration. 
This research has benefitted from the SpeX Prism Spectral Libraries, maintained by Adam Burgasser at \url{http://www.browndwarfs.org/spexprism}.
This research has made use of the SIMBAD database \citep{2000AAS..143....9W},
the Aladin sky atlas \citep{2000AAS..143...33B},
and the VizieR catalogue access tool
developed and operated at CDS, Strasbourg, France.
The authors recognize and acknowledge the significant cultural role
and reverence that the summit of Maunakea has with the
indigenous Hawaiian community, and that the W. M. Keck
Observatory stands on Crown and Government Lands that the
State of Hawai’i is obligated to protect and preserve for future
generations of indigenous Hawaiians. 
Portions of this work
were conducted at the University of California, San Diego,
which was built on the unceded territory of the Kumeyaay
Nation, whose people continue to maintain their political
sovereignty and cultural traditions as vital members of the San
Diego community.}
\end{acknowledgments}

\vspace{5mm}
\facilities{Keck Observatory(NIRES), Infrared Telescope Facility(SpeX), Apache Point Observatory(TripleSpec), {Magellan(FIRE), Gemini-North(GNIRS)}}

\software{
astropy \citep{2013AandA...558A..33A,2018AJ....156..123A,2022ApJ...935..167A},  
Matplotlib \citep{2007CSE.....9...90H},
NumPy \citep{2011CSE....13b..22V},
pandas \citep{mckinney-proc-scipy-2010},
SciPy \citep{2020NatMe..17..261V},
SpeXTool \citep{2004PASP..116..362C},
SPLAT \citep{2017ASInC..14....7B}
}

\appendix

% \section{T Subdwarf Candidates\label{app:candidates}}

% The following Table provides a list of all 95 candidate T subdwarfs identified following the selection criteria listed in Section~\ref{sec:selection}.
% We list the CATWISE2020 designation (J2000 sexigesimal coordinates at epoch 2015 May 28);
% $JKW1W2$ photometry from 2MASS, the UKIDSS Galactic Plane Survey \citep{2007MNRAS.379.1599L} and Hemisphere Survey \citep{2018MNRAS.473.5113D}, the
% Vista Via Lactae and Hemisphere Surveys \citep{2013Msngr.154...35M}, and CATWISE2020.
% astrometry from CATWISE2020 or Gaia DR3 \citep{2023AandA...674A...1G};
% calculated reduced $J$-band proper motion ($H_J$);
% measured or estimated distance from \citet{2021ApJS..253....7K};
% and estimated tangential velocity.
% We also note any sources that have been previously reported in the literature.

% \adamb{do we keep this or just point to "future discovery papers"?}

% \include{table_candidates}

\section{IRTF/SpeX Comparison Sample\label{app:splat}}

The following Tables provide the measured spectral indices (Table~\ref{tab:indices_splat}) and LOWZ model fit parameters (Table~\ref{tab:splat_modelfit}) for a sample of 184 late-L and T dwarfs with low-resolution near-infrared spectra obtained with the IRTF/SpeX instrument \citep{2003PASP..115..362R} and curated in the SPLAT archive \citep{2017ASInC..14....7B}. 
Data references for the individual spectra are included in the Table endnotes.
The model fits for these data were conducted exclusively on the individual grid models without interpolation.

\begin{longrotatetable}
\begin{deluxetable}{llcccccccccccl}
\tablecaption{Index Measurements for Dwarf Late-L and T Dwarf Sample\label{tab:indices_splat}} 
\tabletypesize{\tiny} 
\tablehead{ 
%  & 
% \multicolumn{1}{c}{Lit.} & & & & & &  & & &
% \multicolumn{1}{c}{Est.} & &
% \multicolumn{1}{c}{Data}
% \\
\colhead{Designation} & 
\colhead{SpT\tablenotemark{a}} & 
\colhead{[CH4-J]} & 
\colhead{[CH4-H]} & 
\colhead{[CH4-K]} & 
\colhead{[H2O-J]} & 
\colhead{[H2O-H]} & 
\colhead{[H2O-K]} & 
\colhead{[Y/J]} & 
\colhead{[K/H]} & 
\colhead{[K/J]} & 
\colhead{[H-dip]} & 
\colhead{$\langle\zeta_T\rangle$} & 
\colhead{Ref.}   \\ 
} 
\startdata 
J00001354+2554180 & T4.5 & 0.549$\pm$0.015 & 0.548$\pm$0.006 & 0.259$\pm$0.005 & 0.327$\pm$0.002 & 0.349$\pm$0.005 & 0.480$\pm$0.004 & 0.476$\pm$0.005 & 0.352$\pm$0.002 & 0.197$\pm$0.001 & 0.738$\pm$0.008 & 1.00$\pm$0.09 &  [3] \\
J00062785+1857288 & L7 & 1.011$\pm$0.021 & 1.056$\pm$0.018 & 1.001$\pm$0.013 & 0.727$\pm$0.018 & 0.781$\pm$0.017 & 0.809$\pm$0.013 & 0.508$\pm$0.013 & 0.711$\pm$0.011 & 0.691$\pm$0.011 & 0.995$\pm$0.012 & 0.93$\pm$0.12 &  [29] \\
J00150206+2959323 & L7 & 0.982$\pm$0.013 & 1.047$\pm$0.016 & 0.924$\pm$0.009 & 0.773$\pm$0.011 & 0.743$\pm$0.012 & 0.770$\pm$0.011 & 0.585$\pm$0.011 & 0.658$\pm$0.008 & 0.583$\pm$0.007 & 0.983$\pm$0.010 & 0.85$\pm$0.10 &  [12] \\
J00164396+2304267 & T0.0 & 0.704$\pm$0.012 & 1.005$\pm$0.013 & 0.798$\pm$0.015 & 0.559$\pm$0.010 & 0.534$\pm$0.011 & 0.675$\pm$0.013 & 0.578$\pm$0.010 & 0.573$\pm$0.009 & 0.449$\pm$0.006 & 1.010$\pm$0.012 & 1.04$\pm$0.06 &  [28] \\
\nodata & \nodata & 0.752$\pm$0.012 & 1.046$\pm$0.015 & 0.796$\pm$0.011 & 0.555$\pm$0.011 & 0.564$\pm$0.012 & 0.668$\pm$0.012 & 0.602$\pm$0.013 & 0.594$\pm$0.008 & 0.465$\pm$0.006 & 1.045$\pm$0.011 & 0.99$\pm$0.07 &  [29] \\
J00185151+5153306 & L7.0 & 0.845$\pm$0.020 & 1.122$\pm$0.019 & 1.011$\pm$0.013 & 0.662$\pm$0.017 & 0.710$\pm$0.018 & 0.798$\pm$0.011 & 0.481$\pm$0.016 & 0.829$\pm$0.017 & 0.893$\pm$0.016 & 1.015$\pm$0.016 & 1.20$\pm$0.15 &  [28] \\
J00311004+5749363 & L8 & 0.900$\pm$0.009 & 1.030$\pm$0.009 & 0.843$\pm$0.007 & 0.653$\pm$0.007 & 0.658$\pm$0.007 & 0.687$\pm$0.007 & 0.582$\pm$0.009 & 0.655$\pm$0.007 & 0.591$\pm$0.005 & 1.028$\pm$0.007 & 0.99$\pm$0.15 &  [20] \\
\nodata & \nodata & 0.940$\pm$0.007 & 1.073$\pm$0.007 & 0.837$\pm$0.005 & 0.653$\pm$0.005 & 0.656$\pm$0.007 & 0.678$\pm$0.005 & 0.600$\pm$0.006 & 0.651$\pm$0.005 & 0.570$\pm$0.004 & 1.021$\pm$0.005 & 0.98$\pm$0.15 &  [34] \\
\enddata 
\tablenotetext{a}{Spectral type as inferred from direct comparison to spectral standards.}
\tablerefs{ 
[1] \citet{2004AJ....127.2856B};[2] \citet{2006AJ....131.2722C};[3] \citet{2006ApJ...637.1067B};[4] \citet{2006ApJ...639.1095B};[5] \citet{2007AJ....133.2320S};[6] \citet{2007AJ....134.1162L};[7] \citet{2007ApJ...654..570L};[8] \citet{2007ApJ...659..655B};[9] \citet{2008ApJ...681..579B};[10] \citet{2009AJ....137..304S};[11] \citet{2010ApJ...710.1142B};[12] \citet{2010ApJS..190..100K};[13] \citet{2011AJ....142...77D};[14] \citet{2011AJ....142..171G};[15] \citet{2011ApJ...732...56G};[16] \citet{2011ApJS..197...19K};[17] \citet{2012ApJ...757..100D};[18] \citet{2013ApJ...772..129B};[19] \citet{2013ApJ...776..126C};[20] \citet{2013ApJ...777...84B};[21] \citet{2013ApJS..205....6M};[22] \citet{2013PASP..125..809T};[23] \citet{2014ApJ...783..122K};[24] \citet{2014ApJ...784...65B};[25] \citet{2014ApJ...787..126L};[26] \citet{2014ApJ...794..143B};[27] \citet{2015AJ....150..182K};[28] \citet{2015ApJ...814..118B};[29] \citet{2016ApJ...817..112S};[30] \citet{2016ApJ...822L...1S};[31] \citet{2016ApJ...830..144R};[32] \citet{2017AJ....153..196S};[33] \citet{2017AJ....154..112K};[34] Burgasser et al. (in prep)
}
\end{deluxetable} 
\end{longrotatetable}

\startlongtable
\begin{deluxetable}{llccccccl}
\tablecaption{LOWZ Fit Parameters for Local Late$-$L and T Dwarfs\label{tab:splat_modelfit}} 
\tabletypesize{\tiny} 
\tablehead{ 
 & & \multicolumn{6}{c}{Best-fit LOWZ Parameters} \\
 \cline{3-8}
\colhead{Designation} & 
\colhead{SpT} & 
\colhead{\teff} & 
\colhead{\logg} & 
\colhead{[M/H]} & 
\colhead{C/O} & 
\colhead{$\log{k_{zz}}$} & 
\colhead{$\chi^2_r$} & 
\colhead{Ref.}   \\ 
 & & (K) & (cm/s$^2$) & (dex) & & (cm$^2$/s) \\
} 
\startdata 
J00001354+2554180 & T4.5 & 1200 & 5.2 & 0.00 & 0.55 & 2 & 111 &  [3] \\
J00062785+1857288 & L7 & 1600 & 4.5 & 0.75 & 0.85 & 10 & 29 &  [28] \\
J00150206+2959323 & L7 & 1600 & 4.5 & 0.25 & 0.85 & 10 & 23 &  [12] \\
J00164396+2304267 & T0.0 & 1600 & 4.5 & 0.00 & 0.55 & 2 & 18 &  [27] \\
J00164397+2304265 & T0 & 1600 & 4.5 & 0.00 & 0.55 & 2 & 31 &  [28] \\
J00185151+5153306 & L7.0 & 1600 & 4.5 & 0.75 & 0.85 & 10 & 69 &  [27] \\
J00311004+5749363 & L8 & 1600 & 4.5 & 0.50 & 0.85 & 10 & 38 &  [20] \\
\nodata & \nodata & 1600 & 4.5 & 0.50 & 0.85 & 10 & 89 &  [33] \\
\enddata 
\tablerefs{ 
[1] \citet{2004AJ....127.2856B}; [2] \citet{2006AJ....131.2722C}; [3] \citet{2006ApJ...637.1067B}; [4] \citet{2006ApJ...639.1095B}; [5] \citet{2007AJ....133.2320S}; [6] \citet{2007AJ....134.1162L}; [7] \citet{2007ApJ...654..570L}; [8] \citet{2007ApJ...659..655B}; [9] \citet{2008ApJ...681..579B}; [10] \citet{2009AJ....137..304S}; [11] \citet{2010ApJ...710.1142B}; [12] \citet{2010ApJS..190..100K}; [13] \citet{2011AJ....142...77D}; [14] \citet{2011AJ....142..171G}; [15] \citet{2011ApJ...732...56G}; [16] \citet{2011ApJS..197...19K}; [17] \citet{2012ApJ...757..100D}; [18] \citet{2013ApJ...772..129B}; [19] \citet{2013ApJ...776..126C}; [20] \citet{2013ApJ...777...84B}; [21] \citet{2013ApJS..205....6M}; [22] \citet{2013PASP..125..809T}; [23] \citet{2014ApJ...784...65B}; [24] \citet{2014ApJ...787..126L}; [25] \citet{2014ApJ...794..143B}; [26] \citet{2015AJ....150..182K}; [27] \citet{2015ApJ...814..118B}; [28] \citet{2016ApJ...817..112S}; [29] \citet{2016ApJ...822L...1S}; [30] \citet{2016ApJ...830..144R}; [31] \citet{2017AJ....153..196S}; [32] \citet{2017AJ....154..112K}; [33] 
Burgasser et al.\ (in prep.)
} 
\end{deluxetable} 

\section{Bolometric Corrections from LOWZ Models\label{app:bc}}

We provide in this section bolometric corrections computed from a subset of the LOWZ model grid \citep{2021ApJ...915..120M}, specifically:  
{\logg} = 5.0 (in cm/s$^2$); $\log{k}_{zz}$ = 2 (in cm$^2$/s); C/O = 0.55; $-$2.5 $\leq$ [M/H] $\leq$ 1.0,
and 500~K $\leq$ {\teff} $\leq$ 1600~K. 
These corrections are for the MKO $YJHK$ and WISE $W1W2$ filters and are computed on the Vega system.
See Section~\ref{sec:temperature} for details on these calculations.

\startlongtable
\begin{deluxetable}{ccccccccc}
\tablecaption{Bolometric Corrections for Brown Dwarfs Based on the LOWZ Models\label{tab:bcs}} 
\tabletypesize{\scriptsize} 
\tablehead{ 
 & & & \multicolumn{6}{c}{Bolometric Correction} \\
\cline{4-9}
\colhead{\teff} & 
\colhead{[M/H]} & 
\colhead{$M_{bol}$} & 
\colhead{$MKO~Y$} & 
\colhead{$MKO~J$} & 
\colhead{$MKO~H$} & 
\colhead{$MKO~K$} & 
\colhead{$WISE~W1$} &
\colhead{$WISE~W2$} \\
\colhead{(K)} & 
\colhead{{(dex)}} & 
\colhead{(mag)} & 
\colhead{(mag)} & 
\colhead{(mag)} & 
\colhead{(mag)} & 
\colhead{(mag)} & 
\colhead{(mag)} & 
\colhead{(mag)} \\
} 
\startdata 
500 & $-$2.50 & 20.86 & $-$0.94 & $-$1.71 & $-$0.13 & $-$5.84 & 4.58 & 6.02 \\ 
500 & $-$2.00 & 20.86 & $-$0.61 & $-$1.29 & 0.11 & $-$5.18 & 4.35 & 6.13 \\ 
500 & $-$1.50 & 20.86 & $-$0.17 & $-$0.56 & 0.36 & $-$4.19 & 4.00 & 6.26 \\ 
500 & $-$1.00 & 20.86 & 0.24 & 0.33 & 0.61 & $-$2.82 & 3.52 & 6.36 \\ 
500 & $-$0.50 & 20.86 & 0.63 & 1.25 & 0.86 & $-$1.34 & 2.99 & 6.33 \\ 
500 & $-$0.25 & 20.86 & 0.83 & 1.62 & 1.08 & $-$0.72 & 2.70 & 6.24 \\ 
500 & 0.00 & 20.86 & 0.93 & 1.91 & 1.22 & $-$0.01 & 2.46 & 6.14 \\ 
500 & 0.25 & 20.86 & 0.92 & 2.08 & 1.30 & 0.61 & 2.26 & 6.00 \\
\enddata 
\tablecomments{Magnitudes are on the Vega system, and evaluation of $M_{bol}$ assumes R = 0.077~R$_\odot$. We further constrain the LOWZ models to have {\logg} = 5 (cm/s$^2$), {\kzz} = 2 (cm$^2$/s), and C/O = 0.55.}
\end{deluxetable}

\bibliography{byw-sdt}{}
\bibliographystyle{aasjournal}

\end{document}